\documentclass{aa}
\bibpunct{(}{)}{;}{a}{}{,} % to follow the A&A style
\usepackage{graphicx}
\usepackage{subcaption}
\usepackage{xspace}
\usepackage{booktabs}
\usepackage{txfonts}
\usepackage[colorlinks=true,citecolor=blue]{hyperref}

\newcommand{\vsini}{\ensuremath{v \sin{i}}\xspace}

\newcommand{\ms}{\ensuremath{\mathrm{m\,s^{-1}}}\xspace}
\newcommand{\kms}{\ensuremath{\mathrm{km\,s^{-1}}}\xspace}
\newcommand{\A}{\ensuremath{\mathrm{\AA}}\xspace}

\newcommand{\Halpha}{\ensuremath{\mathrm{H\alpha}}\xspace}
\newcommand{\CaHK}{\ion{Ca}{II}\,H\&K\xspace}

\newcommand{\pEWHalpha}{pEW\ensuremath{'(\Halpha)}\xspace}
\newcommand{\IHalpha}{\ensuremath{I_{\Halpha}}\xspace}
\newcommand{\logLHalphaLbol}{\ensuremath{\log(L_{\Halpha}/L_{\mathrm{bol}})}\xspace}

\newcommand{\NaD}{\ensuremath{\mathrm{\ion{Na}{I}\,D}}\xspace} % doublet
\newcommand{\Prot}{\ensuremath{P_{\mathrm{rot}}}\xspace}
\newcommand{\Prothalf}{\ensuremath{\frac{1}{2}P_{\mathrm{rot}}}\xspace}

\newcommand{\caracal}{\texttt{caracal}\xspace}
\newcommand{\serval}{\texttt{serval}\xspace}
\newcommand{\raccoon}{\texttt{raccoon}\xspace}

\begin{document}

\title{The CARMENES search for exoplanets around M dwarfs}

\subtitle{Line-by-line sensitivity to activity in M dwarfs\thanks{Tables containing information about the sensitivity of the different lines  to activity are available in electronic form at the CDS via anonymous ftp to cdsarc.u-strasbg.fr (130.79.128.5) or via \url{ https://cdsarc.cds.unistra.fr/cgi-bin/qcat?J/A+A/}}}

\author{
M.~Lafarga \inst{1,2,3,4},
I.~Ribas \inst{1,2},
M.~Zechmeister \inst{5},
A.~Reiners \inst{5},
\'A.~L\'opez-Gallifa\inst{6,7},
D.~Montes \inst{6},
A.~Quirrenbach \inst{8},
P.\,J.~Amado \inst{9},
J.\,A.~Caballero\inst{10},
M.~Azzaro\inst{11},
V.\,J.\,S.~B\'ejar\inst{12,13},
A.\,P.~Hatzes\inst{14},
Th.~Henning\inst{15},
S.\,V.~Jeffers\inst{16},
A.~Kaminski\inst{8},
M.~K\"urster\inst{15},
P.~Sch\"ofer\inst{9},
A.~Schweitzer\inst{17},
H.\,M.~Tabernero\inst{7},
M.\,R.~Zapatero~Osorio\inst{7}
}

\institute{
Institut de Ci\`encies de l'Espai (ICE, CSIC), Campus UAB, c/ de Can Magrans s/n, 08193 Cerdanyola del Vall\`es, Spain
\and %2
Institut d'Estudis Espacials de Catalunya (IEEC), c/ Gran Capit\`a 2-4, 08034 Barcelona, Spain
\and %3
Department of Physics, University of Warwick, Gibbet Hill Road, Coventry CV4 7AL, United Kingdom
\and %4
Centre for Exoplanets and Habitability, University of Warwick, Coventry, CV4 7AL, UK
\and %5
Institut f\"ur Astrophysik, Georg-August-Universit\"at, Friedrich-Hund-Platz 1, 37077 G\"ottingen, Germany
\and %6
Departamento de F\'isica de la Tierra y Astrof\'isica \& IPARCOS-UCM (Instituto de F\'isica de Partículas y del Cosmos de la UCM),
Facultad de Ciencias F\'isicas, Universidad Complutense de Madrid, 28040 Madrid, Spain
\and %7
Centro de Astrobiolog\'ia, CSIC-INTA, Carretera de Ajalvir km 4, 28850 Torrej\'on de Ardoz, Madrid, Spain
\and %8
Landessternwarte, Zentrum f\"ur Astronomie der Universt\"at Heidelberg, K\"onigstuhl 12, 69117 Heidelberg, Germany
\and %9
Instituto de Astrof\'isica de Andaluc\'ia (IAA-CSIC), Glorieta de la Astronom\'ia s/n, 18008 Granada, Spain
\and %10
Centro de Astrobiolog\'ia, CSIC-INTA, ESAC, Camino Bajo del Castillo s/n, 28692 Villanueva de la Ca\~nada, Madrid, Spain
\and %11
Centro Astron\'onomico Hispano en Andaluc\'ia, Observatorio de Calar Alto, Sierra de los Filabres, 04550 G\'ergal, Spain
\and %12
Instituto de Astrof\'isica de Canarias, V\'ia L\'actea s/n, 38205 La Laguna, Tenerife, Spain
\and %13
Departamento de Astrof\'isica, Universidad de La Laguna, 38026 La Laguna, Tenerife, Spain
\and %14
Th\"uringer Landesstenwarte Tautenburg, Sternwarte 5, 07778 Tautenburg, Germany
\and %15
Max-Planck-Institut f\"ur Astronomie, K\"onigstuhl 17, 69117 Heidelberg, Germany
\and %16
Max-Planck-Institut für Sonnensystemforschung, D-37077, Göttingen, Germany
\and %17
Hamburger Sternwarte, Gojenbergsweg 112, 21029 Hamburg, Germany
}

\titlerunning{Sensitivity to activity of M dwarf spectral lines}

\authorrunning{Lafarga et al.}

\date{Received, 01 December 2022 / Accepted, 09 February 2023}

\abstract
{
Radial velocities (RVs) measured from high-resolution stellar spectra are routinely used to detect and characterise orbiting exoplanet companions. The different lines present in stellar spectra are created by several species, which are non-uniformly affected by stellar variability features such as spots or faculae. Stellar variability distorts the shape of the spectral absorption lines from which precise RVs  are measured, posing one of the main problems in the study of exoplanets.}
{In this work we aim to study how the spectral lines present in M dwarfs are independently impacted by stellar activity.}
{We used CARMENES optical spectra of six active early- and mid-type M dwarfs to compute line-by-line RVs and study their correlation with several well-studied proxies of stellar activity.}
{We are able to classify spectral lines based on their sensitivity to activity in five M dwarfs displaying high levels of stellar activity. We further used this line classification to compute RVs with activity-sensitive lines and less sensitive lines, enhancing or mitigating stellar activity effects in the RV time series. For specific sets of the least activity-sensitive lines, the RV scatter decreases by $\sim2$ to 5 times the initial one, depending on the star. Finally, we compare these lines in the different stars analysed, finding the sensitivity to activity to vary from star to star.}
{Despite the high density of lines and blends present in M dwarf stellar spectra, we find that a line-by-line approach is able to deliver precise RVs. Line-by-line RVs are also sensitive to stellar activity effects, and they allow for an accurate selection of activity-insensitive lines to mitigate activity effects in RV. However, we find stellar activity effects to vary in the same insensitive lines from star to star.}


\maketitle
%---------------------------------------------------------------------

\section{Introduction}

High-resolution stellar spectra are routinely used to study and characterise exoplanet companions orbiting stars through the Doppler spectroscopy or radial velocity (RV) technique. Stellar spectra are affected by the intrinsic variability of the stellar hosts. Stars are not quiet, homogeneous bodies, but they display variability on different timescales and amplitudes, including the effects of oscillations \citep[e.g.][]{bedding2001oscillations,bazot2012oscillations,kunovachodzic2021piMen}, granulation \citep[e.g.][]{meunier2015granulation,cegla2018convection}, and magnetically active regions such as spots and faculae \citep[e.g.][]{saardonahue1997activity,desort2007activity,lagrange2010spot}. These features distort the stellar spectra, introducing biases in the measured RVs that can be large enough to mimic or hide the signal caused by a planet. Magnetically active regions are specially important because they co-rotate with the star and hence have timescales of the order of the stellar rotation period (similar to the orbital periods of close-in planets) and they impact the RVs on the \ms level.

The different absorption lines observed in stellar spectra are created by the different atomic or molecular species present in the stellar photosphere. Different atoms and molecules have different sensitivities to temperature, magnetic field strength, and convection pattern. These are parameters affected by photospheric stellar activity features: spots and faculae possess strong magnetic fields that inhibit convective motions and change the temperature in these regions. We therefore expect that changes in these parameters due to stellar activity will affect the profile of different absorption lines in different ways, depending on the sensitivity of the lines to these parameters. Line profile changes affect not only RV measurements used to study exoplanets, but also the determination of stellar properties and chemical abundances, especially in young, active stars \citep[e.g.][]{reiners2016sun,meunier2017VariabilityGranualtionConvective,passegger2019carmenesPhotParam,shulyak2019carmenesMagnetic,spina2020activity,abia2020rubidium,shan2021vanadium,liebing2021convectiveblueshift}.

Usual methods to determine RVs \citep[either by cross-correlation or template-matching schemes, e.g.][]{baranne1996elodie,pepe2002coralieSaturns,anglada-escude2012harpsTerra,zechmeister2018serval} yield a global RV measurement for the full spectrum. This means that these RV measurements average over the different asymmetries and shifts experienced by individual lines. Consequently, information related to the different effects of activity on different spectral regions is lost. Some spectroscopic activity indicators are also determined from the entire spectral range of the observations. The full-width-at-half-maximum (FWHM) of the cross-correlation function (CCF) or bisector asymmetries (such as the bisector inverse slope, BIS) are measured from a CCF that averages a large number of lines \citep[e.g.][]{baranne1996elodie,queloz2001noplanet,lafarga2020carmenesccf}. The chromatic index (CRX) or the differential line width (dLW) also come from a template-matching method that takes into account wide spectral regions at the same a time \citep{zechmeister2018serval}. Therefore, as occurs with RVs, the activity information that these indicators contain is also averaged among many absorption lines that may show different activity effects. Other indicators such as those measuring the emission from the core of chromospheric lines, such as the \CaHK or the \Halpha lines \citep[e.g][]{noyes1984rotation,lovis2011magnetic,schofer2019carmenesActInd}, probe activity in the chromosphere, and hence, may not be perfectly correlated with the photospheric activity, which is what causes changes in the RVs.

Recently, a number of studies have started to focus on how line profiles change due to the effects of activity. \citet{davis2017activityPCA} used simulated time series of disc-integrated spectra with spots, faculae, and Doppler shifts due to planetary companions to study their different signatures. By applying a principal component analysis (PCA) on the simulated spectra, the authors found that spots and faculae induce variability in the spectral lines different from that introduced by pure Doppler shifts.

Several works have found line profile variations correlated with activity indices in HARPS observations \citep[High-Accuracy Radial velocity Planetary Searcher,][]{mayor2003harps} of the nearby dwarf $\alpha$\,Cen\,B, a moderately active K1\,V star that shows a clear activity modulation in its RVs and activity indicators. By comparing spectra of high- and low-activity states of the star, \citet{thompson2017acenBactivity} were able to identify lines whose profile changes depending on the activity level. The specific morphology of the variations differs on a line-to-line basis, but several lines show depth variations. The pseudo-equivalent widths measured from some of these features are rotationally modulated, and show correlations with the $\log R'_\mathrm{HK}$ activity index.

\citet{wise2018activelines} also studied line profile variations in HARPS observations of $\alpha$\,Cen\,B and $\epsilon$\,Eri, an active K2\,V star. They found that the depth (or core flux) of about 40 absorption lines is correlated with the $S$ index derived from the \CaHK lines, and they periodically change with the stellar rotation period. \citet{ning2019linesbayesselection} extended the previous work with an automated method to identify activity-sensitive lines with a Bayesian variable selection method, which accounts for dependencies between lines and uses different activity indicators ($S$ index, \NaD, \Halpha, CCF BIS, and FWHM) to trace activity changes in the RVs. \citet{lisogoroskyi2019activityalphacenb} used the $\alpha$\,Cen\,B dataset to measure equivalent widths and asymmetries and compute their correlation with the $S$ index, finding almost 350 activity-sensitive lines, which include the 40 lines compiled by \citet{wise2018activelines}. Methods such as these could be used to find activity indicators derived from the properties of photospheric absorption lines for stars of other spectral types and for observations at different wavelength ranges.

Instead of studying line profile variations, \citet{dumusque2018indivline} measured the RV of individual absorption lines present in stellar spectra \citep[following the method described in][]{bouchy2001photonnoise} and correlated them with an activity indicator. Similarly to previous studies, the stars used are relatively early-type cool dwarfs (G1\,V to K1\,V, including $\alpha$\,Cen\,B) observed with HARPS. In the case of $\alpha$\,Cen\,B, the author used the global RV as the activity indicator since, in principle, the RV variation of this star is solely due to activity. Different correlation strengths between the line-by-line RVs and the activity indicator were interpreted as the lines having different sensitivities to stellar activity. This work also showed that a judicious selection of the lines used to compute the total RV of a spectrum (taking into account the sensitivity of the lines to activity) can result in measurements where the activity signal is mitigated or amplified depending on the lines selected. 

\citet{cretignier2020indivline} continued the work presented in \citet{dumusque2018indivline} by refining the method used to measure RVs from individual lines and studying the RV relation with the line properties. The authors show that, in $\alpha$\,Cen\,B, lines with different depths display different effects due to activity; in particular, the RV effect is inversely proportional to the line depth. This agrees with the fact that shallow lines, which are formed deeper in the stellar photosphere where the convection velocity is larger, are more affected by the inhibition of this convection in the presence of an activity feature, while deep lines, which formed in the outer regions of the photosphere where the convection velocity is lower, show a diminished effect. \citet{cretignier2020indivline} also propose a new activity indicator based on the RV difference between deep and shallow spectral lines. \citet{siegel2022depths} used line-by-line RVs to build a novel activity indicator -- the depth metric -- based on the depth variations of activity-sensitive lines. The authors used this metric to study the effects of activity in the HARPS RVs of  $\alpha$\,Cen\,B and HD 13808, an active K2\,V star hosting two Neptune-mass planets, finding it to be efficient at mitigating stellar activity in these Sun-like stars. Within each individual line, \citet{almoulla2022lbl} further showed that RVs measured from different line segments (different parts of the lines formed at different temperatures) correlate with stellar activity in the Sun and $\alpha$\,Cen\,B, observed with HARPS-N \citep{cosentino2012harpsn} and HARPS, respectively. Several line-by-line approaches were also tested within the EXPRES Stellar Signals Project \citep[][]{jurgenson2016expres,zhao2022express} in four G and K dwarf stars, including the method we present here.

The aforementioned studies focused on Sun-like stars but they did not include M dwarf stars. M dwarfs display a spectrum with a higher density of features, including atomic lines and molecular bands, which makes it difficult to separate individual lines due to blending and the presence of the molecular pseudo-continuum. Despite that, a method such as the CCF, which uses a mask built from selecting `individual' lines, is still able to deliver precise RVs \citep[e.g.][]{lafarga2020carmenesccf}, so it is expected that it could be possible to study different activity effects on individual lines. Moreover, convective blueshift, which is affected by the presence of active regions, is different in M dwarfs and Sun-like stars \citep[it is decreased for M dwarfs e.g.][]{beeck2013convection,baroch2020carmenescrxYZCMi,liebing2021convectiveblueshift}. Therefore, lines in M dwarf spectra could show different activity-related effects.

Recently, \citet{bellotti2022linelists} studied the effect of using different combinations of lines to compute least squares deconvolution (LSD) profiles in three M dwarf stars (EV~Lac, AD~Leo, and DS~Leo) observed with ESPaDonS \citep[Echelle SpectroPolarimetric Device for the Observation of Stars,][]{donati2003espadons} and NARVAL \citep[][]{auriere2003narval}. A line selection based on several line parameters (depth, wavelength, and Landé factor) does not result in stellar activity effects seen in the RVs derived from the LSD being mitigated. However, a randomised algorithm is able to find a subset of lines that minimises the RV scatter, without computing line-by-line RVs. \citet{artigau2022lbl} applied a line-by-line approach similar to the one presented by \citet{dumusque2018indivline} to compute precise RVs of the M dwarf stars Proxima Cen, observed with HARPS, and Barnard's star, observed in the near-infrared with SPIRou \citep[SpectroPolarim\`etre InfraRouge,][]{donati2020spirou}, finding similar RV precision as template-matching techniques. From their line-by-line framework, the authors also introduce an activity indicator similar to the differential line width implemented by \citet{zechmeister2018serval}. \citet{martioli2022TOI1759}, \citet{gan2022TOI2136}, and \citet{cadieux2022TOI1452} also used the same line-by-line framework to measure precise RVs of M dwarf stars observed with SPIRou, and \citet{radica2022K2-18} applied the same method to measure precise RVs from HARPS and CARMENES  \citep[Calar Alto high-Resolution search for M dwarfs with Exo-earths with Near-infrared and optical Echelle Spectrographs,][]{quirrenbach2016carmenes,quirrenbach2018carmenes}  spectra of K2-18. 

Inspired by the method proposed by \citet{dumusque2018indivline}, in this work we apply a similar approach to observations of M dwarfs obtained with the high-resolution spectrograph CARMENES. In a sample of six early to mid M dwarfs, and different activity levels, we compute line-by-line RVs, classify lines according to their sensitivity to activity, and use this classification to compute RVs affected by activity to different degrees. We also study how the activity sensitivity of the same lines varies for the different stars studied.
This article is structured as follows. In Sect. \ref{sec:indlintargets}, we present the stars analysed. In Sect. \ref{indlinmethodrv} we explain the method followed to determine line-by-line RVs, including our initial line selection. Sect. \ref{sec:indlinact} deals with the classification of lines based on their sensitivity to activity, and in Sect. \ref{sec:indlinrvselected} we use this classification to compute RVs in which the changes induced by activity have been removed or enhanced. We compare the activity sensitivity of the lines in the selected stars in Sect. \ref{sec:indlincomparisonstar}. Finally, we discuss and summarise our findings in Sect. \ref{sec:indlinsummary}.

%---------------------------------------------------------------------

\section{Targets} \label{sec:indlintargets}

%---------------------------------------------------------------------
%\begin{table*}
\begin{sidewaystable}
\centering
\caption{Main properties of the six  analysed stars.}
\label{tab:indivlinobj}
\tabcolsep=0.06cm
%{\scriptsize % --- Smaller text
{\small
\begin{tabular}{llccccccccccccc}
\hline\hline
Karmn&Name&RA\tablefootmark{(a)}&DEC\tablefootmark{(a)}&$d$\tablefootmark{(a)} & Sp.                   &$J$\tablefootmark{(c)}&\vsini\tablefootmark{(d)}&\Prot\tablefootmark{(e)}&\pEWHalpha\tablefootmark{(f)}&\logLHalphaLbol\tablefootmark{(f)}& [Fe/H]\tablefootmark{(g)}& \# obs &std RV\\
     &    &(J2000)              &(J2000)               &[pc]                   &type\tablefootmark{(b)}&[mag]                 &[\kms]                   & [d]                    & [\AA]                       &                                  &                          & VIS    &[\ms] \\
\midrule
J07446+035& YZ CMi, \object{GJ 285}        &07:44:40.17&+03:33:08.9&5.9889$\pm$0.0012  & M4.5\,V & $6.581\pm0.024$ & $4.0\pm1.5$ & $2.78\pm0.01$    & $-7.281\pm0.024$ & $-3.6099\pm0.0014$ & \ldots         & 51 & 88.2\\
J05019+011& \object{1RXS J050156.7+010845} &05:01:56.66&+01:08:42.9&25.288$\pm$0.023   & M4.0\,V & $8.526\pm0.026$ & $6.5\pm1.5$ & $2.12\pm0.02$    & $-6.356\pm0.019$ & $-3.5805\pm0.0013$ & $-0.15\pm0.16$ & 19 & 90.3\\
J22468+443& EV Lac, \object{GJ 873}        &22:46:49.73&+44:20:02.4&5.05159$\pm$0.00056& M3.5\,V & $6.106\pm0.030$ & $3.5\pm1.5$ & $4.38\pm0.03$    & $-4.983\pm0.021$ & $-3.6497\pm0.0018$ & $-0.19\pm0.16$ & 107& 50.2\\
J10196+198& AD Leo, \object{GJ 388}        &10:19:36.28&+19:52:12.0&4.96509$\pm$0.00073& M3.0\,V & $5.449\pm0.027$ & $2.4\pm1.5$ & $2.2399\pm0.0006$& $-4.520\pm0.040$ & $-3.6140\pm0.0030$ & \ldots         & 26 & 18.4\\
J15218+209& OT Ser, \object{GJ 9520}       &15:21:52.93&+20:58:39.9&11.4515$\pm$0.0028 & M1.5\,V & $6.610\pm0.021$ & $4.3\pm1.5$ & $3.37\pm0.01$    & $-2.878\pm0.018$ & $-3.7657\pm0.0028$ & $-0.34\pm0.16$ & 53 & 36.7\\
J11201--104& \object{LP 733-099}           &11:20:06.10&-10:29:46.7&18.930$\pm$0.014   & M2.0\,V & $7.814\pm0.026$ & $3.6\pm1.5$ & $5.643\pm0.005$  & $-1.877\pm0.016$ & $-3.9580\pm0.0040$ & $-0.21\pm0.16$ & 29 & 18.3\\
\hline
\end{tabular}
} % --- Smaller text
\tablefoot{
Values taken from the Carmencita database \citep{caballero2016carmencita}. We also show the number of CARMENES VIS observations obtained, and their RV scatter, measured as the standard deviation (std) of the \serval RVs (instrumental drift and nightly average corrected). Stars are sorted with decreasing activity level as measured from \pEWHalpha.
\tablefoottext{a}{\citet{gaia2021edr3}.}
\tablefoottext{b}{\citet{hawley1996PMSU}, except J05019+011, J10196+198 \citet{alonsofloriano2015carmenesinlowres}, J11201--104 \citet{riaz2006mdwarf}.}
\tablefoottext{c}{\citet{skrutskie20062mass}.}
\tablefoottext{d}{\citet{reiners2018carmenes324}, J10196+198 \citet{kossakowski2022adleo} \citep[computed as in][]{reiners2018carmenes324}.}
\tablefoottext{e}{\citet{diezalonso2019carmenesRotPhot}, except J10196+198 \citet{morin2008mdwarfmag}, and J11201--104 \citet{revilla2020thesis,shan2022prot}.}
\tablefoottext{f}{\citet{schofer2019carmenesActInd}.}
\tablefoottext{g}{\citet{schweitzer2019carmenesMR}.}
}
%\end{table*}
\end{sidewaystable}
%---------------------------------------------------------------------

We used observations obtained as part of the CARMENES \citep[][]{quirrenbach2016carmenes,quirrenbach2018carmenes} main survey (guaranteed-time observations – GTO programme). CARMENES is installed at the 3.5\,m telescope at Calar Alto Observatory in Almer\'ia, Spain, and consists of a pair of cross-dispersed, fibre-fed echelle spectrographs with complementary wavelength coverage, which allow simultaneous observations in the visible and the near-infrared wavelength range. The visible (VIS) channel covers the spectral range $\lambda$ = 5200--9600\,\A at a resolution of $R=94\,600$, with an average sampling of 2.5 pixels per spectral resolution element. The near-infrared (NIR) channel covers the range $\lambda$ = 9600--17100\,\A at a resolution of $R=80\,400$, and has an average sampling of 2.8 pixels per spectral element. The CARMENES survey has been ongoing since 2016. It monitors over 300 M dwarfs across all spectral subtypes with the main goal of detecting orbiting exo\-planets with the Doppler method \citep{reiners2018carmenes324}.

Here, we are interested in seeing the effect of activity in individual spectral lines, for different types of stars in the CARMENES sample. Therefore, we selected stars of different spectral types and different activity levels, as measured from the average pseudo-equivalent width of their \Halpha line \citep[\pEWHalpha,][]{schofer2019carmenesActInd}. To select the appropriate targets, we considered the following criteria. To be able to properly characterise the individual lines in the spectrum, we limited our targets to bright stars ($J\leq$\,9\,mag, to have high S/N per line), with low rotational velocity ($\vsini\leq7\,\kms$, to avoid strong line blending, which difficults the identification of lines and reduces the number of lines available). We also selected only targets for which we had $\gtrsim\,20$\,observations, which allowed us to derive reliable correlations between the RV of the individual lines and the activity indicators. In our early tests, we found that another limiting factor was the RV scatter of the observations. With the method that we used to measure line-by-line RVs, the error on the RV of each line is on average about 300\,\ms for bright targets. If we then average the RV values of about 1000 lines, we have a maximum precision of about 10\,\ms in the RV of each observation. For this reason, we also excluded stars with RV scatter (std) smaller than $\sim$15\,\ms (that is, stars showing small RV variability).

These criteria left us with the six targets shown in Table \ref{tab:indivlinobj}: two early M dwarfs, J15218+209 (OT~Ser) and J11201--104 (LP~733-099), and four mid M dwarfs, J07446+035 (YZ~CMi), J05019+011 (1RXS J050156.7+010845), J22468+443 (EV~Lac), and J10196+198 (AD~Leo). The table shows the main properties of the selected stars together with the number of observations available and their RV scatter. Due to the constraint on the RV scatter, the selected stars have relatively high activity level (from $\pEWHalpha\leq-1.8\,\AA$, up to $\pEWHalpha\sim-7\,\AA$). And due to the constraint on brightness, our sample only includes early and mid spectral types.

%---------------------------------------------------------------------

\section{RV computation} \label{indlinmethodrv}

\subsection{Line selection: CCF mask} \label{sec:indlin_linelist}

After the target selection, the second step in our analysis was to identify and select individual lines\footnote{For clarity, throughout this work we use the word `lines' to refer to individual absorption features or minima in the spectrum, even though these features are not true atomic lines but blends of several lines or a feature in a molecular band.} in the spectrum. To do that, we made use of the \raccoon pipeline, which we previously developed to create CCF binary masks \citep{lafarga2020carmenesccf}. To select lines, we looked for minima in a \serval spectral template, characterised them by fitting a Gaussian, and chose the features with depth, FWHM, and contrast between certain cut values. \serval \citep{zechmeister2018serval} is the default CARMENES pipeline to estimate RVs, based on a template-matching approach. It creates a high S/N stellar template by co-adding observations, and subsequently uses this template to compute a least-squares fit to each of the observations, from which the RV time series is obtained iteratively. Before co-adding, the observations are corrected for the corresponding barycentric motion of the Earth and any other known drift so that the stellar lines are optimally aligned. The templates have a similar format as the observations, and echelle orders are considered individually. Here, the templates were built with CARMENES VIS observations, and the cut values on depth, FWHM, and contrast were the same as the ones used for the standard masks \citep[see Table 1 in ][]{lafarga2020carmenesccf}. We also took into account the position of telluric lines and the varying position of the spectra on the CCD due to the barycentric movement of the Earth as explained in \citet{lafarga2020carmenesccf}. In summary, to account for tellurics, we broadened the features of a telluric mask by the maximum barycentric Earth RV (BERV) of the observations, and removed from the line list those overlapping with telluric features, taking into account the absolute RV of the target star. We also removed lines at the order extremes which are not always present in the observed spectra due to the varying BERV of different observations. In this work, we only need the wavelength position of the lines, that is, we do not need the full information of a binary mask, which includes wavelength and weight. The wavelength positions are given by the minimum of a Gaussian fit to the line.

We used two different line lists, depending on the spectral type of the star. For J15218+209 and J11201--104 (spectral types M1.5\,V and M2.0\,V, and similar \vsini) we used a line list created from the \serval template of J15218+209, with 1712 lines. For the other four stars (spectral types between M3.0\,V and M4.5\,V, and also similar \vsini), we used a line list obtained from the J07446+035 template, with 2207 lines. Both J07446+035 and J22468+443 are relatively bright and have over 20 observations, so the line list could have been build from either star. We selected J07446+035 because it has a \vsini close to the mean \vsini of the four stars, which can affect the lines present in the spectrum \citep[e.g.][]{lafarga2020carmenesccf}, however, given the uncertainties in the \vsini, we expect a line list made from a template of J22468+443 to yield similar results as the template used here.

%---------------------------------------------------------------------

\subsection{Line-by-line RV}

%---------------------------------------------------------------------
\begin{figure}
\centering
\includegraphics[width=\linewidth]{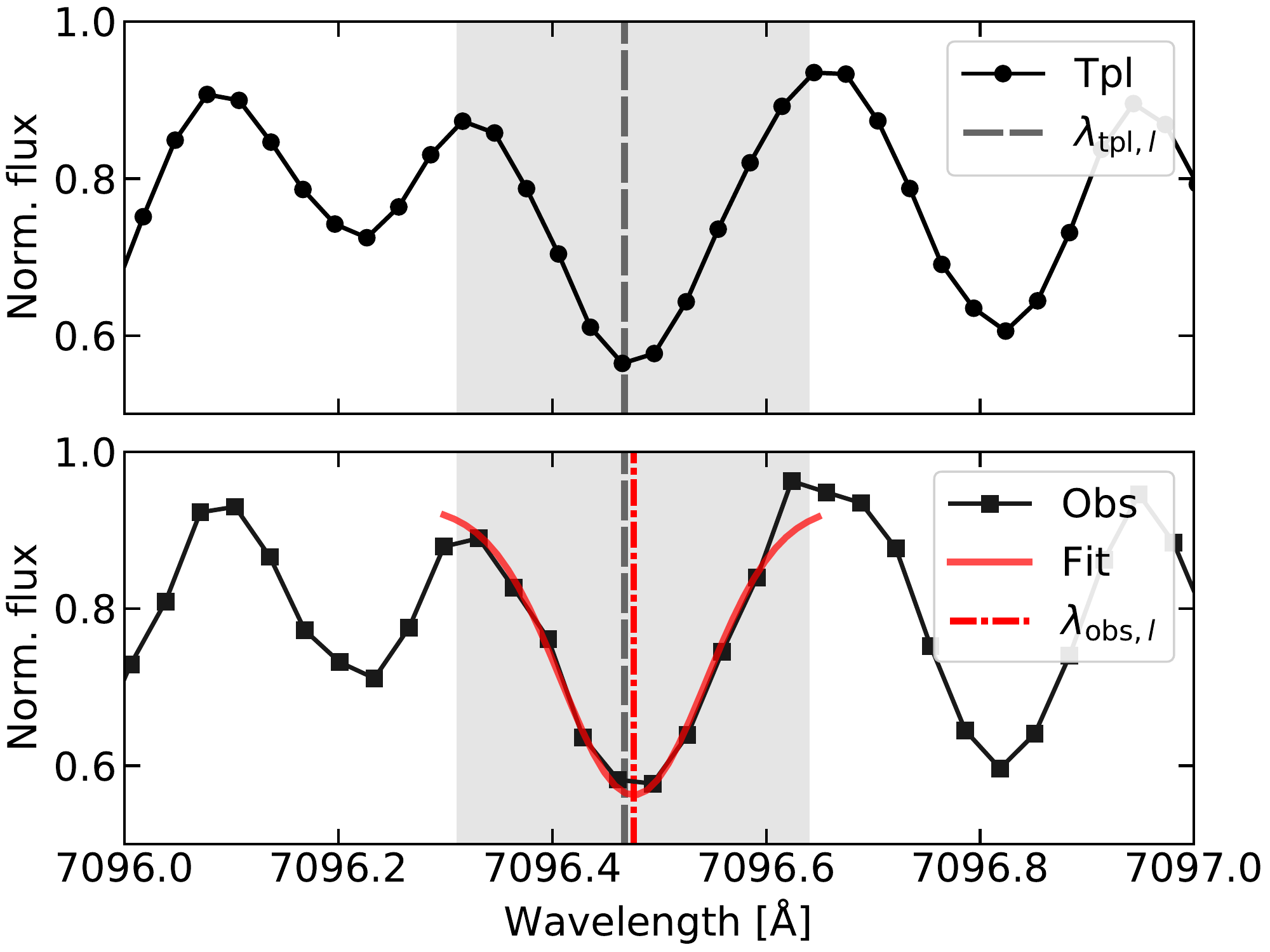}
\caption{RV computation of a representative individual line. The dashed grey vertical line represents the wavelength of the line $l$ in the line list, as measured in the spectrum template (black dots, in this example, $\lambda_{\mathrm{tpl},l}=7096.4675\,\AA$).
The dash-dotted red vertical line represents the wavelength of the line $l$ measured in an observation (black squares) with a Gaussian fit to the line (solid red line, in this example, $\lambda_{\mathrm{obs},l}=7096.4761\,\AA$). The shaded grey region marks the data points considered in the Gaussian fit to the spectrum. The Doppler shift computed with Eq. \ref{eq:dopplershift} in this case is of 363\,\ms.
}
\label{fig:illustrationindivlinefit}
\end{figure}
%---------------------------------------------------------------------

%---------------------------------------------------------------------
\begin{figure*}
    \centering
    \includegraphics[width=\linewidth]{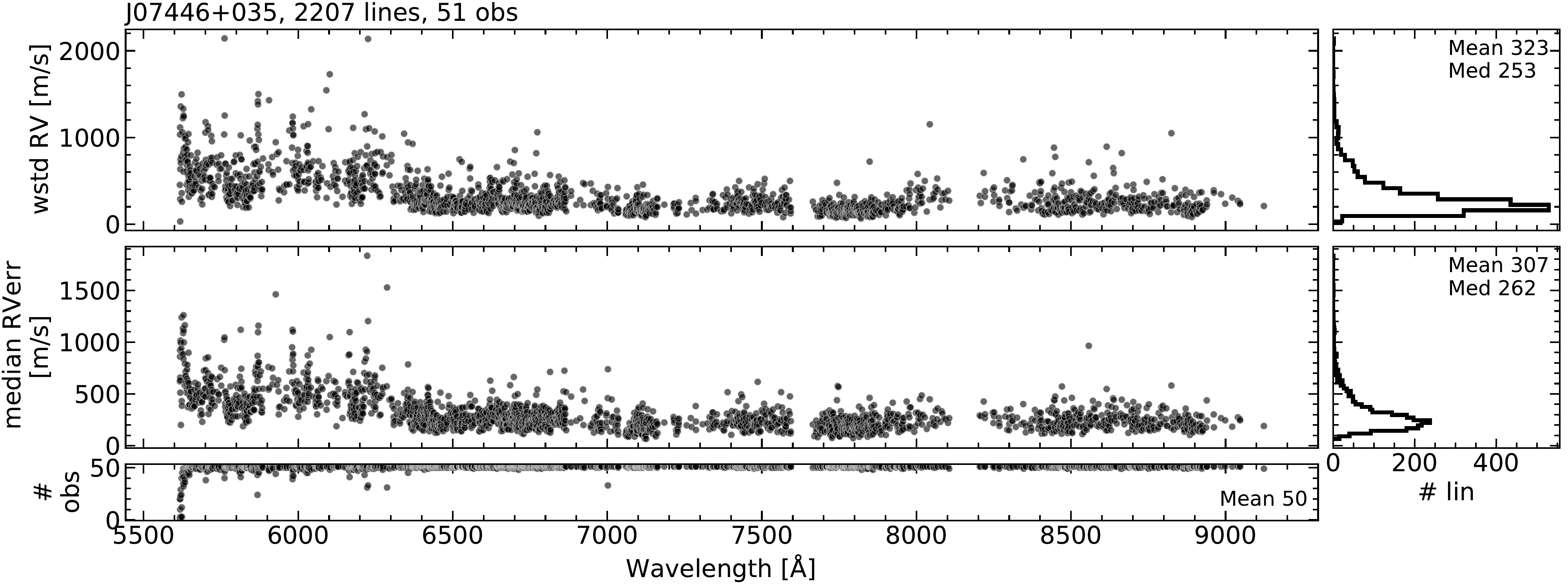}
    \caption{RV scatter, median RV error and number of observations used, as a function of the individual line wavelength, for J07446+035 (YZ~CMi).}
    \label{fig:statsrvlinerrJ07446+035}
\end{figure*}
%---------------------------------------------------------------------

Next, for all the available observations of each target, we computed an RV for each of the lines in the line list, that is, a line-by-line RV. We used observations reduced with \caracal, the standard CARMENES reduction pipeline \citep{caballero2016carmenes}, which reduces the spectra by flat-relative optimal extraction \citep{zechmeister2014fox} and outputs the different spectral orders in vacuum wavelength (throughout this paper all wavelengths are in vacuum). The spectra were corrected for the BERV, secular acceleration, and instrumental drifts, as measured by the standard CARMENES pipelines \citep{zechmeister2018serval,trifonov2018carmenes1st,tal-or2019systematichires}.

There are several ways to measure line-by-line radial velocities. To obtain the RV of a specific line $l$ in one of the observed spectra, we could compute its CCF using a single-line mask, or apply a template-matching algorithm using only a small predefined region around the line. We opted for a more straightforward method. First, we obtained the position of the line $l$ in the observed spectrum, $\lambda_{\mathrm{obs},\,l}$, by fitting a Gaussian to the region around the line in the observed data (similarly to the process of line characterisation when building CCF masks mentioned above). The line position is given by the minimum of the best fit. Then, we computed the RV of the line as the Doppler shift between the position of the line in the observation, $\lambda_{\mathrm{obs},\,l}$, and the position of the line in the line list, $\lambda_{\mathrm{tpl},\,l}$ (i.e. as measured in the template, also from a Gaussian fit)
\begin{equation}\label{eq:dopplershift}
\mathrm{RV}_l = c \left( 1 - \frac{\lambda_{\mathrm{tpl},\,l}}{\lambda_{\mathrm{obs},\,l}} \right),
\end{equation}
where $c$ is the speed of light. This is illustrated in Fig. \ref{fig:illustrationindivlinefit}. 
To estimate the uncertainty on the individual RV measurements, we used the formal error of the Gaussian fit to the observation. We did not consider the uncertainty of the initial Gaussian fit to the high S/N template (i.e. we did not propagate this uncertainty) because we are interested in the relative RV measurement.
Although we are fitting lines that are not completely Gaussian-shaped, this error gives an indication of the goodness of the fit for the different lines, which reflects the S/N of the different regions of the observed spectra.

The spectral region used in the Gaussian fit is constrained by the adjacent local maxima at each side of the line minimum, as measured in the \serval template (shaded grey area in Fig. \ref{fig:illustrationindivlinefit}). In this way, we made sure to always use the same region around each line. This may not happen if, instead, we measured the maxima in each different observation, because their position could change depending on the S/N, activity effects, or tellurics which may not have been considered by our mask. For  J11201--104, we used the line limits obtained from the template of J15218+209, instead of using its own template, which did not have a high S/N because the star is relatively faint ($J$ > 7\,mag) and we did not have a large number of observations available. J05019+011 is the faintest star in our sub-sample ($J$ > 8\,mag) and has only 19 observations, therefore we also tried to use the line limits from the template of another star, in this case J07446+035. However, for J05019+011, we obtained better results (i.e. a better match between the template and observed lines) using the line limits obtained from its own template, which could be due to the fact that this star has a slightly larger \vsini than J07446+035.

During the creation of the line list (Sect. \ref{sec:indlin_linelist} above), we already took into account the regions contaminated by tellurics affecting the template. However, depending on the absolute RV of the target star and the BERV correction of the target observations, regions different than the ones in the template can be affected by tellurics. Therefore, before computing the RVs, we further removed the affected lines following the steps described above.

Since we are using observations of an echelle spectrograph, the observed spectrum is divided into different orders. For most orders, a wavelength region larger than the free spectral range falls on the detector and is extracted by the pipeline (i.e. there is a wavelength overlap between consecutive orders).
For the lines in the overlap regions, we only used the redder part of the bluer order, as opposed of the bluer part of the redder order in each overlap. This is because in general, we found that the redder part of most of the orders has better S/N than the bluer part of the next order.

As an example, in Fig. \ref{fig:statsrvlinerrJ07446+035} we show, for each line, the scatter of the RVs of all the observations of J07446+035, measured as the weighted standard deviation of the RVs of each epoch. We also show the median RV error of each line. We see that most lines show a scatter close to 300\,\ms, and the typical error in the individual line RV is also of about 300\,\ms.
Lines located in the bluer region of the spectral range, with wavelengths shorter than $\sim$~6400\,\A, are the ones that show a larger RV scatter and error. This happens because the bluer part of the spectral range is where observations of M dwarfs have lower S/N, which makes it difficult to correctly identify the lines and measure their RV. We obtain similar values, about 300--400\,\ms, for the rest of stars.

%---------------------------------------------------------------------

\subsection{Total RV} \label{sec:indlinmethodtotalrv}

We averaged the line-by-line RVs to compute a total RV per observation and compare it to the RVs obtained with standard methods that consider all the lines simultaneously: the CCF obtained with \raccoon and the template-matching scheme from \serval. In the following, we refer to the method of computing the total RV by averaging the line-by-line RVs as the lines average (LAV) method.
The LAV RV uncertainties are given by the standard error of the mean. Before computing their mean, we discarded some data points. We did not use line-by-line RVs corresponding to bad Gaussian fits to the spectra, which we identified as those with RV errors larger than 1000 \ms and smaller than 20 \ms (i.e. points where the fit was clearly not successful).  This procedure typically removed less than 1\% of the total lines. We note that these cuts work for the stars in our sample but in general, they will depend on the properties of each specific star, such as the S/N or the rotational velocity. A more general way to remove data points with bad Gaussian fits would be to directly reject outliers in the $\chi^2$ distribution of all the Gaussian fits. After removing these data points, we performed, in each observation separately, a $4 \sigma$ clipping on the RVs of all the lines to discard outliers. This procedure typically discards about 1\% of all lines. Finally, we discarded lines that did not have a reliable RV measurement in more than ten observations. This process mainly removed lines in the bluer part of the spectrum, where the S/N is lowest, and weak lines, which can be properly identified and characterised in a co-added spectrum template such as the one produced by \serval, but not in single observations that have much lower S/Ns.

In Figs. \ref{fig:rvtsallcompareJ07446+035} to \ref{fig:rvtsallcompareJ11201--104} (in Appendix \ref{sec:app_totalrv_all}), we show the average RV of all the lines compared to the RV obtained with the CCF method and \serval's template matching, for the targets under analysis. The masks used to compute the CCFs were the same as the ones used to define the individual line lists. That is, masks created from \serval templates of J15218+209 (used in the CCF of J15218+209 and J11201--104) and J07446+035 (used in the CCF of J07446+035, J05019+011, J22468+443, and J10196+198). The \serval RVs of each target are computed using a template made by co-adding the observations of the target itself. We obtain comparable RV values for the three different methods (LAV, CCF, and \serval, but see next paragraph on the RV uncertainties), but the LAV RVs are in general closer to the CCF RVs than to those obtained with \serval. This is probably due to the fact that the line list used to compute the individual line RVs is obtained from the mask used in the CCF method, while \serval considers the entire wavelength domain, not just a set of lines. We also note that \serval includes the weakest lines in the spectrum, which are excluded in the CCF masks. Therefore, \serval might result in higher RV scatter for active stars. For simplicity, in the following we only use the CCF RVs and not the \serval ones, but the results obtained are comparable.

Despite the LAV RV values being similar to those obtained with the CCF method, and also despite using the same line list as the CCF, the LAV RV uncertainties (of the order of 10\,\ms) are larger than the CCF ones ($\sim$4\,\ms). This difference could indicate that the individual line RV uncertainties are overestimated. As explained above, we used the formal uncertainty of the Gaussian fit to the individual line, and not an estimate of the actual RV content of the line, which could lead to smaller uncertainties. Using the formal uncertainty of the fit allows us to compare different lines, that is, the formal uncertainty is adequate for relative measurements, but is not ideal when comparing with other methods such as the CCF. It is also possible that the average of the Gaussian fits of the individual lines introduces noise and is less reliable than computing the CCF with all the lines and then fitting a Gaussian to the averaged profile, an effect that would be enhanced if the individual line RV uncertainties are not accurate. Hence, we only obtained comparable RV values between the LAV and the CCF RVs because the dispersion due to the stellar activity of the stars is significantly large. Considering the uncertainty, the actual precision of the LAV RVs is worse than the CCF RVs.

There are some observations for which the difference between the LAV RVs and the other two datasets (i.e. CCF and \serval RVs) is significantly larger than the rest. These observations are the ones with the lowest S/N within each time series. We can observe this difference of S/N in the RV errors of the three datasets, which are significantly larger for these observations, and also, in the number of lines used to compute the average RV, which is significantly lower than for the rest of epochs (again see Figs. in Appendix  \ref{sec:app_totalrv_all}). For instance, in J07446+035 (Fig. \ref{fig:rvtsallcompareJ07446+035}), the LAV RV of the fourth observation starting from the end of the time series deviates from the CCF and \serval values, and uses less lines than the rest of the epochs. As another example, in J15218+209 (Fig. \ref{fig:rvtsallcompareJ15218+209}), there is an observation near BJD 2457800 with a LAV RV deviating from the CCF and \serval ones, larger uncertainties, and using less lines than the rest of data points. This decrease in the number of lines is related to the lower S/N of these observations, which makes it difficult to correctly identify the lines in the observed spectrum and obtain a reliable Gaussian fit of the line. This seems to indicate that computing total RVs by averaging the RVs of individual lines (as computed here) is more sensitive to the S/N of the data than the other two methods. In the following analysis, we did not consider these observations with low S/N (specifically, we discarded observations with $\mathrm{S/N}<25\,\mathrm{to}\,50$ in the CARMENES VIS reference order 82, centred at about 7500\,\AA, depending on the average S/N of the different stars).

We also discarded observations with strong flares. Flares add continuum flux over the whole spectral range, and can be easily identified by an increase in the \Halpha core emission. This change in flux can have a strong effect on RVs, introducing drifts of several hundred \ms \citep[see e.g.][]{reiners2009CNLeoflare}. Aside from stronger than average \Halpha emission, flares also affect the other activity indicators, which show extreme values in their time series, or clear outliers. Therefore, to avoid RV biases due to strong flares when computing the correlations with the activity indicators, we discarded observations with strong \Halpha emission by performing a 3 sigma clipping on the \Halpha index, \IHalpha, which measures the ratio of flux around the centre of the \Halpha line to the flux in reference bandpasses on either sides of the line \citep[as defined in][]{zechmeister2018serval}. We observed that other activity indicators such as the contrast of the CCF and the differential line width \citep[dLW, which accounts for changes in the line widths of the observed spectrum compared to a spectrum template, also defined in][]{zechmeister2018serval}, show clear outliers corresponding to strong flare events, but that do not have an \IHalpha value large enough to be removed by the sigma clipping procedure. Therefore, we also performed a 3 sigma clipping on the dLW time series.

%---------------------------------------------------------------------

\section{Activity effect on individual lines} \label{sec:indlinact}

\subsection{Correlation between line RV and activity indicators}\label{sec:corrlinact}

%---------------------------------------------------------------------
\begin{figure*}
\centering
\includegraphics[width=0.85\linewidth]{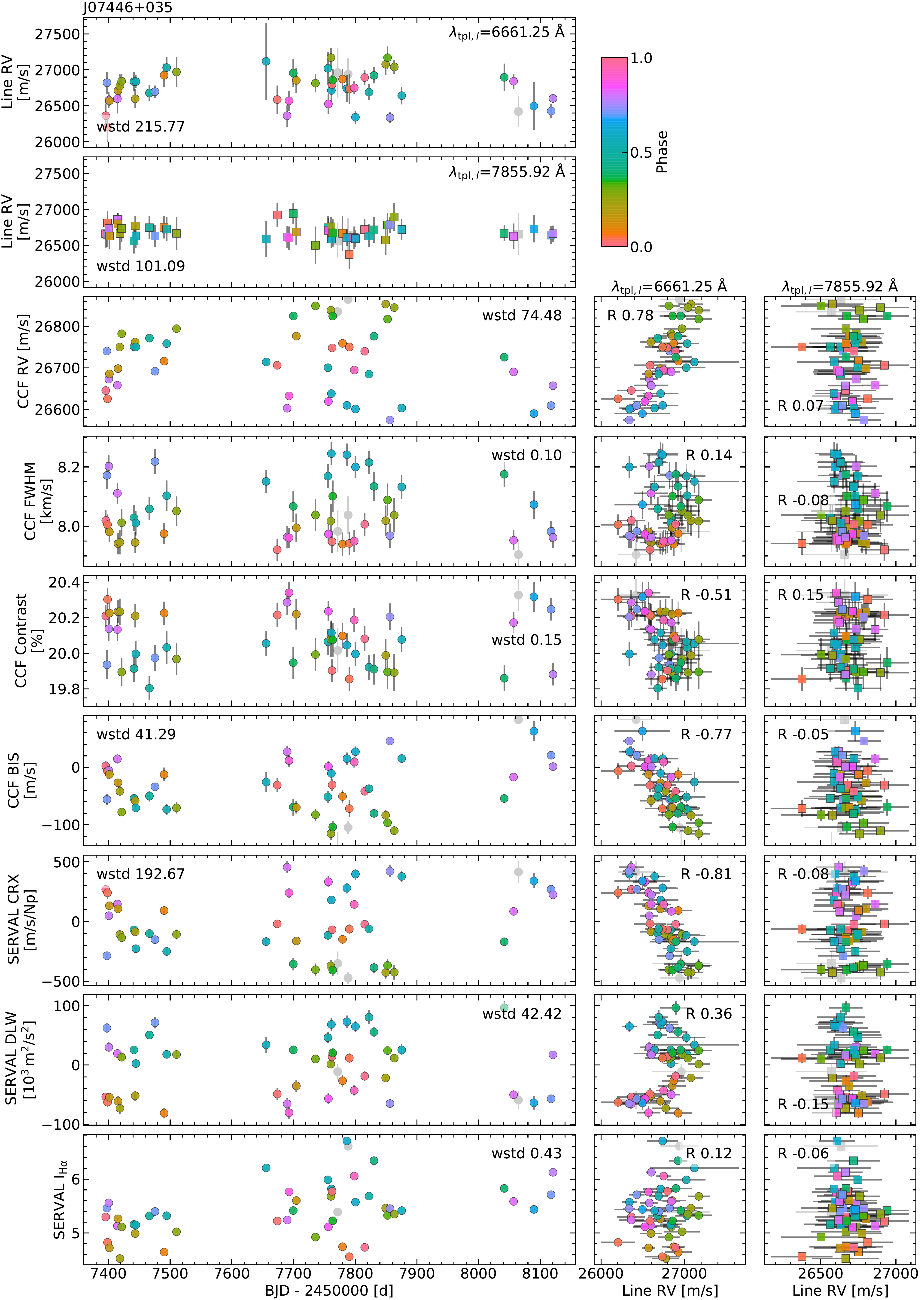}
\caption{Time series RV of an `active' ($\lambda_{\mathrm{tpl},l}=6661.25\,\A$, \emph{top left}, dots) and `inactive' ($\lambda_{\mathrm{tpl},l}=7855.92\,\A$, \emph{top left}, squares) line for J07446+035 (YZ~CMi). The panels \emph{below} show the time series of the CCF RV, FWHM, contrast, and BIS, and the CRX, dLW, and \IHalpha computed by \serval (\emph{left}), together with their correlation with the RV of the individual line (\emph{middle} and \emph{right}). The text in the time series panels shows the scatter of the data, computed as their weighted standard deviation (wstd). The text in the correlation panels shows the Pearson's correlation coefficient (R).
All data points are colour-coded with the stellar rotation phase. Observations removed as explained in Sect. \ref{sec:indlinmethodtotalrv} are marked in grey.}
\label{fig:indivlinrvtscorrJ07446+035}
\end{figure*}
%---------------------------------------------------------------------

A way to study how stellar activity affects different lines is to check for correlations between the RV of each individual line and an activity indicator. Strong correlations would indicate that a certain line is highly affected by activity, while no correlation could mean that the line is not very sensitive to activity effects \citep{dumusque2018indivline}.

We checked the linear correlation between the line-by-line RVs and several activity indicators: CCF FWHM, contrast, and BIS \citep[computed with \raccoon as in ][]{lafarga2020carmenesccf}, and CRX, dLW and \IHalpha \citep[computed with \serval as in ][]{zechmeister2018serval}. To quantify the correlations, we computed the Pearson's correlation coefficient R. A value of R close to 1 indicates a strong linear correlation, $-1$, a strong anti-correlation, and R close to 0 indicates no correlation. In Fig. \ref{fig:indivlinrvtscorrJ07446+035} we show, as an example, the time series and correlations of two different lines of J07446+035: an `active' line that shows strong correlations with several activity indicators, $\lambda_{\mathrm{tpl},l}=6661.25\,\AA$, and an `inactive' line, for which we do not observe any clear correlation, $\lambda_{\mathrm{tpl},l}=7855.92\,\AA$.

BIS and CRX are known to show clear anti-correlations with the total RV of the spectrum if it is affected by activity (\citealp[see e.g.][]{zechmeister2018serval,tal-or2018carmenesRVloud,lafarga2021indicators}, \citealp[but also][for an example of a positive correlation]{kossakowski2022adleo}). Here, we also observe a strong anti-correlation with the individual RVs of several lines, such as the example line $\lambda_{\mathrm{tpl},l}=6661.25\,\AA$ shown in Fig.  \ref{fig:indivlinrvtscorrJ07446+035}. For the other indicators, such as FWHM and dLW, we observe some loop- or circular-like shapes following the phase of the stellar rotation modulation (the circular-like shape being due to phase shift) but not a clear linear positive or negative correlation. The same applies to the correlation between total RV and these indicators \citep[again see e.g.][]{zechmeister2018serval,lafarga2021indicators,jeffers2022EVLac}. Indicators such as the FWHM or the dLW measure the width of the lines (i.e. second moment of the line profile), as opposed to other proxies such the BIS, which are sensitive to line asymmetries (third moment of the line profile). Therefore, linear correlations between the line or total RV and the FWHM or dLW are not necessarily expected \citep[see e.g.][]{jeffers2022EVLac,cardona2022young}. For these indicators, other types of correlations should be investigated.

In addition to the usual activity indicators, we also computed the correlation between the line-by-line RVs and the total RV obtained from the CCF (also in Fig. \ref{fig:indivlinrvtscorrJ07446+035}). In very active stars, where the modulations observed in the total RV are clearly caused by activity, the total RV itself can also be considered an activity indicator. This seems to be the case for the targets under study (but see Sect. \ref{sec:indlin_resultsJ10196+198} about J10196+198), and, as expected, several lines show a clear strong correlation. It is important to note that if the target star hosts exoplanet companions, the total RV will also contain the Doppler shifts due to the gravitational pull of these companions. Hence, in such cases, the total RV is not a good proxy for stellar activity, and any correlations can be biased by the modulation caused by the orbiting companions, unless this modulation does not significantly contribute to the RV. 

%---------------------------------------------------------------------

\subsection{Correlation difference between activity indicators}

%---------------------------------------------------------------------
\begin{figure*}
\centering
\begin{subfigure}[]{0.41\linewidth}
\centering
\includegraphics[width=\textwidth]{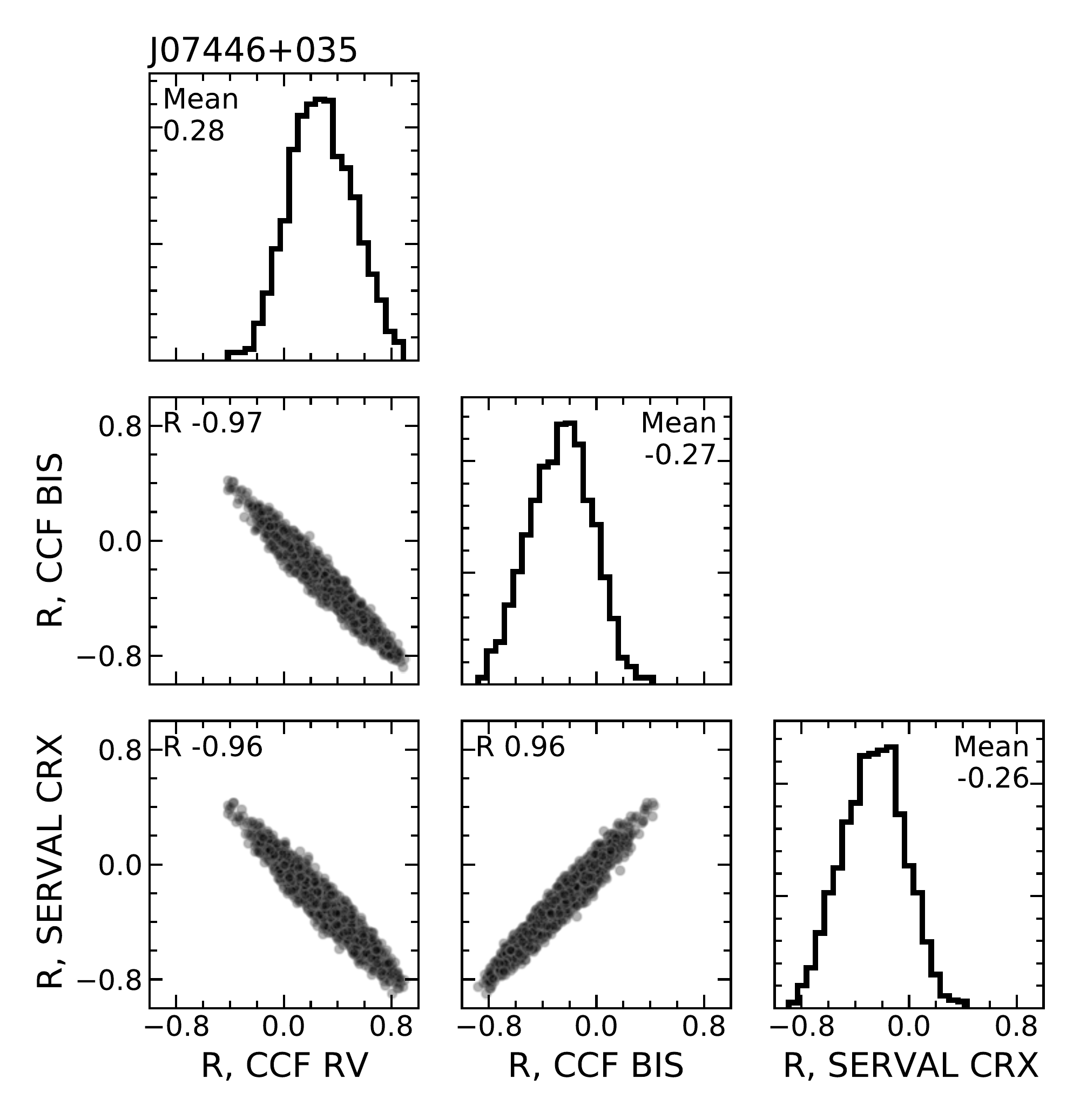}
\end{subfigure}
\,
\begin{subfigure}[]{0.41\linewidth}
\centering
\includegraphics[width=\textwidth]{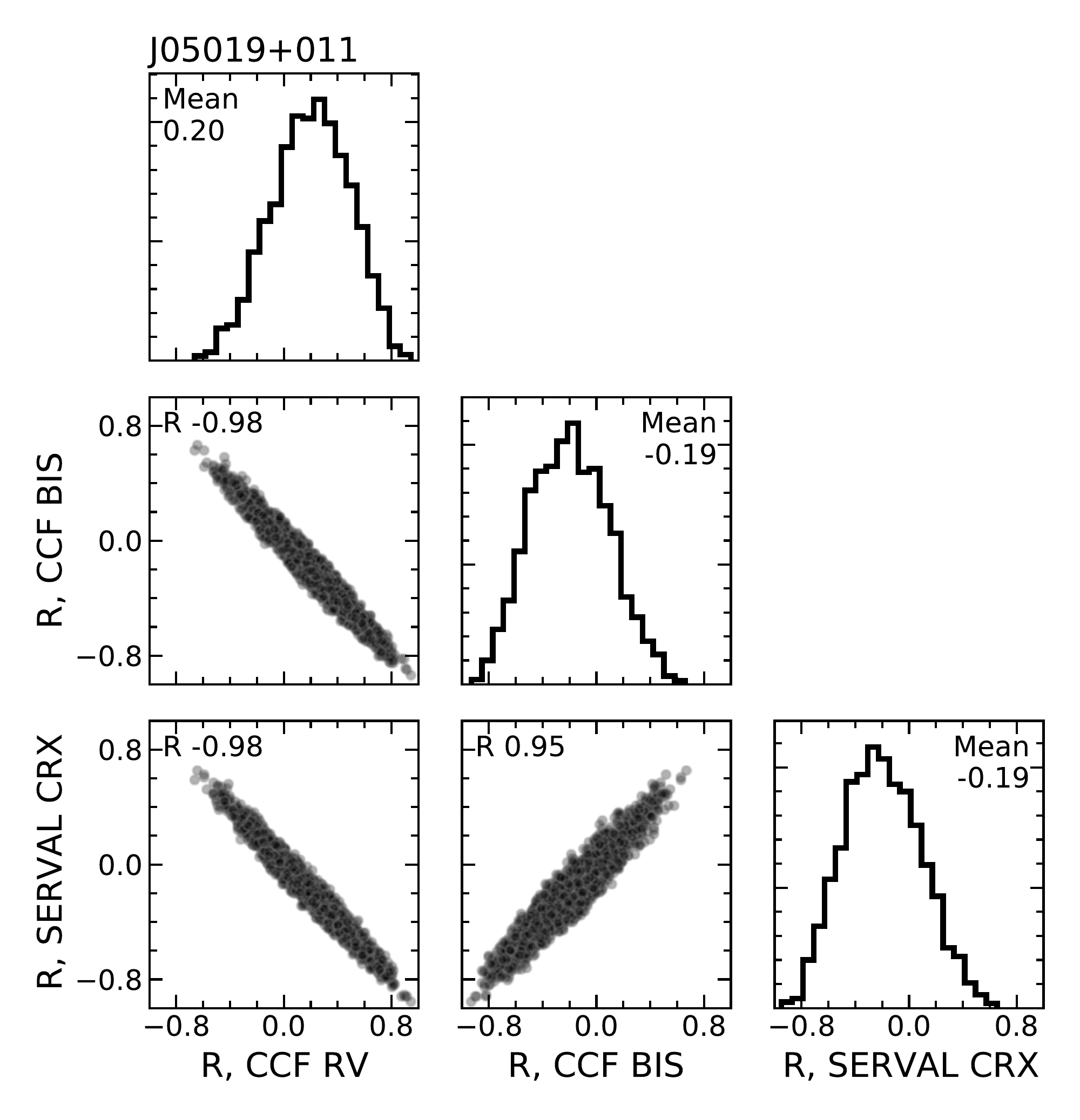}
\end{subfigure}
\\
\begin{subfigure}[]{0.41\linewidth}
\centering
\includegraphics[width=\textwidth]{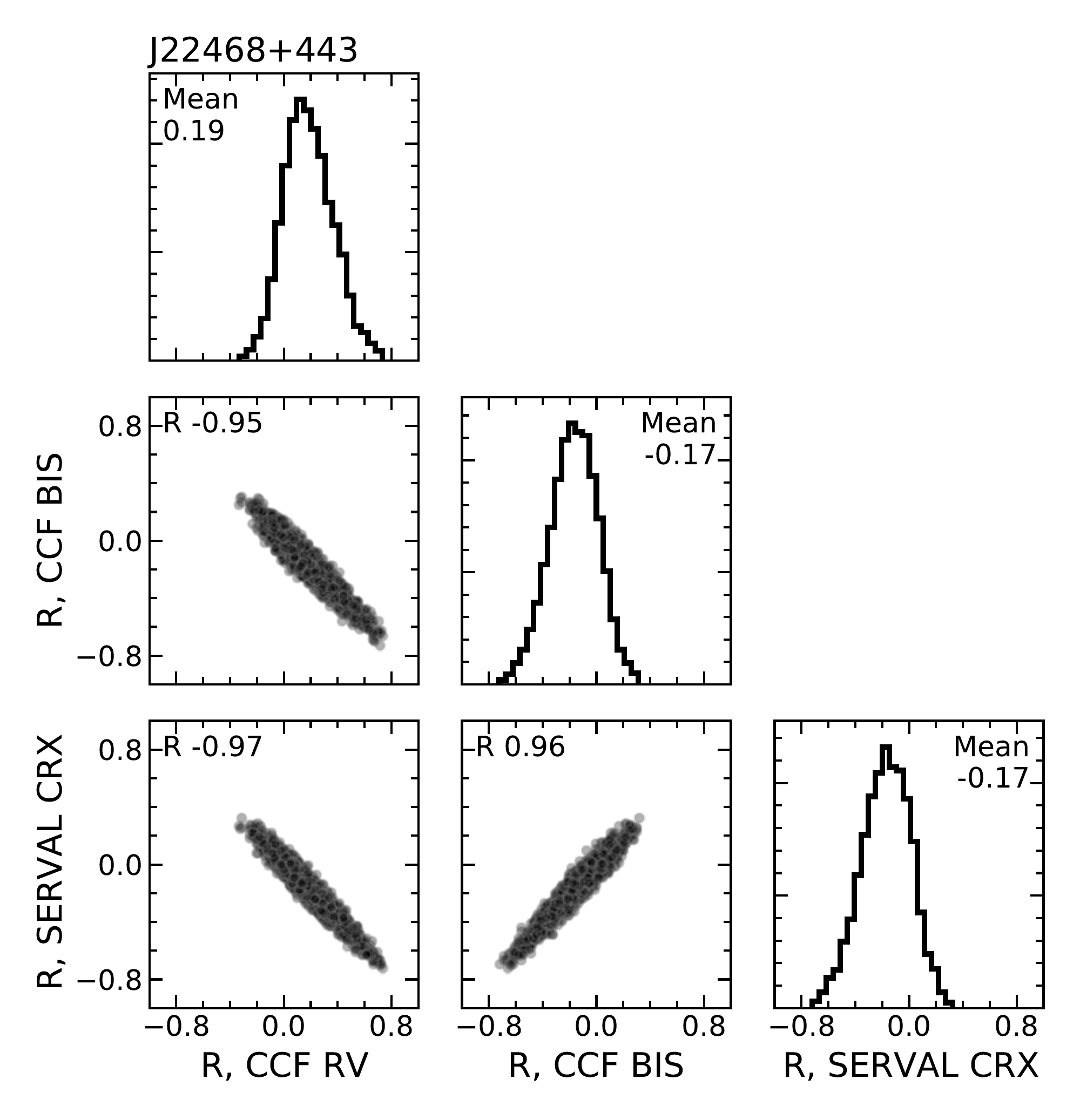}
\end{subfigure}
\,
\begin{subfigure}[]{0.41\linewidth}
\centering
\includegraphics[width=\textwidth]{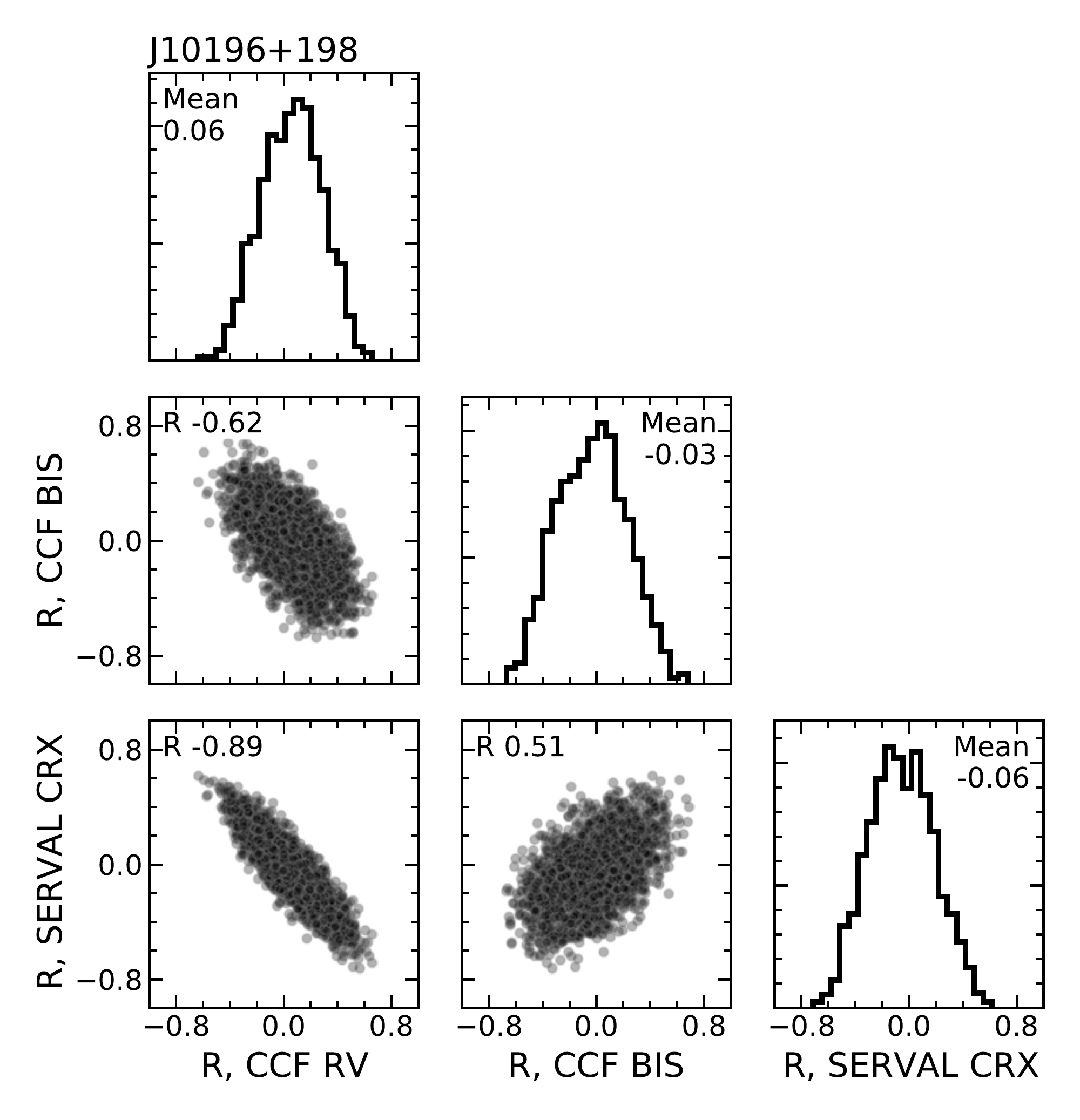}
\end{subfigure}
\\
\begin{subfigure}[]{0.41\linewidth}
\centering
\includegraphics[width=\textwidth]{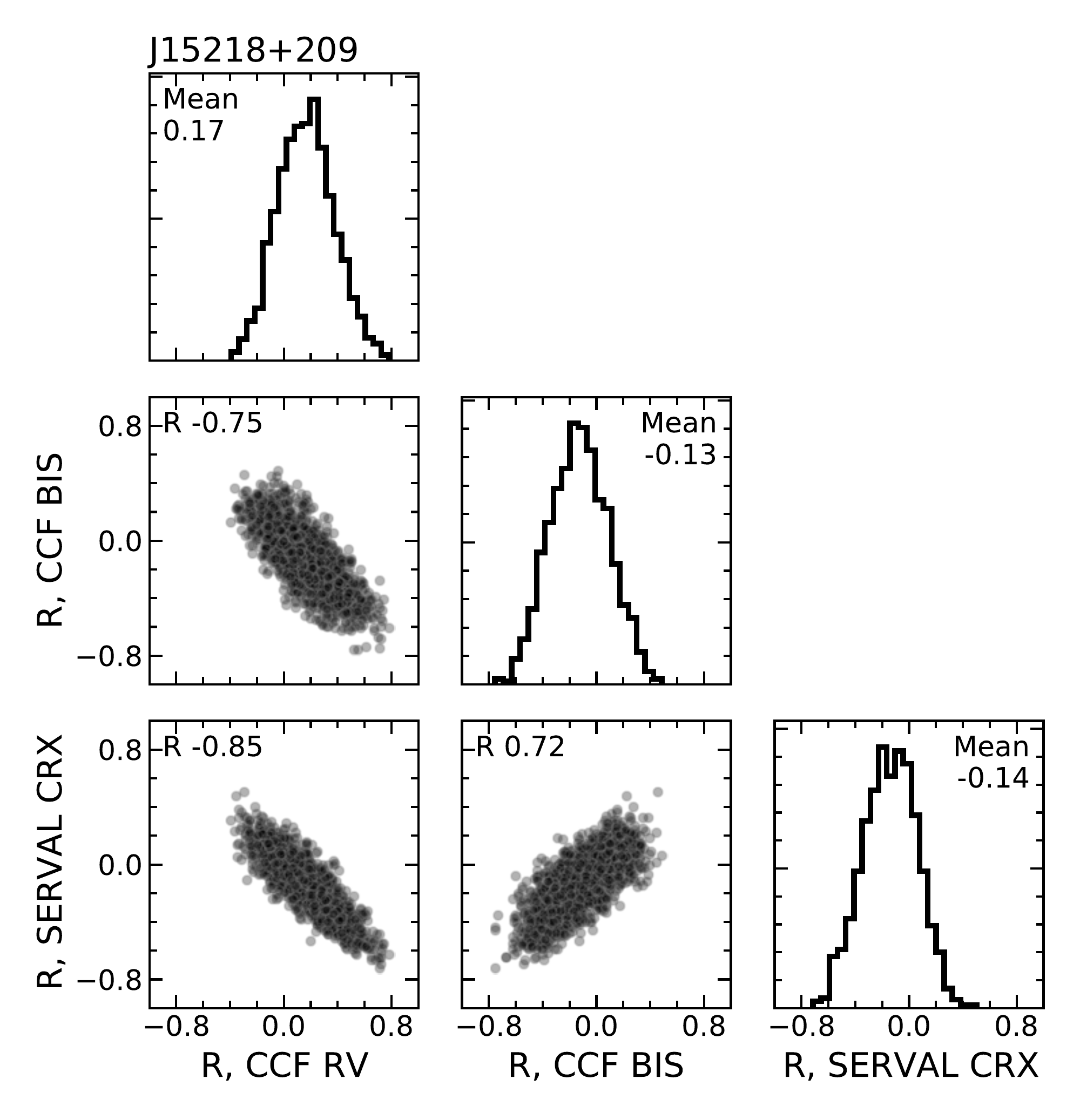}
\end{subfigure}
\,
\begin{subfigure}[]{0.41\linewidth}
\centering
\includegraphics[width=\textwidth]{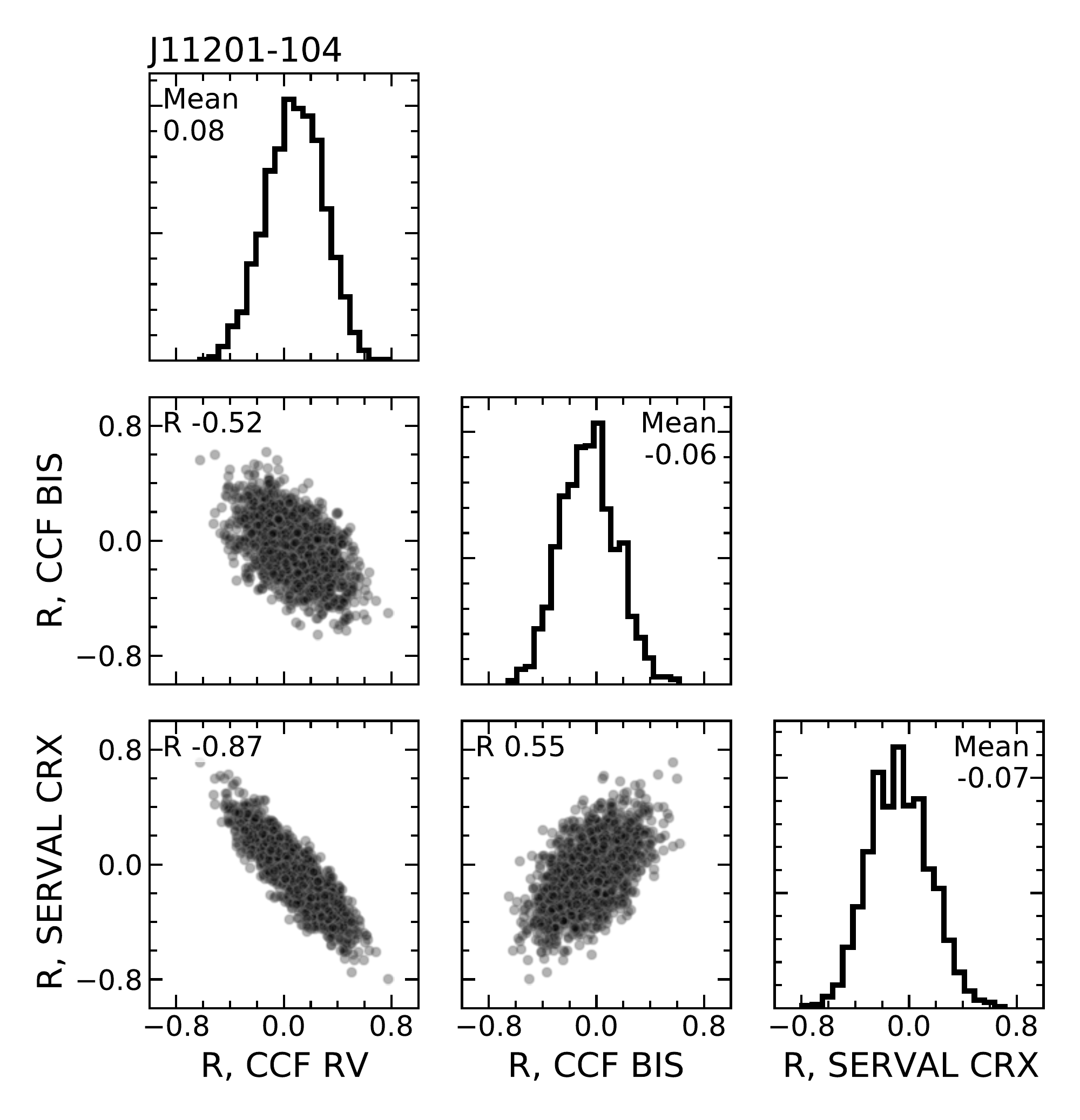}
\end{subfigure}
\caption{Comparison of the R values obtained for the correlation between the line RVs and CCF RV, BIS, and CRX, for J07446+035 (\emph{top left}), J05019+011 (\emph{top right}), J22468+443 (\emph{middle left}), J10196+198 (\emph{middle right}), J15218+209 (\emph{bottom left}), and J11201--104 (\emph{bottom right}). The histogram panels also show the mean value of the distribution.}
\label{fig:cornerRparams}
\end{figure*}
%---------------------------------------------------------------------

In the following, we focus the analysis on the three indicators that show a simple, approximately linear, correlation with the line-by-line RVs: total CCF RV, CCF BIS, and \serval's CRX. Aside from the correlation, BIS and CRX are the most effective indicators in tracing activity in very active, early- and mid-type M dwarfs as those analysed here, at least within the CARMENES GTO sample \citep{lafarga2021indicators}. To see if the correlation strength of different lines is consistent among these three activity indicators, we compared the R values obtained for the correlation between the line-by-line RVs and the indicators (see Fig. \ref{fig:cornerRparams}).

For J07446+035, J05019+011, and J22468+443 (the stars with the largest average \pEWHalpha and largest RV scatter), the three indicators show similar R scatter and similar values for all the lines, that is, lines with a strong correlation with CCF RV also show a strong correlation (anti-correlation in these cases) with BIS and CRX. Therefore, we expect subsets of active and inactive lines selected based on the correlation with these indicators to be similar. J10196+198 and the earlier type stars, J15218+209 and J11201--104, show less well-defined correspondence between the R values of the three indicators (i.e. the data points in Fig. \ref{fig:cornerRparams} show a larger spread than in the three previously-discussed stars). This means that the three activity indicators result in correlations of slightly different strength (i.e. different R) for the same line.

%---------------------------------------------------------------------

\subsection{Correlation strength as a function of the line wavelength}

\begin{figure*}
\centering
\includegraphics[width=\linewidth]{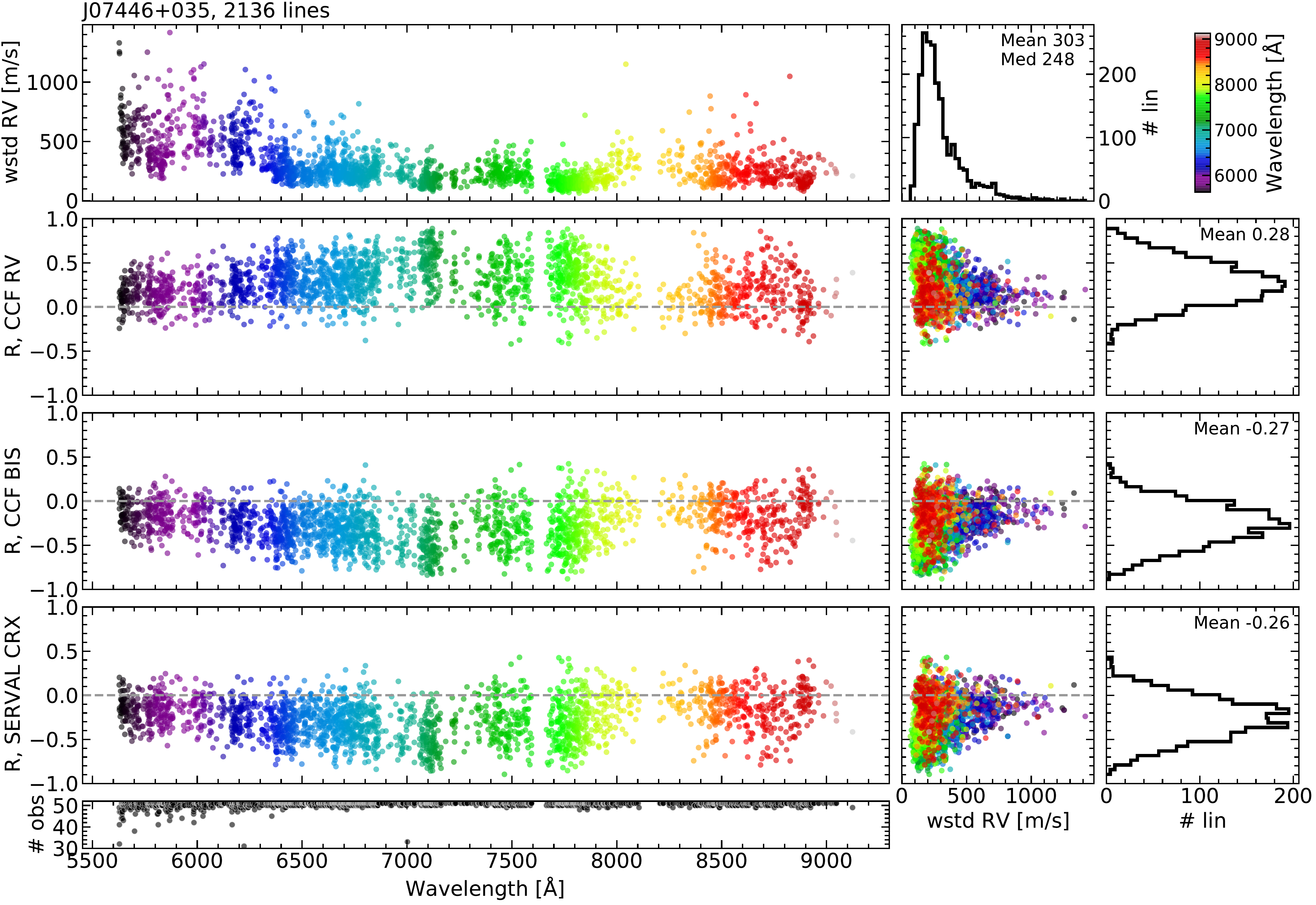}
\caption{RV scatter of each line (computed as the weighted standard deviation of the line RV of all the observations, wstd RV), and Pearson's correlation coefficient R as a function of the wavelength, for the correlation with CCF RV, BIS, and CRX (\emph{top} to \emph{bottom left panels}) for J07446+0355 (YZ~CMi). All data points are colour-coded with the wavelength. The \emph{middle panels} show the correlation between the corresponding R and the wstd RV. The histograms on the \emph{right} show the distribution of R. The \emph{top middle} histogram shows the distribution of wstd RV. The \emph{bottom left panel} shows the number of observations used per line.}
\label{fig:RcomparisonparamsJ07446+035}
\end{figure*}

Next, we study the distribution of R values as a function of wavelength. In Fig. \ref{fig:RcomparisonparamsJ07446+035}, we show the distribution of R values of all the lines obtained from the correlations with the three selected indicators (CCF RV, BIS, and CRX) as a function of the line wavelength, for J07446+035. In the same figure, we also show the correlation of the R values with the RV scatter of each line, measured as the weighted standard deviation of the line RV in all the observations (wstd RV, as in Fig. \ref{fig:statsrvlinerrJ07446+035}).

For the three indicators, the distribution of R values is not constant in wavelength. The average value of |R| increases from short to long wavelengths, peaking for lines between $\lambda\sim7000$ and 8000\,\A, and decreasing again for redder wavelengths. Regarding the RV scatter of each line, its average value decreases from the bluest wavelengths up to $\sim7500\,\AA$, and remains constant for longer wavelengths. Figure \ref{fig:RcomparisonparamsJ07446+035} only shows data of J07446+035, but we observe a similar behaviour for the other five stars.

The increase of |R| with wavelength in the blue part of the spectrum seems to be related to the decrease in RV scatter. In the blue, the spectrum has lower S/N than in the red. Due to this, the RV of the bluest lines has large uncertainties and is not as precise as the RV of lines at longer wavelengths. Hence, this decrease in precision due to low S/N results in an increase of the RV scatter. The low S/N also drives the decrease in |R|. If the line RV measurements are not precise enough, it is not possible to detect any correlation with activity, resulting in |R| values close to 0. Indeed, we see that blue lines with large RV scatter have |R| $\sim0$. We conclude that these |R| values close to 0 do not mean that these lines are not sensitive to activity, but instead reflect the lack of precision in their RV measurement. In other words, the R coefficient seems to be biased at low S/N (which is what we observe in the blue part of the spectrum).

Regarding the red part of the spectrum, for wavelengths longer than $\sim7500\,\AA$, we do not expect the decrease in |R| to be caused by a decrease in precision or S/N, because the RV scatter of these lines remains approximately constant. This decrease in |R| could be due to wavelength-dependent physical effects, such as the temperature contrast effect. Activity features such as spots have cooler temperatures (and are hence darker) than the surrounding `quiet' photosphere. This temperature contrast or difference in flux breaks the symmetric flux contribution from the blue-shifted and red-shifted hemispheres of the star, which results in a distortion of the spectral line profiles as the spot covers different parts of the rotating stellar disc. This temperature contrast effect decreases with wavelength, because the contrast in flux is less pronounced at redder wavelengths. That is, the effect on the spectral lines should be smaller at longer wavelengths. Therefore, this decrease in temperature contrast at red wavelengths could be the cause of the decrease in |R| with wavelength in the red part of the spectrum. This effect should be similar to what indicators such as the CRX trace \citep{zechmeister2018serval}. The decrease of |R| with wavelength is not seen bluewards of $\lambda\sim7500\,\AA$, perhaps because the decrease in S/N (and hence the loss in RV precision) dominates.

Another explanation for lines with no correlation ($\mathrm{|R|}\sim0$) and large RV scatter could be that these lines are affected by activity in different ways, for instance, there could be a chromospheric component. Therefore, these lines would still show significant RV scatter, but would not be correlated with photospheric activity indicators such as those used here. Due to the fact that most lines with large RV scatter also show large RV uncertainties (implying a low S/N as the cause of the scatter), we do not expect such effects to affect a significant number of lines, but this is something that requires further work.

%---------------------------------------------------------------------
%---------------------------------------------------------------------

%---------------------------------------------------------------------

\subsection{Selection of active and inactive lines}\label{sec:lineselection}

The strength of the correlation with the activity indicator R allows us to classify the lines based on their sensitivity to activity. We have selected several sets of inactive and active lines depending on their R value. We considered as inactive lines with R values around 0, $|\mathrm{R}| \leq \mathrm{R_{cut}}$, for several $\mathrm{R_{cut}}$ from 0.1 to 0.4. As active lines, we selected lines with extreme R values. In the case of the correlation with the total RV, we selected lines with R close to 1, $R > \mathrm{R_{cut}}$, for several positive $\mathrm{R_{cut}} $, such as 0.4, 0.6, 0.8. For the correlations with BIS and CRX, we expect an anti-correlation if there is a modulation due to activity, therefore we selected lines with $R < \mathrm{R_{cut}}$, for several negative $\mathrm{R_{cut}}$, such as $-$0.4, $-$0.6, $-$0.8.

We also selected lines based on their RV scatter. As mentioned above, lines with the largest RV dispersion have an R close to 0 (Fig. \ref{fig:RcomparisonparamsJ07446+035}), probably due to the fact that they do not contain enough RV information to yield a precise measurement, and not because they are in fact less sensitive to activity. Therefore, when selecting inactive lines, we also performed some cuts discarding lines with large RV scatter (wstd RV).

Another way to select correlated or uncorrelated lines reliably would be to take into account the line-by-line RV uncertainty, or the S/N of the line. Similarly to the cuts based on the RV scatter above, we could have discarded lines with a median RV uncertainty above a specific value, which could have helped us avoid selection biases. Avoiding selection biases could be relevant especially for the active line selection, since we did not use the scatter of the line-by-line RVs (because active lines would have a larger scatter), and hence, our only selection criteria is based solely on the value of R, which can be biased at low S/N. We performed several tests of active line selection by limiting the median RV uncertainty of the lines to several values around the overall median RV uncertainty ($\sim$200--300\,\ms). These tests show that, despite the limited number of active lines used, the LAV RV scatter and the periodicities present in the LAV RVs are very similar to those obtained without this extra cut on the RV uncertainty. Hence, we decided to only include here the simpler case of a single cut on R value.

To test if the correlations obtained are statistically significant (i.e. to know if the R values are expected from random fluctuations), we computed their p-value. We see that for all stars and indicators, $|\mathrm{R}|$ values $\geq0.3$ have p-values close to 0 (<0.05), and that there are no high $|\mathrm{R}|$ values (i.e. close 1 or $-1$) with a high p-value. This indicates that the correlation is statistically significant. Also as expected, the smaller the $|\mathrm{R}|$ value, the higher is the p-value. Therefore, when performing cuts on the distribution of R values, we are indeed selecting or rejecting lines with a significant correlation with the activity indicators.

Tables containing information about the sensitivity to activity of the different lines are available online, one table for each of the six stars studied in this work. Specifically, each table includes the central wavelength of the line as measured in the spectral template used, the scatter of the line RV, and the Pearson's correlation coefficient R obtained for the correlation between the line RV and the three activity indicators considered (that is, three different R values, one per indicator). An example of the information contained in these tables can be seen in Figure \ref{fig:RcomparisonparamsJ07446+035}. We note again that even though we use the term line, these lines correspond to minima in the spectrum \citep[as identified in][]{lafarga2020carmenesccf} and are the result of blends of true atomic lines or features in molecular bands.

%---------------------------------------------------------------------

\section{Total RV computation with selected lines} \label{sec:indlinrvselected}

We used the different sets of active and inactive lines to recompute the total RV of each observation. As before, we used the LAV method (average of the individual line RVs). In this section, we show the results obtained for the six targets studied. To evaluate the effect of activity in the new total RVs, we analysed the change in the time series RV scatter, as well as the change of the activity-related signals present in the generalised Lomb-Scargle periodogram \citep{zechmeister2009gls}. Here we only include figures of the results obtained with the correlation with the total RV for the first target analysed, J07446+035. For clarity and completeness, figures obtained with the correlation with the BIS and CRX for J07446+035, and all figures of the rest of the targets, can be found in Appendix \ref{sec:app_rvs_periodograms}.

%---------------------------------------------------------------------

\subsection{J07446+035 (YZ CMi, GJ 285)}

J07446+035 (YZ CMi, GJ 285) is a mid-type M dwarf and one of the most active stars in our sample. It shows a large RV scatter, of almost 90\,\ms, due to stellar activity. Global RVs and both CRX and BIS show modulations at the stellar rotation period, and these two activity indicators show strong linear correlations with the RVs \citep[see][]{zechmeister2018serval,tal-or2018carmenesRVloud,lafarga2021indicators,schofer2022carmenes4stars}. \citet{baroch2020carmenescrxYZCMi} used the deviations from a straight linear correlation to constrain starspot and convective motion parameters for this star. As mentioned above, for this star we used an initial line list derived from a template of observations of J07446+035 itself, with $\sim2000$ lines.

\paragraph{Inactive lines}

%---------------------------------------------------------------------
\begin{figure*}
\centering
\begin{subfigure}[]{0.32\linewidth}
\centering
\includegraphics[width=\textwidth]{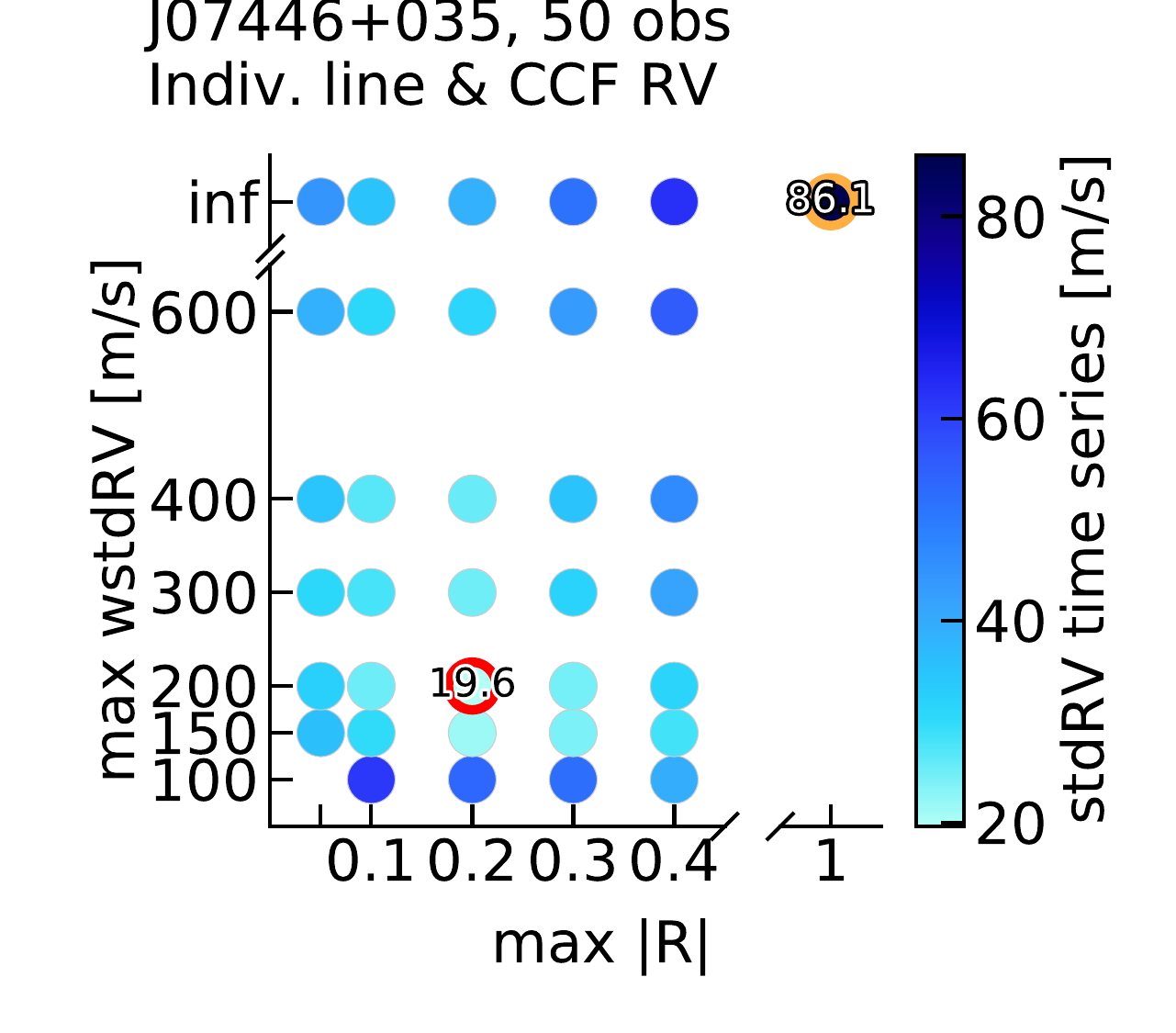}
\end{subfigure}
\,
\begin{subfigure}[]{0.32\linewidth}
\centering
\includegraphics[width=\textwidth]{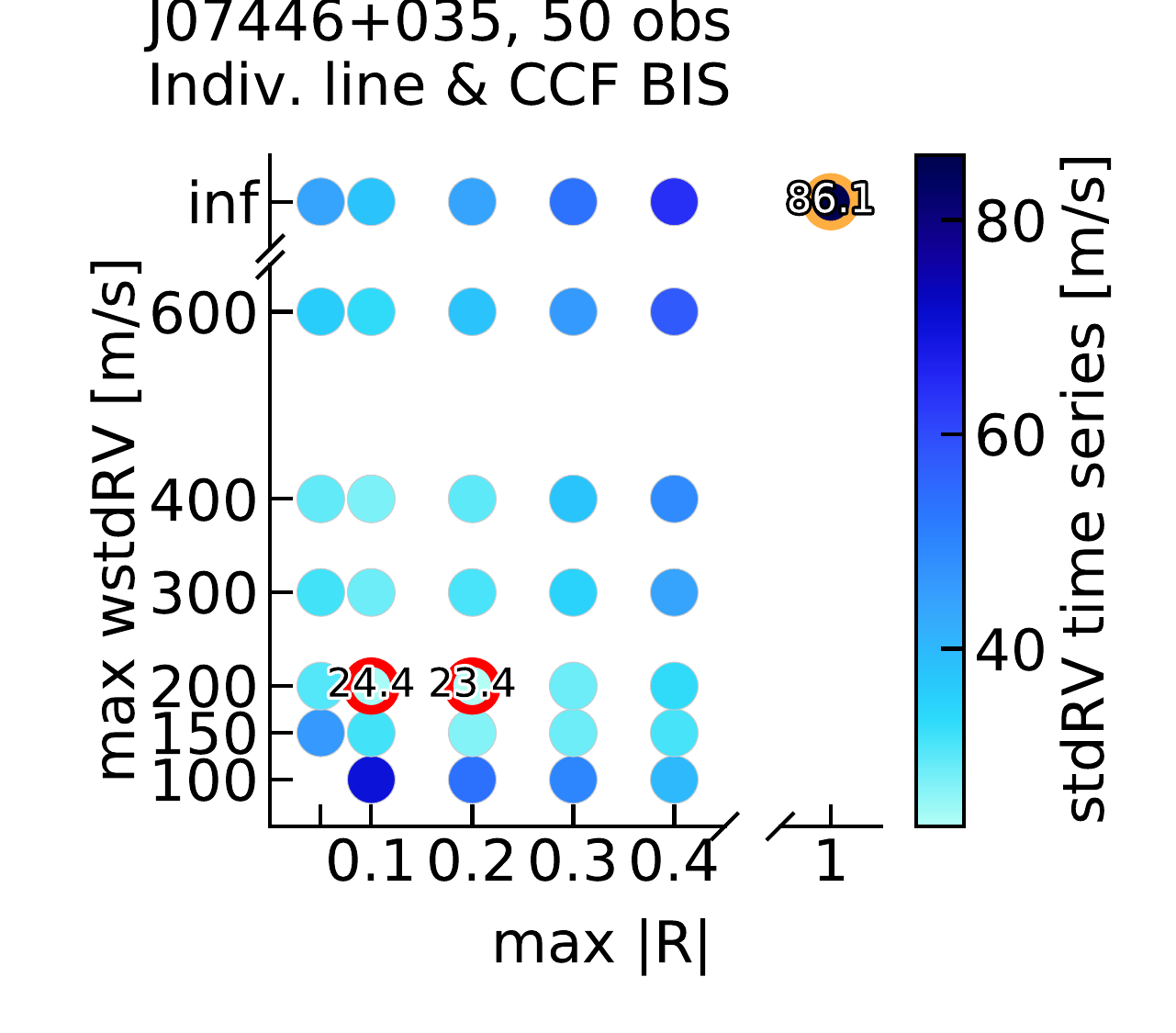}
\end{subfigure}
\,
\begin{subfigure}[]{0.32\linewidth}
\centering
\includegraphics[width=\textwidth]{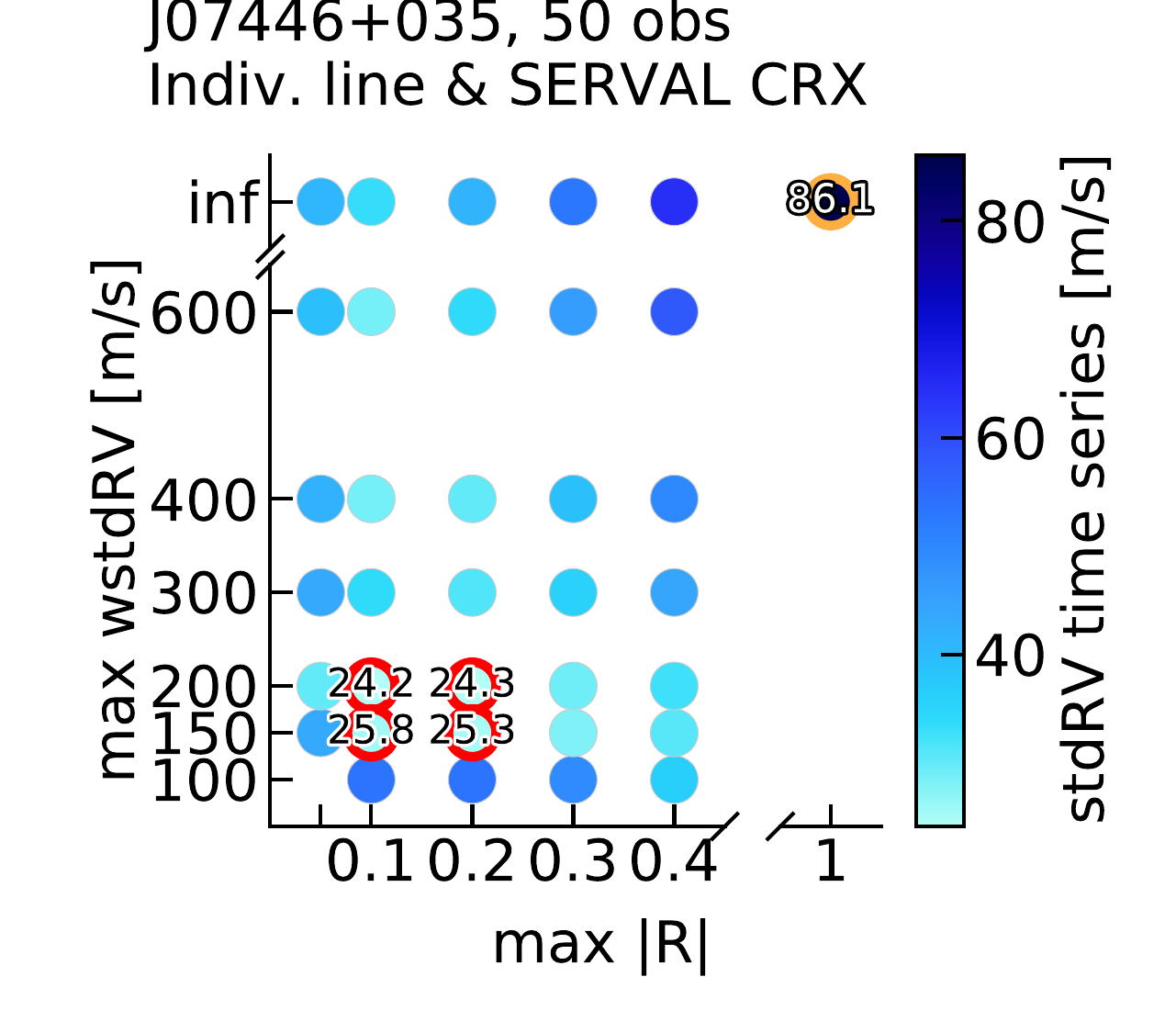}
\end{subfigure}
\caption{Scatter of the LAV RV time series of several sets of inactive lines for J07446+0355 (YZ~CMi). The different line sets are selected by limiting the line RV scatter (wstdRV, y-axis) and the strength of the correlation between the line RV and an activity indicator (Pearson's correlation coefficient R, x-axis). Each data point corresponds to a different set of selected lines. The colour of the data points indicates the scatter of the average RV time series obtained for each line set. We show the results obtained using the correlation with three activity indicators: total RV of the spectrum (\emph{left}), BIS (\emph{middle}), and CRX (\emph{right}). The data points highlighted in red indicate the datasets with the smallest RV time series scatter (values within 10\% of the absolute minimum). The data point highlighted in orange indicates the initial dataset that uses all lines to compute the RV. The numbers in these points indicate the scatter obtained using these sets of lines (i.e. they reflect the colour of the data point).}
\label{fig:cuts2dinactiveJ07446+035}
\end{figure*}
%---------------------------------------------------------------------

%---------------------------------------------------------------------
\begin{figure*}
\centering
\includegraphics[width=0.93\linewidth]{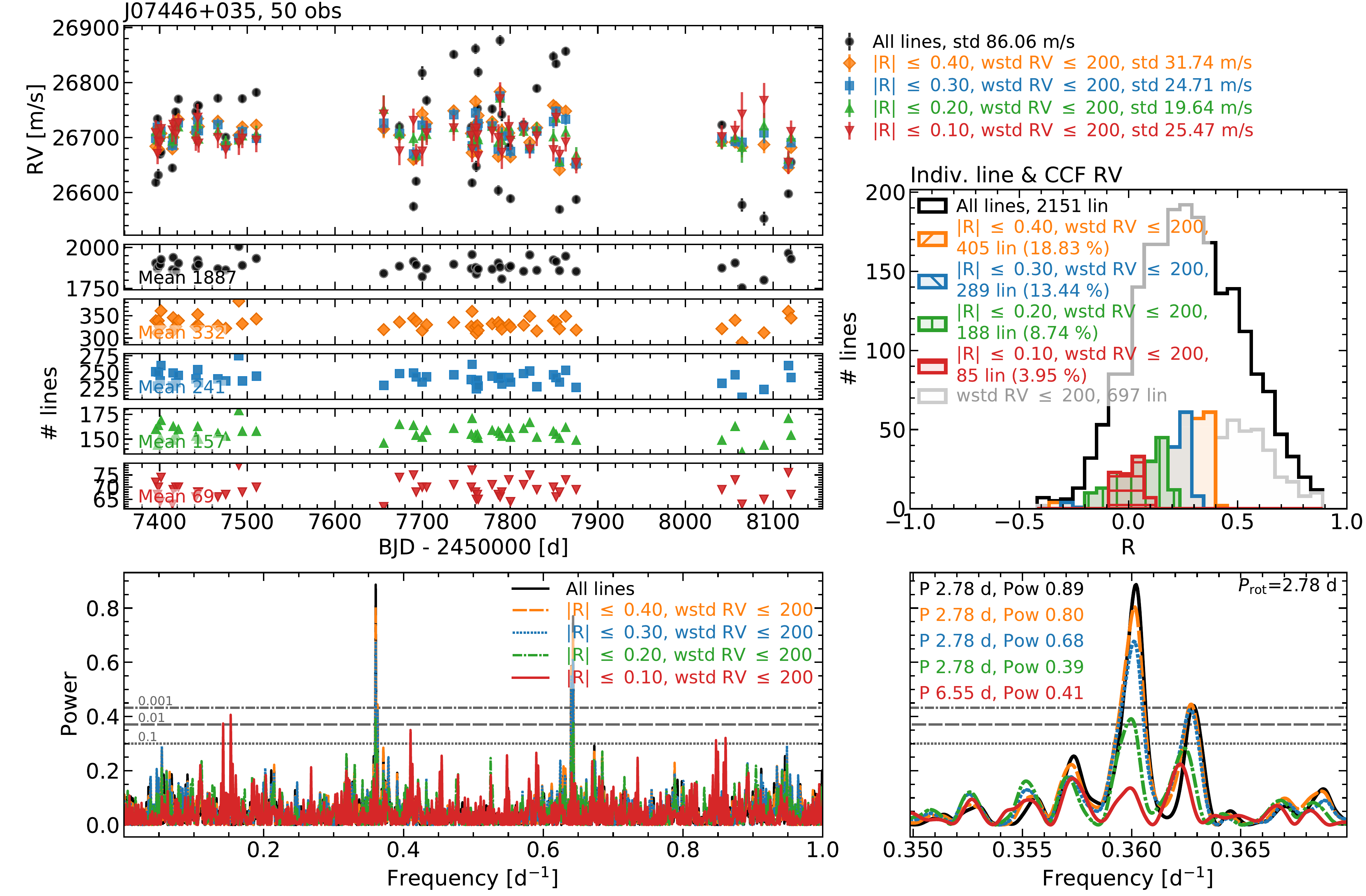}
\caption{Line-by-line results obtained using inactive lines for J07446+035 (YZ~CMi). \emph{Top left}: LAV RV time series obtained using RVs of selected inactive individual lines (colour), together with the LAV RV time series obtained using all the lines (black). The lines are selected based on the correlation of their RVs with the total spectrum RV, and their scatter. \emph{Middle left}: Number of lines used to compute the average RV in each observation, for the different selection of lines used in the \emph{top left panel}. \emph{Top right}: Distribution of the Pearson's correlation coefficient R from the correlation between the individual line RVs and the total spectrum RV. The regions in colour indicate the different selection of lines used to compute the RVs of the \emph{top left panel}. Additionally, we show in grey the distribution of R values after applying only the cut in line RV scatter (wstd). \emph{Bottom left}: Periodograms of the RV datasets plotted on the \emph{top left panel}. The horizontal lines correspond to the 0.1, 0.01, and 0.001 FAP levels, respectively. \emph{Bottom right}: Zoom in of the periodogram region around the peak corresponding to the rotation period of the star. The text indicates the period and power of the highest peak of each periodogram.}
\label{fig:tsnhpccfrvinactivebestJ07446+035}
\end{figure*}
%---------------------------------------------------------------------

We show a summary of the LAV RV time series obtained for the different sets of inactive lines tested in Fig. \ref{fig:cuts2dinactiveJ07446+035}. As we decrease the range of the R values of the averaged lines towards zero, and remove lines with large RV dispersion (wstd RV), the scatter of the total RV time series decreases. For the three indicators, the smallest RV scatter occurs when the RVs are computed using lines with $|\mathrm{R}|\leq0.1$ or $\leq0.2$ and $\mathrm{wstd~RV\lesssim200}\,\ms$, but the scatter is not minimised exactly for the same cuts. For the correlation with the total RV, the scatter becomes $\sim5$ times smaller than the initial one, and for the correlation with BIS and CRX, $\sim4$ times smaller. This decrease in RV scatter occurs up to a specific point. After that, if we further decrease the maximum $|\mathrm{R}|$ and/or wstd~RV of the selected lines, the scatter starts to increase. As we decrease the number of averaged lines, the uncertainty of the averaged RV tends to increase too, due to the fact that we have less RV content. However, when the number of averaged inactive lines is smaller than $\sim100$, the uncertainty in the averaged RV can be larger than the dispersion of the data points (see e.g. \ref{fig:tsnhpccfrvinactivebestJ05019+011} and other figures in Appendices \ref{sec:app_rvs_periodograms} and \ref{sec:app_indlincomparisonstar}). This increase could indicate that the LAV RV errors are overestimated if a small number of lines is used to compute the average, and a more robust or accurate method of determining the errors would be needed.

To further analyse the recomputed RVs and their modulations, we computed the periodogram of the different RV datasets. Fig. \ref{fig:tsnhpccfrvinactivebestJ07446+035} shows the LAV RV time series, the number of lines used, and the corresponding periodograms for four sets of lines selected using the correlation with the total RV. The different RV datasets shown correspond to those obtained with lines having $\mathrm{wstd~RV\lesssim200}\,\ms$ and four different maximum values of $|\mathrm{R}|$, 0.4, 0.2, 0.1, and 0.05, which include some of the datasets with the smallest time series RV scatter. For comparison, we also show the original RV obtained using all the lines. Figs. \ref{fig:tsnhpccfbisinactivebestJ07446+035} and \ref{fig:tsnhpservalcrxinactivebestJ07446+035} show the same but for the LAV RVs obtained with BIS and CRX, respectively. We observe a similar behaviour for the lines selected based on the correlations with the three indicators, which was expected, since for J07446+035 all three indicators show very similar R values for the same lines (as seen in Fig. \ref{fig:cornerRparams}).

We see that as the RV scatter decreases when restricting the lines used, the power of the periodogram peak at \Prot is also reduced. This seems to indicate that, by rejecting lines with large $|\mathrm{R}|$, we are effectively reducing the activity signal present in the RV measurements. If all the lines were equally affected by activity, we would expect the RV scatter to become larger as we restrict the R of the averaged lines. This is because, as we decrease the range of allowed R values, we are decreasing the number of lines used, and hence, we are degrading the precision of the RV measurement. Despite that, since the RV scatter decreases as we reduce the number of lines used, it seems that, for this star, reducing the red noise caused by activity has a larger effect than the increase in white photon noise due to the reduction in RV content, as also found by \citet{dumusque2018indivline} for $\alpha$\,Cen\,B. We note that some of the smallest RV scatters are obtained using datasets with only $\sim200$ lines, about 10\% of the original $\sim2000$. With more restrictive cuts, the RV scatter starts to increase. This probably occurs because the number of lines is too small to obtain sufficient precision in the RV measurements, and hence the photon noise dominates. For these sets of lines, the periodogram peak at \Prot also becomes less significant, and the highest peak in the periodogram is not related to \Prot any more. This could reflect the fact that the RVs of the datasets with the smallest number of lines contain mostly noise.

\paragraph{Active lines}

%---------------------------------------------------------------------
\begin{figure*}
\centering
\includegraphics[width=0.93\linewidth]{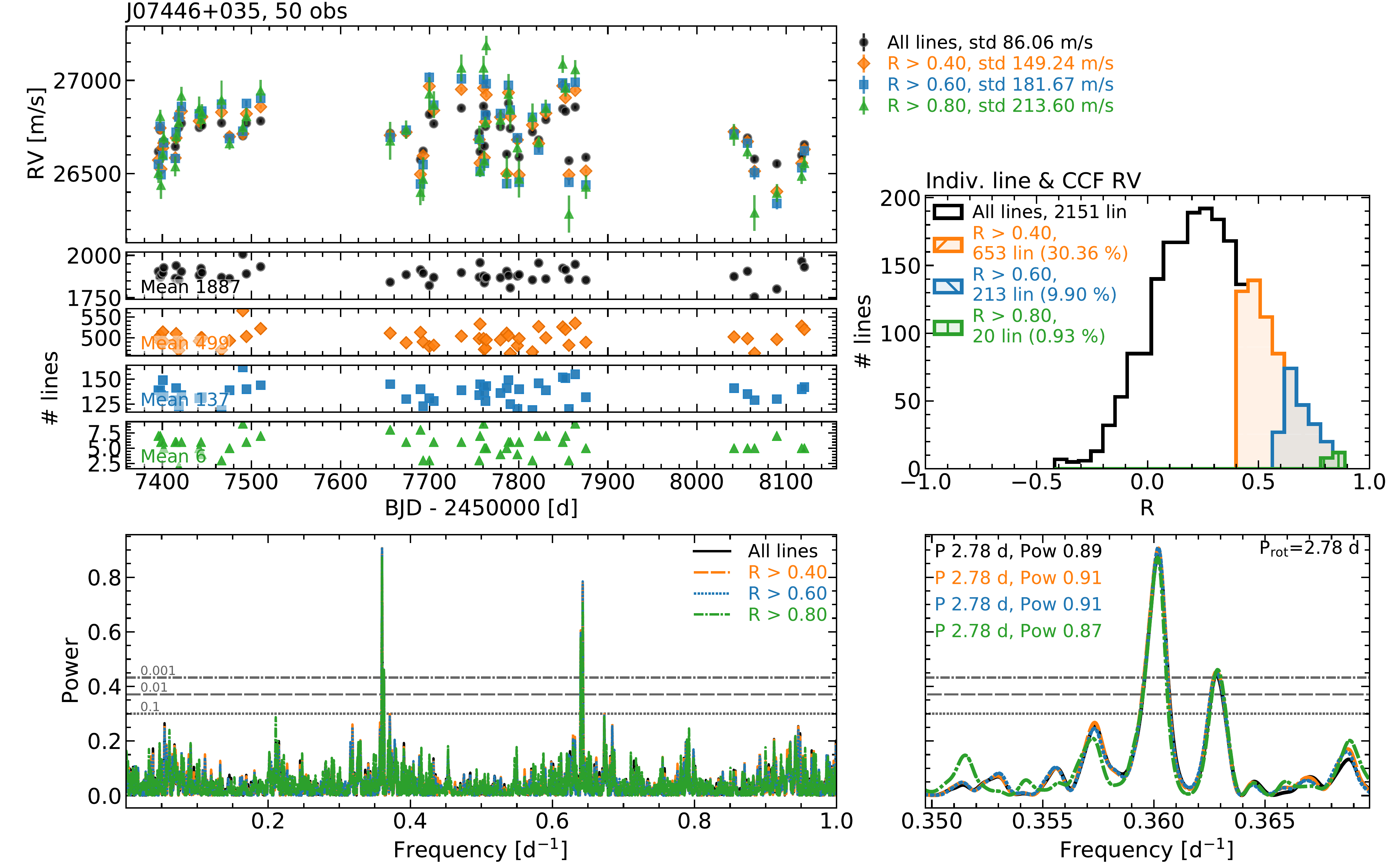}
\caption{Same as Fig. \ref{fig:tsnhpccfrvinactivebestJ07446+035}, but for three sets of active lines.}
\label{fig:tsnhpccfrvactiveJ07446+035}
\end{figure*}
%---------------------------------------------------------------------

Figures \ref{fig:tsnhpccfrvactiveJ07446+035}, \ref{fig:tsnhpccfbisactiveJ07446+035}, and \ref{fig:tsnhpservalcrxactiveJ07446+035} show the LAV RV time series, number of lines, and periodograms of three sets of active lines, selected using the correlations with the total RV, BIS, and CRX, respectively. Here, the scatter increases significantly as we restrict the minimum value of R towards 1 for the correlation with the total RV, or towards $-1$ in the case of BIS and CRX, becoming more than 2 times larger for the most extreme sets. This increase could be due to the fact that we are only using the most active lines, but also to an increase in the photon noise, because we are reducing the number of lines used. In all cases, the periodogram shows the highest peak at the stellar rotation period, all with similar power. We note that for $\mathrm{R_{cut}}=0.8$, using 10 to 20 lines (0.5 to less than 1\,\% of the initial set of lines), depending on the activity indicator, the periodogram still clearly displays a signal due to the stellar rotation, very similar to the periodogram obtained using the $\sim2000$ initial lines.

%%%%%%%%%%%%%%%%%%%%%%%%%%%%%%%%%%%%%%%%%%%%%%%%%%
%%%%%%%%%%%%%%%%%%%%%%%%%%%%%%%%%%%%%%%%%%%%%%%%%%

\subsection{J05019+011 (1RXS J050156.7+010845)}

J05019+011 (1RXS J050156.7+010845) is, as one can expect from its name, a relatively strong X-ray emitter \citep{haakonsen2009rosat2mass}. Based on its kinematics and activity indicators, it has repeatedly proposed as a member of the young $\beta$~Pictoris moving group \citep[e.g.][]{schlieder2012youngstars,alonsofloriano2015betapic}. In the CARMENES data, it has an initial RV scatter of about 90\,\ms from the \serval RVs. The LAV method results in a larger scatter, close to 106\,\ms. We only have 19 observations for this target, however this is sufficient to see that the global RVs, BIS, and CRX show a significant modulation at the stellar rotation period, and that the activity indicators and the RVs are linearly correlated, as in J07446+035 \citep[see][]{lafarga2021indicators}. For this star we used the line list built from the J07446+035 template.

\paragraph{Inactive lines}

Fig. \ref{fig:cuts2dinactiveJ05019+011} shows the RV scatter obtained for the different sets of inactive lines considered. Figs. \ref{fig:tsnhpccfrvinactivebestJ05019+011}, \ref{fig:tsnhpccfbisinactivebestJ05019+011}, and \ref{fig:tsnhpservalcrxinactivebestJ05019+011} show the RV time series, number of lines, and periodograms of four datasets, obtained using the correlation with the total RV, BIS, and CRX. In this case, the RV scatter is minimised for the datasets that include lines with $|\mathrm{R}|\leq0.3$ or $\leq0.4$ and $\mathrm{wstd~RV\leq300}\,\ms$, for the correlation with the total RV, BIS, and CRX. The decrease in RV scatter is of $\sim2.5$ times compared to the initial set of lines. The periodogram shows a significant peak at 2.09\,d, close to but not exactly at \Prot, 2.12\,d, whose power decreases as we restrict the lines used. We note here that \citet{revilla2020thesis} find a \Prot of 2.088\,d using \emph{TESS} data, which is closer to the value derived from the RVs here.

\paragraph{Active lines}

Regarding the active lines (Figs. \ref{fig:tsnhpccfrvactiveJ05019+011}, \ref{fig:tsnhpccfbisactiveJ05019+011}, and \ref{fig:tsnhpservalcrxactiveJ05019+011}), the RV scatter increases as we restrict the number of lines used, and the periodogram power of the peak at 2.09\,d remains significant. Contrary to the case of J07446+035, the periodogram of the RVs obtained by the average of the lines with $\mathrm{R>0.80}$ for the total RV, or $\mathrm{R<-0.80}$ for BIS and CRX, is less clear than the previous two cuts, and the peak at \Prot is not as significant.

%%%%%%%%%%%%%%%%%%%%%%%%%%%%%%%%%%%%%%%%%%%%%%%%%%
%%%%%%%%%%%%%%%%%%%%%%%%%%%%%%%%%%%%%%%%%%%%%%%%%%

\subsection{J22468+443 (EV Lac, GJ 873)}

J22468+443 (EV Lac, GJ 873) is well known flaring mid M dwarf. In the data analysed here, the periodograms of the RVs and the indicators show signals at both \Prot (4.38~days) and \Prothalf, with the strongest signal at \Prothalf, and the indicators and the RVs show linear correlations \citep{lafarga2021indicators,schofer2022carmenes4stars,jeffers2022EVLac,cardona2022young}. The stronger signal at \Prothalf is probably due to the fact that RVs and some indicators (including BIS and CRX) show a modulation with a double dip structure within one rotation period, which favours \Prothalf over \Prot. Interestingly, for the same set of observations, other indicators show a single dip structure. This could be due to different indicators tracing different moments of the line profile \citep{lafarga2021indicators,schofer2022carmenes4stars,jeffers2022EVLac}. CCF and LAV RVs have a scatter of $\sim40\,\ms$, about 10\,\ms (1.2 times) smaller than the one obtained with \serval (Fig. \ref{fig:rvtsallcompareJ22468+443}). We used the line list built from the J07446+035 template. Similarly to the two previous stars, the correlations obtained with the total RV, BIS, and CRX are very similar, and hence, we obtain similar results for the three indicators (Fig. \ref{fig:cornerRparams}).

\paragraph{Inactive lines}

The smallest RV scatter occurs when using lines with $|\mathrm{R}|\leq0.1$ or $\leq0.2$ and $\mathrm{wstd~RV}\lesssim200$ or $\lesssim400\,\ms$ (Fig. \ref{fig:cuts2dinactiveJ22468+443}). The minimum scatter is $\sim18-19\,\ms$, $2.2-2.4$ times smaller than the initial $\sim40\,\ms$, depending on the indicator. Regarding the activity modulation present in the RVs (Figs. \ref{fig:tsnhpccfrvinactivebestJ22468+443}, \ref{fig:tsnhpccfbisinactivebestJ22468+443}, and \ref{fig:tsnhpservalcrxinactivebestJ22468+443}), the periodogram shows a very significant signal at \Prothalf, 2.19\,d, and a less significant one at \Prot, 4.38\,d. As we restrict the lines used, the peak at  \Prothalf decreases in power, becoming non-significant for the datsets that result in some of the smallest time series RV scatter (datset with lines with $|\mathrm{R}|\leq0.1$ and $\mathrm{wstd~RV\leq200}\,\ms$). We see that for some datsets, for instance the one with lines with $|\mathrm{R}|\leq0.2$ and $\mathrm{wstd~RV\leq200}\,\ms$, the peak at \Prot increases its power, becoming significant.

\paragraph{Active lines}

For this star, the strength of the correlations between the individual line RVs and the activity indicators does not reach values as large as $\mathrm{R}=0.8$, as for J07446+035 or J05019+011, so the most restrictive cut performed to select active lines is at $\mathrm{R}>0.60$ (Figs. \ref{fig:tsnhpccfrvactiveJ22468+443}, \ref{fig:tsnhpccfbisactiveJ22468+443}, and \ref{fig:tsnhpservalcrxactiveJ22468+443}). As with the previous stars, the RV scatter increases as we restrict the lines used. The power of the periodogram peak at \Prothalf slightly increases, too. We observe a similar behaviour for the correlations obtained with the three indicators.

%%%%%%%%%%%%%%%%%%%%%%%%%%%%%%%%%%%%%%%%%%%%%%%%%%
%%%%%%%%%%%%%%%%%%%%%%%%%%%%%%%%%%%%%%%%%%%%%%%%%%

\subsection{J10196+198 (AD Leo, GJ 388)} \label{sec:indlin_resultsJ10196+198}

J10196+198 (AD Leo, GJ 388) has been the subject of recent, deep analyses to disentangle if the origin of the RV signal is due either to a planetary companion or to stellar activity \citep[e.g.][]{carleo2020giarps,kossakowski2022adleo}. It shows an activity level similar to that of J22468+443 (similar \pEWHalpha and \logLHalphaLbol), but has an RV scatter significantly smaller than J22468+443 or the other two previous stars ($\sim18\,\ms$ for J10196+198, while for the previous stars it is $>40\,\ms$). The number of observations is relatively small, 26, but the periodograms of the RVs, BIS, and CRX, show a significant peak at $\sim$2.24, close to \Prot, so the number of observations and sampling are sufficient for observing activity-related modulations.

The different RV amplitudes between J10196+198 and J22468+443 could be caused by different visible spot configurations. The spin axis of J10196+198 has a relatively low inclination \citep[$i\sim14\degr$,][]{kossakowski2022adleo} in comparison to J22468+443 or the previous stars \citep[which have $i\geq60\degr$, see e.g.][]{morin2008mdwarfmag}. This close to pole-on orientation could cause any visible co-rotating spots to induce a smaller modulation in the RVs, since they would not abruptly appear and disappear as the star rotates. Also, the photosphere of J10196+198 could be more homogeneously spotted, which would also induce smaller RV modulations.

As seen in Sect. \ref{sec:corrlinact}, Fig. \ref{fig:cornerRparams}, for this star there are no lines whose RV shows a very strong correlation with the activity indicators. The correlation coefficients R do not reach very large values, contrary to the findings for the three previous stars. This means that, by using these correlations, we are not able to identify lines that have a strong contribution to activity, even though the activity indicators show significant peaks at \Prot. Aside from this, the R values obtained for each line depend on the activity indicator used to compute the correlation. All this could indicate that the correlations that we find do not have much information related to the activity of the star.

As mentioned before, this star has been speculated to host an exoplanet with an orbital period similar to the stellar rotation, $\sim$2.23\,d, in a 1:1 spin-orbit resonance \citep{tuomi2018adleo}, although this claim has been challenged by further studies \citep{carleo2020giarps,robertson2020activityHPF}. However, since the stellar rotation and the hypothetical planet have the same period, it is difficult to completely rule out the planet's existence \citep{kossakowski2022adleo}. Since the RVs of this star could potentially contain the signal induced by the presence of an orbiting planet, using the total RV as an activity indicator is not a good choice here, because the correlations with the line RVs would not solely reflect the effect of activity.  It could be argued that the planet amplitude is much smaller than the modulation due to activity. However, it would be very challenging to discern the amplitude of said planet from the residual activity present in the RVs computed with inactive lines.
We presented an initial analysis of J10196+198 in \citet{kossakowski2022adleo}, where we studied activity-insensitive lines. Here we summarise it and show in addition the effect of selecting activity-sensitive lines.

\paragraph{Inactive lines}

As we restrict the lines used to those with R closer to 0, we do not see a significant decrease in the RV scatter, but an increase (Fig. \ref{fig:cuts2dinactiveJ10196+198}). Only for a few datsets that contain almost the same lines as the initial one, obtained with the correlation with RV and CRX, the scatter decreases, but not significantly (about 1\,\ms less than the initial 16.6\,\ms). For the datsets obtained using the correlation with BIS, none shows a decrease in the RV scatter. We note that the initial RV scatter obtained with the LAV method is lower than that obtained with the SERVAL RVs, 18.4\,\ms, but larger than the one obtained with the CCF RVs, 15.0\,\ms, which is close to the smallest value obtained with the inactive lines datset.

The periodograms show a peak at the \Prot of the star, 2.24\,d, which decreases in power as we restrict the lines used (Figs. \ref{fig:tsnhpccfrvinactivebestJ10196+198}, \ref{fig:tsnhpccfbisinactivebestJ10196+198}, and \ref{fig:tsnhpservalcrxinactivebestJ10196+198}). This applies to all three indicators, but for the BIS the decrease is smaller than for the total RV and the CRX. Since the RV scatter does not decrease significantly, but remains the same or increases, we attribute this decrease in the significance of the peak at \Prot to the increase in photon noise due to the smaller number of lines used in the datsets.

\paragraph{Active lines}

As we restrict the line selection to those showing the stronger correlations, we see that the RV scatter increases significantly (Figs. \ref{fig:tsnhpccfrvactiveJ10196+198}, \ref{fig:tsnhpccfbisactiveJ10196+198}, and \ref{fig:tsnhpservalcrxactiveJ10196+198}). The activity signal present in the RVs, however, does not remain constant. The peak at \Prot shows a decrease in power for the datsets using the lines with the stronger correlations, and even completely disappears in the RV datsets computed using lines that show strong correlations with BIS. This could indicate that the lines that we identified as active are in fact not related to activity, which agrees with the fact that the strength of the correlations between the individual line RVs and the activity indicators were not large (Fig.~\ref{fig:cornerRparams}).

%%%%%%%%%%%%%%%%%%%%%%%%%%%%%%%%%%%%%%%%%%%%%%%%%%
%%%%%%%%%%%%%%%%%%%%%%%%%%%%%%%%%%%%%%%%%%%%%%%%%%

\subsection{J15218+209 (OT Ser, GJ 9520)}

J15218+209  (OT Ser, GJ 9520) is one of the two early M dwarfs analysed. It has a large RV scatter, 37\,\ms. RV, BIS, and CRX periodograms show a peak at \Prot, 3.37\,d, but it is only significant (FAP < 0.1\,\%) in the case of the RVs \citep{lafarga2021indicators}. For this star we used the line list built from the J15218+209 template itself.

\paragraph{Inactive lines}

The smallest scatter of the RV time series occurs when using lines with $|\mathrm{R}|\leq0.1-0.2$ and $\mathrm{wstd~RV\leq100-300}\,\ms$ (Fig. \ref{fig:cuts2dinactiveJ15218+209}). The RV scatter attains values of 13, 19, and 17\,\ms, 2.7, 2.0, and 2.2 times smaller than the initial 37.2\,\ms, for the correlations with the total RV, BIS, and CRX, respectively. The periodogram of the initial datset shows a significant peak at \Prot, 3.37\,d (Figs. \ref{fig:tsnhpccfrvinactivebestJ15218+209}, \ref{fig:tsnhpccfbisinactivebestJ15218+209}, and \ref{fig:tsnhpservalcrxinactivebestJ15218+209}). For the datsets computed from the correlations with the total RV and CRX, the \Prot peak decreases in power and reaches a FAP of 0.1, which corresponds to a datset with one of smallest RV scatters. For the BIS datsets, the decrease in power is not as clear, and for the datset with the smaller RV scatter, the peak has a similar power as using all the lines.

\paragraph{Active lines}

The time series RV scatter increases as we restrict the lines towards those with stronger correlations (Figs. \ref{fig:tsnhpccfrvactiveJ15218+209}, \ref{fig:tsnhpccfbisactiveJ15218+209}, and \ref{fig:tsnhpservalcrxactiveJ15218+209}). For the datsets obtained with the correlations with the total RV and CRX, the power at the \Prot peak remains significant but decreases slightly. For the BIS datsets, the power decreases significantly. This seems to indicate that the BIS correlations are not as reliable as those obtained from the total RV or CRX.

%%%%%%%%%%%%%%%%%%%%%%%%%%%%%%%%%%%%%%%%%%%%%%%%%%
%%%%%%%%%%%%%%%%%%%%%%%%%%%%%%%%%%%%%%%%%%%%%%%%%%

\subsection{J11201--104 (LP 733-099)}

J11201--104 (LP 733-099) is the other early-type star analysed. The average \logLHalphaLbol indicates that J11201--104 is less active than J15218+209, and it shows a smaller scatter in its RV time series, of about 18\,\ms, in the RVs obtained with \serval, the CCF, and the LAV method (Fig. \ref{fig:rvtsallcompareJ11201--104}). It has a \Prot of 5.643$\pm$0.005 \citep{revilla2020thesis,shan2022prot}, however its periodograms do not show  significant signals. CRX and BIS show linear correlations with the RVs, but they are less clear than in the previous stars. For this star, we used the set of lines derived from the J15218+209 template.

\paragraph{Inactive lines}

The RV time series with the smallest scatters are obtained for the line sets with $|\mathrm{R}|\leq0.1-0.3$ and $\mathrm{wstd~RV\leq200}\,\ms$ (Fig. \ref{fig:cuts2dinactiveJ11201--104}), in the case of the correlations with the total RV and CRX. The scatter decreases from $\sim19\,\ms$ to $\sim11\,\ms$ for the RV datsets and to $\sim13\,\ms$ for the CRX ones, 1.7 and 1.5 time smaller, respectively. In the case of BIS, there are several datsets that show a small scatter close to the minimum one, which is about $\sim15\,\ms$, 1.3 times smaller than the initial one. The periodogram does not show any significant peaks (Figs. \ref{fig:tsnhpccfrvinactivebestJ11201--104}, \ref{fig:tsnhpccfbisinactivebestJ11201--104}, and \ref{fig:tsnhpservalcrxinactivebestJ11201--104}).

\paragraph{Active lines}

Regarding the active lines, the scatter increases significantly, but the periodogram does not shown any significant peaks for any of the datsets used (Figs. \ref{fig:tsnhpccfrvactiveJ11201--104}, \ref{fig:tsnhpccfbisactiveJ11201--104}, and \ref{fig:tsnhpservalcrxactiveJ11201--104}).

%%%%%%%%%%%%%%%%%%%%%%%%%%%%%%%%%%%%%%%%%%%%%%%%%%
%%%%%%%%%%%%%%%%%%%%%%%%%%%%%%%%%%%%%%%%%%%%%%%%%%

\subsection{Overview of the results}

\paragraph{J07446+035, J05019+011, J22468+443, and J15218+209}

For these stars, the scatter of the LAV RV time series significantly decreases when restricting the lines used towards those whose RV shows no correlation with the activity indicators ($|\mathrm{R}|\sim0$), and by removing lines with large RV scatter (limited wstd~RV), that is, when using only inactive lines to compute the RV. The activity-related signals in the RV periodogram also lose significance when using only inactive lines. This indicates that the modulation due to activity present in the RVs is mitigated. These four stars are those with the largest time series RV scatter, from $\sim37$ to $\sim100\,\ms$, and those whose line RVs show the strongest correlation with the activity indicators. When using sets of lines in which the conditions are more restrictive than those mentioned above (i.e. sets with less lines), the RV scatter increases, probably because the photon noise starts to dominate over the 
activity-driven variability.

There are specific sets of lines for which the RV scatter is minimised. These sets change depending on the star and the activity indicator used, but in general, the minimum scatter occurs when using lines with $|\mathrm{R}|\lesssim0.1$ or $\lesssim0.3$ and $\mathrm{wstd~RV\lesssim150}$ or $\lesssim300\,\ms$. For J07446+035, the scatter can be decreased $\sim5$\,times with respect to the initial one. For the other three stars, the maximum decrease is between $\sim2$ and 3\,times. The number of lines in the `best' sets of lines is $\sim100$ to 200 for the mid-type stars J07446+035, J05019+011, and J22468+443, and $\sim400$ for the early-type J15218+209.

For J07446+035, J22468+443, and J15218+209, RVs computed using line sets based on the correlation with the total RV are those that result in the lowest scatter, compared to RVs obtained from line sets based on the correlation with the other two indicators (BIS and CRX). Between BIS and CRX, for J15218+209, the CRX line sets result in smaller RV scatters than those of BIS. For J07446+035 and J22468+443, both BIS and CRX datsets result in similar minimum scatters. For J05019+011, the datsets of the three activity indicators reach similar minimum values.

Doing the same but selecting lines that show a strong correlation with an activity indicator ($\mathrm{R}\sim1$ or $-1$, depending on the activity indicator used), the time series RV scatter increases significantly. This could be due to the fact that we are enhancing the activity signal, but also to the increase in photon noise caused by the decrease in the number of lines (RV content) used. For most of the line sets tested, the activity-related signals in the RV periodograms show similar power as in the periodogram computed from the original RVs. However, for the most restrictive sets of lines ($\mathrm{R}\geq0.6\,\mathrm{to}\,0.8$ or $-0.6$ to $-0.8$, depending on the indicator), the signal loses significance, which could reflect the fact that the photon noise has increased due to the low number of lines used.

We based our choice of R and wstd RV by minimising the LAV RV. We  note that this might not be the optimal criterion to select the cuts in these parameters, because by minimising LAV RV we might eliminate or underestimate the amplitude of other astrophysical signals such as planets. A way to overcome this caveat could be to adopt general thresholds when selecting inactive lines based on the typical R and wstd RV cuts obtained for similar stars without planets.

\paragraph{J10196+198}

This star shows similar activity level as J22468+443, however, its initial RV scatter is significantly smaller ($\sim18\,\ms$ compared to $\sim40\,\ms$ for J22468+443), which may in part be due to the low estimated inclination of the star or to an homogeneously spotted photosphere. The line RVs are less correlated with the activity indicators than in the four previous targets (the correlations between the individual line RVs and the activity indicators show smaller strengths, i.e. R values less close to 1 or $-1$). The different datsets tested did not result in a significant decrease in RV scatter, probably due to the fact that the correlations between the line RVs and the indicators are not sufficiently strong. The periodogram shows a decrease of power at \Prot, probably due to an increase of photon noise in the recomputed RVs. Regarding the sets of active lines, power at \Prot decreases significantly, which agrees with the fact that the correlation between the line RVs and the indicators is not strong. We then also attribute this decrease in power to increasing photon noise.

\paragraph{J11201--104}

This object shows a similar initial scatter as J10196+108 ($\sim 19\,\ms$) and, as in the case of J10196+198, the correlations between the line RVs and the activity indicators are not strong. Despite that, there is a decrease in the RV scatter for some sets of lines obtained with similar selection criteria as the sets that minimised the scatter in the four stars J07446+035, J05019+011, J22468+443, and J15218+209. The maximum scatter decrease in this case is $\sim1.7$\,times the initial one, with a set of about 500 lines. The RV periodogram shows no significant signal.

%%%%%%%%%%%%%%%%%%%%%%%%%%%%%%%%%%%%%%%%%%%%%%%%%%
%%%%%%%%%%%%%%%%%%%%%%%%%%%%%%%%%%%%%%%%%%%%%%%%%%

%---------------------------------------------------------------------

\section{Lines in different stars} \label{sec:indlincomparisonstar}

Next, we investigate if the sensitivities to activity of the different lines (i.e. their R values) are similar in different stars. We performed pairwise comparisons of the stars in two groups: the mid-type stars J07446+035, J05019+011, and J22468+443, which used the same initial line list created from a J07446+035 template, and the early-types J15218+209 and J11201--104, which used the initial line list created from a J15218+209 template. We exclude J10196+198 from this analysis because we were not able to find a set of lines that mitigated the activity signal present in the RVs. For clarity and completeness, we include the corresponding RV time series and periodogram figures in Appendix~\ref{sec:app_indlincomparisonstar}.

%---------------------------------------------------------------------

\subsection{J07446+035 and J05019+011}

%---------------------------------------------------------------------
\begin{figure*}
\centering
\includegraphics[width=\linewidth]{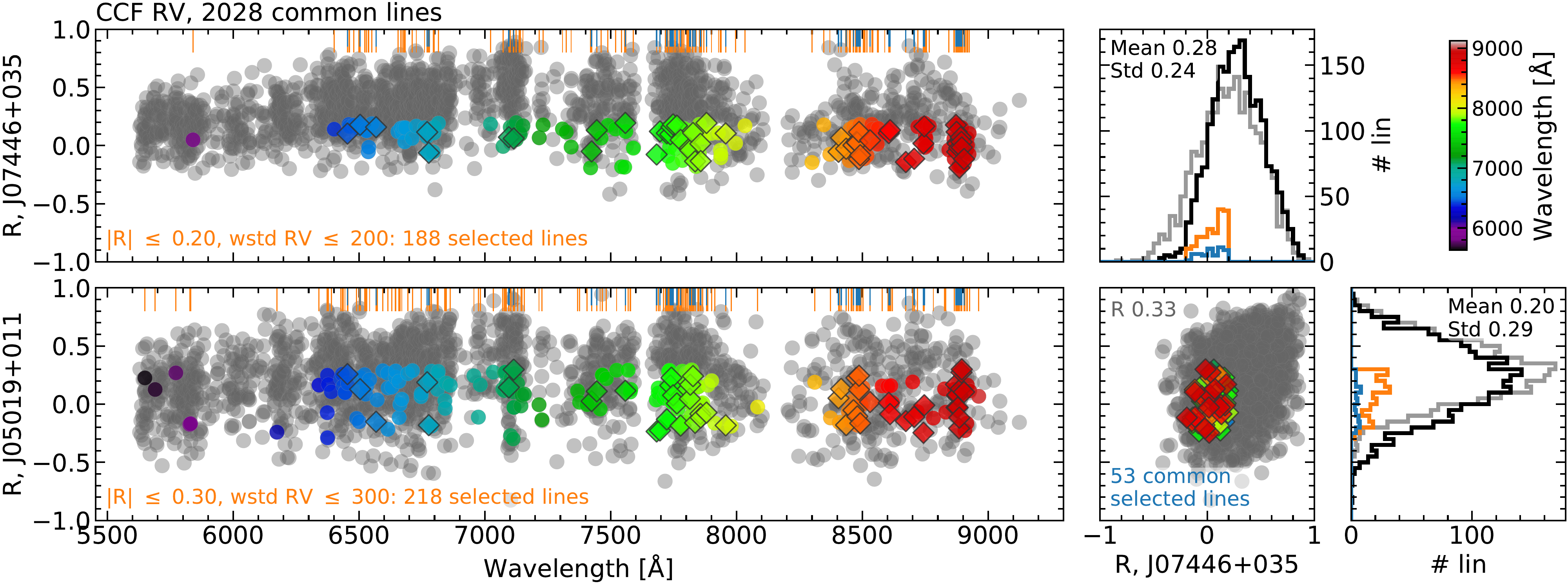}
\caption{Pearson's correlation coefficient R of the lines analysed as a function of their wavelength for J07446+035 (\emph{top left}) and J05019+011 (\emph{bottom left}). Grey dots show lines not selected, coloured dots show lines selected by the specific cut indicated in the orange text, and coloured diamonds indicate selected lines common to both targets. Data points are colour-coded with wavelength. Vertical orange lines at the top of the panels indicate the position of the selected lines, and blue lines indicate those common in both stars.
The \emph{bottom middle panel} shows the correlation between J07446+035 and J05019+011 R values. Grey dots show all lines not selected in any of the targets plus lines selected but not common (i.e. grey and coloured dots in the \emph{left panels}), and coloured diamonds indicate selected lines common in both stars (same as in the \emph{left panels}). Histograms at the \emph{right} show the distribution of R values of each star (J07446+035 \emph{top}, J05019+011 \emph{bottom}). The black histogram shows the distribution of all lines, the orange one, lines selected by the cut indicated in the \emph{left panels}, and the blue one, selected lines common in both stars. The grey histogram shows the distribution of all lines for the other star, for comparison.}
\label{fig:linselcomparisonJ07446+035_J05019+011}
\end{figure*}
%---------------------------------------------------------------------

We compared the distribution of R values of J07446+035 and J05019+011 as a function of the line wavelength in Fig. \ref{fig:linselcomparisonJ07446+035_J05019+011}. The R values shown are those obtained from the correlation between the individual line RVs and the total RV. We show the correlation with this indicator as an example, but we obtained similar results for the correlations with CRX and BIS. Of the 2207 initial lines of the J07446+035 line list, we show here 2028 lines, which are those for which we were able to measure a reliable RV for both stars (i.e. after removing those with low S/N and non-common lines due to different overlap with tellurics or order ends). The distribution of R values of J07446+035 is slightly narrower and shifted towards 1 with respect to that of J05019+011. Many of the lines show different R values in the two stars, since there is only a weak correlation between the two sets ($\mathrm{R\sim0.3}$).

We also compare the datsets of lines for which we obtained the smallest RV scatter: lines with $|\mathrm{R}|\leq0.2$ and $\mathrm{wstd~RV\leq200}\,\ms$ for J07446+035, and $|\mathrm{R}|\leq0.3$ and $\mathrm{wstd~RV\leq300}\,\ms$ for J05019+011, which include 188 and 218 lines, respectively. Of these lines, 53 are common in both datsets. This represents $24-28\,\%$ of the selected lines, and 2.6\,\% of the initial 2028 lines.

Next, we recomputed the RV of each star using the selected lines of the other star, as well as using only the common selected lines. Fig. \ref{fig:tslinselcomparisonJ07446+035_cutJ05019+011} shows the RV time series, lines used, and periodograms of the RVs of J07446+035, computed using these datsets. The RVs recomputed using the best datset of J05019+011 (blue data points in the figure) have a scatter smaller than the initial one, $\sim54\,\ms$ compared to $\sim86\,\ms$, but the periodogram shows a peak at \Prot almost as significant as in the initial datset. The RVs recomputed using the 53 common lines (green data points) show a scatter similar to those obtained using the best data of J07446+035 (orange data points) but slightly larger, of $\sim23\,\ms$ compared to $\sim20\,\ms$, and a periodogram with a peak at \Prot with a low significance.

Figure \ref{fig:tslinselcomparisonJ05019+011_cutJ07446+035} shows the same as Fig. \ref{fig:tslinselcomparisonJ07446+035_cutJ05019+011} but for J05019+011. For the best datset of J07446+035 (orange), the scatter decreases from the initial $\sim106$ to $\sim60\,\ms$, and the peak close to \Prot decreases in power significantly. The RVs obtained from the common lines have a scatter very similar to the one obtained with the best lines of J05019+011 (blue and green), and both periodograms show no significant peaks.

%---------------------------------------------------------------------

\subsection{J07446+035 and J22468+443}

%---------------------------------------------------------------------
\begin{figure*}
\centering
\includegraphics[width=\linewidth]{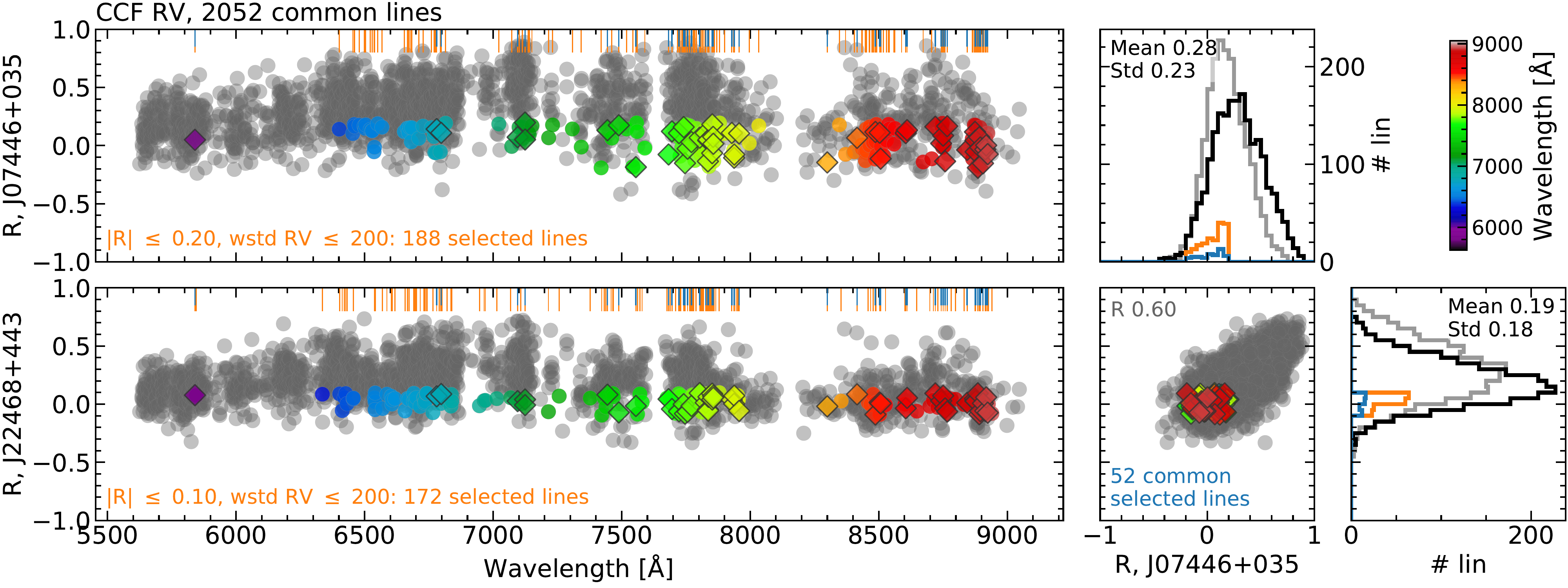}
\caption{Same as Fig. \ref{fig:linselcomparisonJ07446+035_J05019+011}, but for J07446+035 and J22468+443.}
\label{fig:linselcomparisonJ07446+035_J22468+443}
\end{figure*}
%---------------------------------------------------------------------

Figure \ref{fig:linselcomparisonJ07446+035_J22468+443} compares the R values of J07446+035 and J22468+443. In this case, the R distribution of J07446+035, which is the more active of the two stars, is wider and reaches values closer to 1 than that of J22468+443. The R values of the same lines for the two stars are more similar than for the previous two stars (J07446+035 and J05019+011), since now the correlation between the two R datsets is stronger ($\mathrm{R}=0.6$). The cuts yielding the lowest RV scatter are $|\mathrm{R}|\leq0.2$ and $\mathrm{wstd~RV\leq200}\,\ms$ for J07446+035 and $|\mathrm{R}|\leq0.1$ and $\mathrm{wstd~RV\leq200}\,\ms$ for J22468+443. This represents 188 lines for J07446+035 and 172 for J22468+443. Of these selected lines, there are 52 common to both stars ($28-30$\,\% of the selected lines, and 2.5\,\% of the initial 2052 lines).

Figures \ref{fig:tslinselcomparisonJ07446+035_cutJ22468+443} and \ref{fig:tslinselcomparisonJ22468+443_cutJ07446+035} show the RV time series and periodograms recomputed using the best set of lines of the other star, and using the common selected lines, for J07446+035 and J22468+443, respectively. In both cases, using the line list that minimises the RV scatter of the other star results in a significantly smaller RV scatter than initially, about $1.8-2.1$ times smaller. For J07446+035 (blue), this decrease in half is far from the minimum scatter obtained with its own best datset (orange), which is about 4.4 times smaller than the initial one, and the periodogram continues to show a very significant peak at \Prot. But for J22468+443 (orange) the decrease is close to the one obtained with its own best datset (blue), which is about 2.4 times smaller, and the periodogram peak at \Prothalf disappears. As with its own best datset, for J22468+443 there is now more power at \Prot than at \Prothalf (but the peak at \Prot does not become significant in this case). Regarding the common selected lines, in both cases the scatter is close to the minimum one obtained with the best datset of each star, and the periodogram does not show significant peaks related to activity.

\subsection{J22468+443 and J05019+011}

%---------------------------------------------------------------------
\begin{figure*}
\centering
\includegraphics[width=\linewidth]{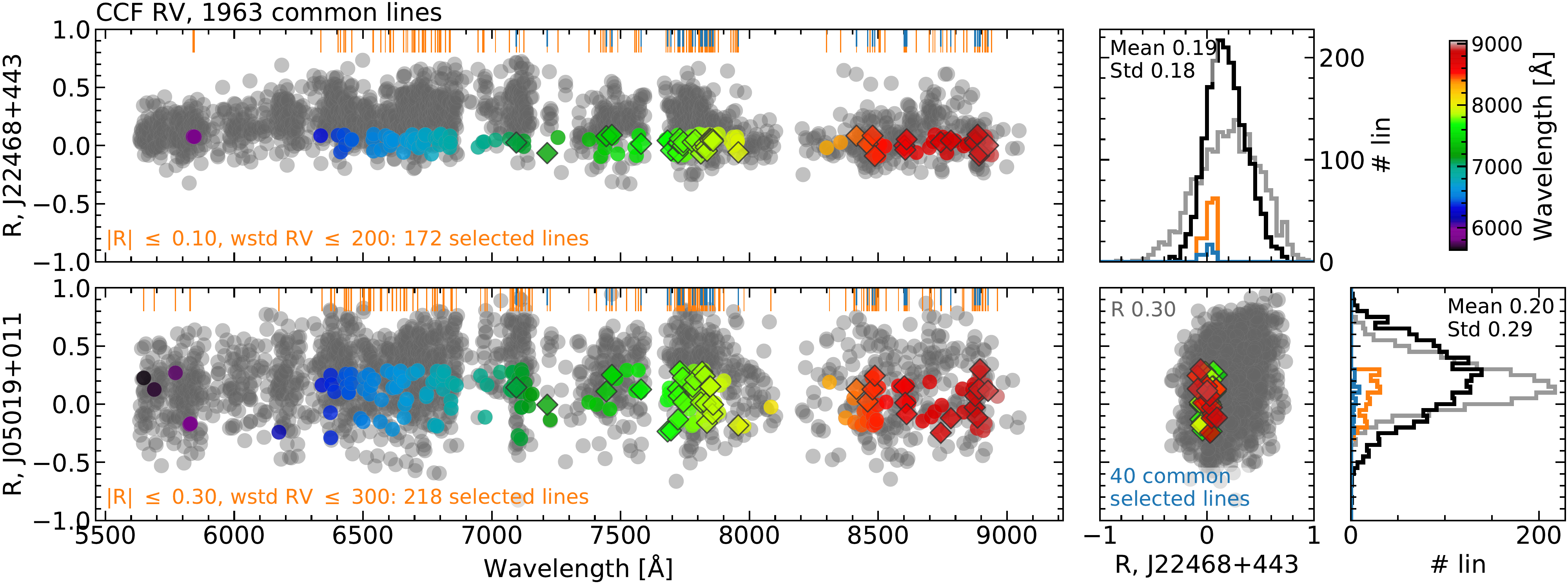}
\caption{Same as Fig. \ref{fig:linselcomparisonJ07446+035_J05019+011}, but for J22468+443 and J05019+011.}
\label{fig:linselcomparisonJ22468+443_J05019+011}
\end{figure*}
%---------------------------------------------------------------------

For J22468+443 and J05019+011, the correlation between R values is low ($\mathrm{R}=0.3$, Fig. \ref{fig:linselcomparisonJ22468+443_J05019+011}). There are 40 common selected lines, $18-23$\% of the 172 and 218 lines that minimise the RV scatter of the stars, which is 2.0\,\% of the initial 1963 lines.

The recomputed RV time series and periodograms are shown in Figs. \ref{fig:tslinselcomparisonJ22468+443_cutJ05019+011} and \ref{fig:tslinselcomparisonJ05019+011_cutJ22468+443}. For J22468+443, using the best datset of J05019+011 (blue) results in a decrease in the RV scatter compared with the initial one (1.4 times smaller), but not as small as the minimum obtained with its own best datset (2.4 times smaller). The periodogram continues to show a significant peak at \Prothalf, although with less power than initially, and in this case, the power at \Prot does not increase. For J05019+011, using the best J22468+443 datset (orange) does not decrease the RV scatter significantly (1.2 times smaller compared to 2.5 times smaller for its own best datset), but the periodogram does no longer show a peak at \Prot. Using the common selected lines, for both stars the RV scatter decreases, but not as much as using their own best datset, and the periodogram does not show any significant peaks.

%---------------------------------------------------------------------

\subsection{J15218+209 and J11201--104}

%---------------------------------------------------------------------
\begin{figure*}
\centering
\includegraphics[width=\linewidth]{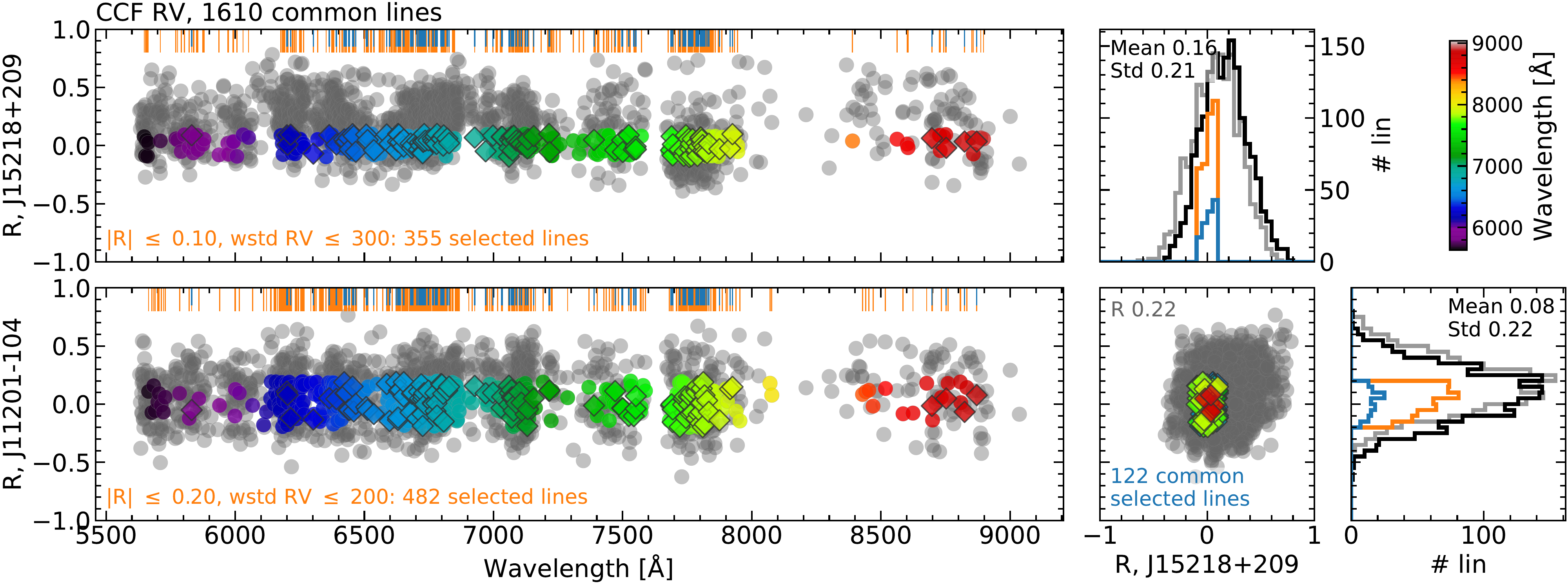}
\caption{Same as Fig. \ref{fig:linselcomparisonJ07446+035_J05019+011}, but for J15218+209 and J11201--104.}
\label{fig:linselcomparisonJ15218+209_J11201--104}
\end{figure*}
%---------------------------------------------------------------------

For these early-type stars, the correlation between R values is even lower than for the mid-type stars ($\mathrm{R}\sim0.2$, Fig. \ref{fig:linselcomparisonJ15218+209_J11201--104}). The selection cuts that result in the smallest RV scatter are $|\mathrm{R}|\leq0.1$ and $\mathrm{wstd~RV\leq300}\,\ms$ for J15218+209 and $|\mathrm{R}|\leq0.2$ and $\mathrm{wstd~RV\leq200}\,\ms$ for J11201--104, which correspond to 355 lines for J15218+209 and 482 lines for J11201--104. Of these, 122 lines are common in both datsets ($25-34$\,\% of the selected lines, and 7.6\,\% of the initial 1610 lines).

In both cases, using the best datset of the other star to recompute the RVs results in a scatter very similar to the initial one (Fig. \ref{fig:tslinselcomparisonJ15218+209_cutJ11201--104}, blue, for J15218+209, and Fig. \ref{fig:tslinselcomparisonJ11201--104_cutJ15218+209}, orange, for J11201--104). In the case of J15218+209, the periodogram also looks similar to the original one, with a significant peak at \Prot. For J15218+209, the RVs computed with the common selected lines (green) have a scatter smaller than the initial one (1.3 times smaller), but larger than the one obtained using the best line set of the star itself (which was 2.7 times smaller, orange datset). In the case of J11201--104, the scatter of the common selected lines RVs is larger than the original one.

%---------------------------------------------------------------------

\subsection{Overview of the results}

From this analysis we conclude that the same lines in different stars show different correlation strengths (i.e. different R values per line). The R values of the pairs of stars J15218+209 and J11201--104, J07446+035 and J05019+011, and J22468+443 and J05019+011, show weak correlations (R from 0.2 to 0.3), while for the pair J07446+035 and J22468+443, the R values are more similar ($R=0.6$).

If we focus on the line sets for which the RV activity signal is most strongly mitigated, in general the number of lines common in the star pairs is low. For the pairs of mid-type stars (J07446+035 and J05019+011, J07446+035 and J22468+443, and J22468+443 and J05019+011), only 2.0\,\% to 2.6\,\% of the initial lines are common in the best sets of the respective stars. For the early-type pair (J15218+209 and J11201--104), this number is  larger, of almost 8\,\% (but their best line sets already have a few hundreds of lines more than those of the mid-type stars). These `common' sets have $\sim50$ lines in the case of the mid-type pairs, and 122 lines for the early-type one.

For the stars in each pair, we used sets of inactive lines based on the correlation of the other star in the pair to compute RVs. This results in small changes in the RV time series compared to using all the lines, due to the lack of correspondence between the correlation strength of the same lines in different stars. In general, the scatter decreases, but not as much as using a set of inactive lines based on the correlations of the star itself, and the periodogram peaks related to \Prot remain significant.

We also recomputed RVs using inactive lines common to the sets that minimised the RV scatter in the two stars of the pair. As mentioned above these sets of common inactive lines have from $\sim50$ to 122 lines, depending on the pair of stars, which represents from 18 to 34\,\% of the lines in the best sets, a significant decrease in the number of lines. In general, using these sets results in RVs and periodograms similar to those obtained using the best set of lines from the star itself, but with slightly larger RV scatters. 

The stars in the two groups used here all show high activity levels, but do not have the exact same properties. The spectral types, rotational velocities, and metallicities are similar but not identical. Based on the results for these stars, we are not able to obtain a general set of lines, even when considering spectral type intervals, for which the effect of activity in the RVs is minimised.

%---------------------------------------------------------------------

\section{Discussion and final remarks} \label{sec:indlinsummary}

We have studied activity effects on individual spectral lines in a set of six active early- and mid-type M dwarfs observed with CARMENES-VIS as part of the CARMENES GTO sample. Here we summarise and put our findings into context. 

We used the \raccoon pipeline to select lines in the stellar spectra and compute line-by-line RVs by comparing the centroid of the line with a reference. By averaging these line-by-line RVs, we computed a global RV per observation, obtaining values comparable to RVs from those resulting from the CCF and template-matching techniques. However, this similarity is only true because the stars selected show large RV dispersions due to stellar activity, which are significantly larger than the RV uncertainties. In fact, the RV uncertainties of the LAV RVs are about one order of magnitude larger than those obtained with the CCF or template-matching methods.

We analysed  the correlation strength between these line-by-line RVs and several spectroscopic activity indicators, following a method analogous to that described in \citet{dumusque2018indivline}. Amongst the different indicators analysed (global RV of the spectrum, CCF FWHM, contrast and BIS, CRX, dLW, and \IHalpha), we find that only the global RV, BIS, and CRX show significant linear correlations with the line-by-line RVs \citep{lafarga2021indicators,jeffers2022EVLac,cardona2022young}. 

Using the strength of these correlations, which we measured with Pearson's correlation coefficient R, we classified the lines according to their sensitivity to activity. We find that the R coefficient is a biased indicator of the correlation with activity when measurements have low precision. This is true for lines located in low S/N regions of the spectrum, especially in the bluer region, which leads to large RV uncertainties.

We then used different sets of lines to compute a new global RV of each observation. With lines having a strong correlation with the activity indicators (i.e. activity-sensitive lines), we obtained activity-dominated RVs, while with sets of lines showing weak correlations (i.e. activity-insensitive lines), we decreased the effect of activity in the RVs. We note here that, despite referring to them as `activity-insensitive' lines, the recomputed RVs still have a significant scatter, larger than the RV uncertainties, and in some cases the RV periodograms still show low-significance signals related to \Prot. Therefore, these lines still contain some contribution from activity, that is, they are the least activity-sensitive lines. We only use activity-insensitive lines for clarity throughout the text. The decrease in activity has been evaluated by analysing the total RV scatter and the presence of significant activity-related signals in the RV periodogram. We have been able to effectively find lines with different correlation strengths and mitigate activity on the RV for five stars of our sample of six: J07446+035, J05019+011, J22468+443, J15218+209, and J11201--104. The maximum decrease in RV scatter obtained using sets of inactive lines is $\sim2$ to 5 times the initial one. These sets of inactive lines have of the order of 100 lines, while initially we started with about 2000 lines.

For J10196+198 (AD~Leo), which shows a lower initial RV scatter and weaker correlations, the method used here is not able to distinguish between active and inactive lines, and hence we did not see an improvement in the RV time series by using subsets of lines \citep[see also][]{kossakowski2022adleo}. This could be because the precision of the individual line RVs is not sufficiently high to deliver global RV time series with smaller scatters. It is also possible that the correlations with the activity indicators are not reliable, either because of the low precision of the RVs or the indicators, or because of the way we quantified the correlations, which could have hampered the classification of lines according to their sensitivity to activity.

We also studied if the same lines show similar correlation strengths between the line-by-line RV and the activity indicators in different stars. By doing this, we aimed to see if a set of inactive lines could be generalised for stars with similar characteristics. We performed pairwise comparisons of the correlation strength of the mid-type M dwarfs, J07446+035, J05019+011, and J22468+443, and, separately, the early-type M dwarfs, J15218+209 and J11201--104. We find that the same insensitive lines in different stars, in general, do not show similar correlation strengths, that is, there is a different activity dependence for the same lines in different stars even if the stars are similar in terms of spectral type, activity level, and rotation. Such a lack of consistency could be due to the fact that most absorption features in M dwarf spectra are line blends rather than well-isolated, single lines. Lines in different stars could be blended to varying degrees, and the combination of the activity-affected profiles of the blended lines could result in different RV variations depending on the specific target star. Further analysis is needed to understand these results.

Using the best set of inactive lines obtained for a specific star to recompute RVs of another target star, even if they have similar properties, is less effective (i.e. results in a less significant reduction of the RV activity signal) than using a line set obtained from the target star itself. Using only the lines common in the best sets of two similar stars achieves better results. However, since the number of common lines is very small, the RV scatter is still large, probably due to the fact that photon noise starts to dominate (i.e. a larger number of lines would be needed to improve the RV precision). Therefore, to achieve the maximum mitigation of activity, the best strategy appears to be selecting lines based on the correlations of each individual star.

The same observations of the six stars analysed here have also been studied by \citet{cardona2022young}. Similarly to \citet{tal-or2018carmenesRVloud}, \citet{cardona2022young} studied the correlation between the total RV and several indicators of stellar activity, but increased the sample of stars used by including most young stars in the CARMENES-GTO sample and analysed a larger number of activity proxies. Moreover, \citet{cardona2022young} went one step further and used the correlation between RV and activity indicators to correct for activity effects in the RVs. \citet{jeffers2022EVLac} also performed a similar de-correlation with the CRX and the centre of light for several sets of J22468+443 (EV~Lac) observations. Table \ref{tab:results} shows a comparison of the results obtained with the line-by-line approach presented here and the de-correlation performed by \citet{cardona2022young}. The decrease in RV scatter obtained by both methods is similar. Similarly, for J22468+443 (EV~Lac), \citet{jeffers2022EVLac} obtained a reduction factor of $\sim$3 to 4, depending on the set of observations used. This was expected because both approaches rely on the use of linear correlations between RV and activity indicators. As opposed to the de-correlation approach, the line-by-line approach can make use of the total RV as a proxy of activity, and hence it does not depend on extra indicators of activity. However, as mentioned above, the total RV modulation can be affected by unknown companions and, for different types of star, other indicators might be better tracers of activity, so using the total RV is not necessarily the best option. On the other hand, the de-correlation approach is simpler and is not limited to bright, relatively slowly rotating M dwarfs showing significant RV scatter.

\begin{table}
\centering
\caption{Initial \serval RV scatter and reduction factor obtained using the method presented here and the de-correlation method by \citet{cardona2022young} of the different stars analysed here.}
\label{tab:results}
\tabcolsep=0.08cm
{\small % --- Smaller text
\begin{tabular}{lcccc}
\hline\hline
Karmn& \multicolumn{2}{c}{This work}             & \multicolumn{2}{c}{\citet{cardona2022young}} \\
\cmidrule(lr){2-3}\cmidrule(lr){4-5}
          & Initial           & Reduction& Initial& Reduction \\
          & std RV [\ms]&factor       &std RV [\ms]         &factor \\
\midrule
J07446+035 &88.2 &4.5 &88.3 &3.6\\
J05019+011  &90.3 &2.1 &82.6 &3.6\\
J22468+443 &50.2 &2.9 &50.6 &2.9\\
J10196+198  &18.4 &1.2 &17.3 &1.9\\
J15218+209 &36.7 &2.7 &36.6 &1.6\\
J11201--104 &18.3 &1.7 &17.9 &1.3\\
\hline
\end{tabular}
}
\tablefoot{The differences in the initial RV scatter between the two methods are due to a slightly different number of observations used \citep[see Sect. \ref{indlinmethodrv} and methods in][]{cardona2022young}.}
\end{table}

Our results from combining lines for different stars are in contrast with those obtained by \citet{bellotti2022linelists} with M dwarf spectra from ESPaDonS and NARVAL. \citet{bellotti2022linelists} were able to achieve a reduction in the total RV scatter of EV~Lac (J22468+443), AD~Leo (J10196+198), and DS~Leo by applying a line list of activity-insensitive lines based on spectra of EV~Lac. These results, however, cannot be immediately compared with ours. \citet{bellotti2022linelists} used a line list based on the VALD database and did not always remove telluric features initially (their algorithm is able to do that afterwards), while our lines have been empirically found in the observed spectra and we have removed any lines overlapping with tellurics. Their initial RV scatters (i.e. using all the lines) are higher than those we obtained with CARMENES-VIS. For EV~Lac, the initial RV scatter of the dataset in \citet{bellotti2022linelists} is of 182\,\ms, and for AD~Leo, 110\,\ms, while with CARMENES-VIS we have 40\,\ms and 17\,\ms, respectively. The final RV scatters (i.e. using a line list that minimises the scatter) obtained by \citet{bellotti2022linelists} are still larger than our starting values. 

One of the reasons for these differences could be the different wavelength ranges of the instruments. The wavelength range of both ESPaDonS and NARVAL include bluer wavelengths than CARMENES-VIS (they start at $\sim3700\,\AA$, while CARMENES-VIS starts at  $5200\,\AA$), and hence, the masks the authors used contain a significantly higher number of lines in the blue than in the red wavelength range. The use of these bluer lines could lead to differences between the RVs measured with different instruments. However, to test variability with wavelength, the authors computed RVs using lines (in this case with tellurics having been removed previously) bluewards and redwards of $5500\,\AA$  separately (i.e. the red set being comparable to the CARMENES-VIS wavelength range), and they found no decrease in the RV scatter with either set of lines. Therefore, it is not clear whether the bluer range of the instruments has a significant impact on the differences mentioned above. Another factor that could be playing a role if the fact that  ESPaDonS and NARVAL have slightly less resolution ($R\sim70\,000$) than CARMENES-VIS ($R=94\,600$), making the identification of single lines harder. Furthermore, the datasets in \citet{bellotti2022linelists} and in this work were observed at very different times (between 2005 and 2016 for EV~Lac and in 2008 for AD~Leo), hence, the stars could have intrinsically different activity levels, which would also contribute to the differences in RV.

In this work we show that for a small sample of active stars with relatively large RV scatters, it is possible to select specific spectral lines to compute RVs that are affected by stellar activity to varying degrees. Our method can be expanded in several ways.
\begin{itemize}
\item The line-by-line RVs were computed by fitting a Gaussian model to the lines, finding their centroids, and comparing them to a list of reference wavelengths. In most cases, our lines do not have clear Gaussian profiles, so we expect to obtain more precise individual RVs by fitting other kinds of functions \citep[e.g. a parabola in the line core such as in][]{reiners2016sun,liebing2021convectiveblueshift} or a model of the spectrum to the region around each individual line \citep[e.g.][]{dumusque2018indivline,artigau2022lbl}. Instead of working on a line-by-line basis, for M dwarfs we could also consider groups of lines, such as those from molecular bands, and measure RVs by template matching. These modifications of the method could improve the precision of line-by-line RVs, and the reliability of the correlations between the line RVs and the activity indicators, allowing for a better classification of the sensitivity of the lines to activity.
\item To quantify the correlations, we used the Pearson’s correlation coefficient R, which measures the strength of linear dependencies (proportional changes between variables). We observed that the R coefficient could be biased due to the low precision of the RV measurements, as it did not provide reliable values in lines with large RV scatter (mainly lines in low-S/N regions). Moreover, we performed cuts on R arbitrarily, without accounting for uncertainties on the exact R value to give an example. For instance, when selecting active lines, a cut on R>$-$0.80 would select the example line at 6661.25\,\AA\,in Fig. \ref{fig:indivlinrvtscorrJ07446+035} for the CRX (R=$-$0.81), but not for the BIS (R=$-$0.77) correlation. A way to account for biases in R could be to use different R limits as a function of the local S/N of the spectrum. One could also use the slope (and its uncertainty) between the line RV and activity indicator as a quantitative indicator for the quality of the correlation. It is also possible to estimate uncertainties on R with Monte Carlo simulations as in \citet{cretignier2020indivline}, which could then be used to further select reliable R values by avoiding selection biases. These approaches could complement the selection based on per-line wstd RV and RV uncertainty presented here.
\item The relationship between the individual line RVs and the three final indicators used (global RV, CCF BIS, and CRX) appears to be linear; however, to further improve the accuracy of the correlations, we could also test if other methods are able to capture these dependencies better, such as the Spearman’s rank correlation coefficient. This correlation coefficient assesses the strength of a monotonic relationship between two variables and hence it is not limited to linear dependencies. The other activity indicators investigated (CCF FWHM, contrast, dLW, and chromospheric lines) clearly show non-linear (circular) relationships with the line-by-line RVs, which seem to arise from phase differences \citep[e.g.][]{bonfils2007gj674,santos2014HD41248,perger2017hadesGJ3942}. If further methods to quantify these dependencies are studied, these indicators could also be used to assess the sensitivity of the different lines to activity. Quantifying the correlation in another way with different indicators could be relevant especially for stars with low levels of stellar activity, because their linear correlations might not be as strong as those presented here, and indicators other than BIS and CRX might be better at tracing activity \citep[see e.g.][]{lafarga2021indicators,cardona2022young}.
\item The method we used to compute the global RVs was to simply average the RVs of the selected individual lines (i.e. compute LAV RVs). However, the selected lines could also be used to build cross-correlation masks \citep[e.g.][]{lafarga2020carmenesccf,rainer2020masks,bellotti2022linelists} or select specific regions in template-matching approaches, which could deliver even more precise RVs (i.e. decrease the uncertainty of the LAV RVs). By doing that, it could also be tested if stellar activity indicators such as those derived from the CCF or the CRX also vary and show a weaker dependence with the recomputed RV and the activity signals, to further probe the presence of activity. By having more precision in the LAV RVs, one could also focus on specific regions of the spectrum, that is, the red part where lines seem to be less active, as opposed to what is presented in this work, where we study the spectrum as a whole. With our approach to average line RVs, we are limited in precision to study the behaviour of RV or CRX computed over a small spectral range.
\item Ideally, the LAV approach would always average the same number of lines. However, depending on the S/N of each line in each observation, the number of averaged lines slightly differs from observation to observation. Rejecting some lines in some observations can lead to a bias in the averaged RV. \citet{artigau2022lbl} accounted for this issue by applying an iterative debiasing process, in which the offset introduced by rejecting specific lines was taken into account when computing the global average RV. We assessed the effect introduced by averaging different lines in our results by computing LAV RVs using only lines common in most observations, as opposed to using as many lines per observation as possible, which is what we did to achieve the results presented above. For all cases tested (i.e. using all lines and applying cuts in R and wstd), the RV scatter decreases by about less than 10\% when using only common lines, compared to using all possible lines. The periodogram structure remains the same, except for cases with a small number of lines (i.e. |R|$\leq$10), probably because in these cases noise dominates. Hence, a debiasing such as the one presented in \citet{artigau2022lbl} would probably slightly decrease the final LAV RVs obtained, and change the exact cut values that minimise the final LAV RV.
\item Our line selection followed an empirical approach, that is we selected  minima present in the spectrum based on their profile and then classified the lines according to their correlation with an activity indicator, without any knowledge of the origin of the lines. By cross-matching the lines selected in our datasets with line databases, one could study dependencies between the sensitivity to activity and physical parameters (such as the excitation potential or the species giving rise to the line), or changes in the line profile \citep[see e.g.][]{wise2018activelines,cretignier2020indivline,bellotti2022linelists}. Spectra observed at a higher resolution could help to characterise shape changes  better and identify line blends.
\item Longer time coverage and/or denser sampling of the observations could also help to characterise short- and long-term changes in activity, including better correlations between the RV and activity proxies.
\end{itemize}

In conclusion, in this work we have presented an analysis of the sensitivity to activity of different individual lines in M dwarf stars, which provides a methodology for exploiting the wealth of information contained in the stellar spectra. We have shown that it is possible to identify lines that correlate with activity to varying degrees in several active M dwarf stars, and that this information can be used to effectively mitigate or enhance the effect of activity on RV measurements. With the current and next generation of high-resolution spectrographs reaching increasingly better RV precisions, stellar activity becomes the ultimate obstacle in RV observations \citep[e.g.][]{crass2021eprv}. Studies about activity effects on spectroscopic observations such as the one presented here will therefore be key in the quest for small Earth-like exoplanets and planets around young stars.
%---------------------------------------------------------------------

\begin{acknowledgements}
We thank the anonymous referee for a constructive and timely report which has helped improve the contents of this article.
CARMENES is an instrument for the Centro Astron\'omico Hispano-Alem\'an de
Calar Alto (CAHA, Almer\'{\i}a, Spain).
CARMENES is funded by the German Max-Planck-Gesellschaft (MPG),
the Spanish Consejo Superior de Investigaciones Cient\'{\i}ficas (CSIC),
the European Union through FEDER/ERF FICTS-2011-02 funds,
and the members of the CARMENES Consortium
(Max-Planck-Institut f\"ur Astronomie,
Instituto de Astrof\'{\i}sica de Andaluc\'{\i}a,
Landessternwarte K\"onigstuhl,
Institut de Ci\`encies de l'Espai,
Institut f\"ur Astrophysik G\"ottingen,
Universidad Complutense de Madrid,
Th\"uringer Landessternwarte Tautenburg,
Instituto de Astrof\'{\i}sica de Canarias,
Hamburger Sternwarte,
Centro de Astrobiolog\'{\i}a and
Centro Astron\'omico Hispano-Alem\'an),
with additional contributions by the Spanish Ministry of Economy,
the German Science Foundation through the Major Research Instrumentation
Programme and DFG Research Unit FOR2544 ``Blue Planets around Red Stars'',
the Klaus Tschira Stiftung,
the states of Baden-W\"urttemberg and Niedersachsen,
and by the Junta de Andaluc\'{\i}a.
Based on data from the CARMENES data archive at CAB (INTA-CSIC).
We acknowledge financial support from the Agencia Estatal de Investigaci\'on 10.13039/501100011033 of the Ministerio de Ciencia e Innovaci\'on and the ERDF ``A way of making Europe'' through projects
PID2019-109522GB-C5[1:4],       % CAB+IAA+IAC+UCM
PGC2018-098153-B-C33,                                   % ICE
PID2021-125627OB-C31,
and the Centre of Excellence ``Severo Ochoa'' and ``Mar\'ia de Maeztu'' awards to the Instituto de Astrof\'isica de Canarias (CEX2019-000920-S), Instituto de Astrof\'isica de Andaluc\'ia (SEV-2017-0709), and Centro de Astrobiolog\'ia (MDM-2017-0737), and the Generalitat de Catalunya/CERCA programme.
ML acknowledges funding from a UKRI Future Leader Fellowship, grant number MR/S035214/1.
\end{acknowledgements}


% for the bibliography, at the end
\bibliographystyle{aa} % style aa.bst
\bibliography{lines} % your references Yourfile.bib
%\bibliography{/home/marina/Dropbox/Zotero/Bibs/2021lines.bib} % your references Yourfile.bib

%---------------------------------------------------------------------

\begin{appendix}

%---------------------------------------------------------------------

\section{Total RV: RV time series comparison}\label{sec:app_totalrv_all}

% ---------------------------------------
% J07446+035 RV TS all lines, CCF, SERVAL
\begin{figure*}
\centering
\includegraphics[width=\linewidth]{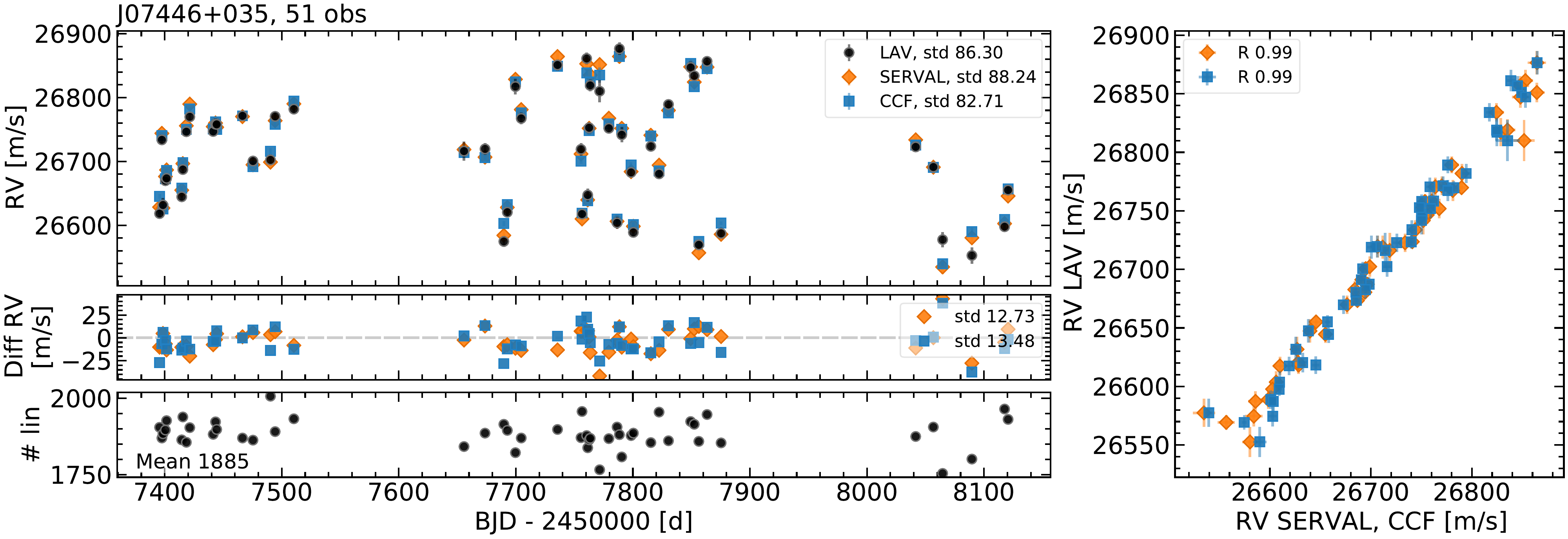}
\caption{Comparison of global RV time series. \emph{Top left}: RV time series of J07446+035 (YZ~CMi, GJ~285) computed with the average of all the individual line RVs (LAV, black dots), \serval (orange diamonds), and CCF (blue squares).
\emph{Middle left}: Difference between the LAV and the \serval (orange diamonds) and CCF (blue squares) RVs. 
\emph{Bottom left}: Number of lines used per observation.
\emph{Right}: Correlation between the LAV and the \serval (orange diamonds) and CCF (blue squares) RVs.} \label{fig:rvtsallcompareJ07446+035}
\end{figure*}

% J05019+011 RV TS all lines, CCF, SERVAL
\begin{figure*}
\centering
\includegraphics[width=\linewidth]{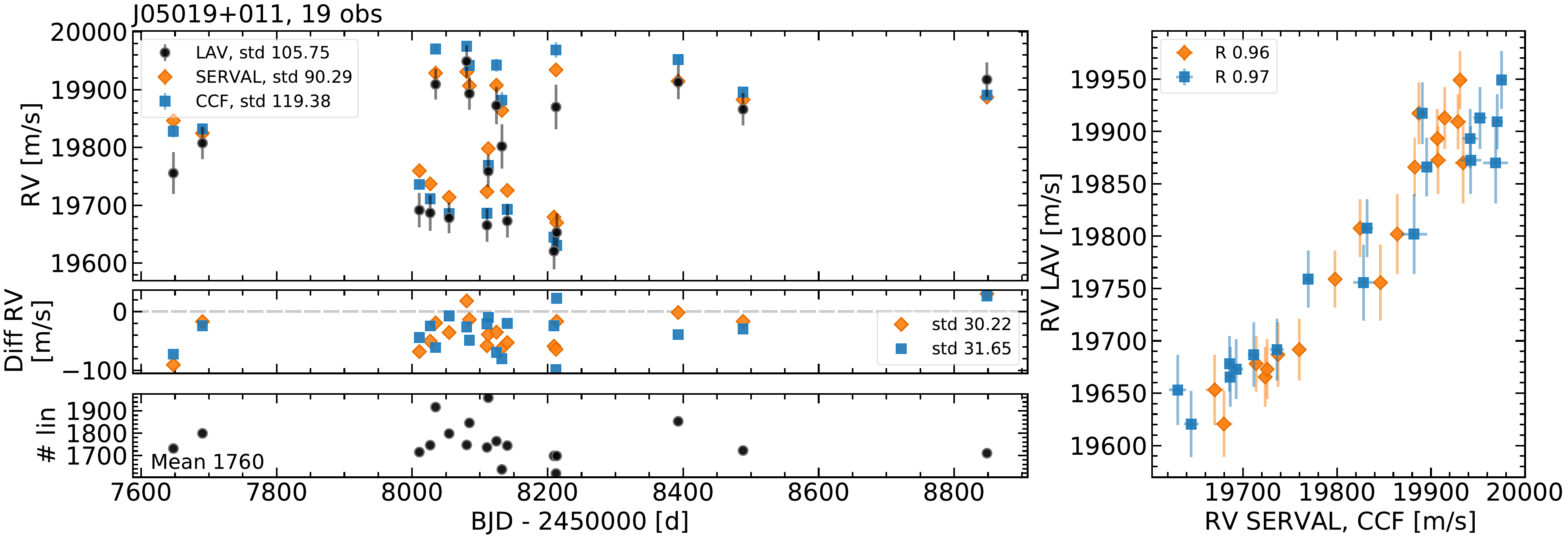}
\caption{Same as Fig. \ref{fig:rvtsallcompareJ07446+035}, but for J05019+011 (1RXS J050156.7+010845).} \label{fig:rvtsallcompareJ05019+011}
\end{figure*}

% J22468+443 RV TS all lines, CCF, SERVAL
\begin{figure*}
\centering
\includegraphics[width=\linewidth]{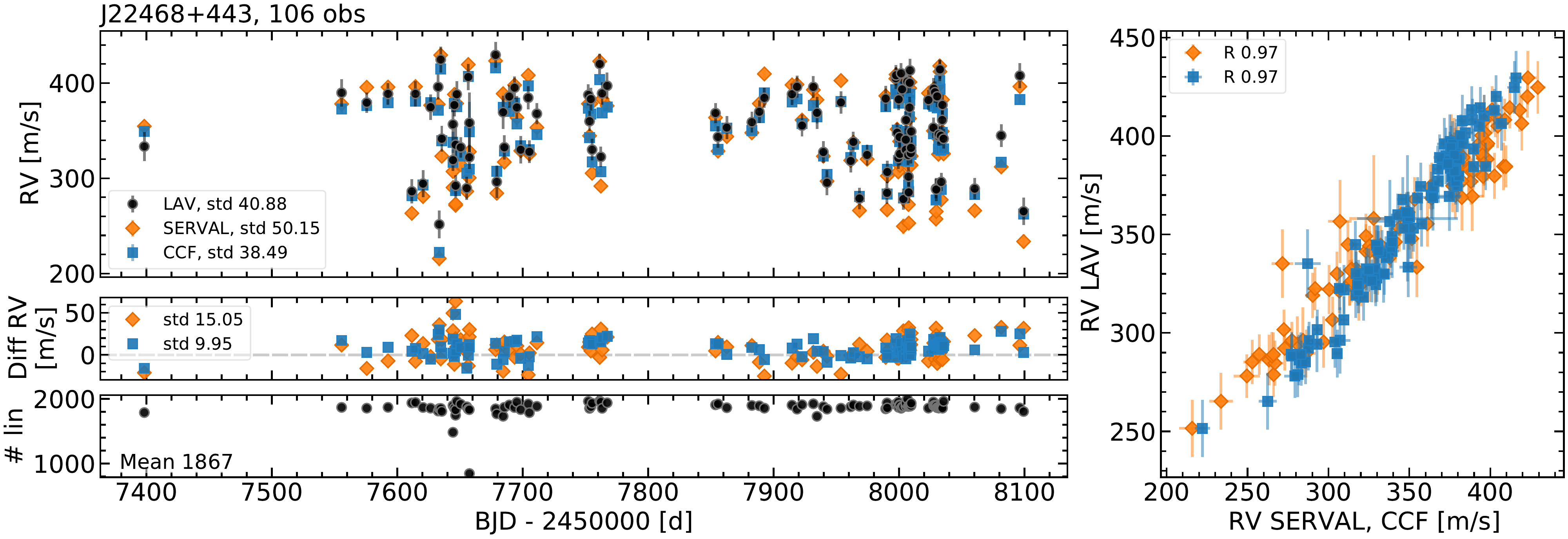}
\caption{Same as Fig. \ref{fig:rvtsallcompareJ07446+035}, but for J22468+443 (EV~Lac, GJ~873).} \label{fig:rvtsallcompareJ22468+443}
\end{figure*}

% J10196+198 RV TS all lines, CCF, SERVAL
\begin{figure*}
\centering
\includegraphics[width=\linewidth]{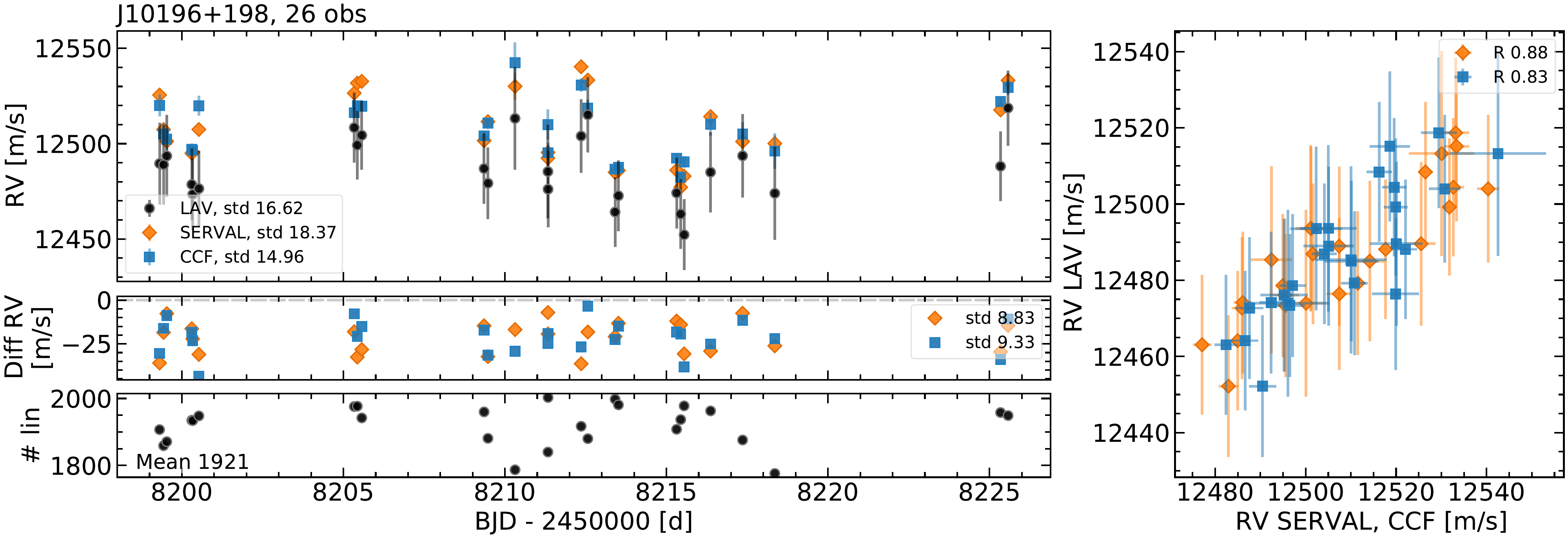}
\caption{Same as Fig. \ref{fig:rvtsallcompareJ07446+035}, but for J10196+198 (AD~Leo, GJ~388).} \label{fig:rvtsallcompareJ10196+198}
\end{figure*}

% J15218+209 RV TS all lines, CCF, SERVAL
\begin{figure*}
\centering
\includegraphics[width=\linewidth]{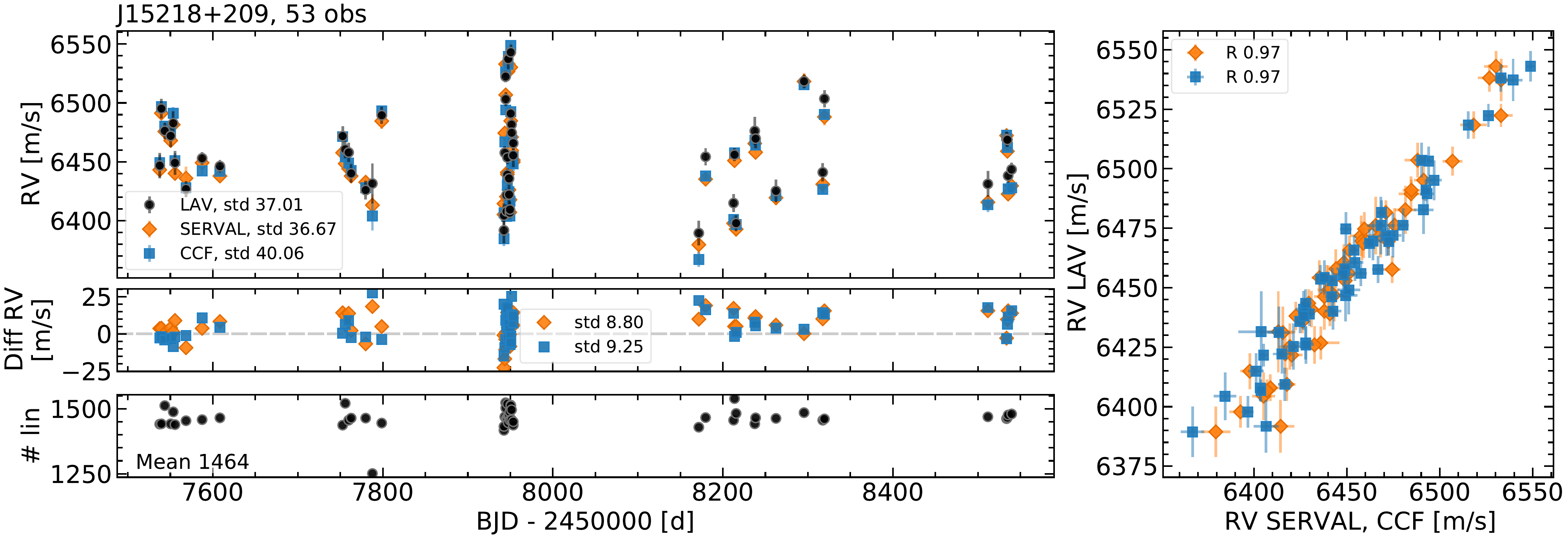}
\caption{Same as Fig. \ref{fig:rvtsallcompareJ07446+035}, but for J15218+209 (OT~Ser, GJ~9520).} \label{fig:rvtsallcompareJ15218+209}
\end{figure*}

% J11201--104 RV TS all lines, CCF, SERVAL
\begin{figure*}
\centering
\includegraphics[width=\linewidth]{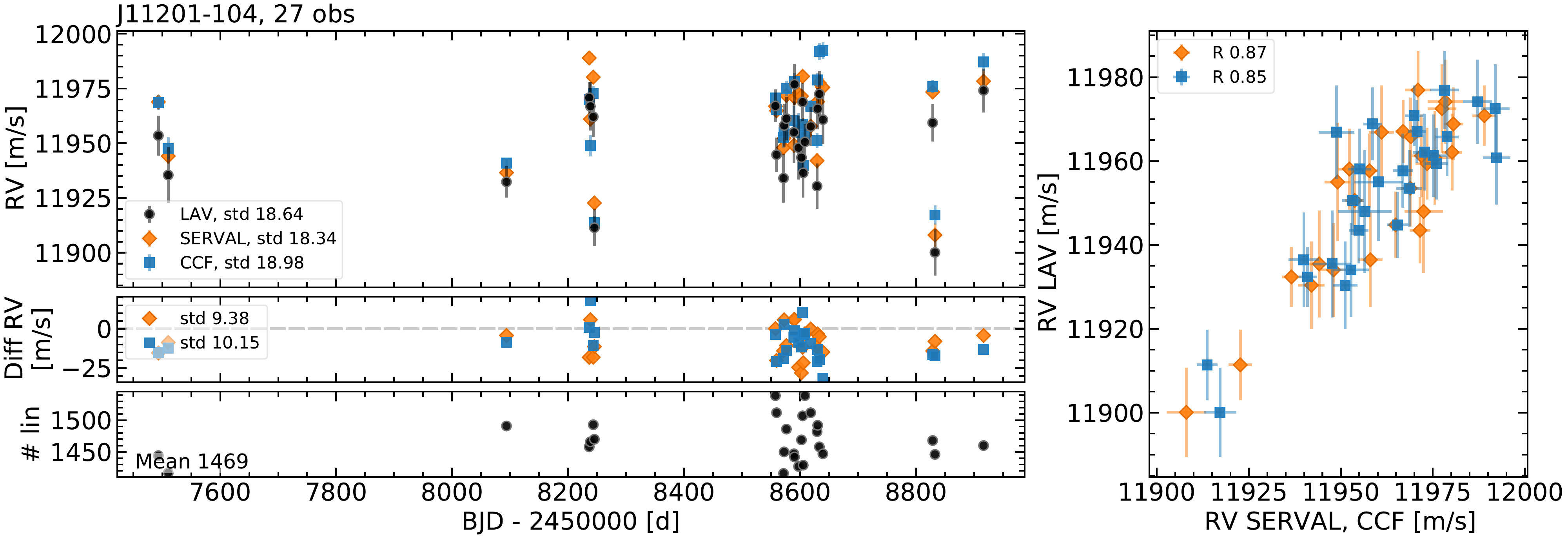}
\caption{Same as Fig. \ref{fig:rvtsallcompareJ07446+035}, but for J11201--104 (LP~733-099).} \label{fig:rvtsallcompareJ11201--104}
\end{figure*}
% ---------------------------------------

%---------------------------------------------------------------------

\section{Total RV computation with selected lines: RV scatters and periodograms}\label{sec:app_rvs_periodograms}

% ------------------------------------------------

% J07446+035 RV TS, Rcoeff periodogram, corr ccfbis
\begin{figure*}
\centering
\includegraphics[width=0.93\linewidth]{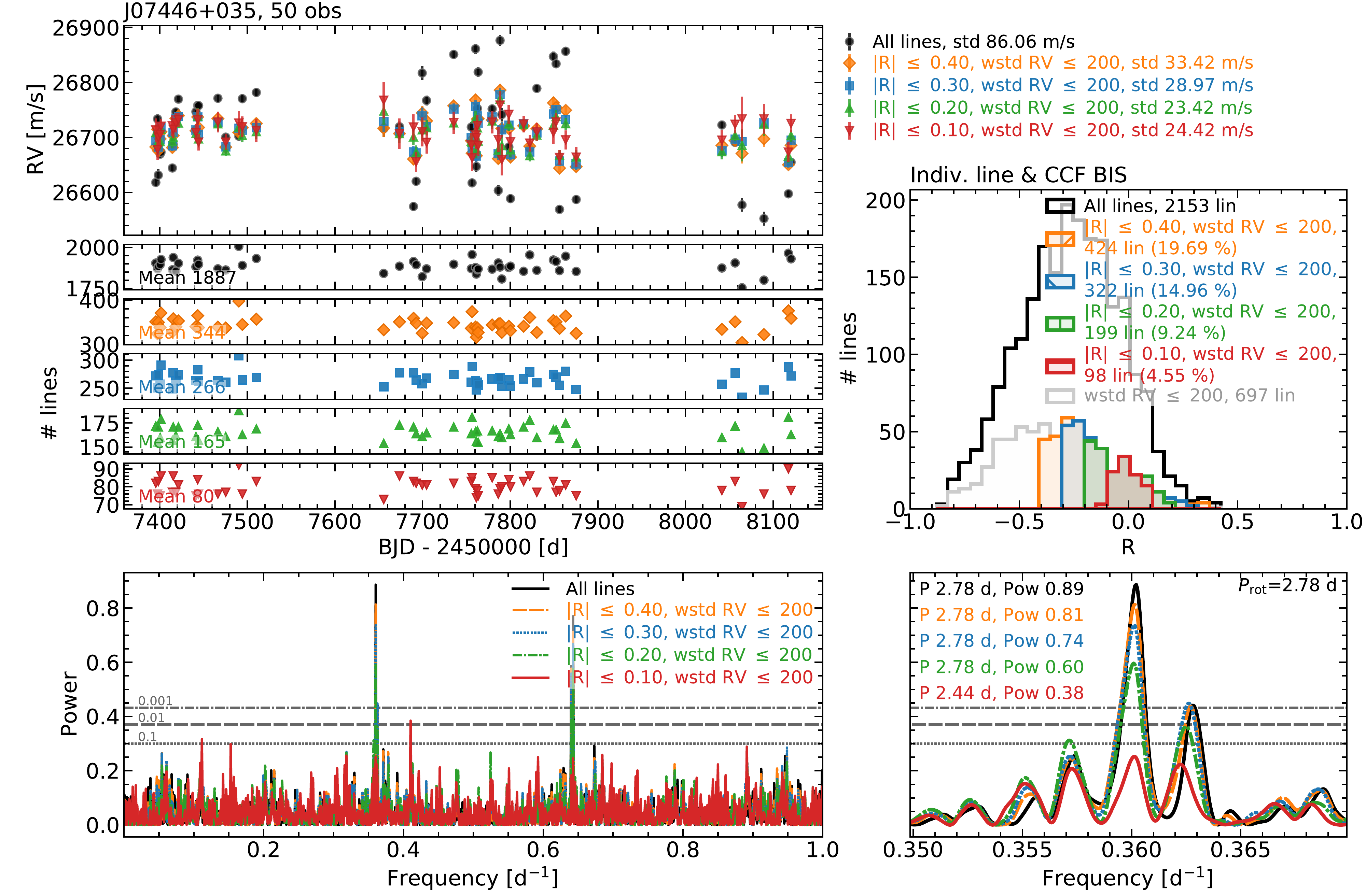}
\caption{Same as Fig. \ref{fig:tsnhpccfrvinactivebestJ07446+035}, but using R values from the correlation between the individual line RVs and CCF BIS.}
\label{fig:tsnhpccfbisinactivebestJ07446+035}
\end{figure*}

% J07446+035 RV TS, Rcoeff, periodogram, corr servalcrx
\begin{figure*}
\centering
\includegraphics[width=0.93\linewidth]{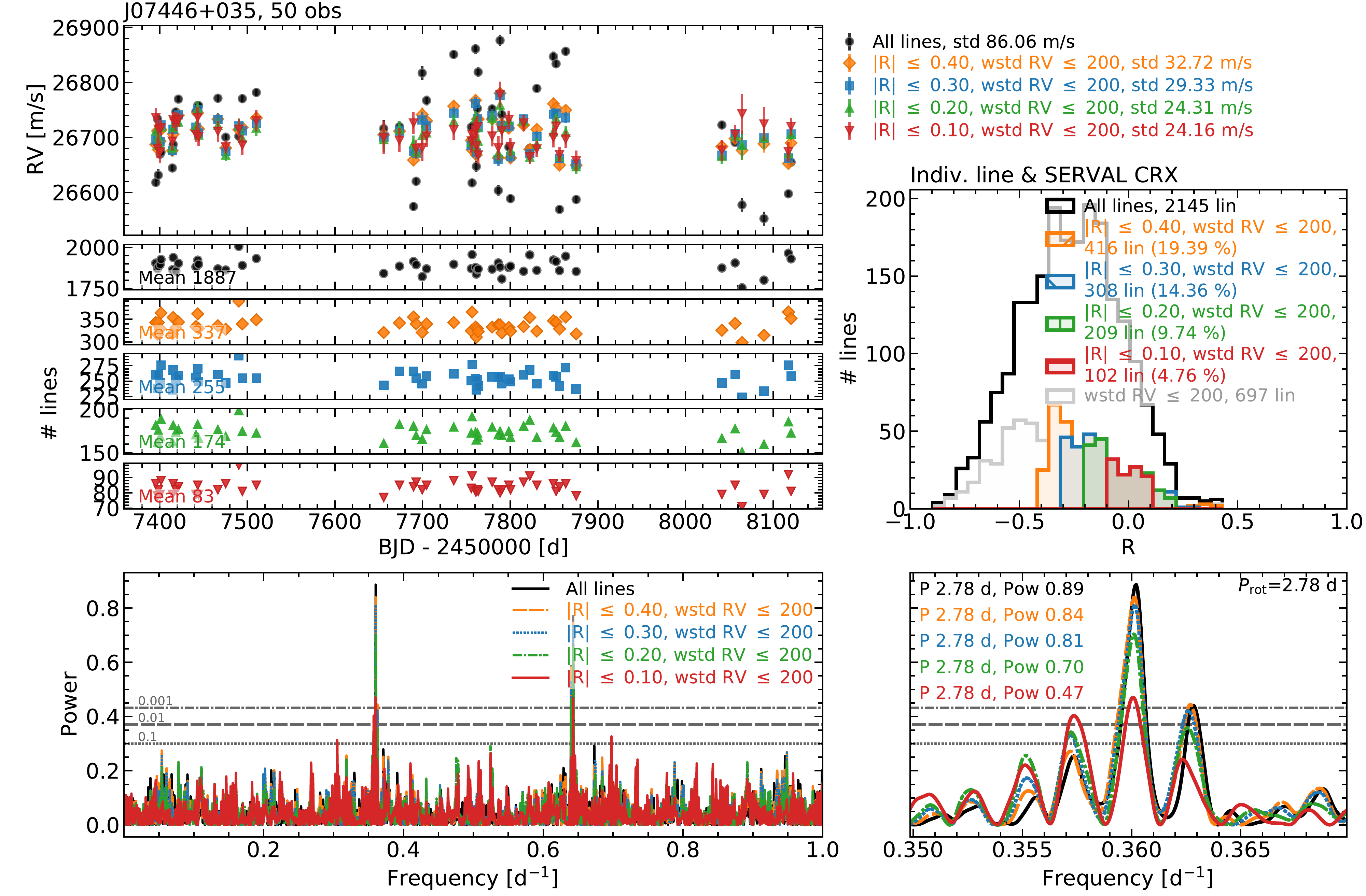}
\caption{Same as Fig. \ref{fig:tsnhpccfrvinactivebestJ07446+035}, but using R values from the correlation between the individual line RVs and CRX.}
\label{fig:tsnhpservalcrxinactivebestJ07446+035}
\end{figure*}

% -------------

% J07446+035 RV TS, Rcoeff periodogram, corr ccfbis, active
\begin{figure*}
\centering
\includegraphics[width=0.93\linewidth]{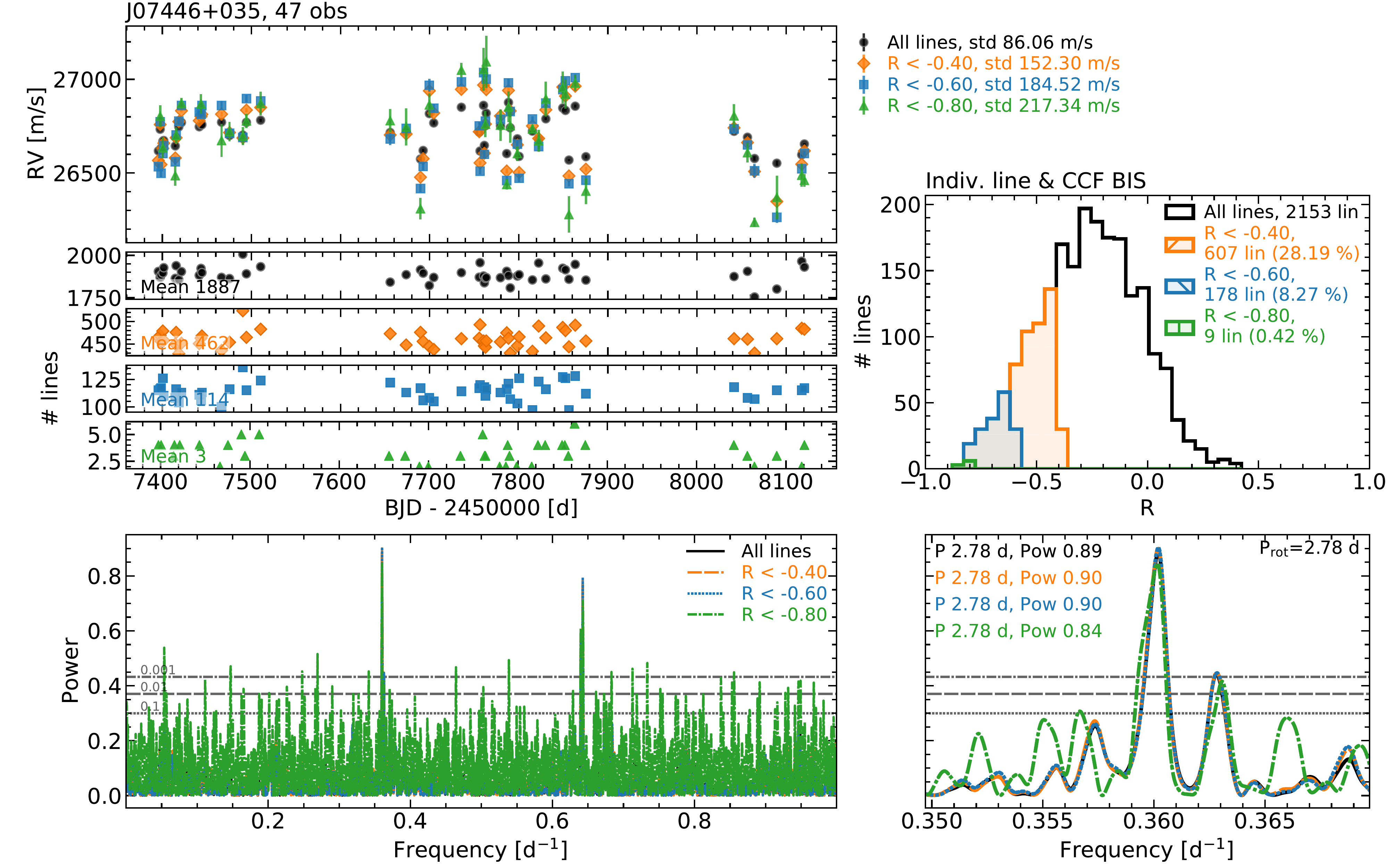}
\caption[]{Same as Fig. \ref{fig:tsnhpccfrvactiveJ07446+035}, but using R values from the correlation between the individual line RVs and CCF BIS.}
\label{fig:tsnhpccfbisactiveJ07446+035}
\end{figure*}

% J07446+035 RV TS, Rcoeff, periodogram, corr servalcrx, active
\begin{figure*}
\centering
\includegraphics[width=0.93\linewidth]{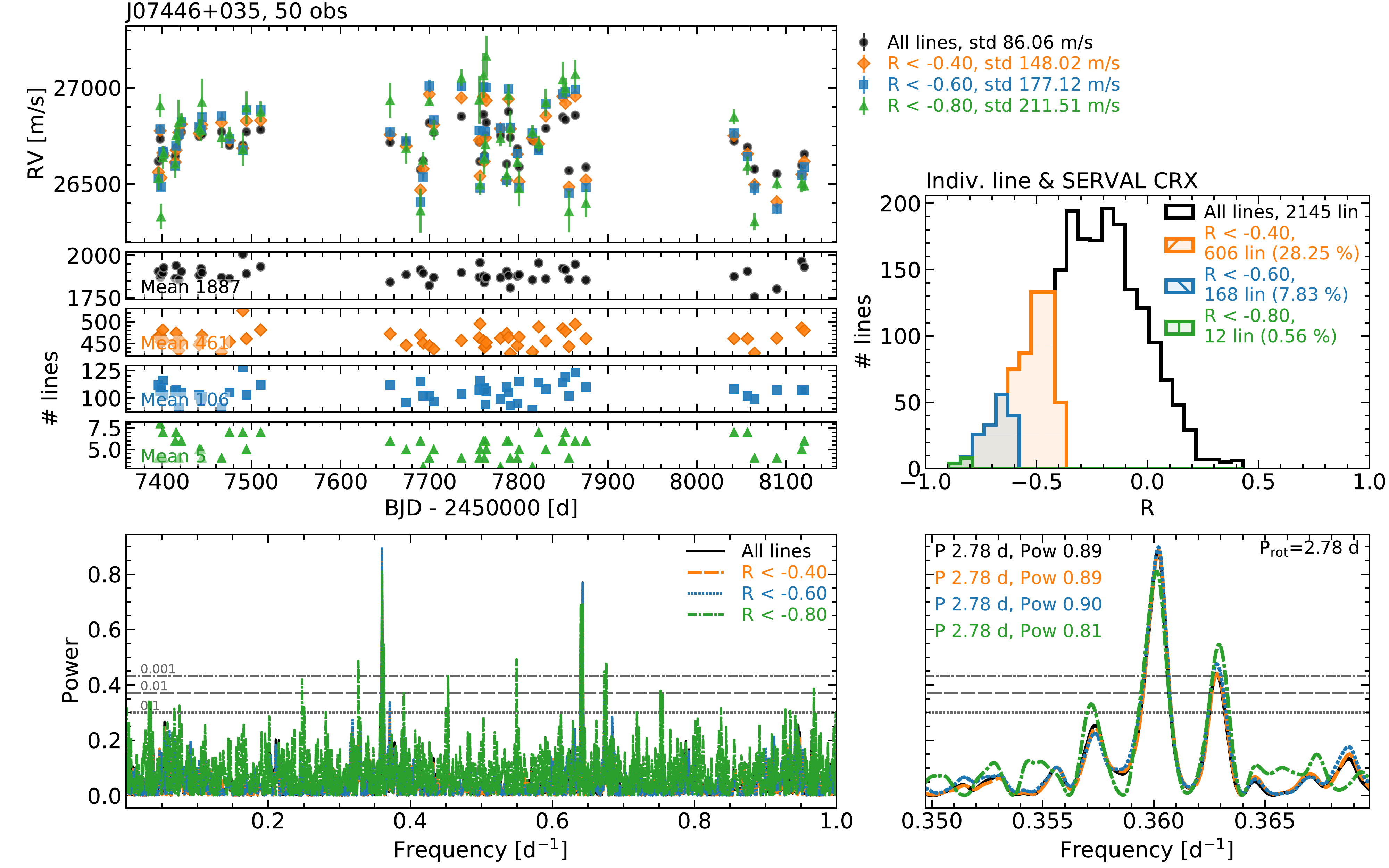}
\caption[]{Same as Fig. \ref{fig:tsnhpccfrvactiveJ07446+035}, but using R values from the correlation between the individual line RVs and CRX.}
\label{fig:tsnhpservalcrxactiveJ07446+035}
\end{figure*}

% ------------------------------------------------

% J05019+011 2dscatter
\begin{figure*}[!b]
\centering
\begin{subfigure}[]{0.32\linewidth}
\centering
\includegraphics[width=\textwidth]{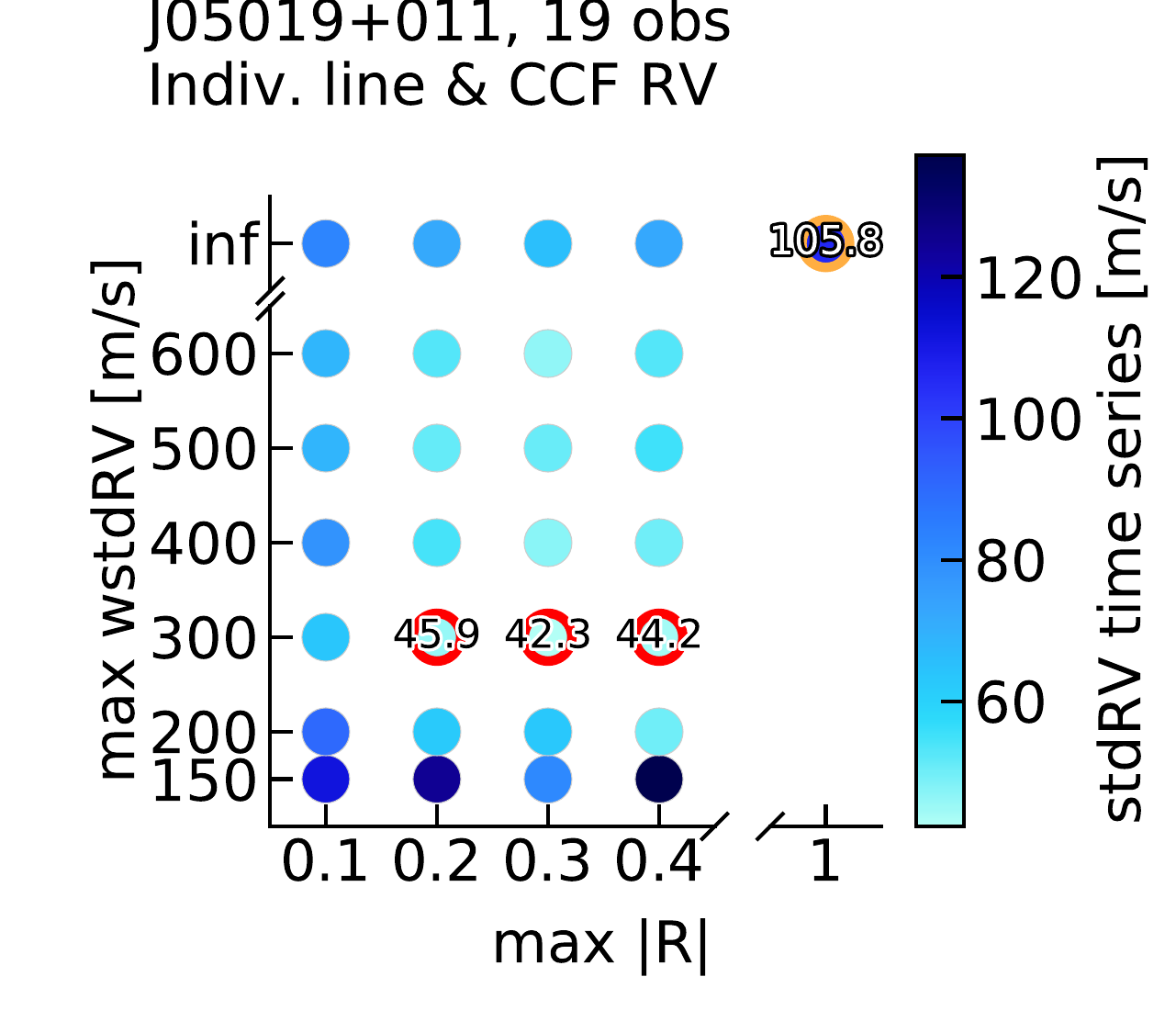}
\end{subfigure}
\,
\begin{subfigure}[]{0.32\linewidth}
\centering
\includegraphics[width=\textwidth]{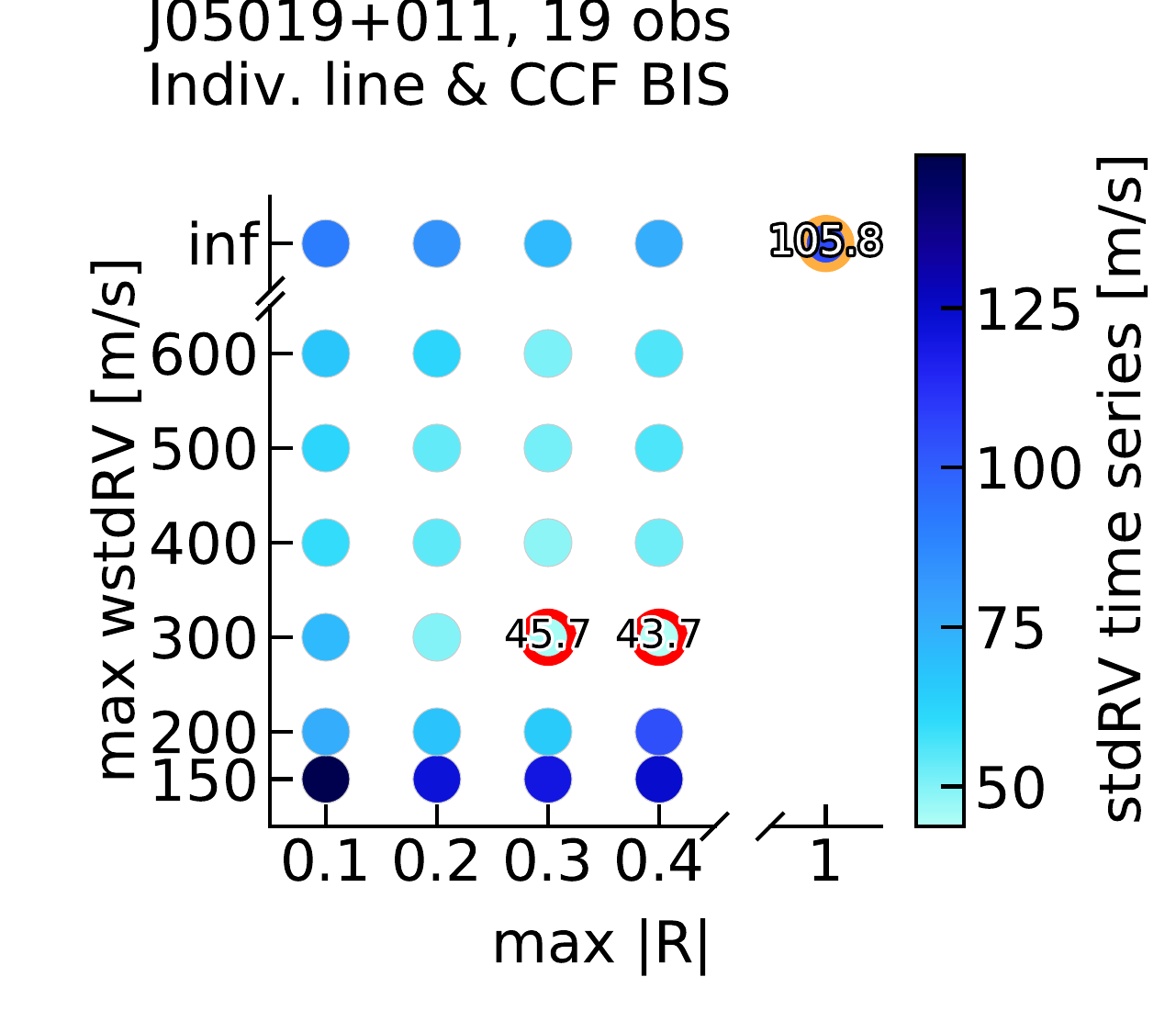}
\end{subfigure}
\,
\begin{subfigure}[]{0.32\linewidth}
\centering
\includegraphics[width=\textwidth]{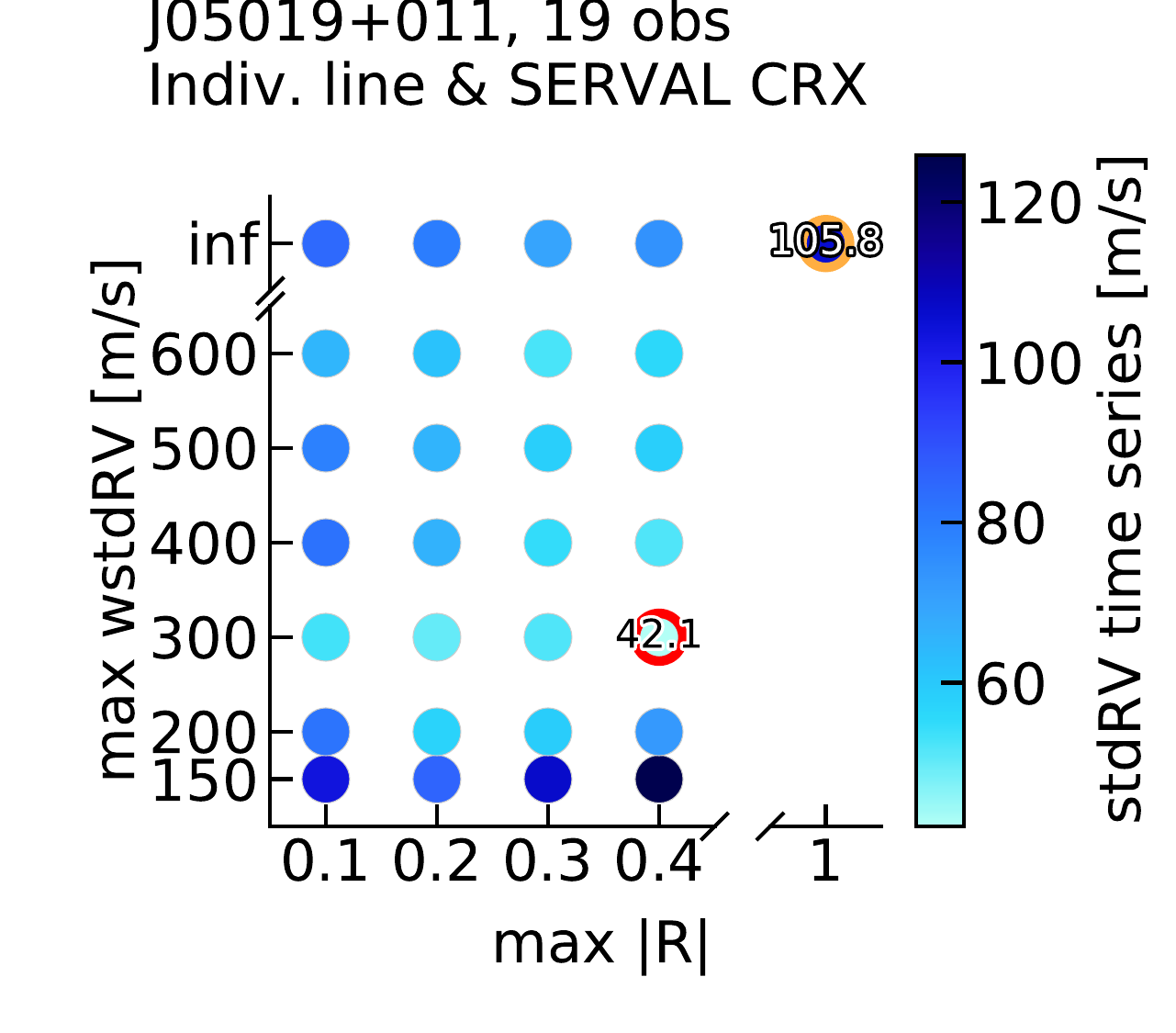}
\end{subfigure}
\,
\caption{Same as Fig. \ref{fig:cuts2dinactiveJ07446+035}, but for J05019+011.}
\label{fig:cuts2dinactiveJ05019+011}
\end{figure*}

% -------------

% J05019+011 RV TS all lines, CCF, SERVAL
\begin{figure*}
\centering
\includegraphics[width=0.93\linewidth]{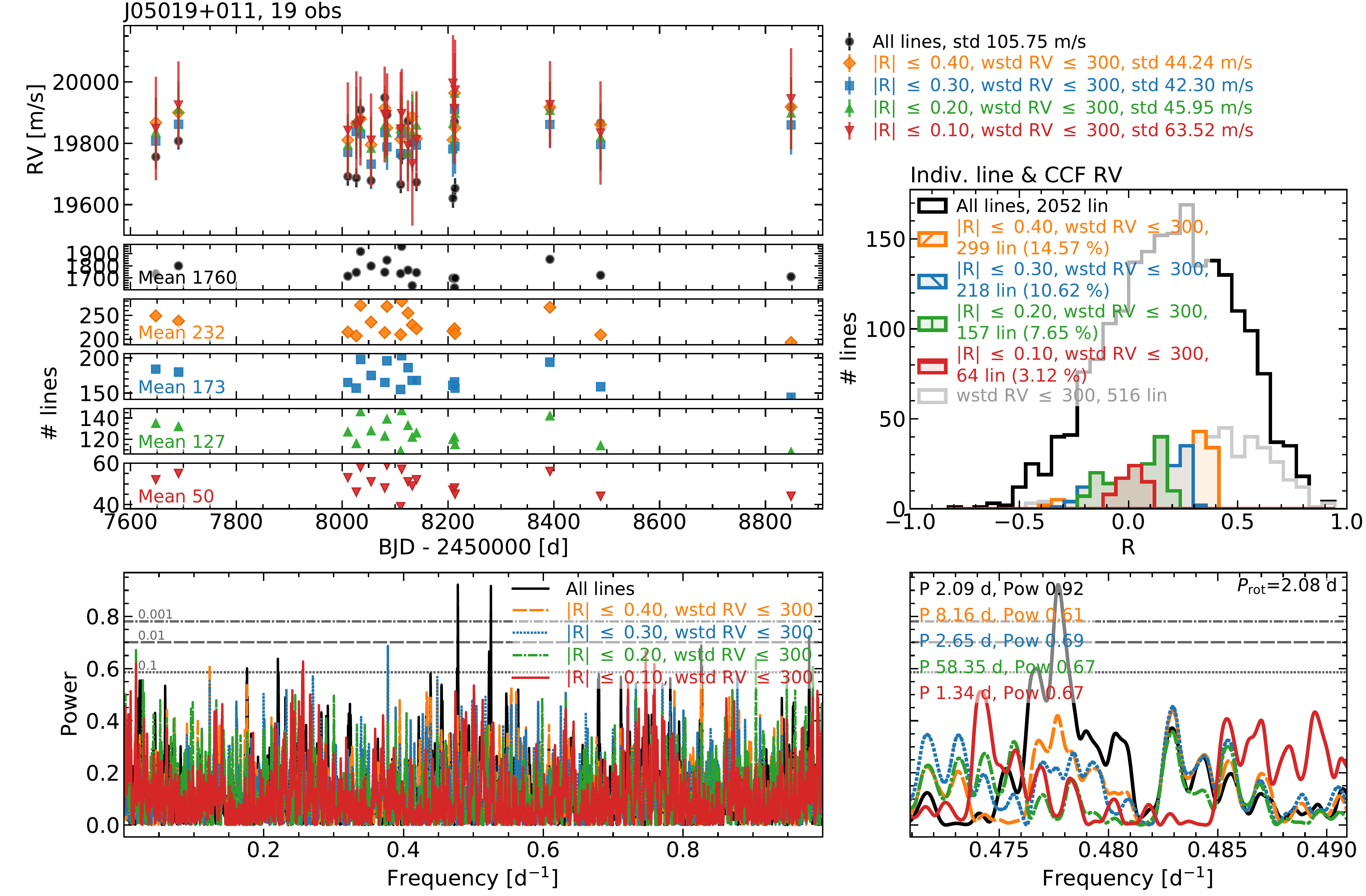}
\caption[Same as Fig. \ref{fig:tsnhpccfrvinactivebestJ07446+035}, but for J05019+011.]{Same as Fig. \ref{fig:tsnhpccfrvinactivebestJ07446+035}, but for J05019+011 and lines with RV scatter $\leq300~\ms$.}
\label{fig:tsnhpccfrvinactivebestJ05019+011}
\end{figure*}

% J05019+011 RV TS, Rcoeff, periodogram, corr ccfbis
\begin{figure*}
\centering
\includegraphics[width=0.93\linewidth]{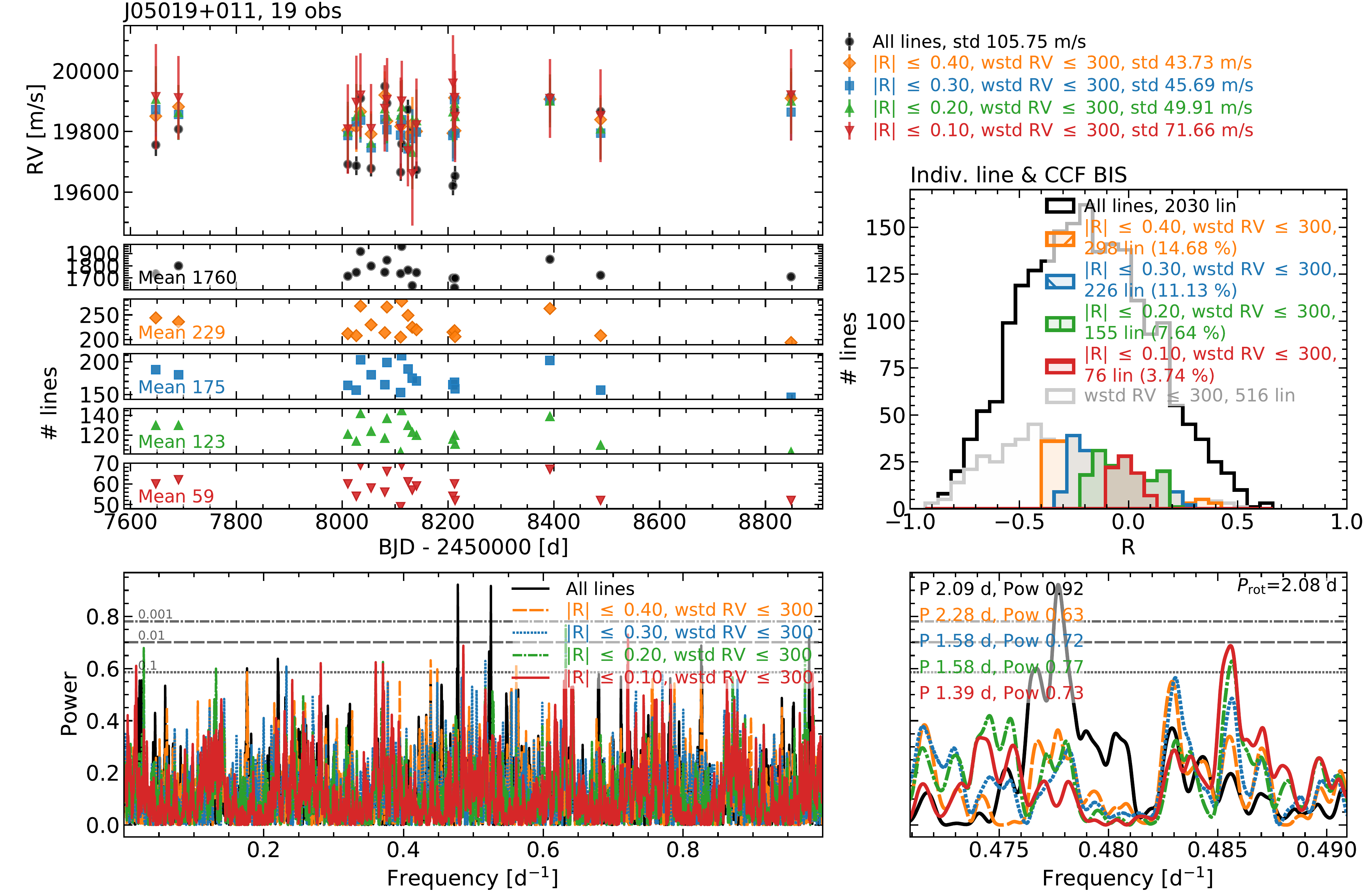}
\caption[Same as Fig. \ref{fig:tsnhpccfbisinactivebestJ07446+035}, but for J05019+011.]{Same as Fig. \ref{fig:tsnhpccfbisinactivebestJ07446+035}, but for J05019+011 and lines with RV scatter $\leq300~\ms$.}
\label{fig:tsnhpccfbisinactivebestJ05019+011}
\end{figure*}

% J05019+011 RV TS, Rcoeff, periodogram, corr servalcrx
\begin{figure*}
\centering
\includegraphics[width=0.93\linewidth]{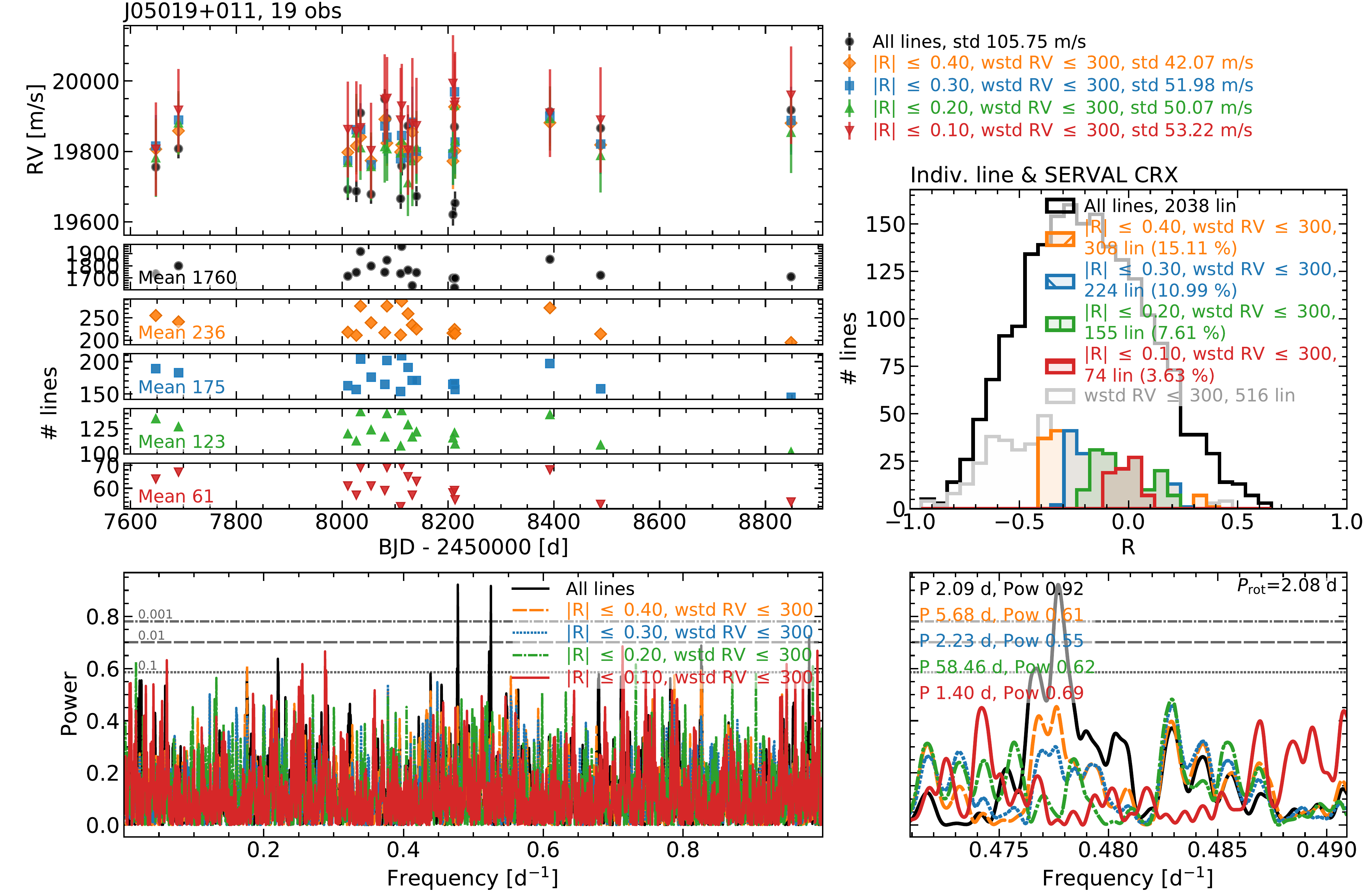}
\caption[Same as Fig. \ref{fig:tsnhpservalcrxinactivebestJ07446+035}, but for J05019+011.]{Same as Fig. \ref{fig:tsnhpservalcrxinactivebestJ07446+035}, but for J05019+011 and lines with RV scatter $\leq300~\ms$.}
\label{fig:tsnhpservalcrxinactivebestJ05019+011}
\end{figure*}

% -------------

% J05019+011 RV TS, Rcoeff, periodogram, corr ccfrv active
\begin{figure*}
\centering
\includegraphics[width=0.93\linewidth]{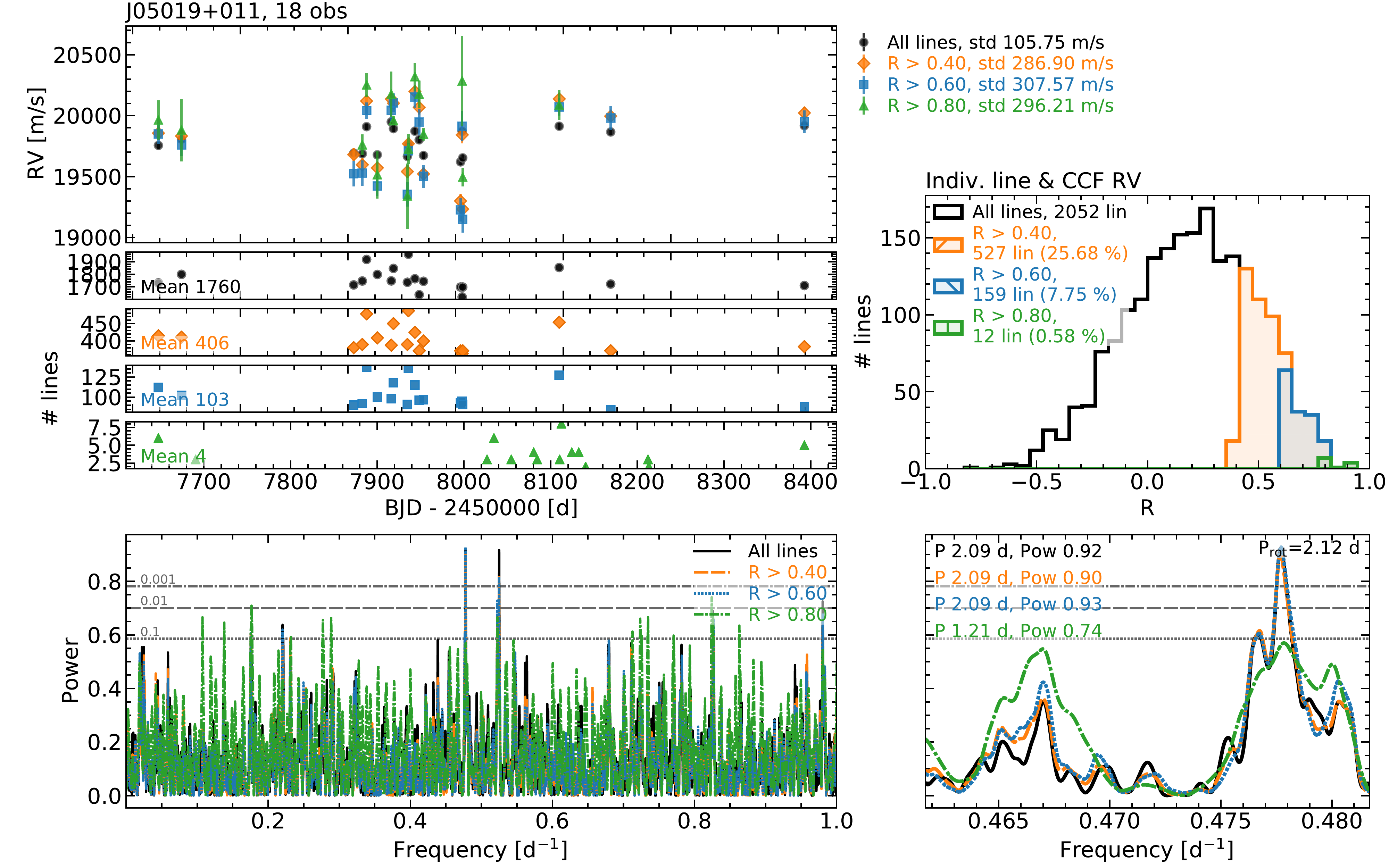}
\caption{Same as Fig. \ref{fig:tsnhpccfrvactiveJ07446+035}, but for J05019+011.}
\label{fig:tsnhpccfrvactiveJ05019+011}
\end{figure*}

% J05019+011 RV TS, Rcoeff, periodogram, corr ccfbis active
\begin{figure*}
\centering
\includegraphics[width=0.93\linewidth]{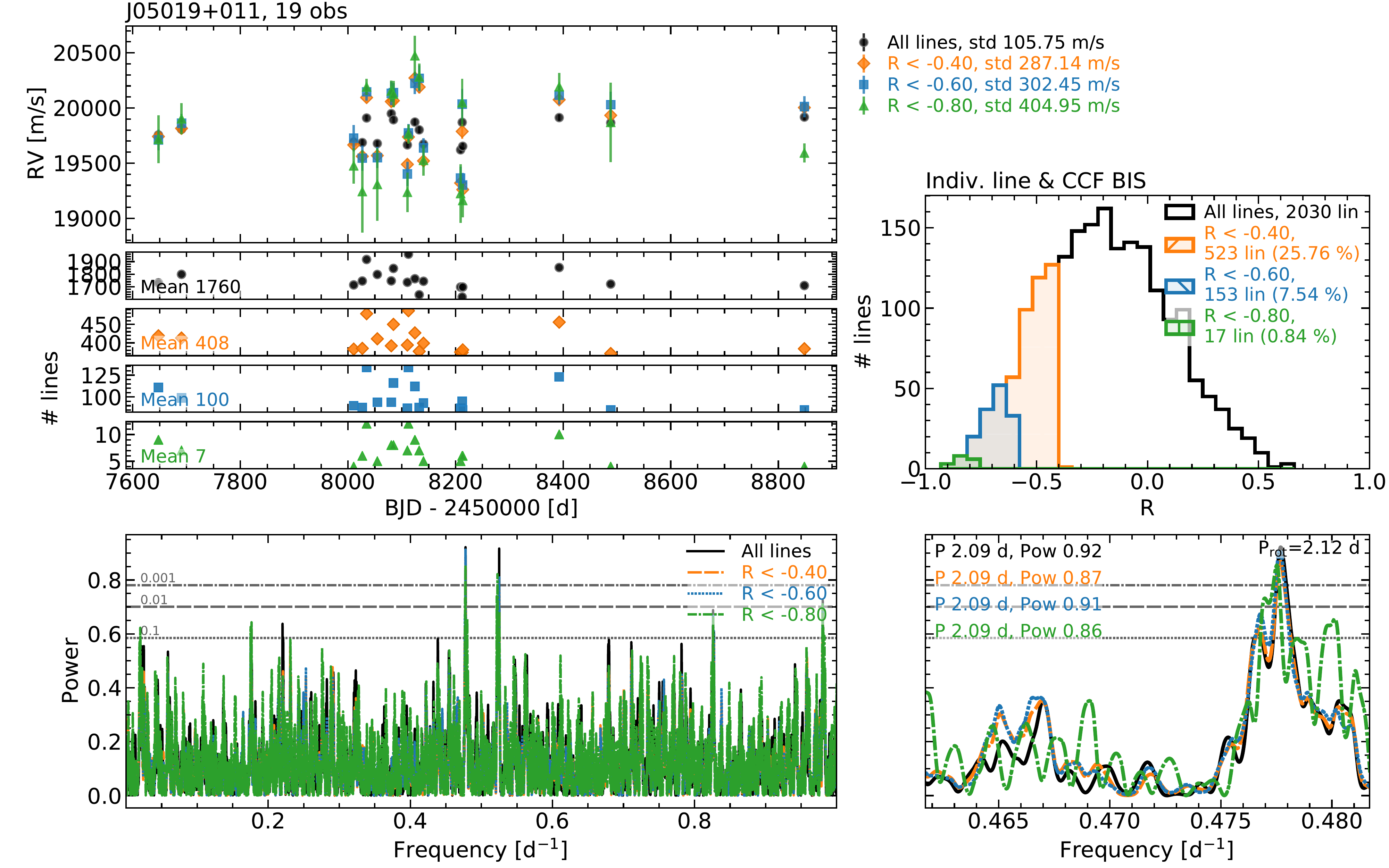}
\caption{Same as Fig. \ref{fig:tsnhpccfbisactiveJ07446+035}, but for J05019+011.}
\label{fig:tsnhpccfbisactiveJ05019+011}
\end{figure*}

% J05019+011 RV TS, Rcoeff, periodogram, corr servalcrx active
\begin{figure*}
\centering
\includegraphics[width=0.93\linewidth]{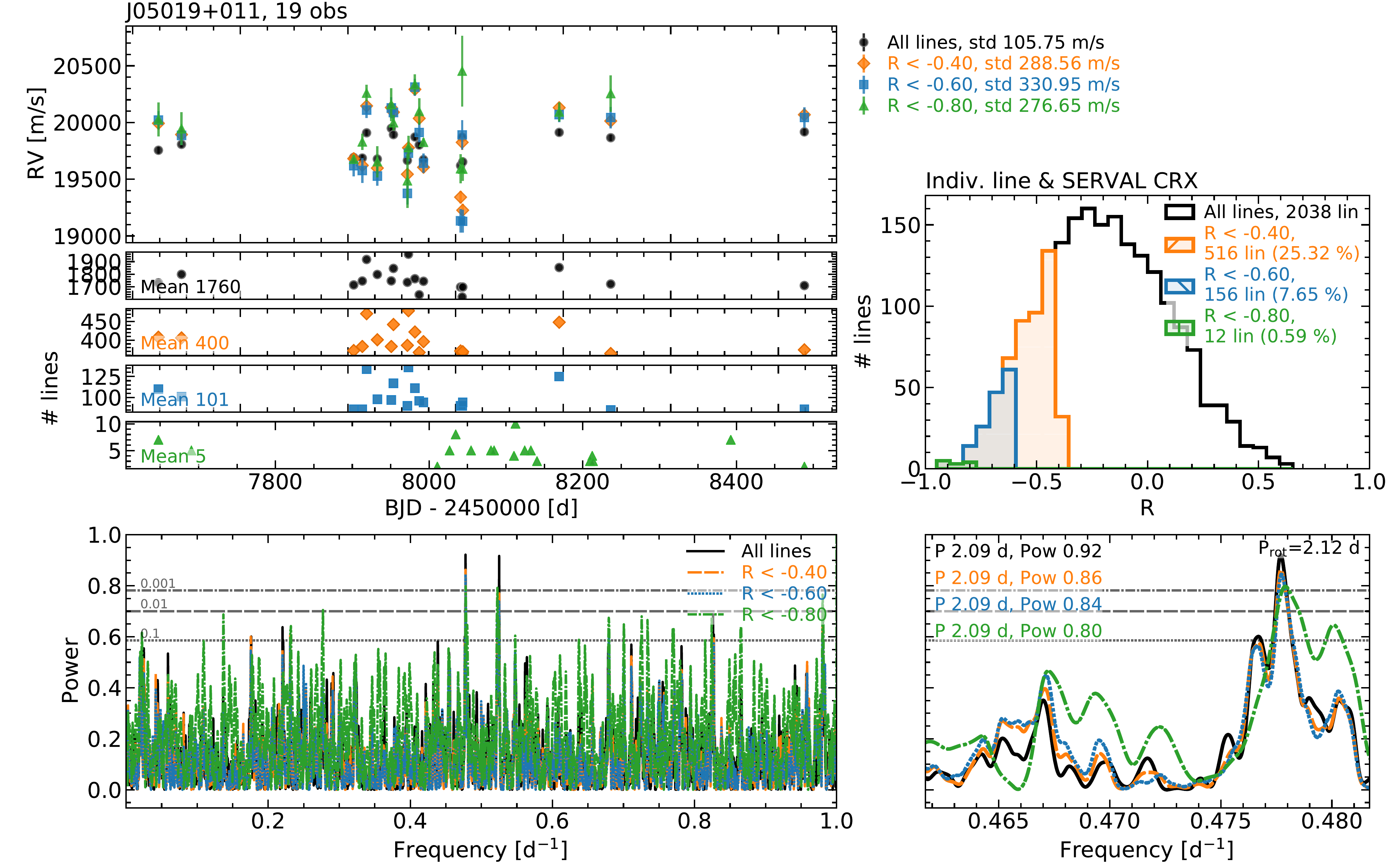}
\caption{Same as Fig. \ref{fig:tsnhpservalcrxactiveJ07446+035}, but for J05019+011.}
\label{fig:tsnhpservalcrxactiveJ05019+011}
\end{figure*}

% ------------------------------------------------

% J22468+443 2dscatter
\begin{figure*}[!b]
\centering
\begin{subfigure}[]{0.32\linewidth}
\centering
\includegraphics[width=\textwidth]{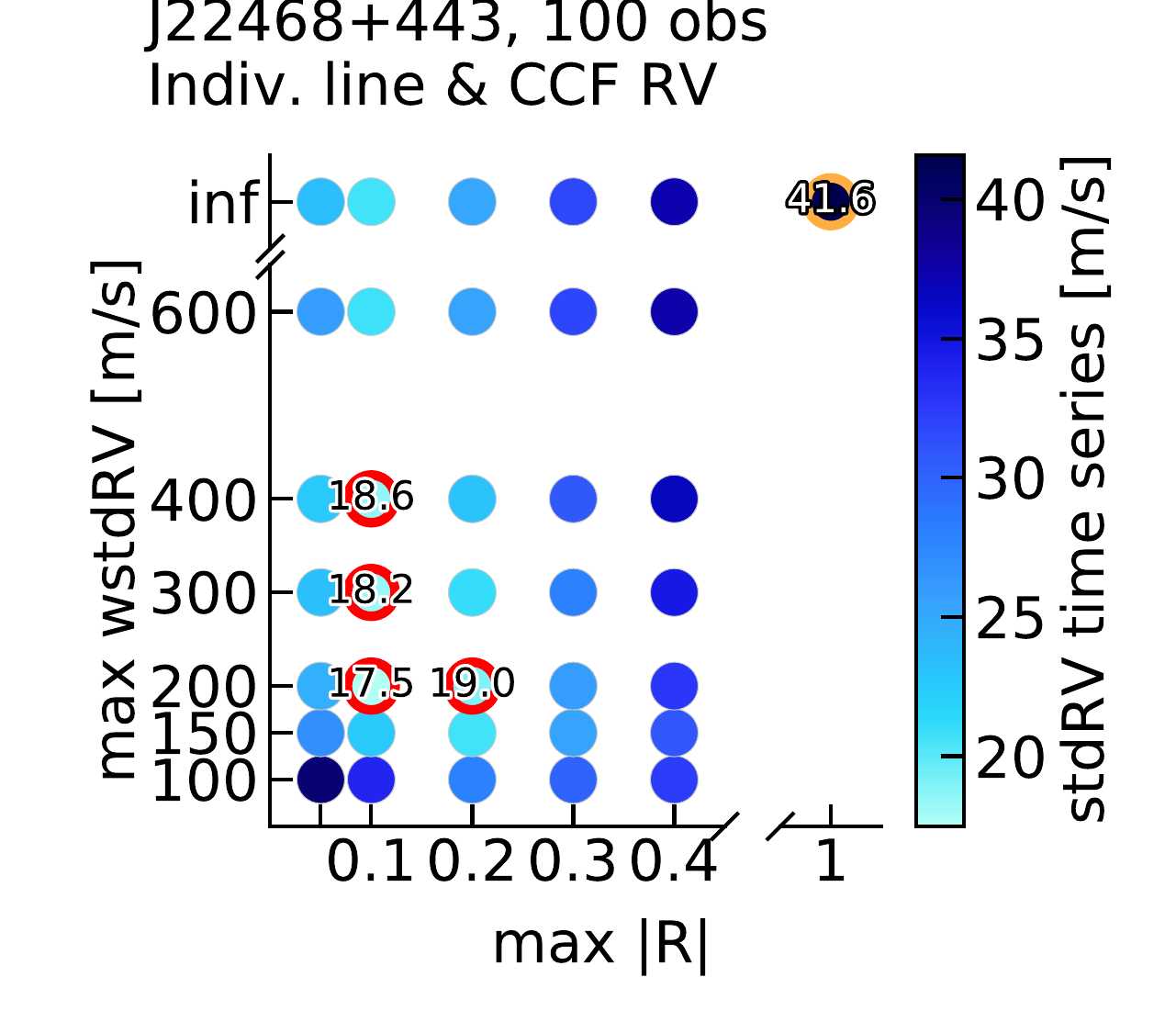}
\end{subfigure}
\,
\begin{subfigure}[]{0.32\linewidth}
\centering
\includegraphics[width=\textwidth]{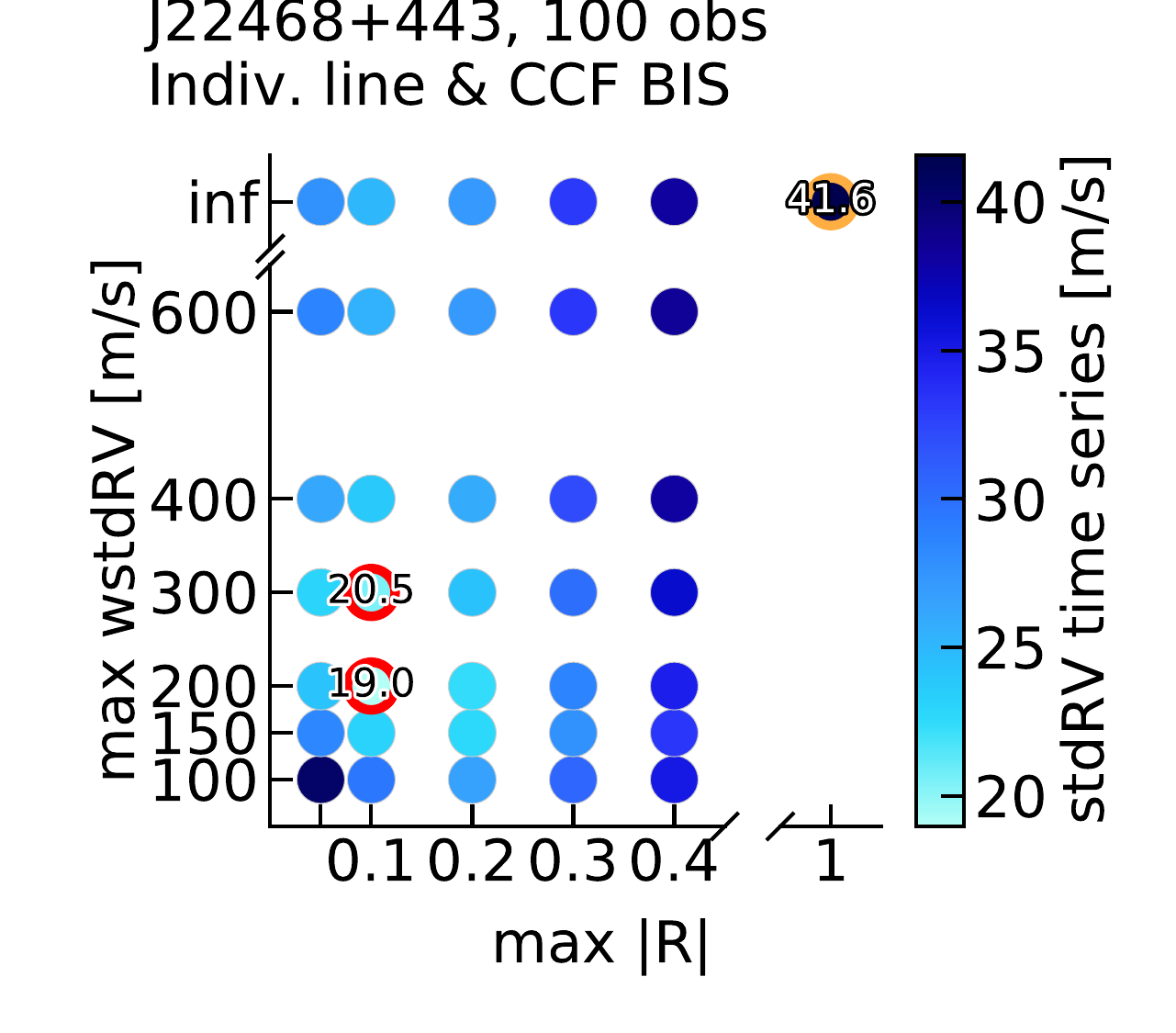}
\end{subfigure}
\,
\begin{subfigure}[]{0.32\linewidth}
\centering
\includegraphics[width=\textwidth]{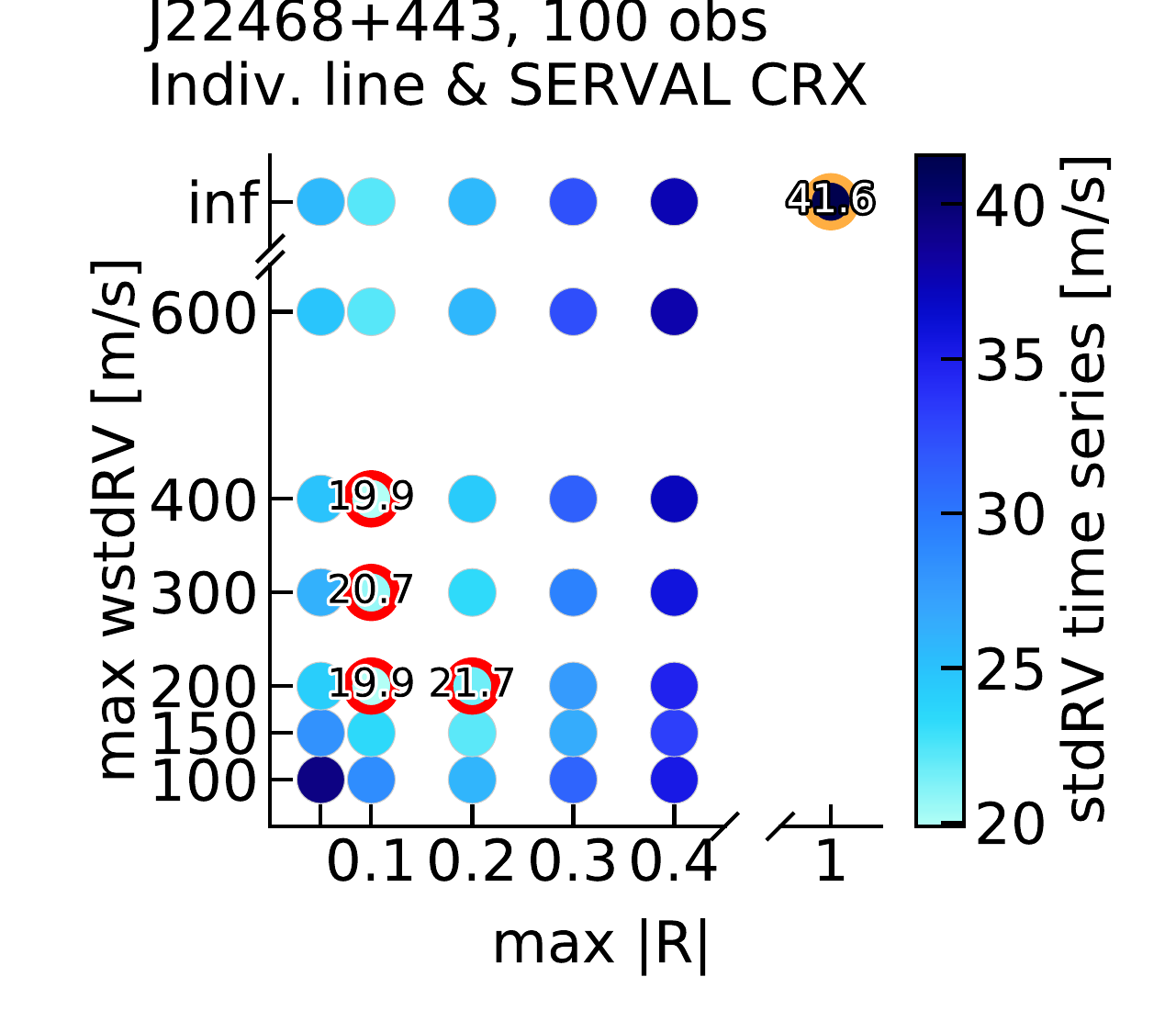}
\end{subfigure}
\,
\caption{Same as Fig. \ref{fig:cuts2dinactiveJ07446+035}, but for J22468+443.}
\label{fig:cuts2dinactiveJ22468+443}
\end{figure*}

% -------------

% J22468+443 RV TS all lines, CCF, SERVAL
\begin{figure*}
\centering
\includegraphics[width=0.93\linewidth]{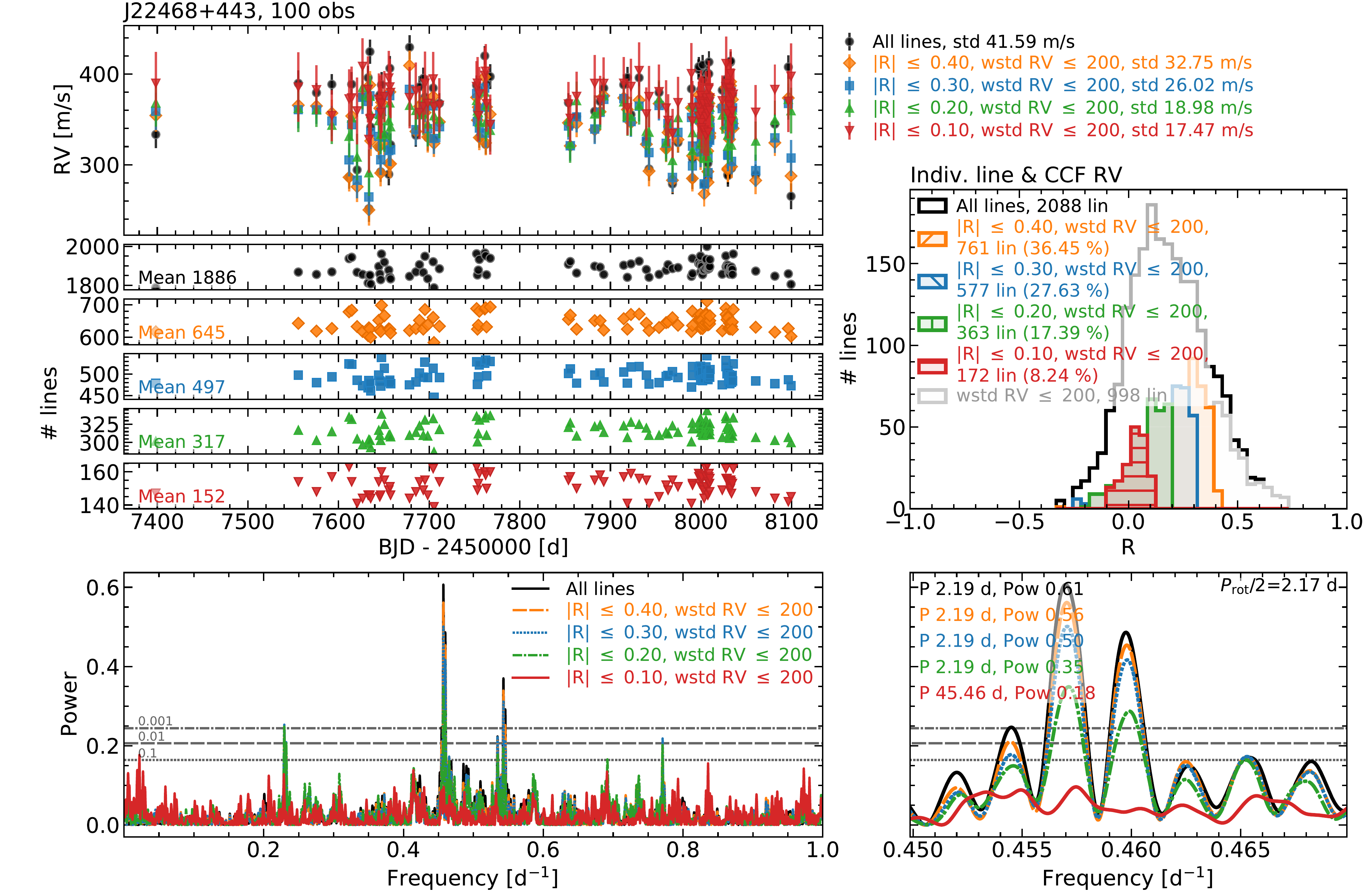}
\caption{Same as Fig. \ref{fig:tsnhpccfrvinactivebestJ07446+035}, but for J22468+443. In this case, the periodogram zoom in on the \emph{right panel} corresponds to half \Prot, which is the highest peak in total RV periodogram.}
\label{fig:tsnhpccfrvinactivebestJ22468+443}
\end{figure*}

% J22468+443 RV TS, Rcoeff, periodogram, corr ccfbis
\begin{figure*}
\centering
\includegraphics[width=0.93\linewidth]{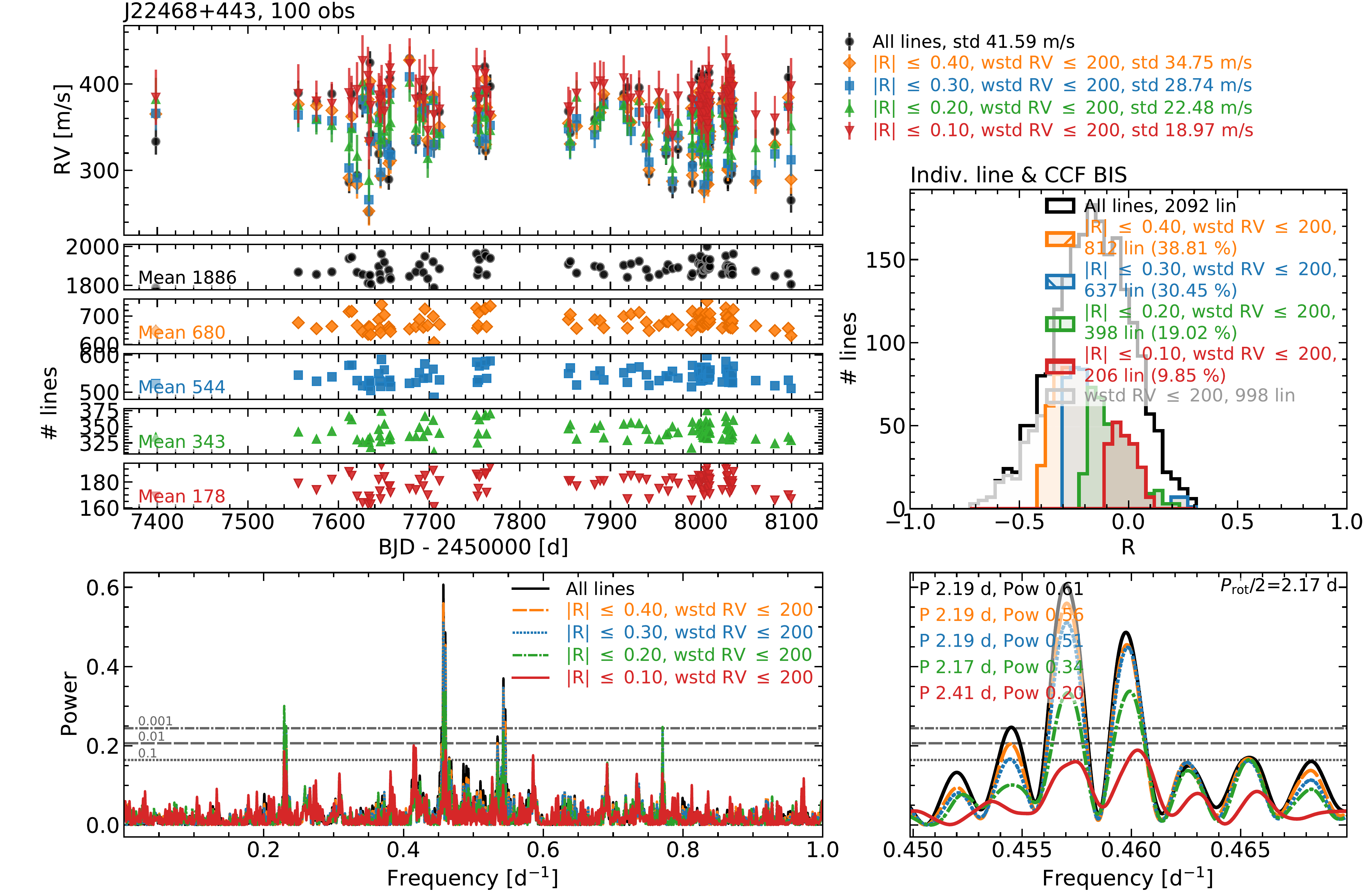}
\caption{Same as Fig. \ref{fig:tsnhpccfbisinactivebestJ07446+035}, but for J22468+443. In this case, the periodogram zoom in on the \emph{right panel} corresponds to half \Prot, which is the highest peak in total RV periodogram.}
\label{fig:tsnhpccfbisinactivebestJ22468+443}
\end{figure*}

% J22468+443 RV TS, Rcoeff, periodogram, corr servalcrx
\begin{figure*}
\centering
\includegraphics[width=0.93\linewidth]{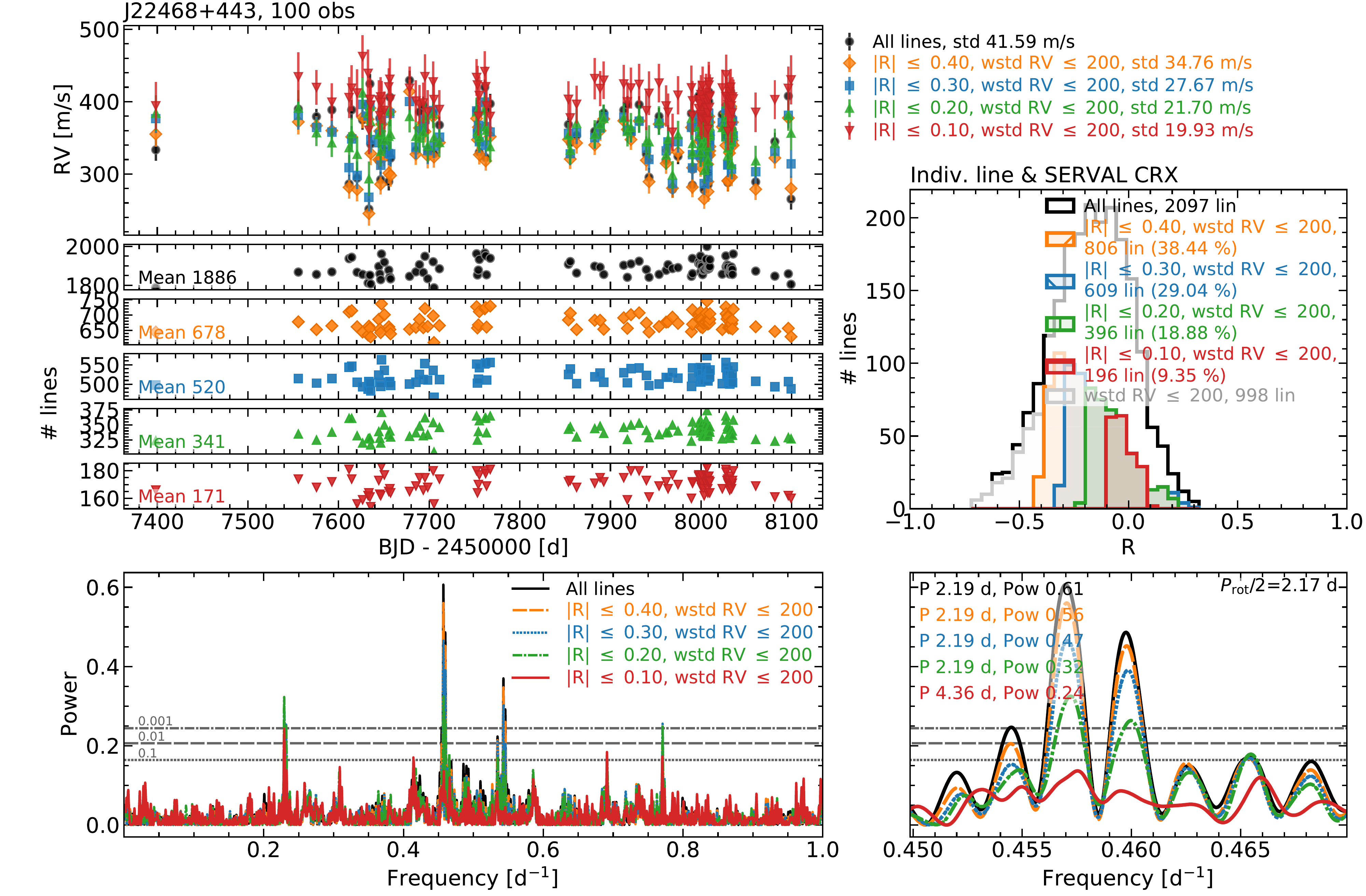}
\caption{Same as Fig. \ref{fig:tsnhpservalcrxinactivebestJ07446+035}, but for J22468+443. In this case, the periodogram zoom in on the \emph{right panel} corresponds to half \Prot, which is the highest peak in total RV periodogram.}
\label{fig:tsnhpservalcrxinactivebestJ22468+443}
\end{figure*}

% -------------

% J22468+443 RV TS, Rcoeff, periodogram, corr ccfrv active
\begin{figure*}
\centering
\includegraphics[width=0.93\linewidth]{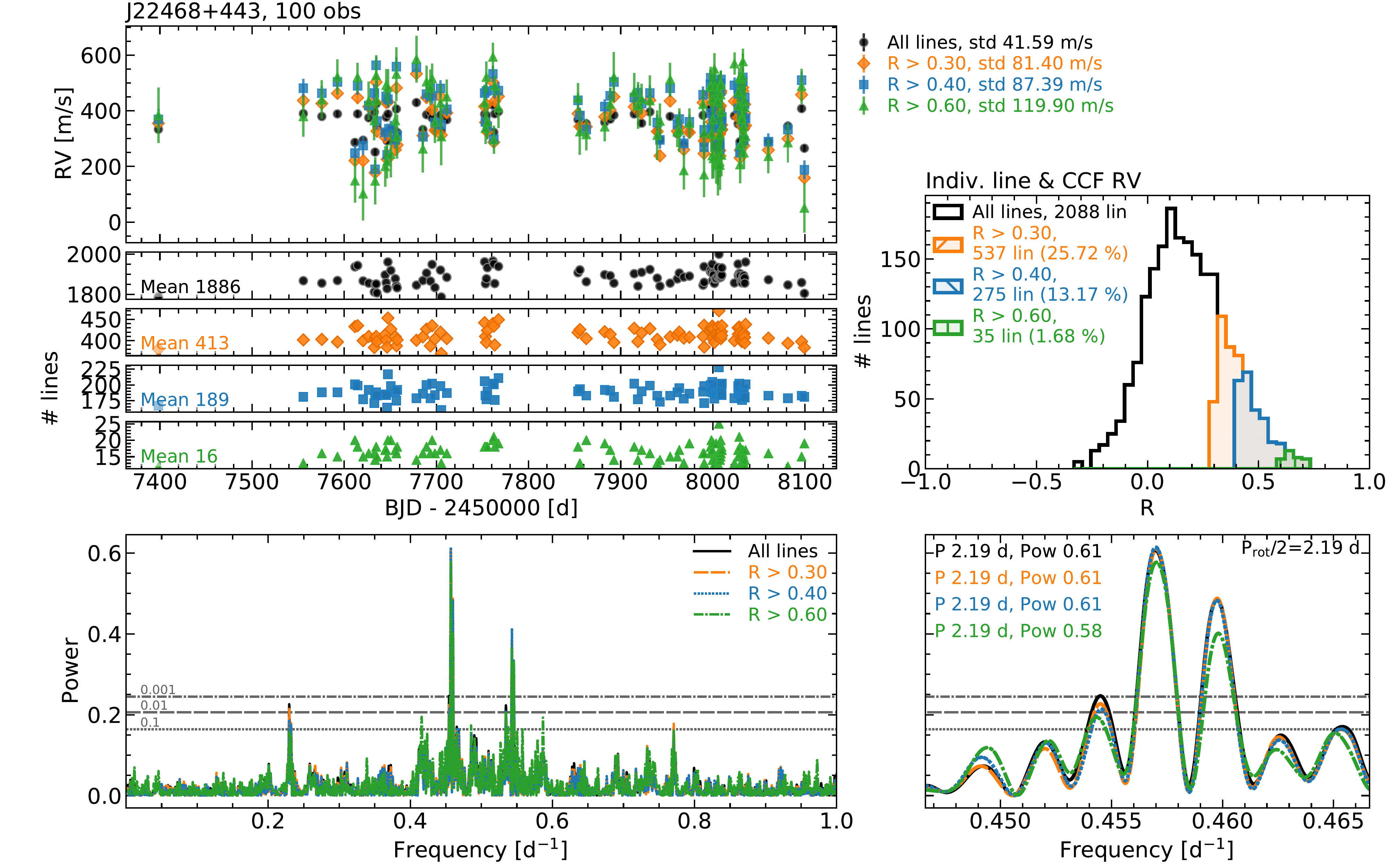}
\caption{Same as Fig. \ref{fig:tsnhpccfrvactiveJ07446+035}, but for J22468+443.}
\label{fig:tsnhpccfrvactiveJ22468+443}
\end{figure*}

% J22468+443 RV TS, Rcoeff, periodogram, corr ccfbis active
\begin{figure*}
\centering
\includegraphics[width=0.93\linewidth]{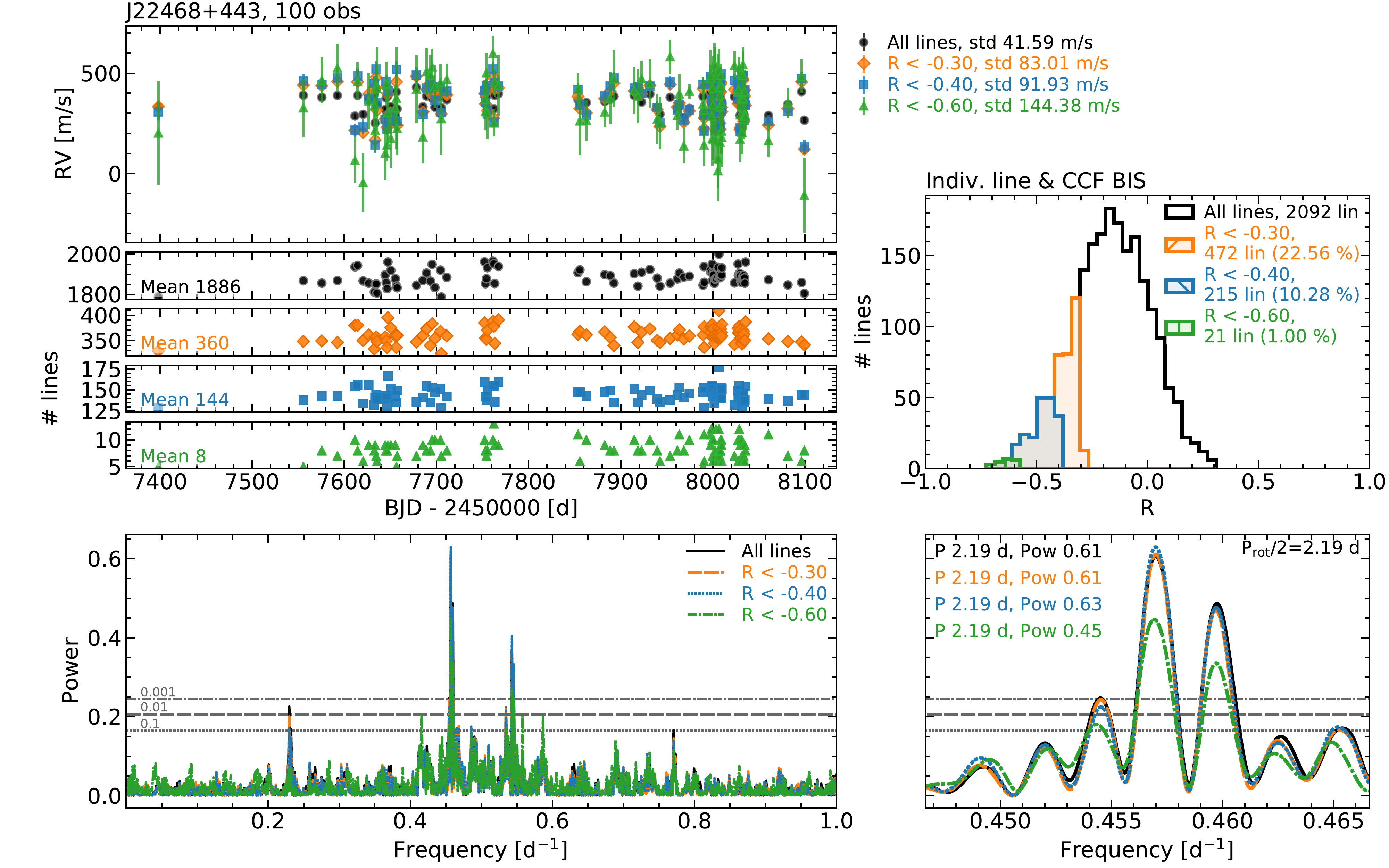}
\caption{Same as Fig. \ref{fig:tsnhpccfbisactiveJ07446+035}, but for J22468+443.}
\label{fig:tsnhpccfbisactiveJ22468+443}
\end{figure*}

% J22468+443 RV TS, Rcoeff, periodogram, corr servalcrx active
\begin{figure*}
\centering
\includegraphics[width=0.93\linewidth]{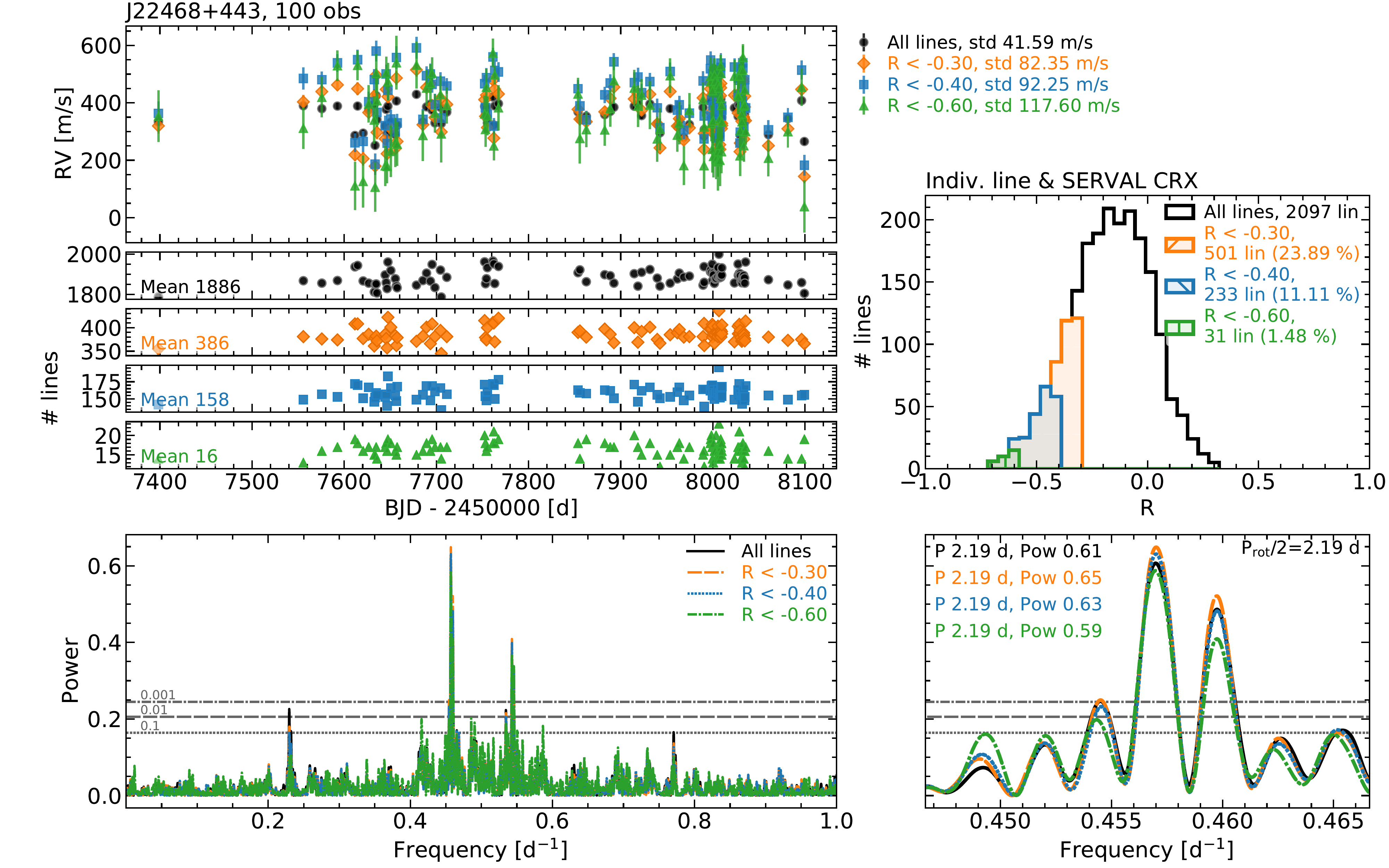}
\caption{Same as Fig. \ref{fig:tsnhpservalcrxactiveJ07446+035}, but for J22468+443.}
\label{fig:tsnhpservalcrxactiveJ22468+443}
\end{figure*}

% ------------------------------------------------

% J10196+198 2dscatter
\begin{figure*}
\centering
\begin{subfigure}[]{0.32\linewidth}
\centering
\includegraphics[width=\textwidth]{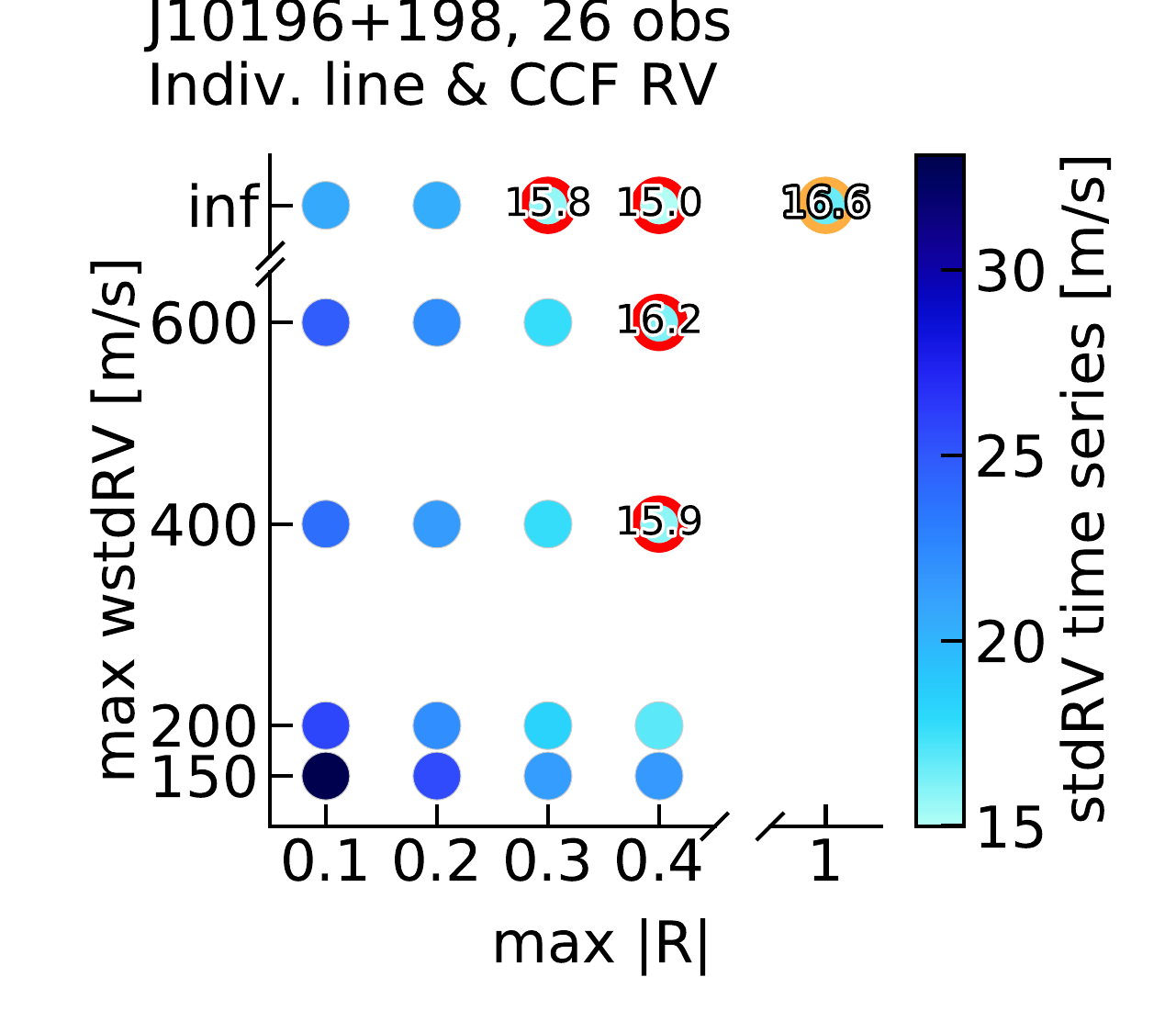}
\end{subfigure}
\,
\begin{subfigure}[]{0.32\linewidth}
\centering
\includegraphics[width=\textwidth]{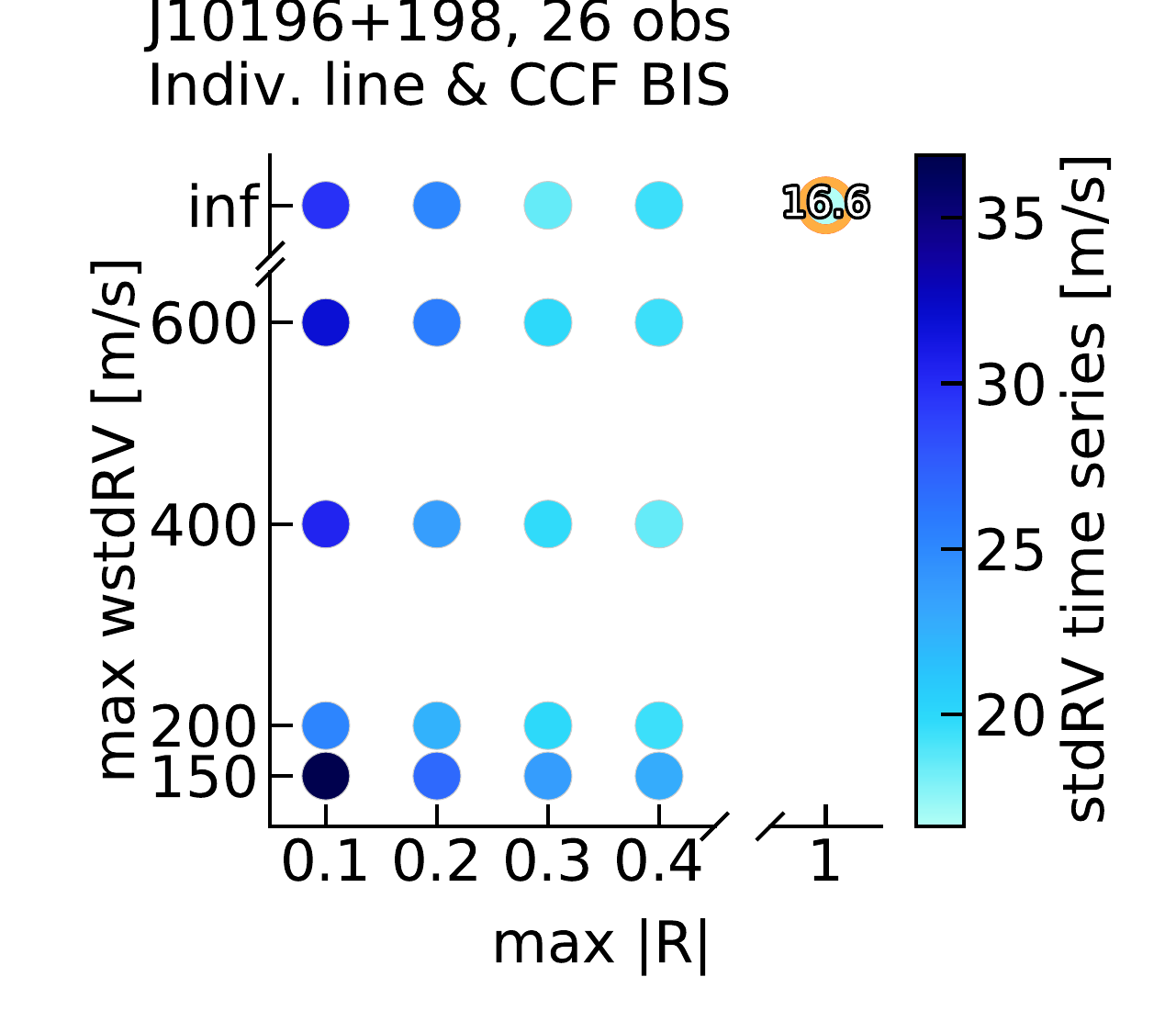}
\end{subfigure}
\,
\begin{subfigure}[]{0.32\linewidth}
\centering
\includegraphics[width=\textwidth]{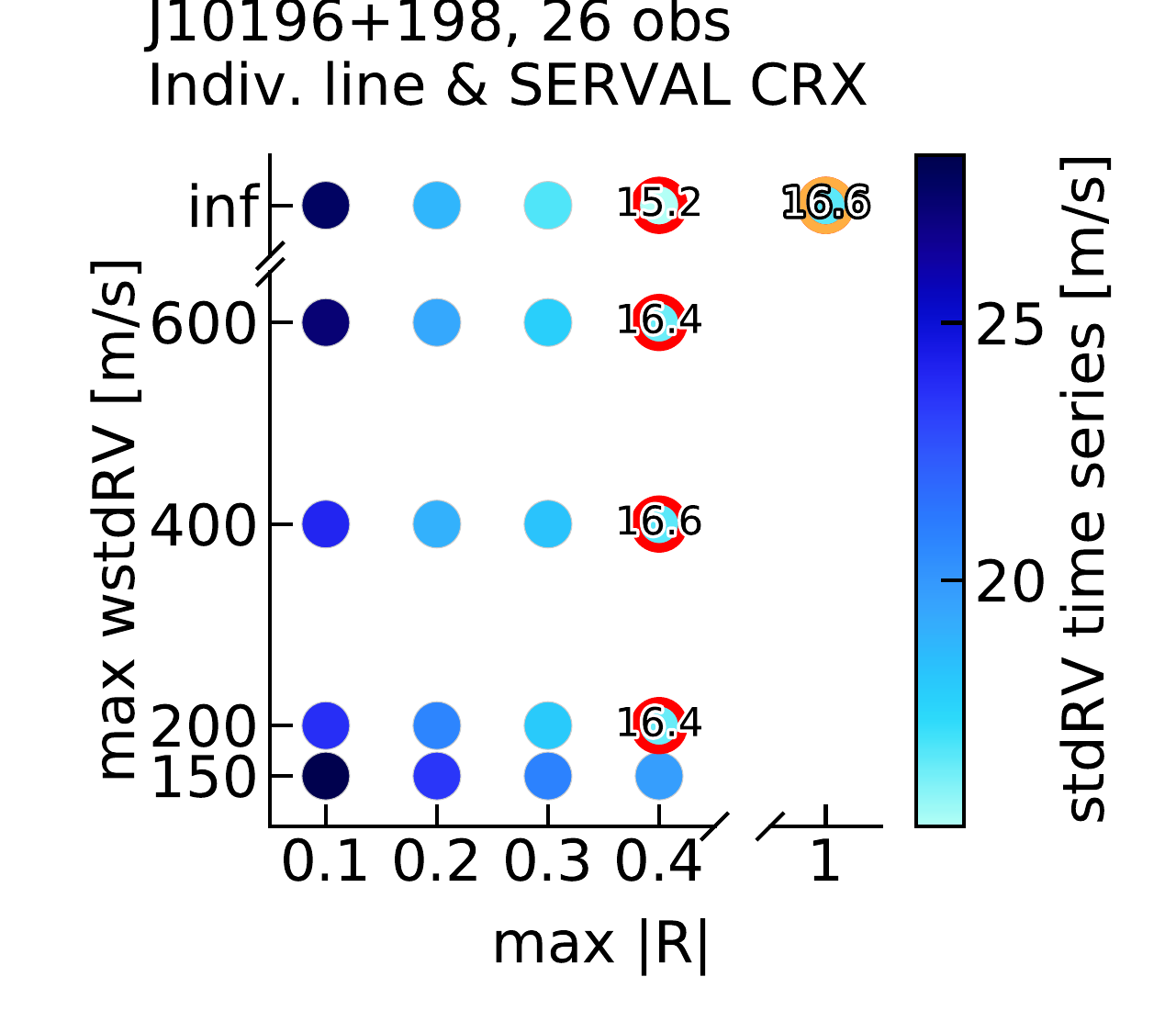}
\end{subfigure}
\,
\caption{Same as Fig. \ref{fig:cuts2dinactiveJ07446+035}, but for J10196+198.}
\label{fig:cuts2dinactiveJ10196+198}
\end{figure*}

% -------------

% J10196+198 RV TS all lines, CCF, SERVAL
\begin{figure*}
\centering
\includegraphics[width=0.93\linewidth]{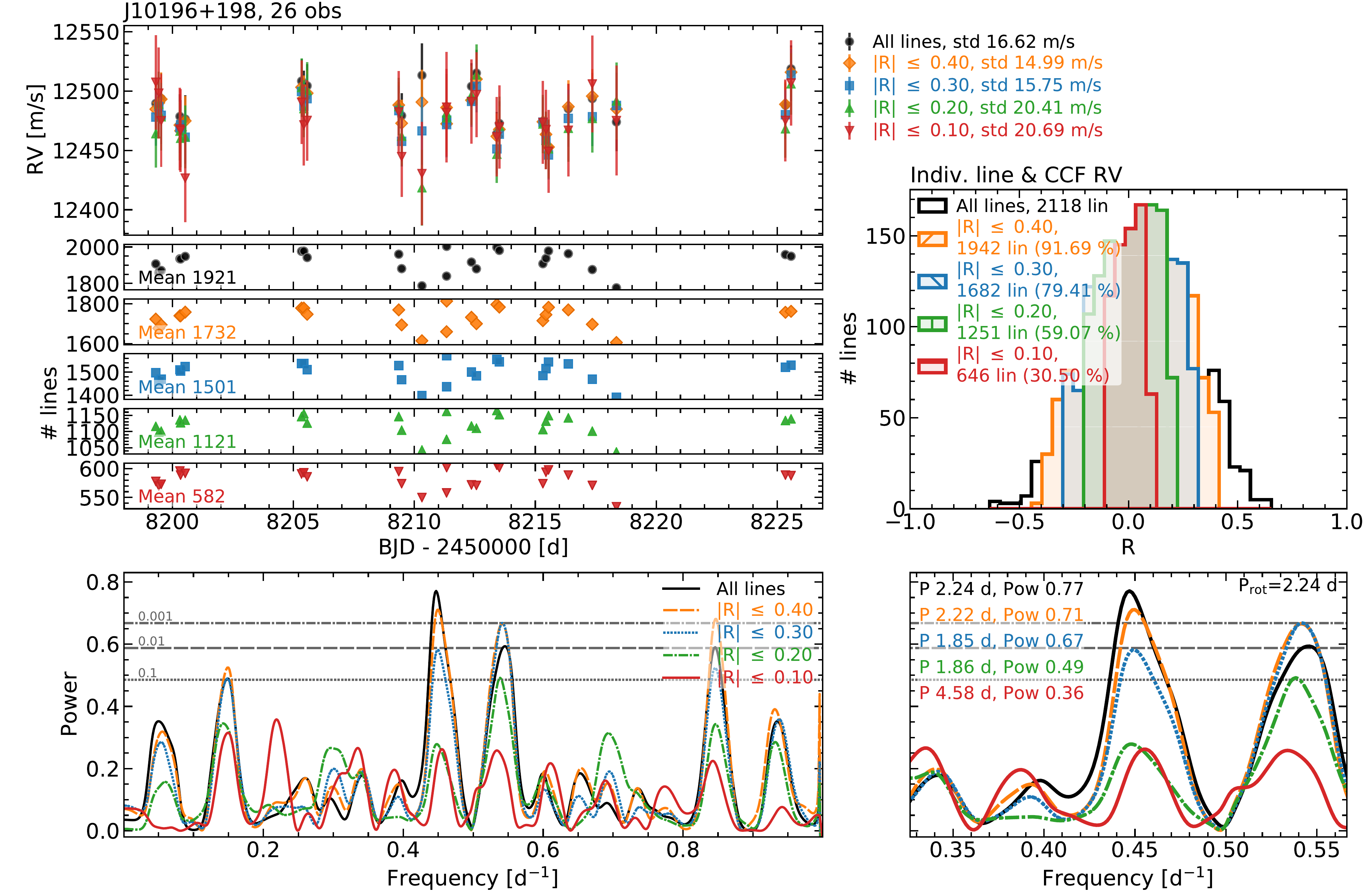}
\caption[Same as Fig. \ref{fig:tsnhpccfrvinactivebestJ07446+035}, but for J10196+198.]{Same as Fig. \ref{fig:tsnhpccfrvinactivebestJ07446+035}, but for J10196+198 and lines with no limit on the RV scatter.}
\label{fig:tsnhpccfrvinactivebestJ10196+198}
\end{figure*}

% J10196+198 RV TS, Rcoeff, periodogram, corr ccfbis
\begin{figure*}
\centering
\includegraphics[width=0.93\linewidth]{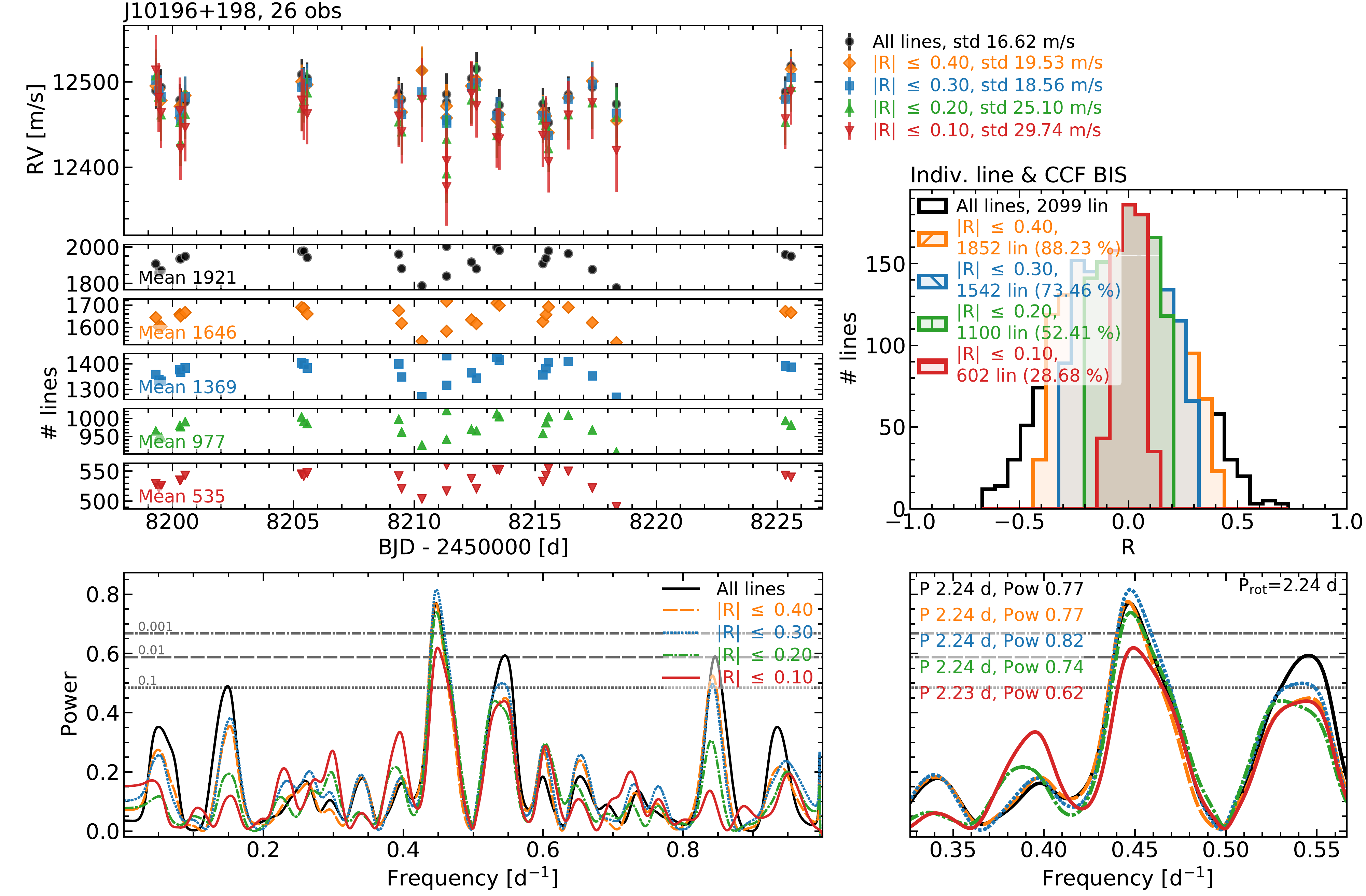}
\caption[Same as Fig. \ref{fig:tsnhpccfbisinactivebestJ07446+035}, but for J10196+198.]{Same as Fig. \ref{fig:tsnhpccfbisinactivebestJ07446+035}, but for J10196+198 and lines with no limit on the RV scatter.}
\label{fig:tsnhpccfbisinactivebestJ10196+198}
\end{figure*}

% J10196+198 RV TS, Rcoeff, periodogram, corr servalcrx
\begin{figure*}
\centering
\includegraphics[width=0.93\linewidth]{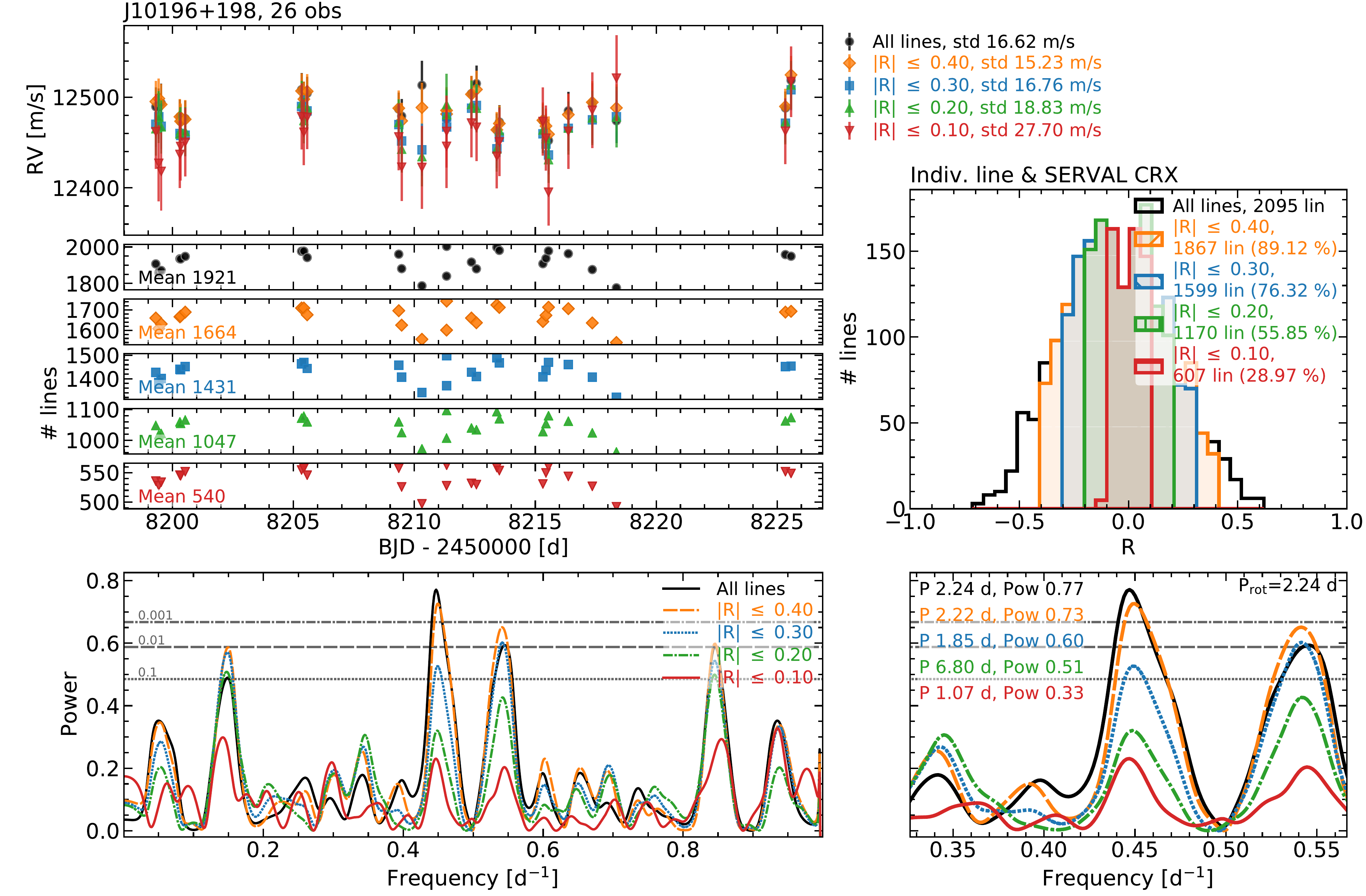}
\caption[Same as Fig. \ref{fig:tsnhpservalcrxinactivebestJ07446+035}, but for J10196+198.]{Same as Fig. \ref{fig:tsnhpservalcrxinactivebestJ07446+035}, but for J10196+198 and lines with no limit on the RV scatter.}
\label{fig:tsnhpservalcrxinactivebestJ10196+198}
\end{figure*}

% -------------

% J10196+198 RV TS, Rcoeff, periodogram, corr ccfrv active
\begin{figure*}
\centering
\includegraphics[width=0.93\linewidth]{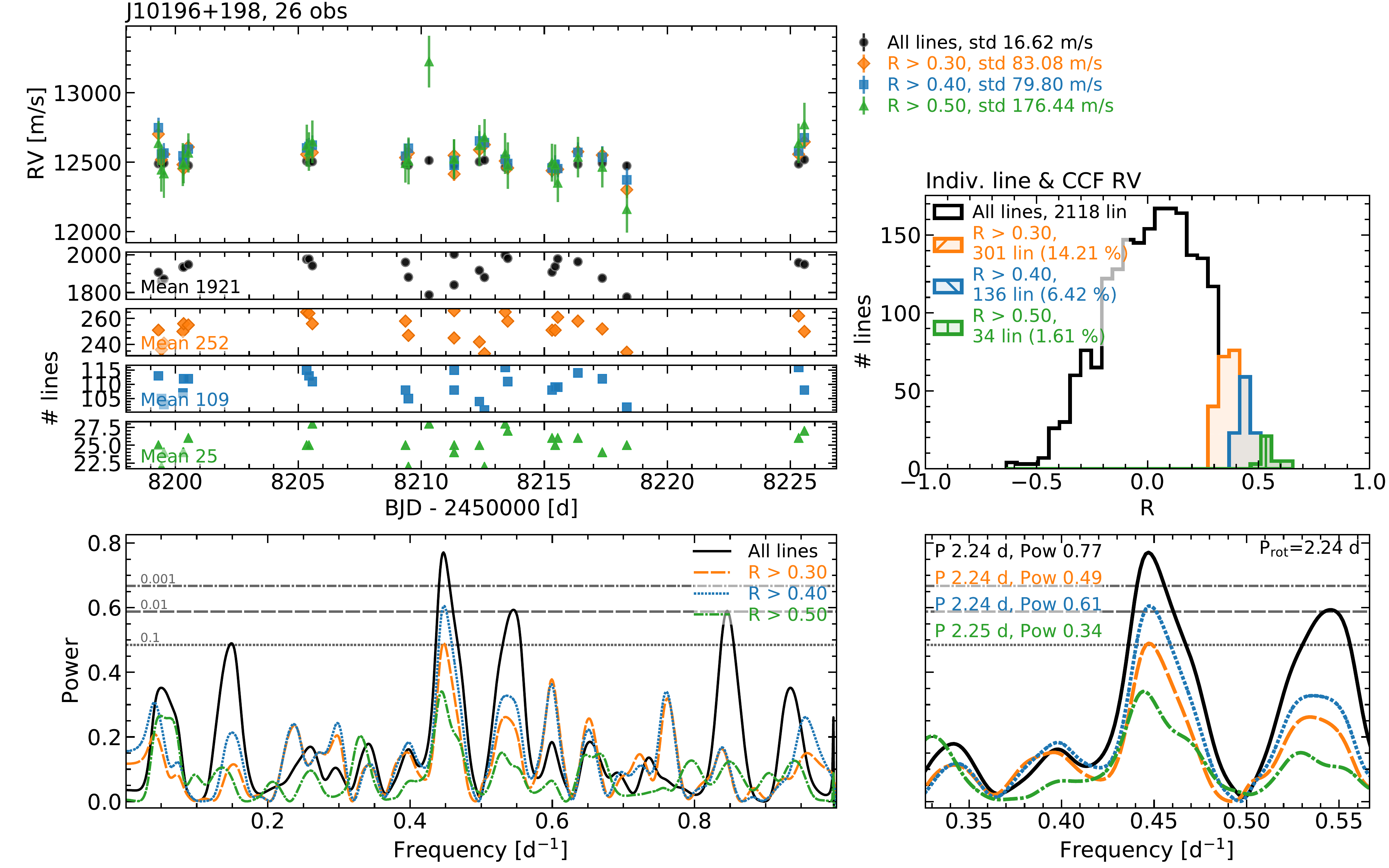}
\caption{Same as Fig. \ref{fig:tsnhpccfrvactiveJ07446+035}, but for J10196+198.}
\label{fig:tsnhpccfrvactiveJ10196+198}
\end{figure*}

% J10196+198 RV TS, Rcoeff, periodogram, corr ccfbis active
\begin{figure*}
\centering
\includegraphics[width=0.93\linewidth]{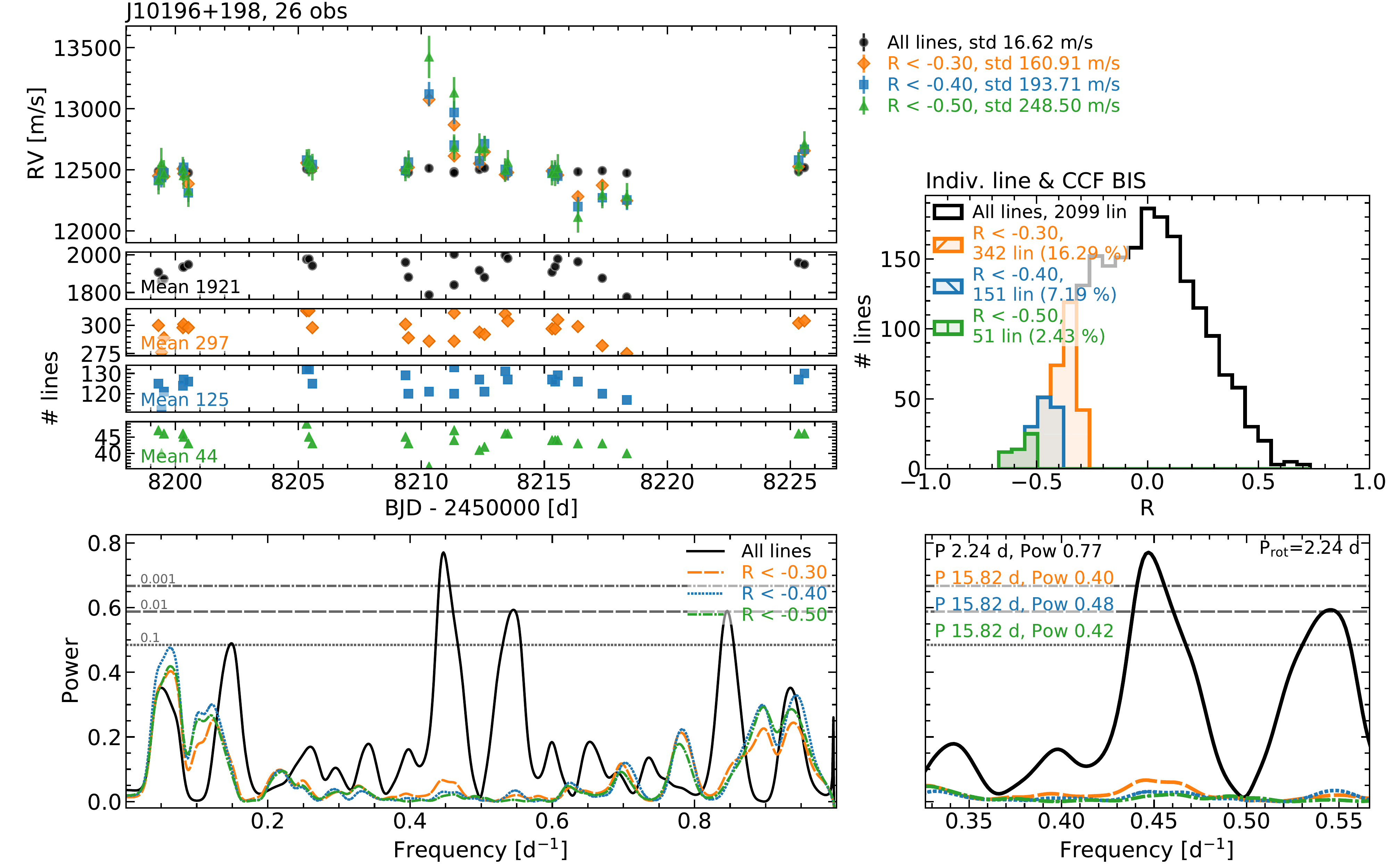}
\caption{Same as Fig. \ref{fig:tsnhpccfbisactiveJ07446+035}, but for J10196+198.}
\label{fig:tsnhpccfbisactiveJ10196+198}
\end{figure*}

% J10196+198 RV TS, Rcoeff, periodogram, corr servalcrx active
\begin{figure*}
\centering
\includegraphics[width=0.93\linewidth]{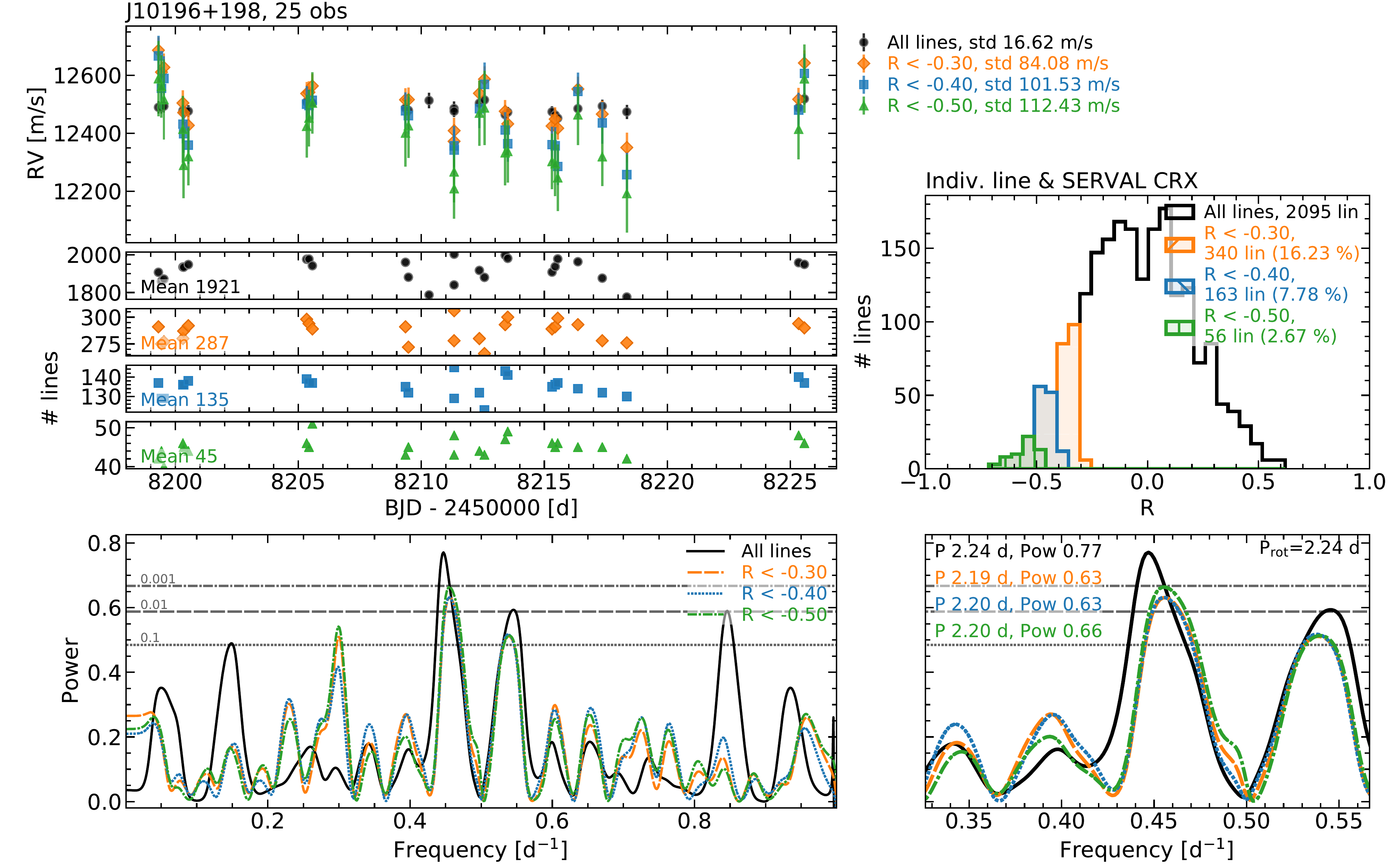}
\caption{Same as Fig. \ref{fig:tsnhpservalcrxactiveJ07446+035}, but for J10196+198.}
\label{fig:tsnhpservalcrxactiveJ10196+198}
\end{figure*}

% ------------------------------------------------

% J15218+209 2dscatter
\begin{figure*}
\centering
\begin{subfigure}[]{0.32\linewidth}
\centering
\includegraphics[width=\textwidth]{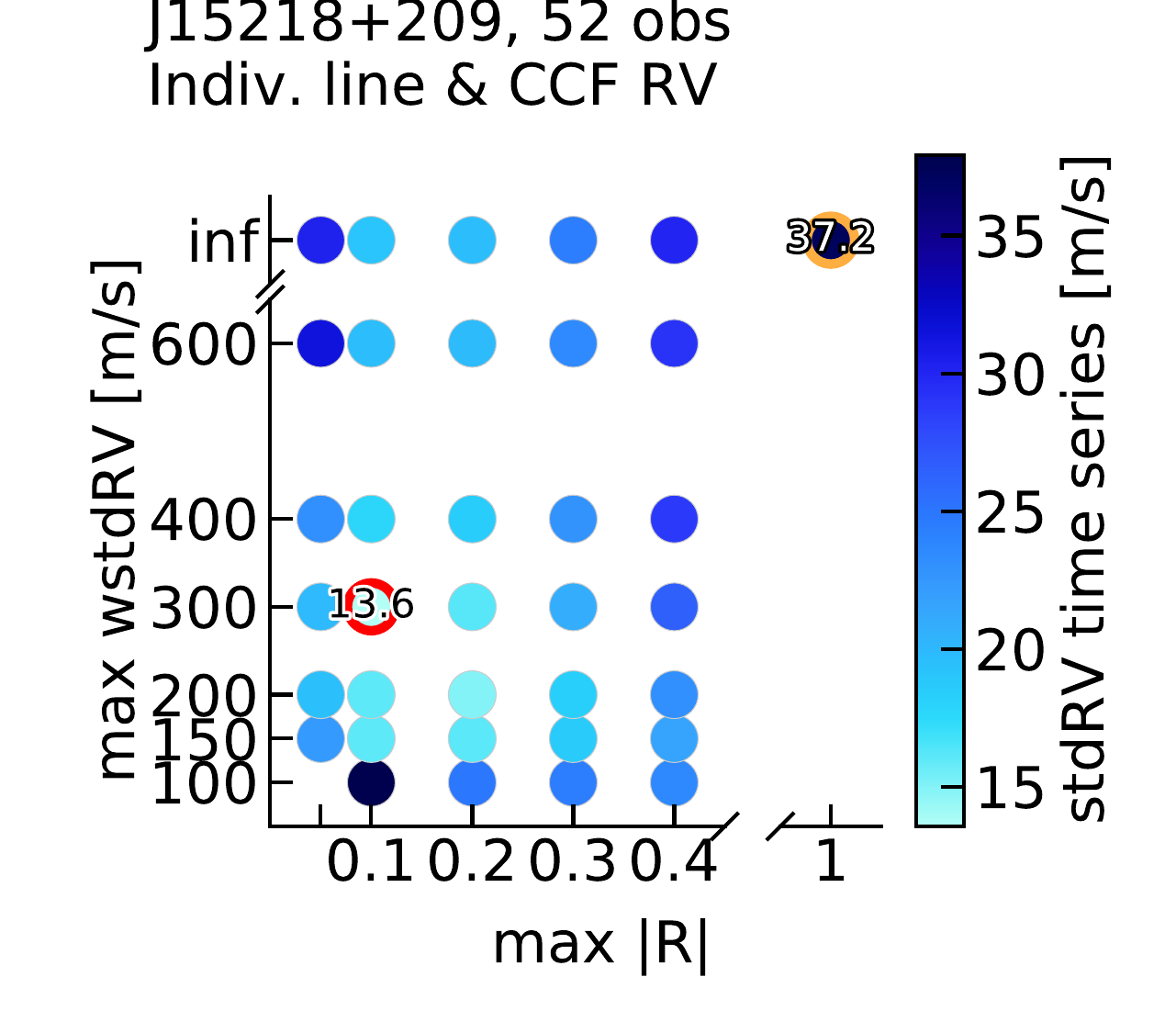}
\end{subfigure}
\,
\begin{subfigure}[]{0.32\linewidth}
\centering
\includegraphics[width=\textwidth]{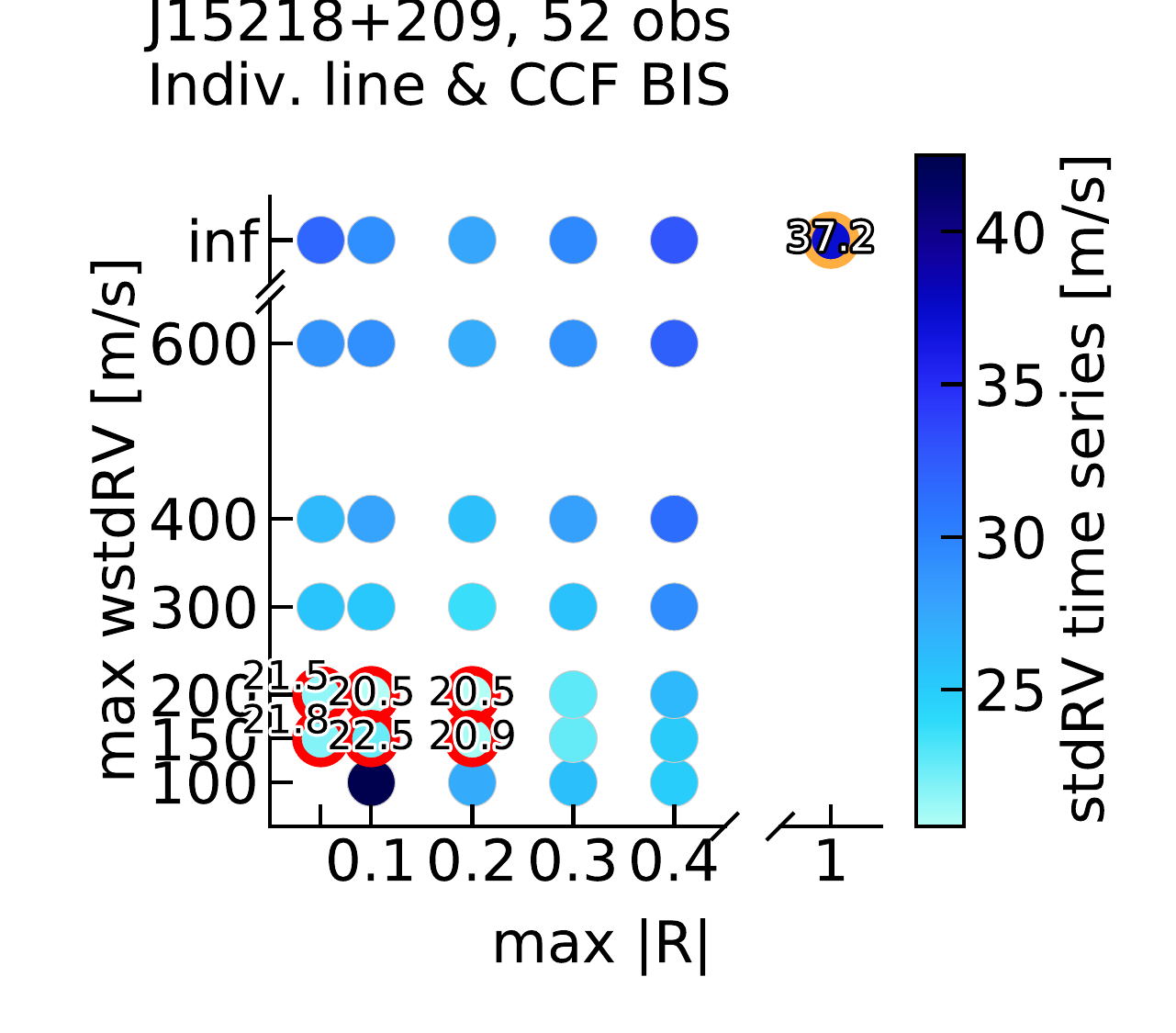}
\end{subfigure}
\,
\begin{subfigure}[]{0.32\linewidth}
\centering
\includegraphics[width=\textwidth]{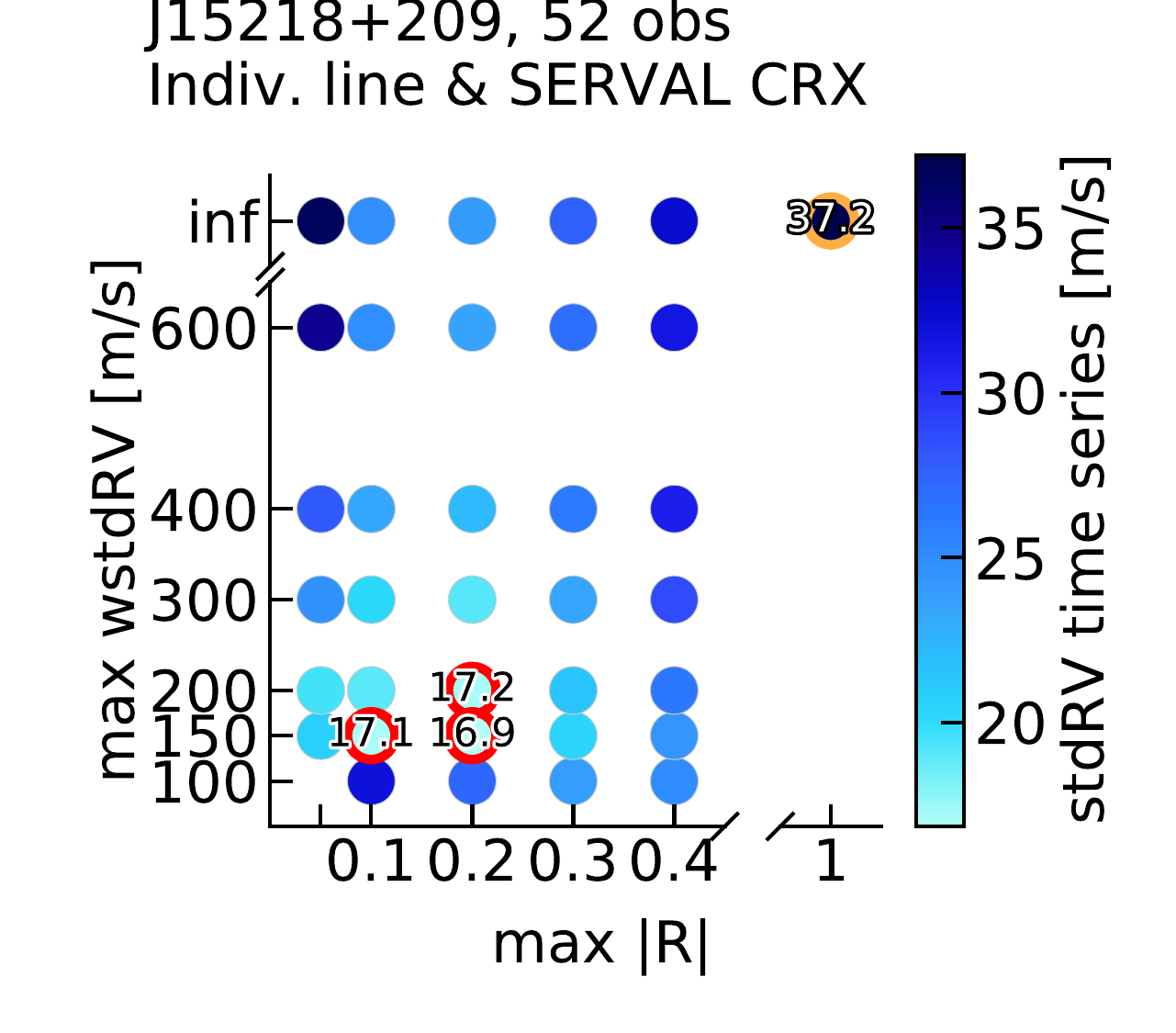}
\end{subfigure}
\,
\caption{Same as Fig. \ref{fig:cuts2dinactiveJ07446+035}, but for J15218+209.}
\label{fig:cuts2dinactiveJ15218+209}
\end{figure*}

% -------------

% J15218+209 RV TS all lines, CCF, SERVAL
\begin{figure*}
\centering
\includegraphics[width=0.93\linewidth]{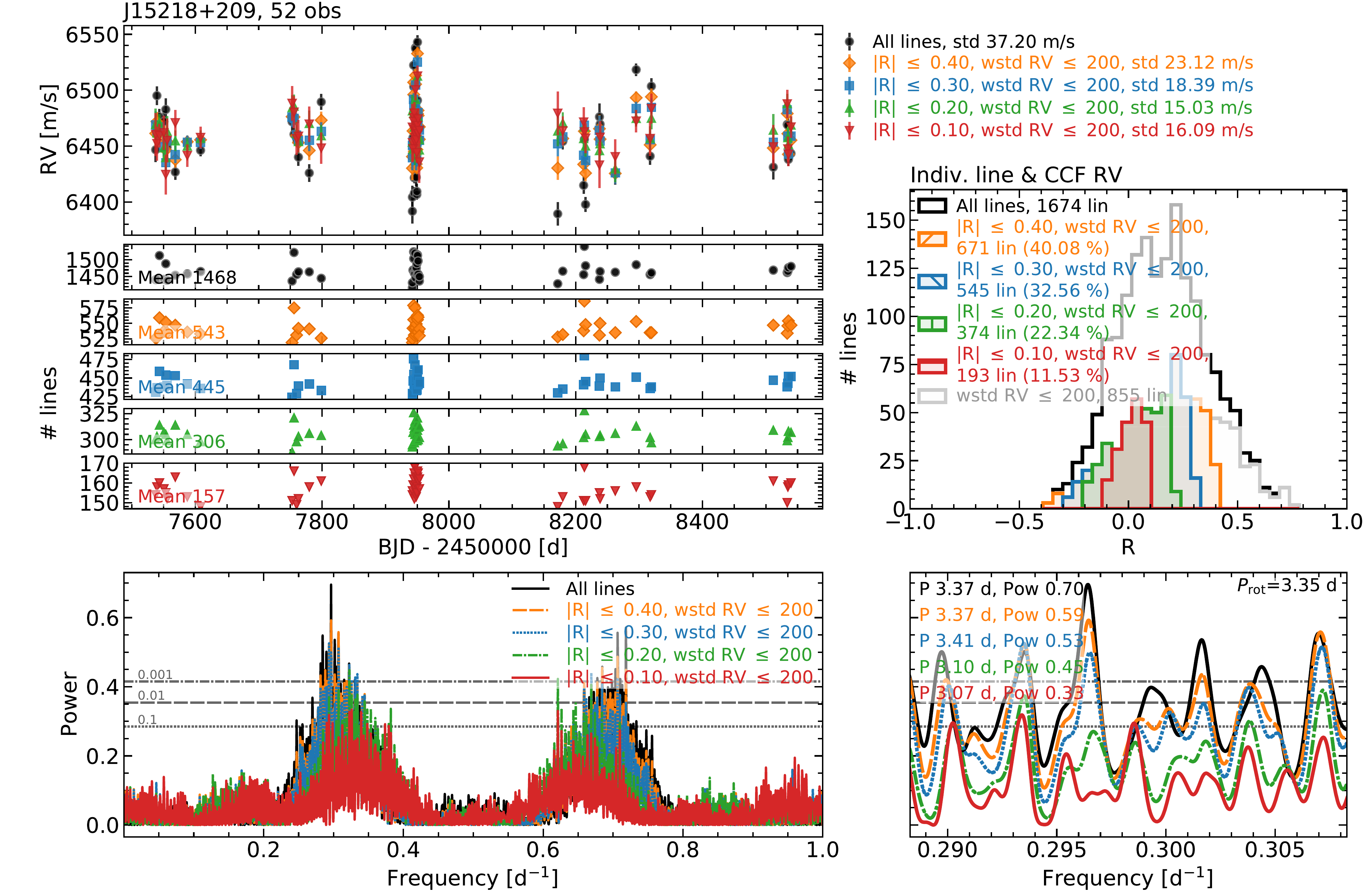}
\caption{Same as Fig. \ref{fig:tsnhpccfrvinactivebestJ07446+035}, but for J15218+209.}
\label{fig:tsnhpccfrvinactivebestJ15218+209}
\end{figure*}

% J15218+209 RV TS, Rcoeff, periodogram, corr ccfbis
\begin{figure*}
\centering
\includegraphics[width=0.93\linewidth]{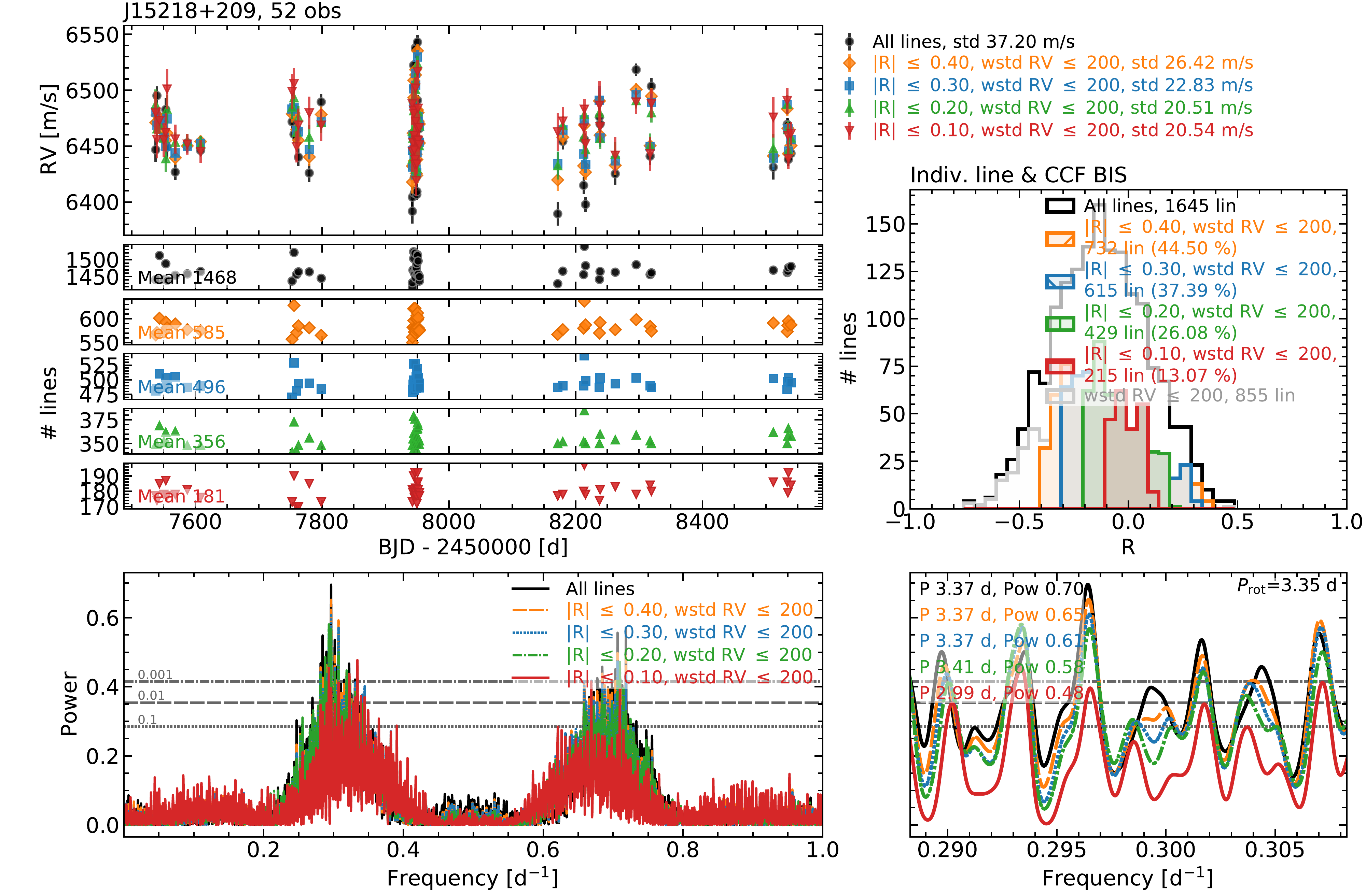}
\caption{Same as Fig. \ref{fig:tsnhpccfbisinactivebestJ07446+035}, but for J15218+209.}
\label{fig:tsnhpccfbisinactivebestJ15218+209}
\end{figure*}

% J15218+209 RV TS, Rcoeff, periodogram, corr servalcrx
\begin{figure*}
\centering
\includegraphics[width=0.93\linewidth]{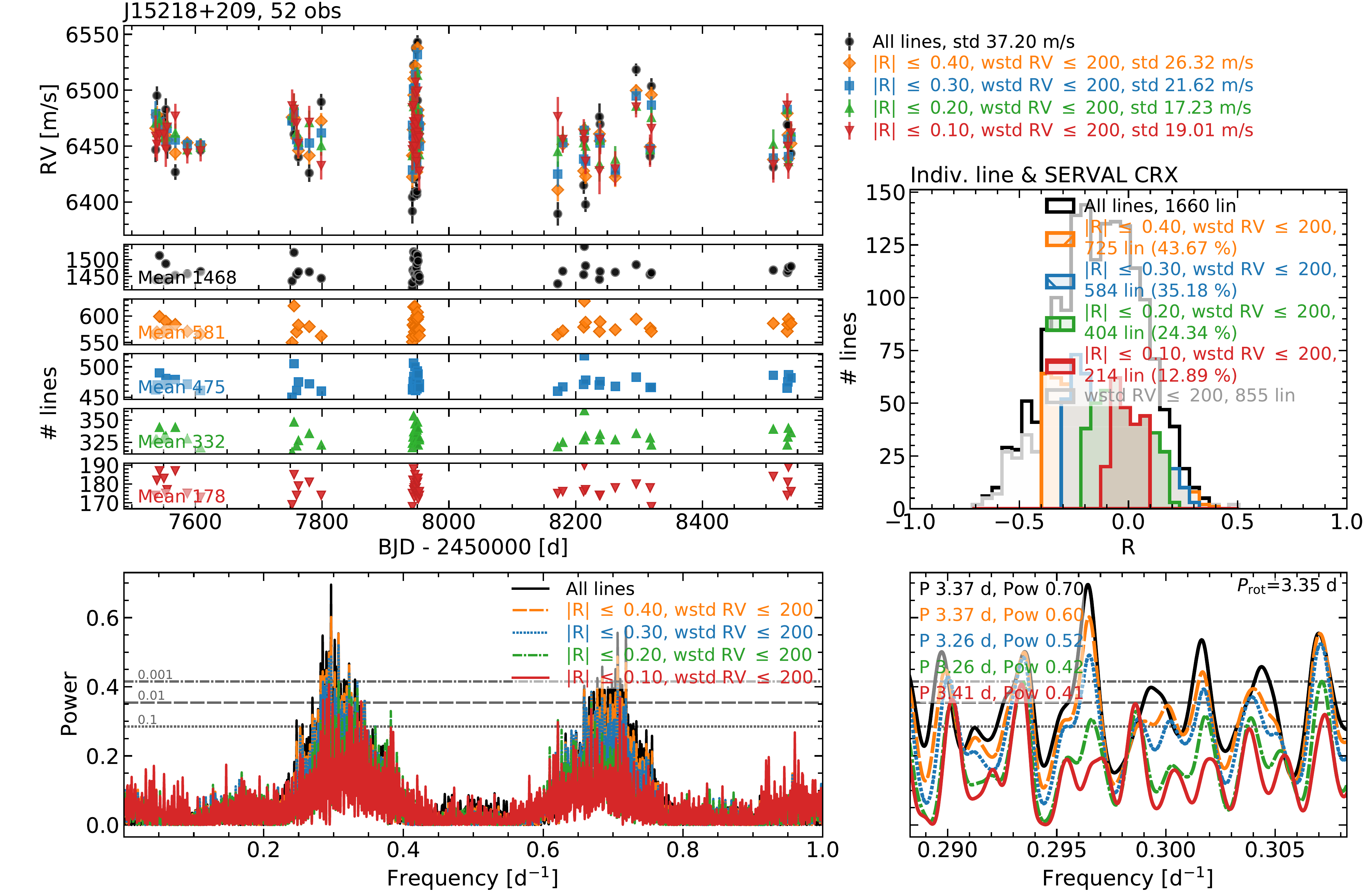}
\caption{Same as Fig. \ref{fig:tsnhpservalcrxinactivebestJ07446+035}, but for J15218+209.}
\label{fig:tsnhpservalcrxinactivebestJ15218+209}
\end{figure*}

% -------------

% J15218+209 RV TS, Rcoeff, periodogram, corr ccfrv active
\begin{figure*}
\centering
\includegraphics[width=0.93\linewidth]{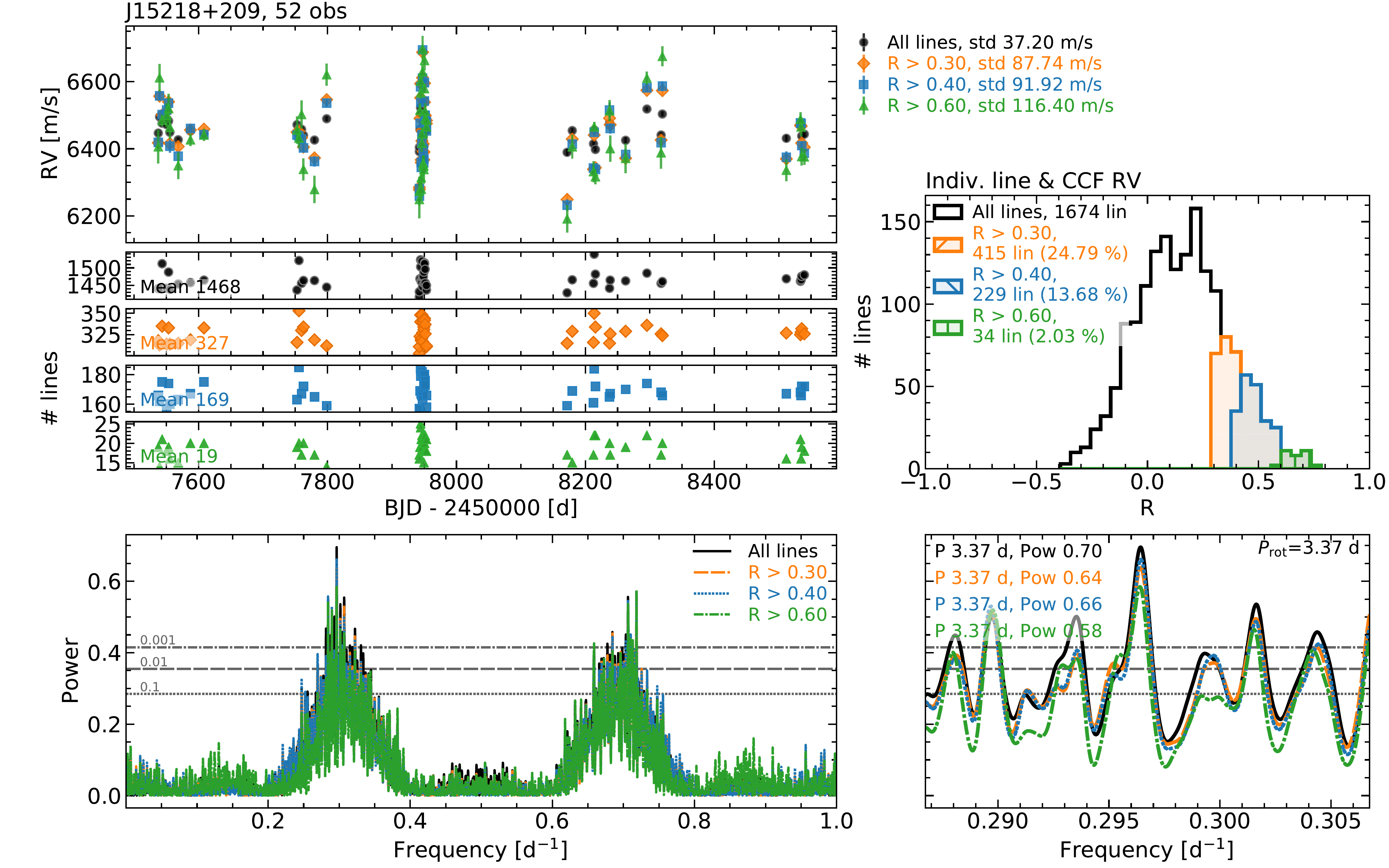}
\caption{Same as Fig. \ref{fig:tsnhpccfrvactiveJ07446+035}, but for J15218+209.}
\label{fig:tsnhpccfrvactiveJ15218+209}
\end{figure*}

% J15218+209 RV TS, Rcoeff, periodogram, corr ccfbis active
\begin{figure*}
\centering
\includegraphics[width=0.93\linewidth]{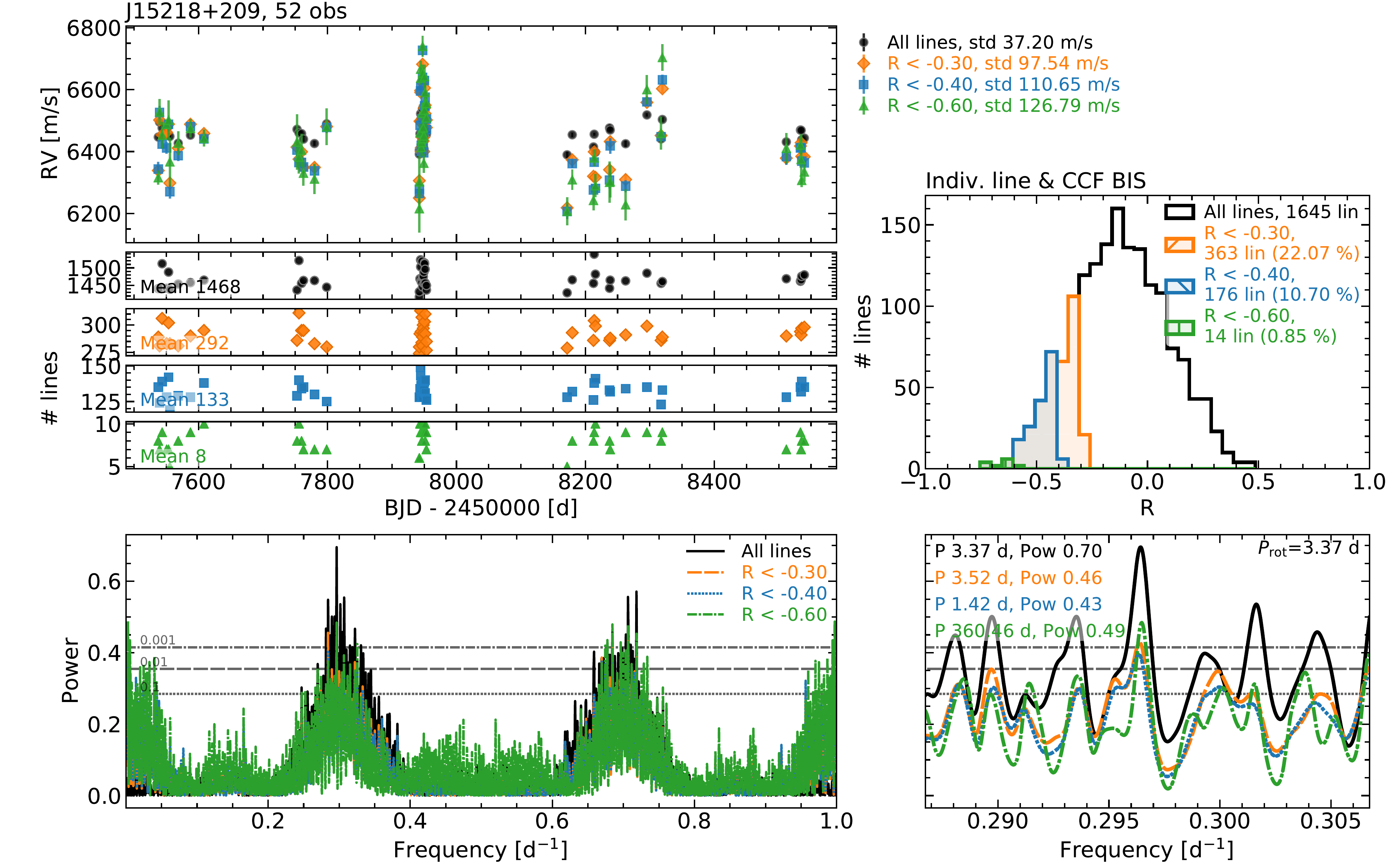}
\caption{Same as Fig. \ref{fig:tsnhpccfbisactiveJ07446+035}, but for J15218+209.}
\label{fig:tsnhpccfbisactiveJ15218+209}
\end{figure*}

% J15218+209 RV TS, Rcoeff, periodogram, corr servalcrx active
\begin{figure*}
\centering
\includegraphics[width=0.93\linewidth]{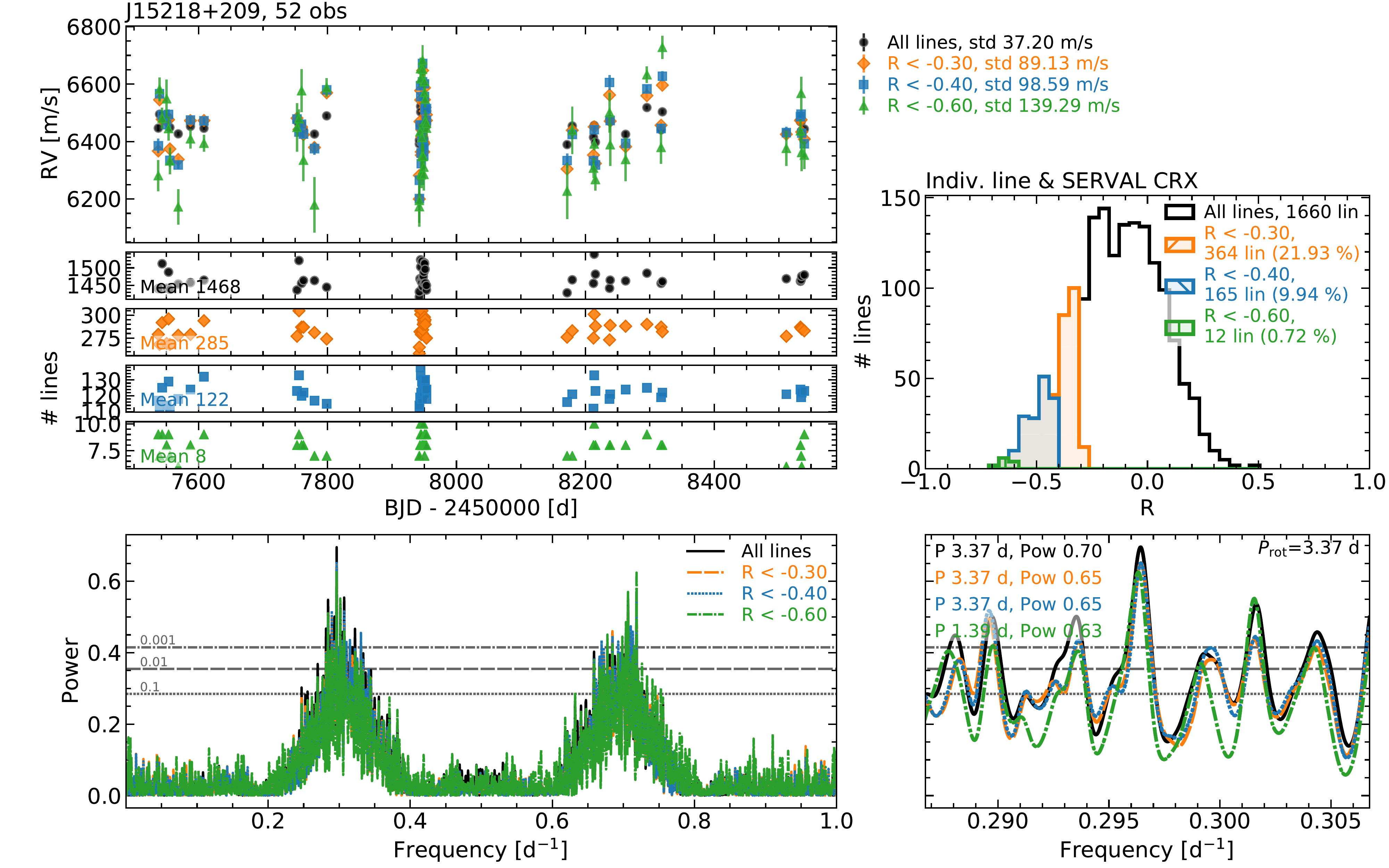}
\caption{Same as Fig. \ref{fig:tsnhpservalcrxactiveJ07446+035}, but for J15218+209.}
\label{fig:tsnhpservalcrxactiveJ15218+209}
\end{figure*}

% ------------------------------------------------

% J11201--104 2dscatter
\begin{figure*}
\centering
\begin{subfigure}[]{0.32\linewidth}
\centering
\includegraphics[width=\textwidth]{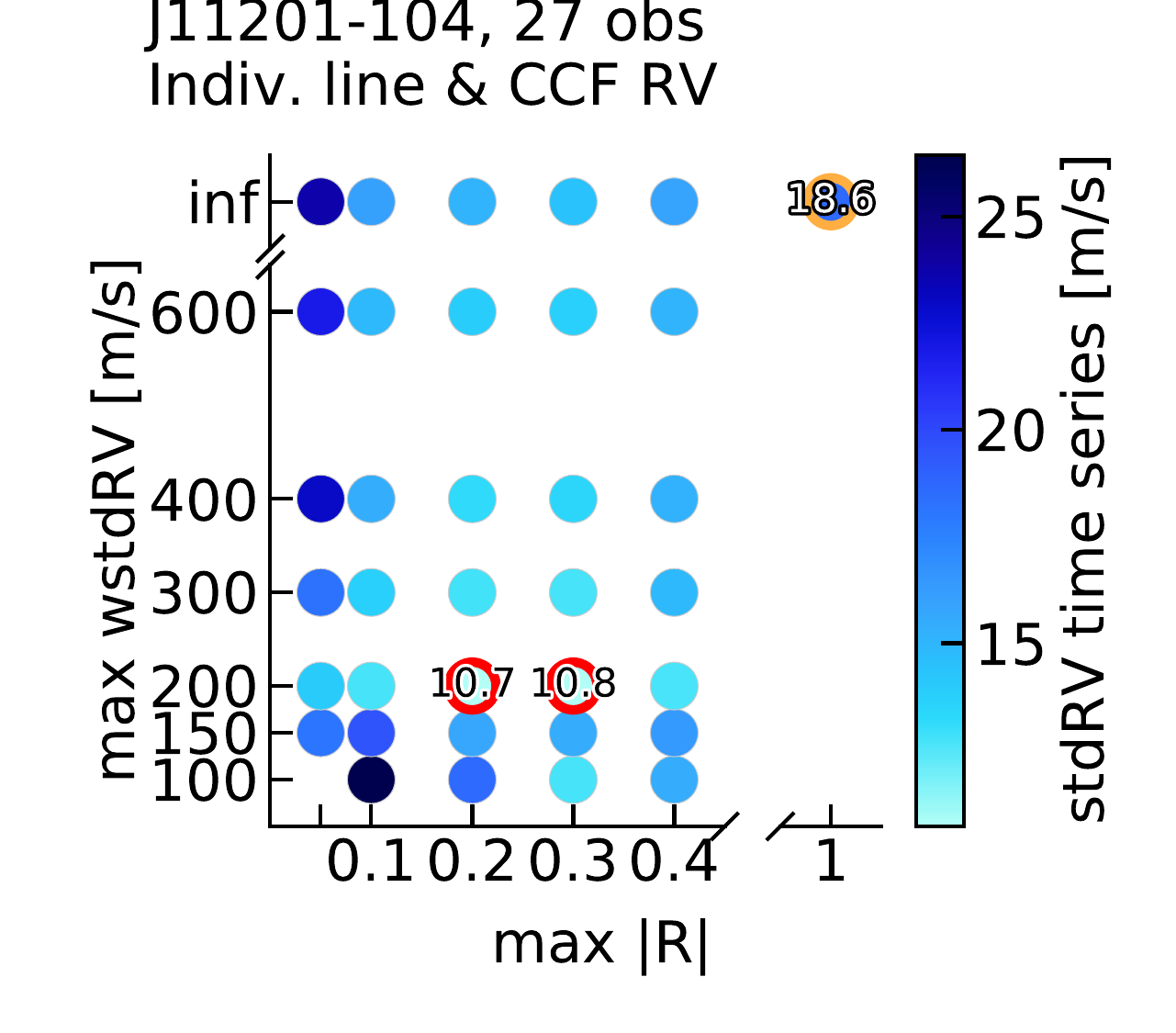}
\end{subfigure}
\,
\begin{subfigure}[]{0.32\linewidth}
\centering
\includegraphics[width=\textwidth]{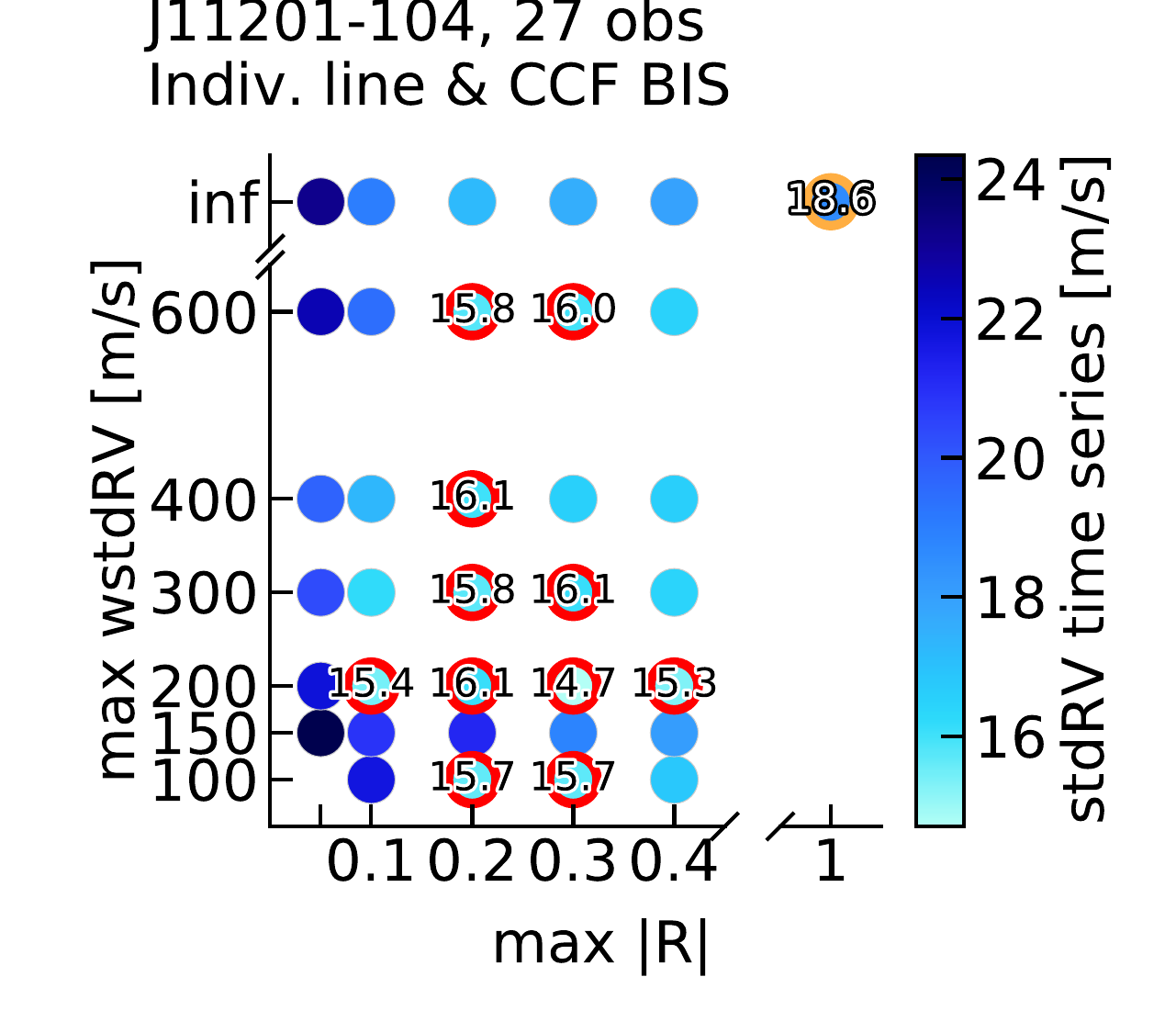}
\end{subfigure}
\,
\begin{subfigure}[]{0.32\linewidth}
\centering
\includegraphics[width=\textwidth]{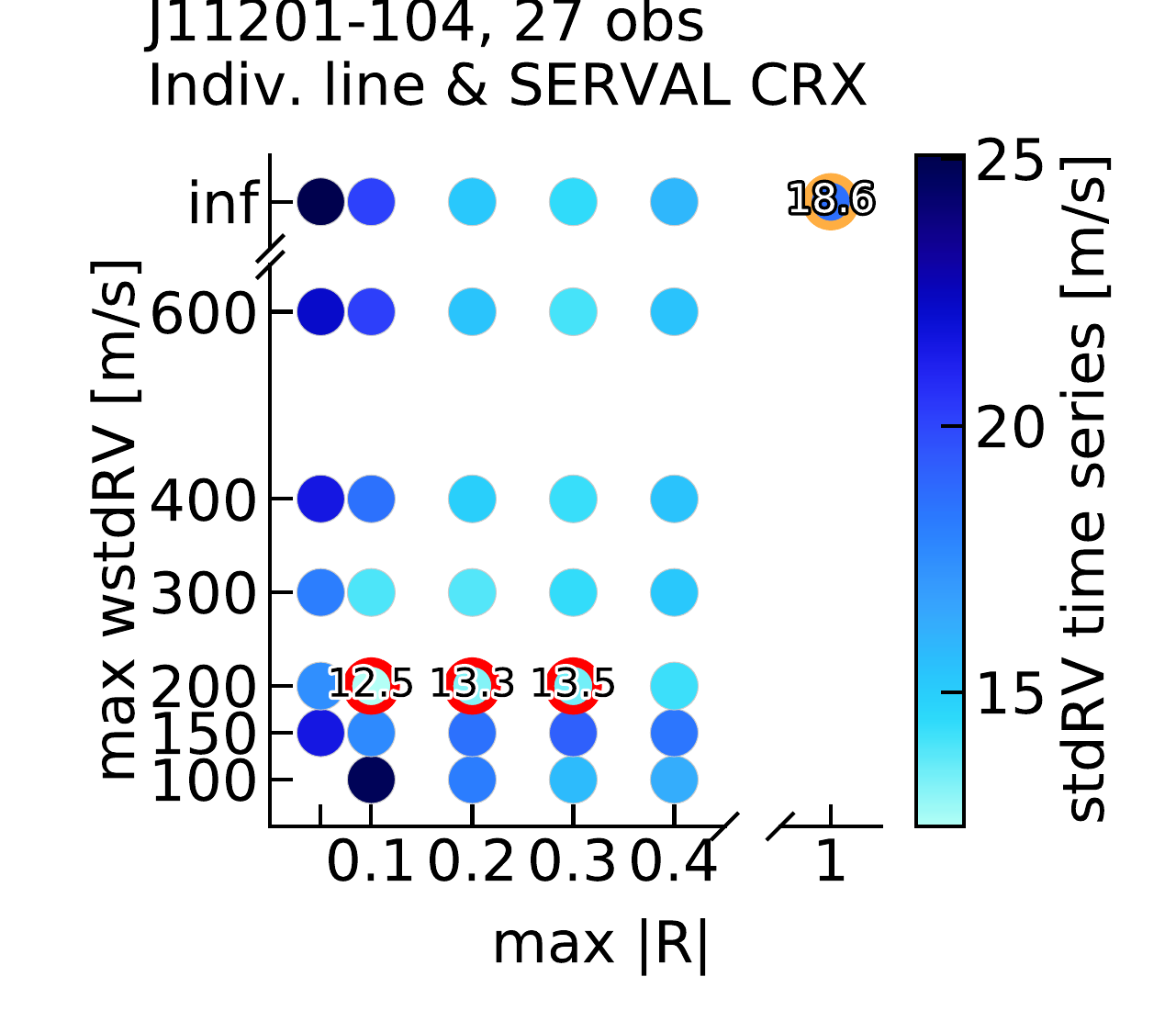}
\end{subfigure}
\,
\caption{Same as Fig. \ref{fig:cuts2dinactiveJ07446+035}, but for J11201--104.}
\label{fig:cuts2dinactiveJ11201--104}
\end{figure*}

% -------------

% J11201--104 RV TS all lines, CCF, SERVAL
\begin{figure*}
\centering
\includegraphics[width=0.93\linewidth]{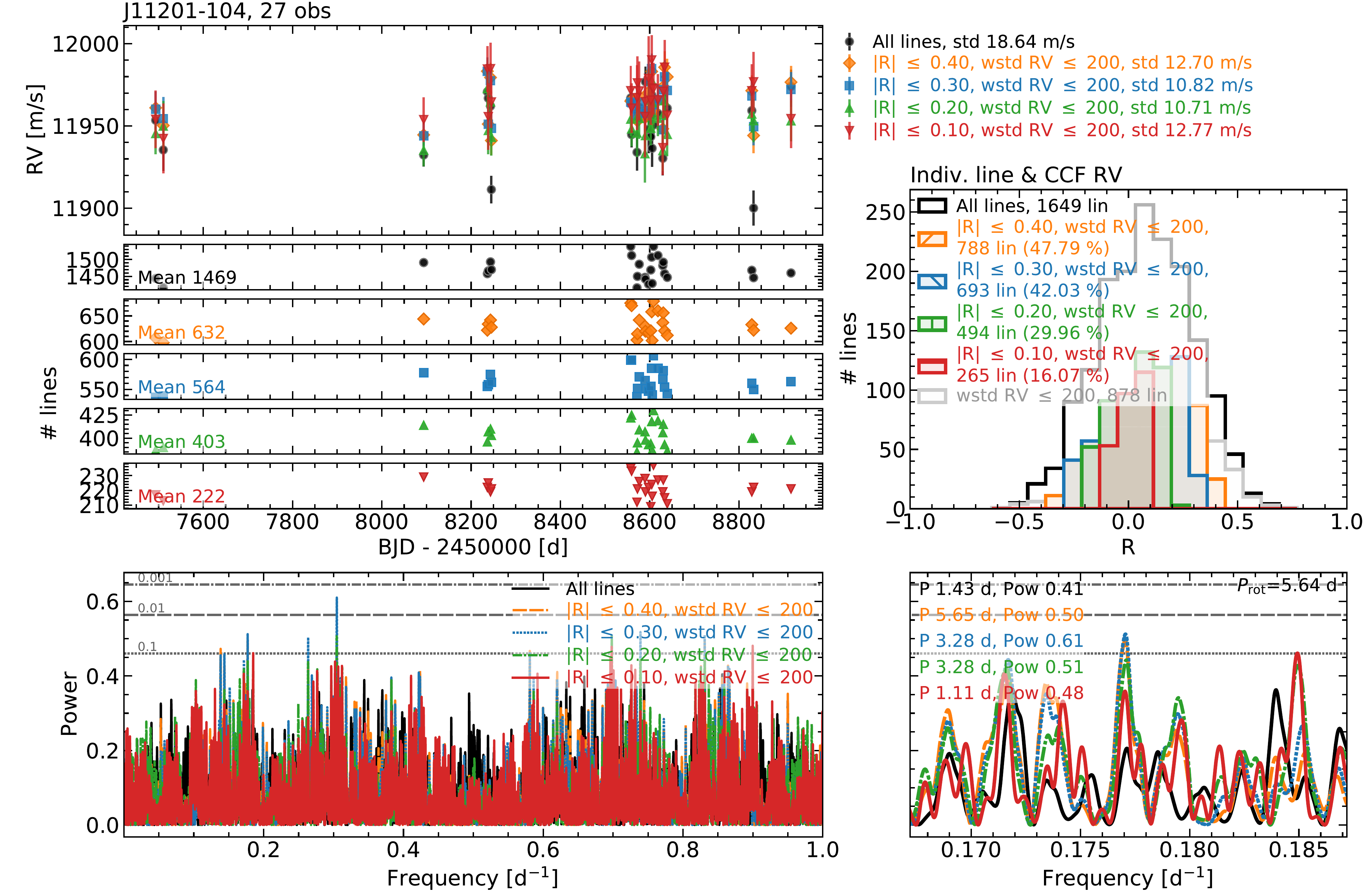}
\caption[Same as Fig. \ref{fig:tsnhpccfrvinactivebestJ07446+035}, but for J11201--104.]{Same as Fig. \ref{fig:tsnhpccfrvinactivebestJ07446+035}, but for J11201--104. Since the rotation period of this star is not known, the periodogram zoom in shows the region around the highest peak in the RV time series obtained using the line averages.}
\label{fig:tsnhpccfrvinactivebestJ11201--104}
\end{figure*}

% J11201--104 RV TS, Rcoeff, periodogram, corr ccfbis
\begin{figure*}
\centering
\includegraphics[width=0.93\linewidth]{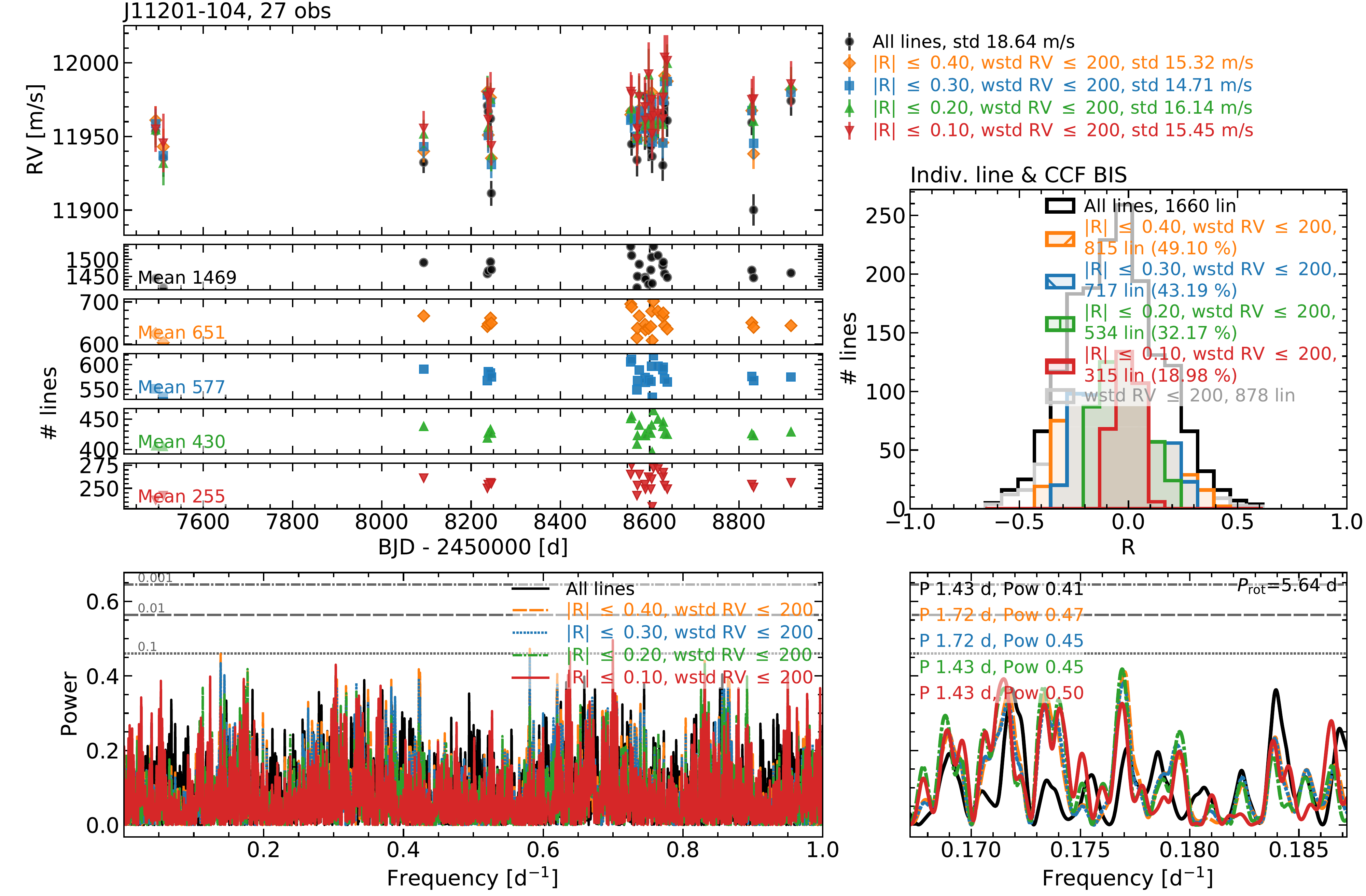}
\caption[Same as Fig. \ref{fig:tsnhpccfbisinactivebestJ07446+035}, but for J11201--104.]{Same as Fig. \ref{fig:tsnhpccfbisinactivebestJ07446+035}, but for J11201--104. Since the rotation period of this star is not known, the periodogram zoom in shows the region around the highest peak in the RV time series obtained using the line averages.}
\label{fig:tsnhpccfbisinactivebestJ11201--104}
\end{figure*}

% J11201--104 RV TS, Rcoeff, periodogram, corr servalcrx
\begin{figure*}
\centering
\includegraphics[width=0.93\linewidth]{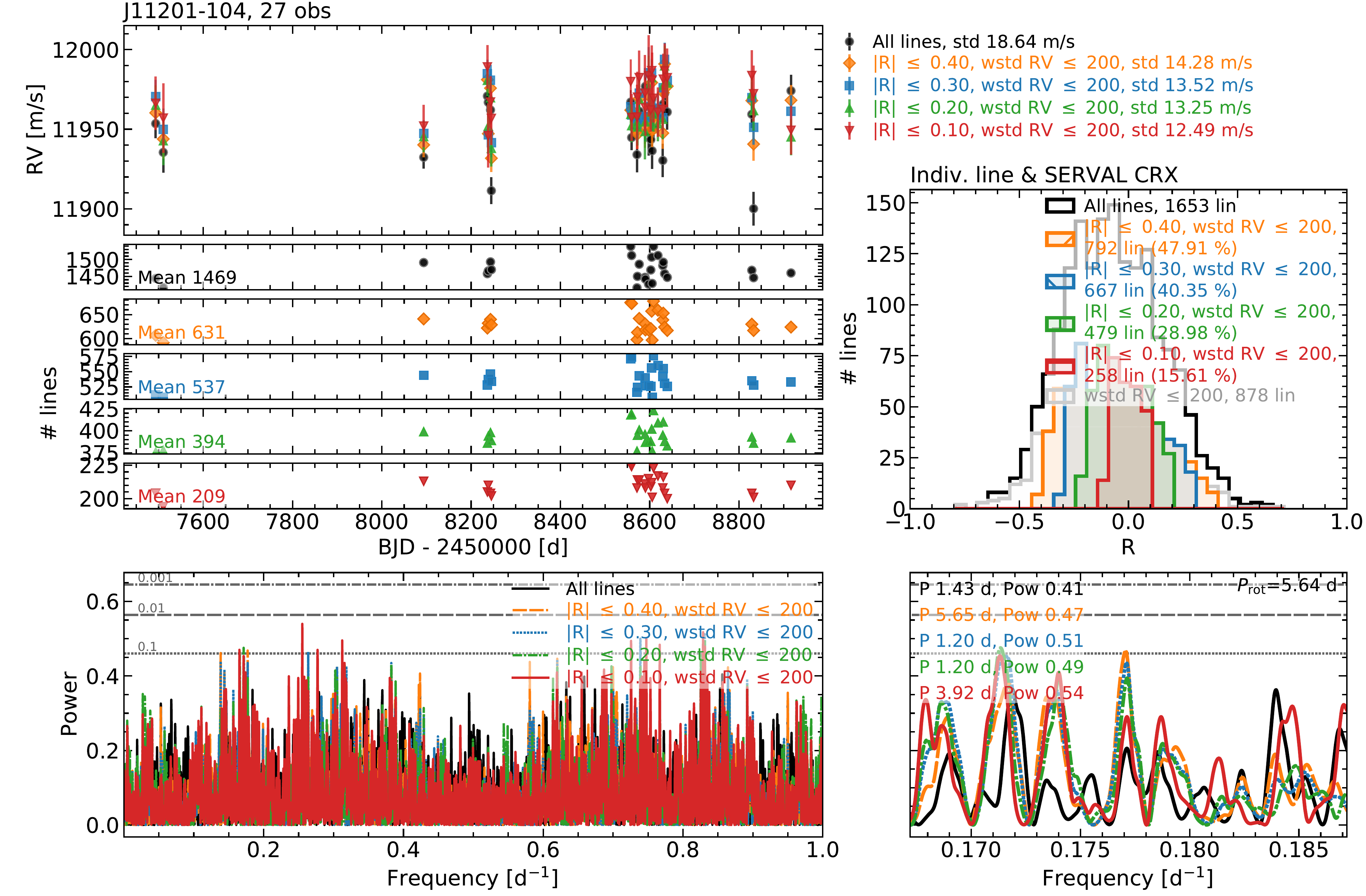}
\caption[Same as Fig. \ref{fig:tsnhpservalcrxinactivebestJ07446+035}, but for J11201--104.]{Same as Fig. \ref{fig:tsnhpservalcrxinactivebestJ07446+035}, but for J11201--104. Since the rotation period of this star is not known, the periodogram zoom in shows the region around the highest peak in the RV time series obtained using the line averages.}
\label{fig:tsnhpservalcrxinactivebestJ11201--104}
\end{figure*}

% -------------

% J11201--104 RV TS, Rcoeff, periodogram, corr ccfrv active
\begin{figure*}
\centering
\includegraphics[width=0.93\linewidth]{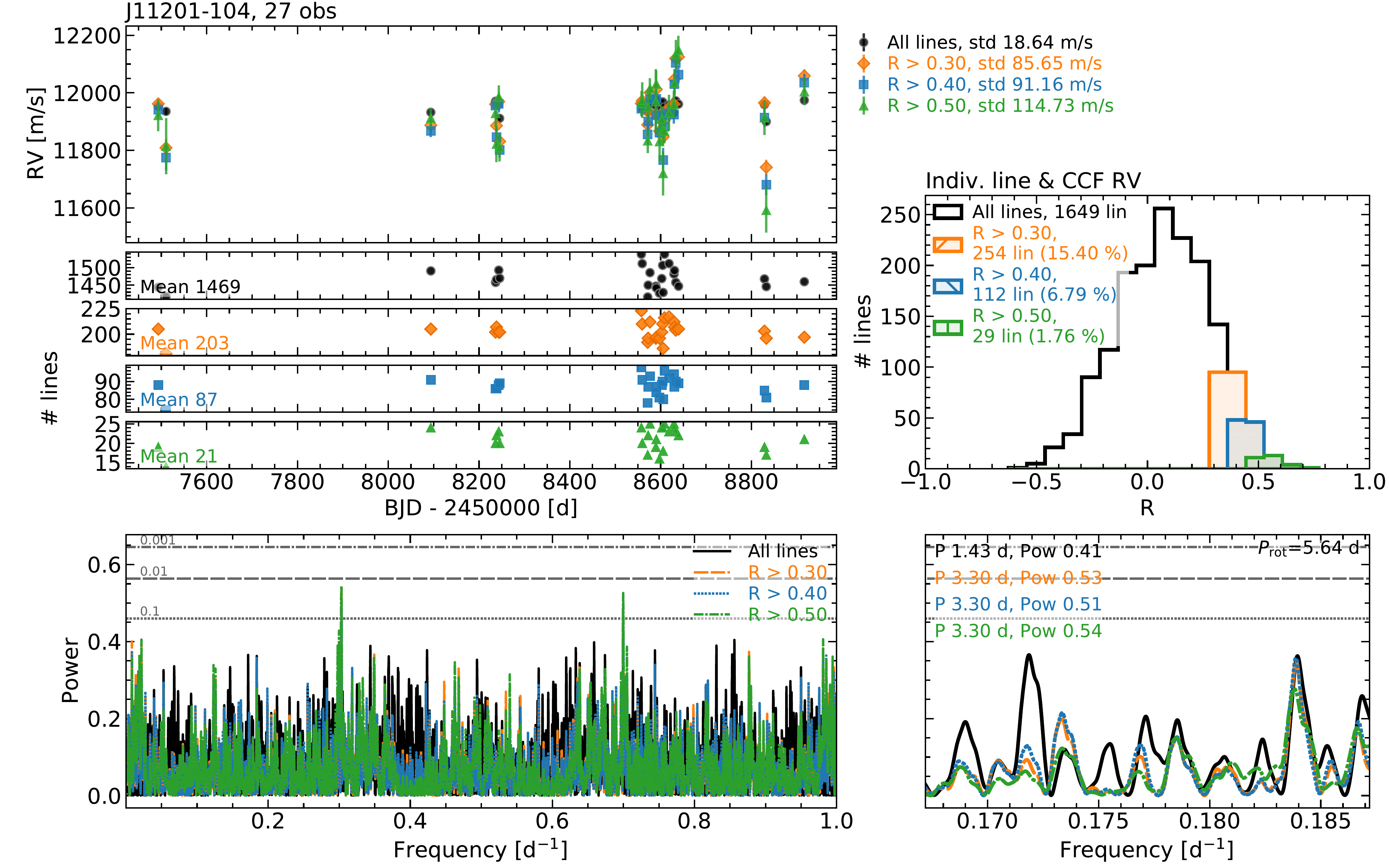}
\caption{Same as Fig. \ref{fig:tsnhpccfrvactiveJ07446+035}, but for J11201--104.}
\label{fig:tsnhpccfrvactiveJ11201--104}
\end{figure*}

% J11201--104 RV TS, Rcoeff, periodogram, corr ccfbis active
\begin{figure*}
\centering
\includegraphics[width=0.93\linewidth]{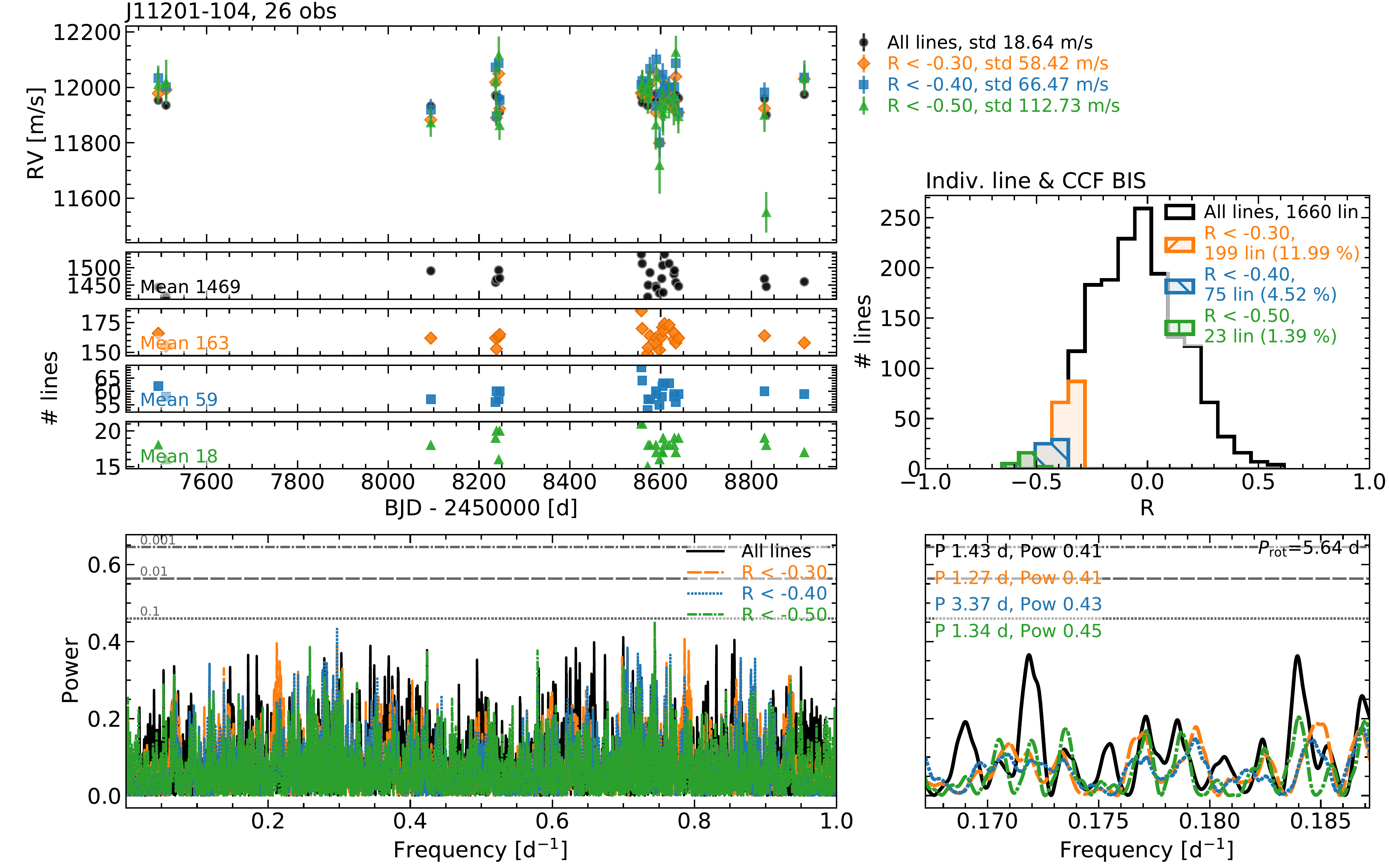}
\caption{Same as Fig. \ref{fig:tsnhpccfbisactiveJ07446+035}, but for J11201--104.}
\label{fig:tsnhpccfbisactiveJ11201--104}
\end{figure*}

% J11201--104 RV TS, Rcoeff, periodogram, corr servalcrx active
\begin{figure*}
\centering
\includegraphics[width=0.93\linewidth]{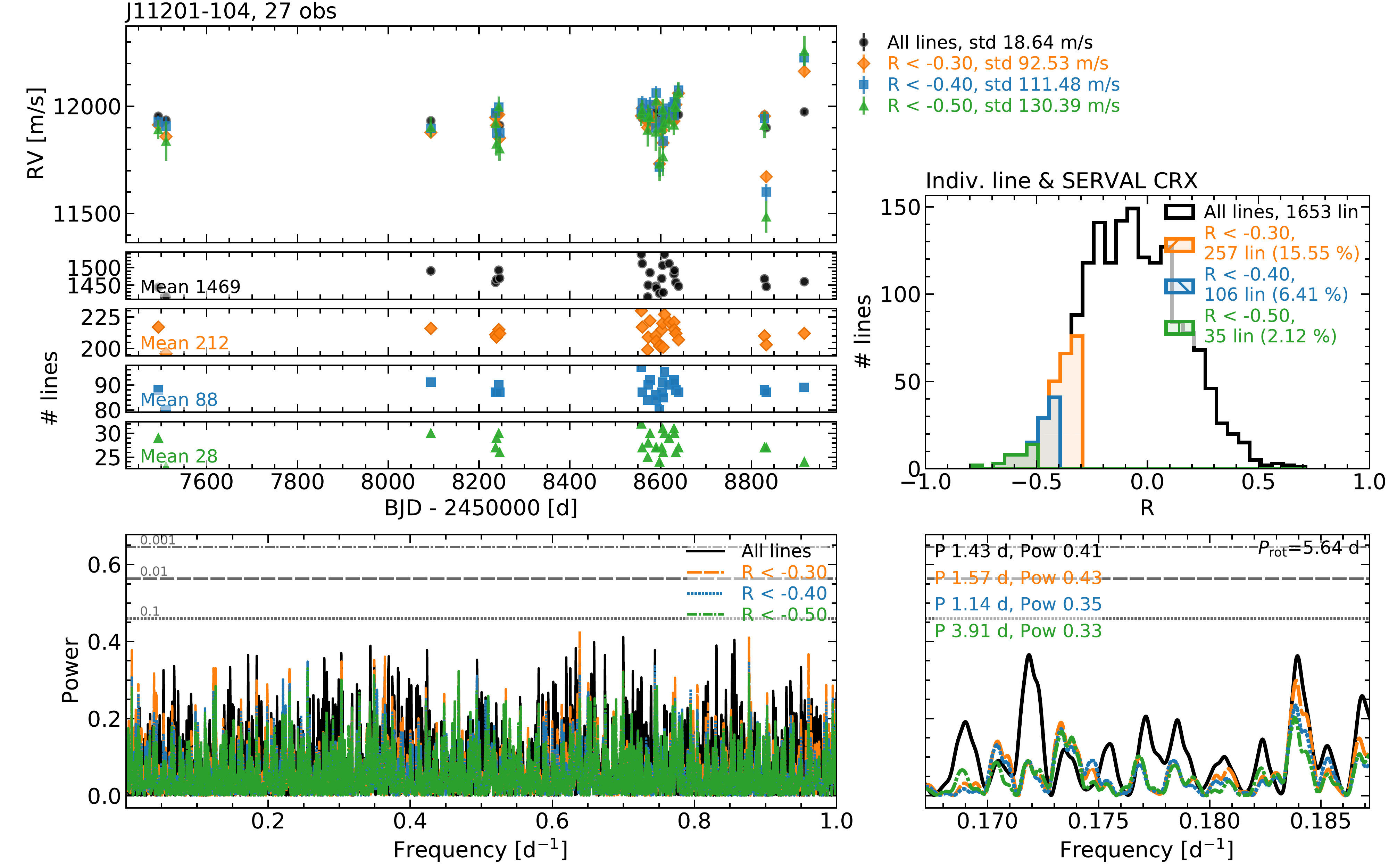}
\caption{Same as Fig. \ref{fig:tsnhpservalcrxactiveJ07446+035}, but for J11201--104.}
\label{fig:tsnhpservalcrxactiveJ11201--104}
\end{figure*}

%---------------------------------------------------------------------

\section{Lines in different stars: RV periodograms}\label{sec:app_indlincomparisonstar}

% J07446+035 and J05019+011
\begin{figure*}
\centering
\includegraphics[width=0.93\linewidth]{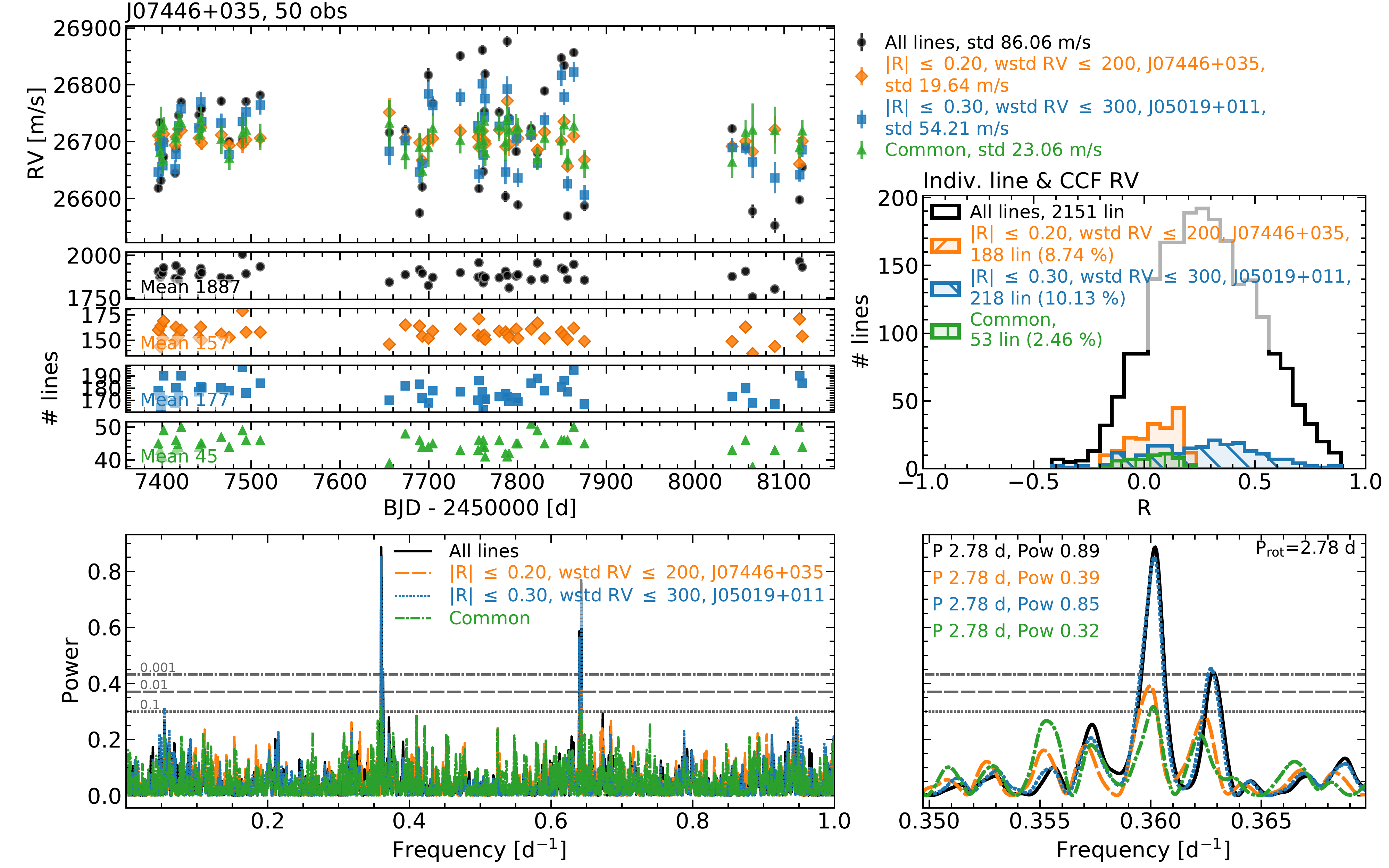}
\caption{Same as Fig. \ref{fig:tsnhpccfrvinactivebestJ07446+035}, but using the following datsets: initial line list (black), lines that minimise the RV scatter of J07446+035 (orange), lines that minimise the RV scatter of J05019+011 (blue) and common lines in the two previous sets (green).}
\label{fig:tslinselcomparisonJ07446+035_cutJ05019+011}
\end{figure*}

\begin{figure*}
\centering
\includegraphics[width=0.93\linewidth]{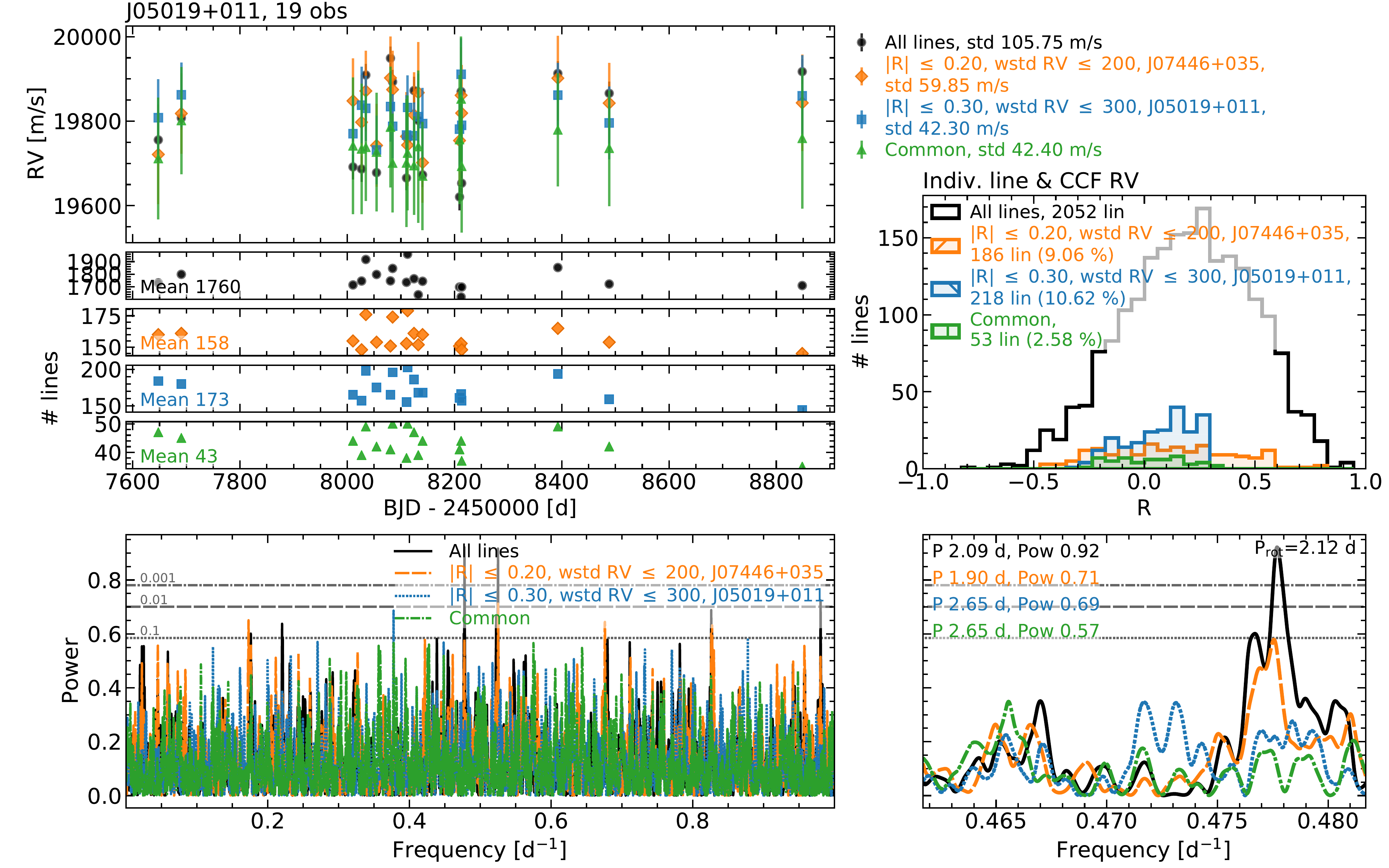}
\caption{Same as \ref{fig:tslinselcomparisonJ07446+035_cutJ05019+011}, but for J05019+011.}
\label{fig:tslinselcomparisonJ05019+011_cutJ07446+035}
\end{figure*}

% ------------------------------------------------

% J07446+035 and J22468+443
\begin{figure*}
\centering
\includegraphics[width=0.93\linewidth]{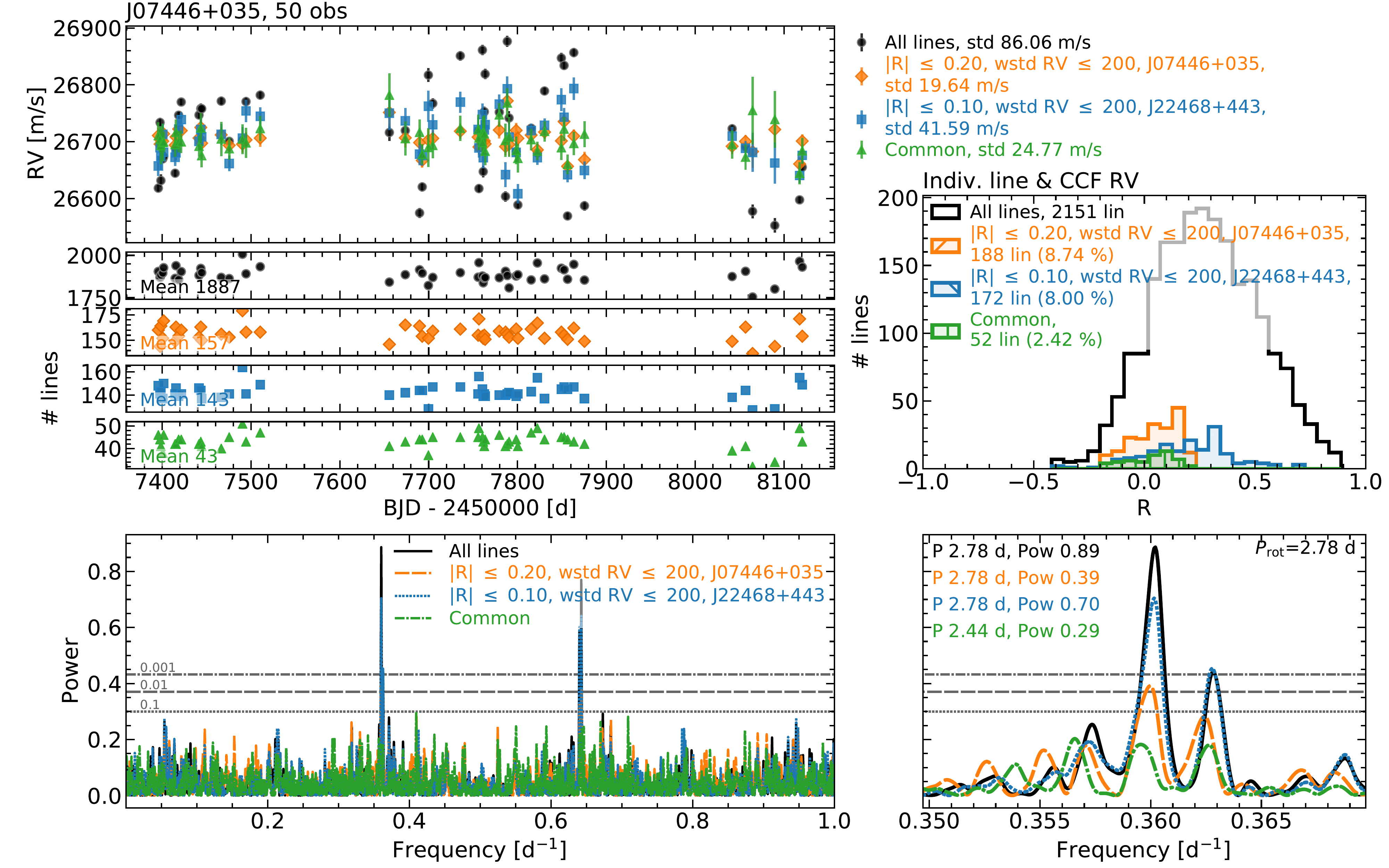}
\caption[Same as \ref{fig:tslinselcomparisonJ07446+035_cutJ05019+011} but using the initial line list, lines that minimise the RV scatter of J07446+035, lines that minimise the RV scatter of J22468+443 and common lines in the two previous sets.]{Same as \ref{fig:tslinselcomparisonJ07446+035_cutJ05019+011}, but using the following datsets: initial line list (black), lines that minimise the RV scatter of J07446+035 (orange), lines that minimise the RV scatter of J22468+443 (blue) and common lines in the two previous sets (green).}
\label{fig:tslinselcomparisonJ07446+035_cutJ22468+443}
\end{figure*}

\begin{figure*}
\centering
\includegraphics[width=0.93\linewidth]{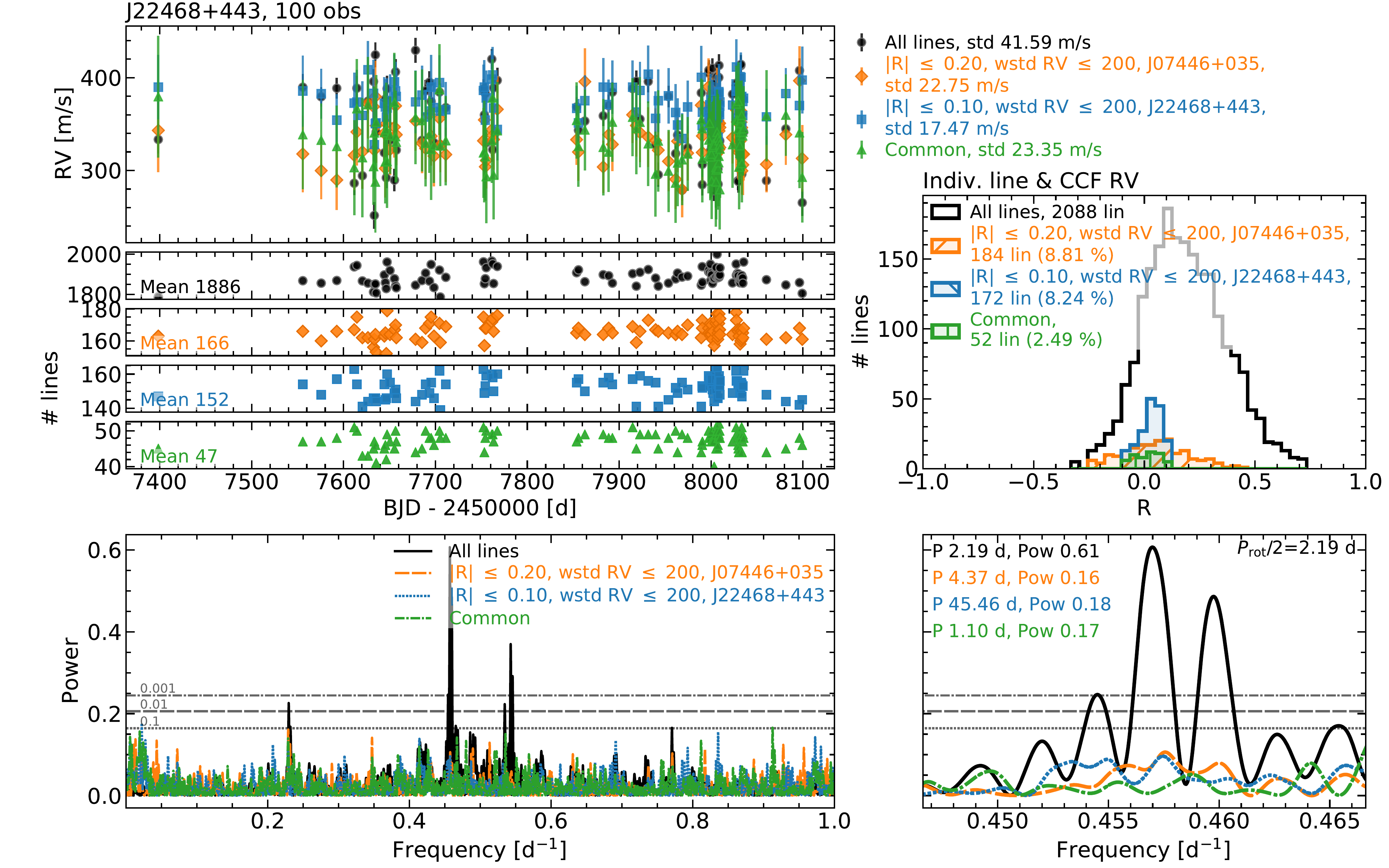}
\caption{Same as Fig. \ref{fig:tslinselcomparisonJ07446+035_cutJ22468+443}, but for J22468+443.}
\label{fig:tslinselcomparisonJ22468+443_cutJ07446+035}
\end{figure*}

% ------------------------------------------------

% J22468+443 and J05019+011
\begin{figure*}
\centering
\includegraphics[width=0.93\linewidth]{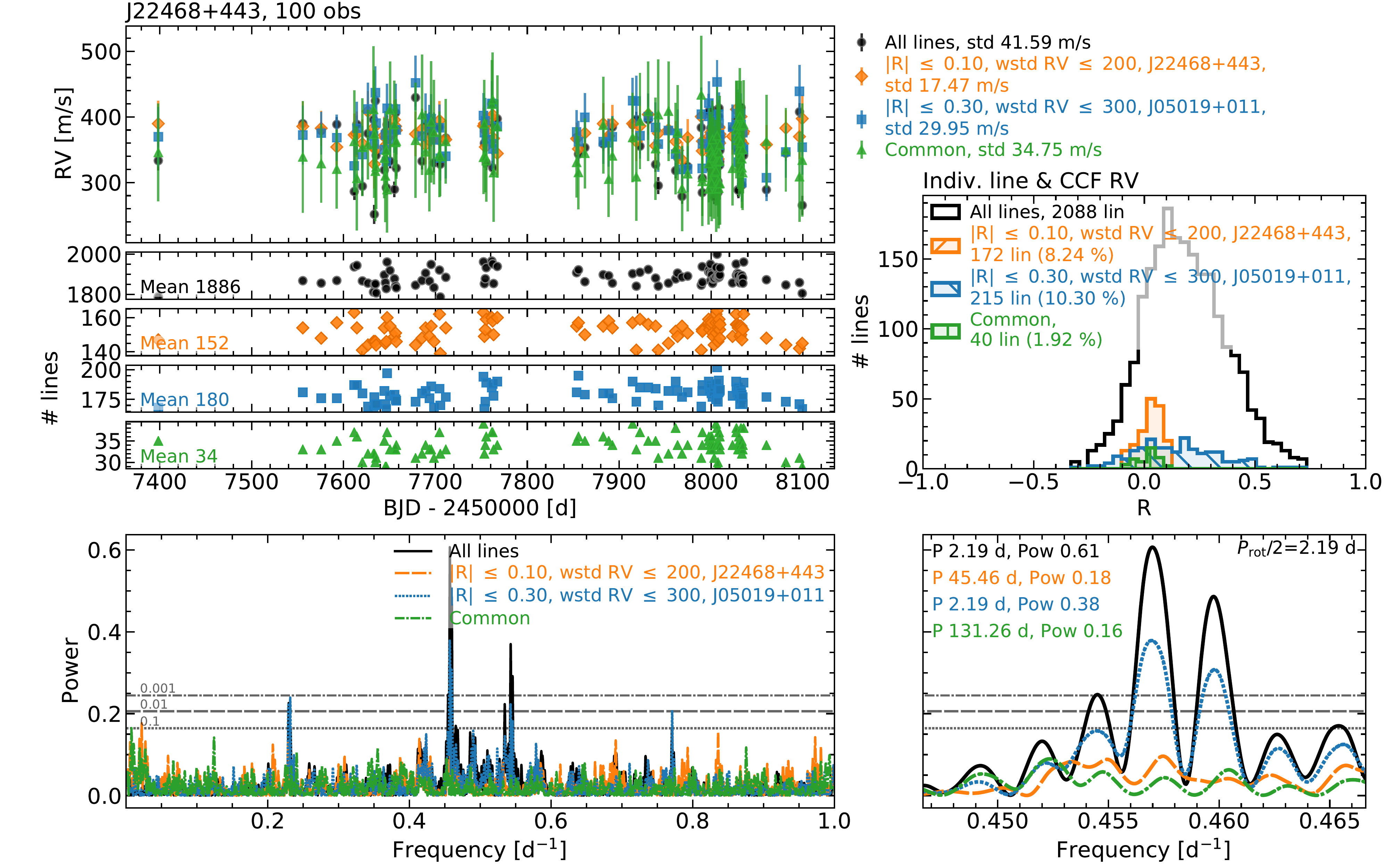}
\caption[Same as \ref{fig:tslinselcomparisonJ07446+035_cutJ05019+011} but for J22468+443 and using the initial line list, lines that minimise the RV scatter of J22468+443, lines that minimise the RV scatter of J05019+011 and common lines in the two previous sets.]{Same as \ref{fig:tslinselcomparisonJ07446+035_cutJ05019+011}, but for J22468+443 and using the following datsets: initial line list (black), lines that minimise the RV scatter of J22468+443 (orange), lines that minimise the RV scatter of J05019+011 (blue) and common lines in the two previous sets (green).}
\label{fig:tslinselcomparisonJ22468+443_cutJ05019+011}
\end{figure*}

\clearpage

\begin{figure*}
\centering
\includegraphics[width=0.93\linewidth]{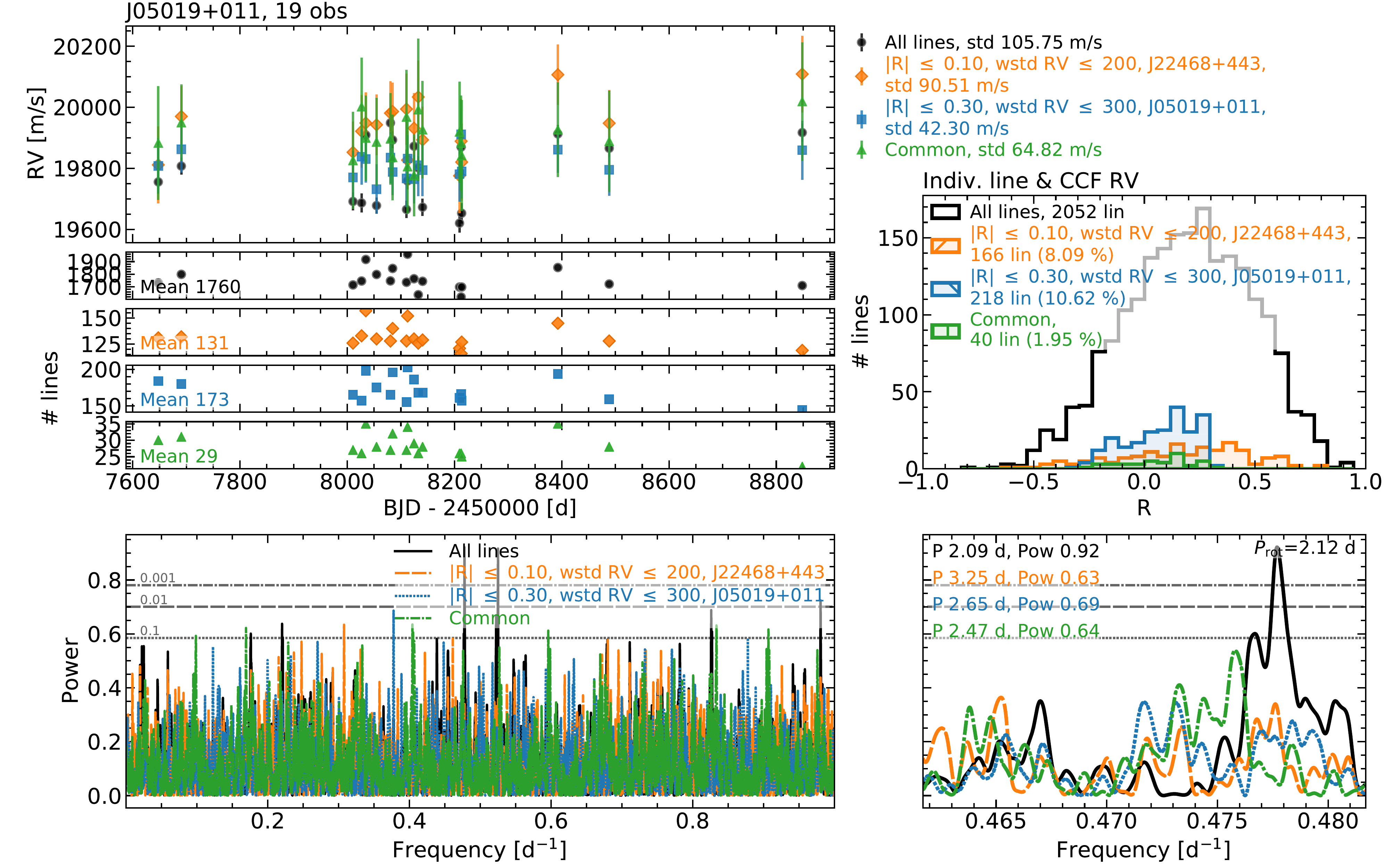}
\caption{Same as \ref{fig:tslinselcomparisonJ22468+443_cutJ05019+011}, but for J05019+011.}
\label{fig:tslinselcomparisonJ05019+011_cutJ22468+443}
\end{figure*}

% ------------------------------------------------

% J15218+209 and J11201--104
\begin{figure*}
\centering
\includegraphics[width=0.93\linewidth]{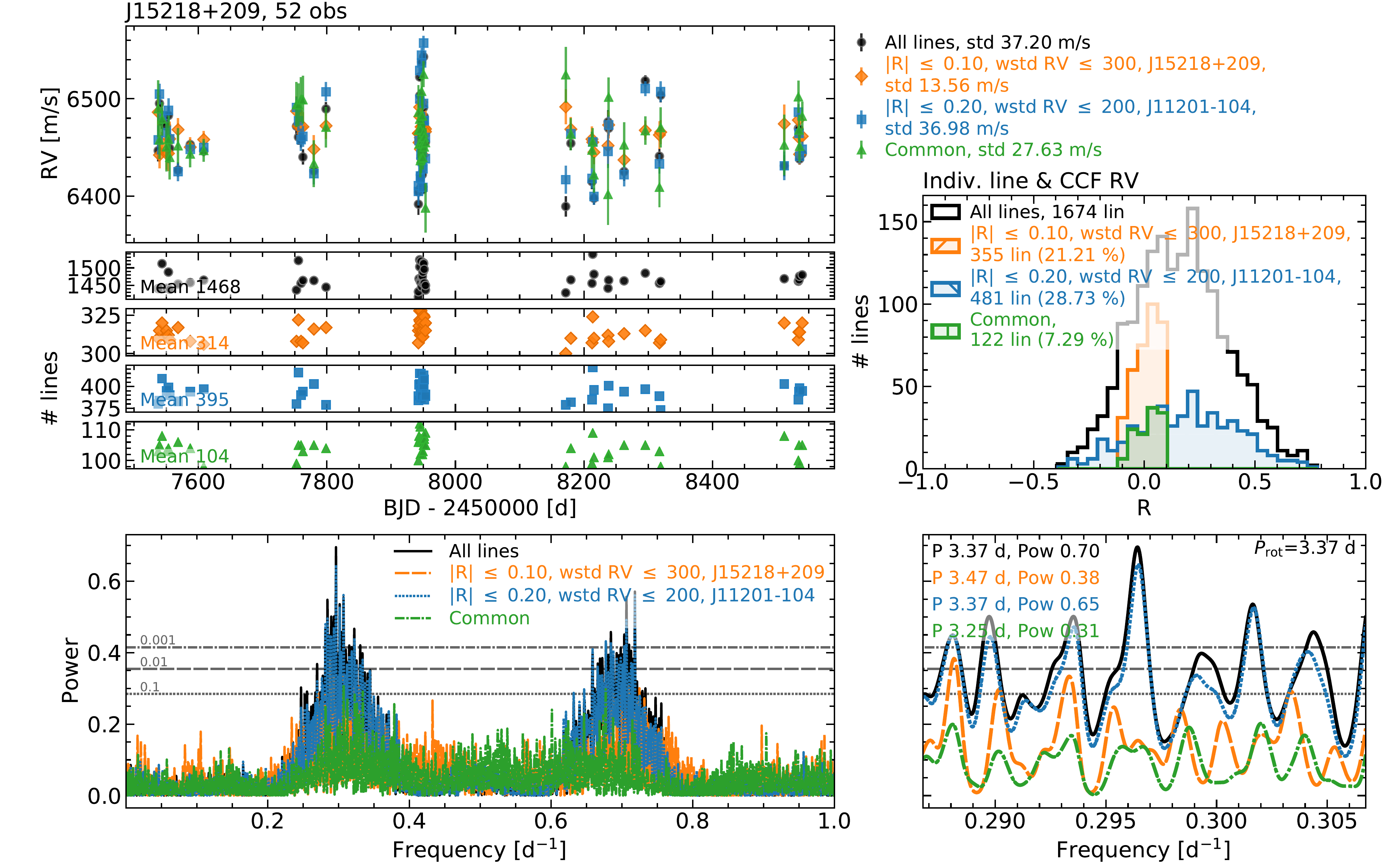}
\caption{Same as \ref{fig:tslinselcomparisonJ07446+035_cutJ05019+011}, but for J15218+209 and using the following datsets: initial line list (black), lines that minimise the RV scatter of J15218+209 (orange), lines that minimise the RV scatter of J11201--104 (blue) and common lines in the two previous sets (green).}
\label{fig:tslinselcomparisonJ15218+209_cutJ11201--104}
\end{figure*}

\begin{figure*}
\centering
\includegraphics[width=0.93\linewidth]{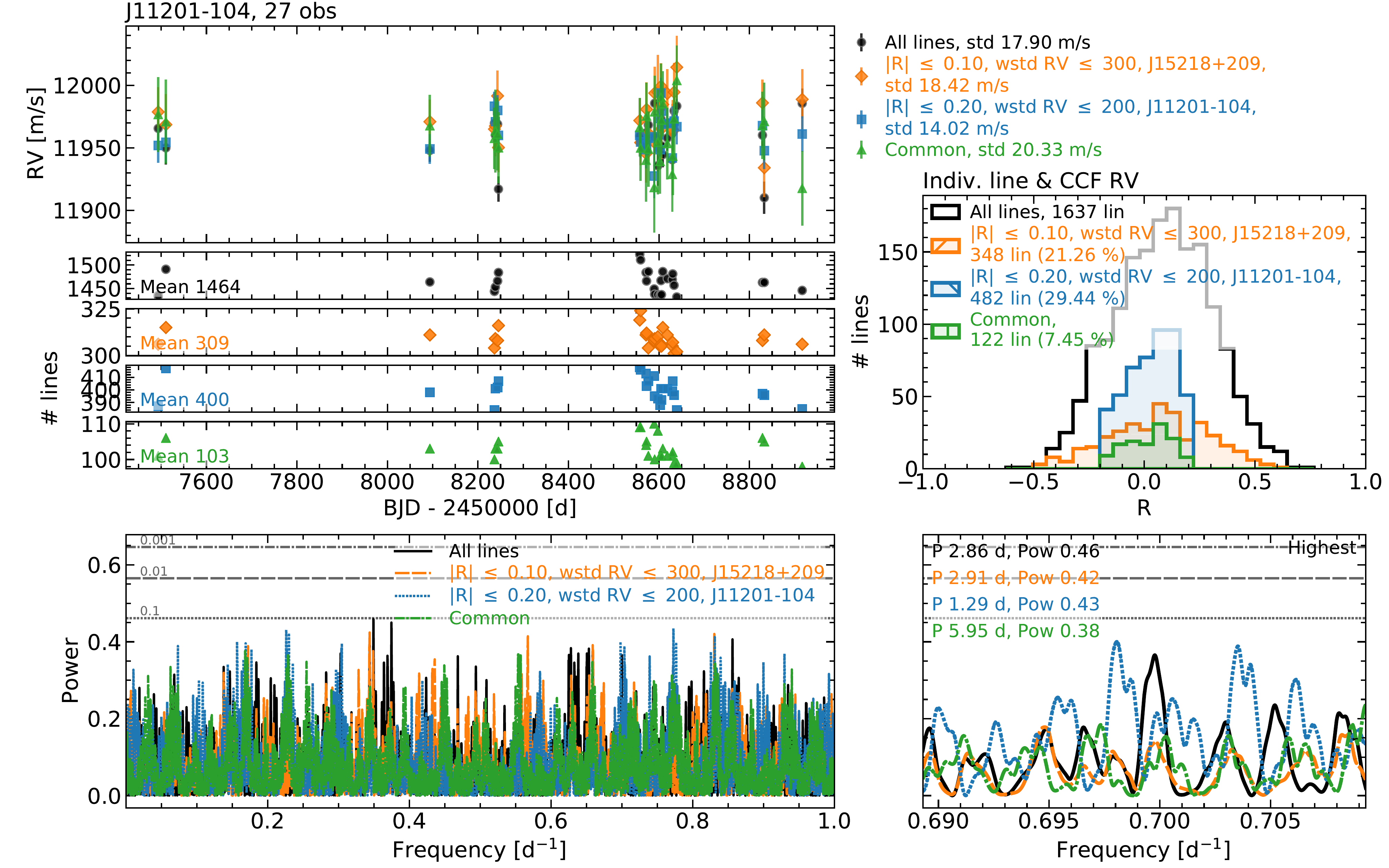}
\caption{Same as Fig. \ref{fig:tslinselcomparisonJ15218+209_cutJ11201--104}, but for J11201--104.}
\label{fig:tslinselcomparisonJ11201--104_cutJ15218+209}
\end{figure*}

%---------------------------------------------------------------------

\end{appendix}

\end{document}